\documentclass[onecolumn,apj]{openjournal}

\usepackage[utf8]{inputenc}
\usepackage{amsmath}
\usepackage{amssymb}
\usepackage{txfonts}

\usepackage[dvipsnames,table]{xcolor}
\definecolor{linkcolor}{rgb}{0.0,0.3,0.5}

\usepackage{graphicx}
\DeclareGraphicsExtensions{.bmp,.png,.jpg,.pdf}
\usepackage{multirow}
\usepackage{makecell}
\usepackage{float}
\usepackage{wrapfig}
\usepackage{sidecap}

\usepackage{soul}
\usepackage{color,colortbl}
\usepackage{textgreek}
\usepackage[autopunct=true,autostyle=false]{csquotes}

\usepackage{fancyhdr}
\usepackage{array}
\usepackage{tabularx}

\usepackage{tensind}
\tensordelimiter{?}

\usepackage{verbatim}
\usepackage[normalem]{ulem}
\usepackage{orcidlink}
\usepackage{mdframed}
\usepackage{tcolorbox}
\usepackage{natbib}
\usepackage{hyperref}
\hypersetup{
    unicode,
    colorlinks=true,
    linkcolor=linkcolor,
    citecolor=linkcolor,
    filecolor=linkcolor,
    urlcolor=blue,
}

\newcommand{\orcidauthor}[3]{\author{\href{http://orcid.org/#1}{#2$^{#3}$}}}
\DeclareUnicodeCharacter{0306}{}

\newcommand\blfootnote[1]{%
  \begingroup
  \renewcommand\thefootnote{}\footnote{#1}%
  \addtocounter{footnote}{-1}%
  \endgroup
}

\makeatletter
\renewcommand{\tablename}{Table}
\renewcommand{\fnum@table}{\tablename\ \thetable}
\def\@makecaption#1#2{%
  \vskip\abovecaptionskip
  \sbox\@tempboxa{#1. #2}%
  \ifdim \wd\@tempboxa >\hsize
    #1. #2\par
  \else
    \global \@minipagefalse
    \hb@xt@\hsize{\hfil\box\@tempboxa\hfil}%
  \fi
  \vskip\belowcaptionskip}
\makeatother
\shorttitle{}
\shortauthors{A. Fontana et al.}
\catcode`\'=12
\begin{document}
\title{Toward the Italian Astronomical Community's Contribution to HWO: Science, Technology, and Instrument Concepts}
% \title{Sticking our necks out on the spectroscopic confirmation of a M$_{UV}=-20.5$ galaxy at z=16.43\vspace{-1.5cm}}
% \title{\objname, a M$_{UV}=-20.5$ source with a plausible  z=16.43 solution based on NIRSpec and NIRCam observations \vspace{-1.5cm}}
% \title{A bright galaxy at z=16.43 plausi, Dbly confirmed by NIRSpec and NIRCam observations \vspace{-1.5cm}}

\orcidauthor{0000-0001-6065-7483}{Fontana$^\dagger$, A.}{1}
\orcidauthor{0000-0003-3758-4516}{Annibali, F.}{2}
\orcidauthor{0000-0003-2888-6133}{Antonietti, N.}{23}
\orcidauthor{0000-0002-4622-4240}{Bianchi, S.}{33}
\orcidauthor{0000-0002-6068-8682}{Biasiotti, L.}{7}
\orcidauthor{0000-0002-1892-2180}{Biazzo, K.}{1}
\orcidauthor{0000-0003-3939-2913}{Bisesi, E.}{7}
\orcidauthor{0000-0002-6177-198X}{Bonomo, A. S.}{5}
\orcidauthor{0000-0003-4590-0136}{Bozza, V.}{21,22}
\orcidauthor{0000-0002-0495-0543}{Briguglio, R.}{3}
\orcidauthor{0000-0002-7704-0153}{Brogi, M.}{4,5}
\orcidauthor{0000-0002-3903-7076}{Busonero, D.}{5}
\orcidauthor{0009-0001-5784-3901}{Cabras, A.}{34}
\orcidauthor{0000-0001-7480-0324}{Cecchi Pestellini, C.}{13}
\orcidauthor{0000-0002-3184-9918}{Cescutti, G.}{7,29}
\orcidauthor{0000-0002-3127-8078}{Ciaravella A.}{13}
\orcidauthor{0000-0002-5281-1417}{Cresci, G.}{3}
\orcidauthor{0000-0003-0378-9249}{Corso, A.J.}{26}
\orcidauthor{0000-0001-9078-5507}{Covino, S.}{10,28}
\orcidauthor{0000-0001-7618-7527}{D'Ammando, F.}{23}
\orcidauthor{0000-0001-5668-6863}{De Rosa, A.}{18}
\orcidauthor{0000-0002-4002-3878}{D’Incecco, P.}{6}
\orcidauthor{0000-0003-3693-3091}{D'Odorico, V.}{7}
\orcidauthor{0000-0002-2662-3762}{D'Orazi, V.}{20,1}
\orcidauthor{0000-0001-8613-2589}{Desidera, S.}{15}
\orcidauthor{0000-0003-3168-2289}{Di Marcantonio, P.}{7}
\orcidauthor{0000-0001-7801-7484}{Di Mauro, M. P.}{18}
\orcidauthor{0000-0002-1381-9125}{Elias-Rosa, N.}{15}
\orcidauthor{0000-0002-2789-816X}{Fineschi, S.}{5}
\orcidauthor{0000-0002-5261-6216}{Fiorellino, E.}{14}
\orcidauthor{0000-0002-1157-0143}{Fiorini, C.}{24}
\orcidauthor{0000-0002-7260-9742}{Fulvio, D.}{31}
\orcidauthor{0000-0002-4266-0643}{Garufi, A.}{23}
\orcidauthor{0000-0002-4480-6909}{Granato, G. L.}{7}
\orcidauthor{0000-0003-1362-8713}{Greco, N.}{31,32}
\orcidauthor{0000-0002-1259-2678}{Guilluy, G.}{5}
\orcidauthor{0000-0002-8068-7695}{Ivanovski, S.}{7}
\orcidauthor{0000-0001-9695-8472}{Izzo, L.}{8}
\orcidauthor{0000-0003-3360-9333}{Jimenez-Escobar, A.}{13}
\orcidauthor{0000-0003-4866-5952}{La Rocca, N.}{9}
\orcidauthor{0000-0001-5570-5081}{Landoni, M.}{10}
\orcidauthor{0000-0001-7819-9003}{Lazzoni, C.}{15}
\orcidauthor{0009-0002-6302-3747}{Liistro, E.}{9}
\orcidauthor{0000-0002-8407-5282}{Majidi, F. Z.}{8}
\orcidauthor{0000-0002-9428-8732}{Mancini, L.}{20,5}
\orcidauthor{0000-0002-2356-3274}{Mangione, A.}{13}
\orcidauthor{0000-0002-4803-2381}{Mannucci, F.}{3}
\orcidauthor{0009-0002-7836-9468}{Marcellino, M.}{11,13}
\orcidauthor{0000-0002-9889-4238}{Marconi, A}{12, 3}
\orcidauthor{0000-0001-9442-2754}{Maris, M.}{7}
\orcidauthor{0000-0001-8892-4301}{Massari, D.}{2}
\orcidauthor{0000-0002-6558-1315}{Melis, A.}{34}
\orcidauthor{0000-0001-8467-1933}{Mesa, D.}{15}
\orcidauthor{0000-0003-1427-2456}{Messa, M.}{2}
\orcidauthor{0000-0002-9900-4751}{Micela, G.}{13}
\orcidauthor{0000-0001-6283-4041}{Milakovi{\'c}, D.}{7,25}
\orcidauthor{0000-0001-8461-4189}{Monai, S.}{7}
\orcidauthor{0000-0002-7616-7136}{Moresco, M.}{14,2}
\orcidauthor{0000-0002-1751-5946}{Munari, E.}{7}
\orcidauthor{0000-0002-5155-130X}{Murante, G.}{7}
\orcidauthor{0000-0002-8951-4408}{Napolitano, L.}{1}
\orcidauthor{0000-0001-9390-0988}{Naponiello, L.}{5}
\orcidauthor{0000-0001-9770-1214}{Nascimbeni, V.}{15}
\orcidauthor{0000-0001-9573-4928}{Pagano, I.}{30}
\orcidauthor{0000-0002-9122-491X}{Palumbo, M. E.}{31}
\orcidauthor{0000-0000-0000-0000}{Pareschi, G.}{9}
\orcidauthor{0000-0003-4209-6533}{Pari, P.}{23}
\orcidauthor{0000-0002-0983-8040}{Pedichini, F.}{1}
\orcidauthor{0000-0002-1383-6750}{Pelizzo, M.G.}{27}
\orcidauthor{0000-0001-8940-6768}{Pentericci, L.}{1}
\orcidauthor{0009-0009-6066-0194}{Pezzotti, C.}{18}
\orcidauthor{0000-0000-0000-0000}{Piazzese, F.}{11,13}
\orcidauthor{0000-0001-7397-8091}{Pilia, M.}{34}
\orcidauthor{0000-0002-1321-8856}{Pino, L.}{3}
\orcidauthor{0000-0002-9937-6387}{Piotto, G.}{16,15}
\orcidauthor{0000-0002-8875-5453}{Piranomonte, S.}{1}
\orcidauthor{0000-0002-7657-7418}{Polychroni, D.}{5}
\orcidauthor{0000-0002-7697-5555}{Ragazzoni, R.}{15,35}
\orcidauthor{0000-0003-1401-6444}{Rigliaco, E.}{15}
\orcidauthor{0000-0003-0243-1464}{Romanato, F.}{16}
\orcidauthor{0000-0003-3997-0488}{Ruffato, G.}{16}
\orcidauthor{0000-0000-0000-0000}{Sacco, G.}{3}
\orcidauthor{0000-0003-4751-7421}{Salvestrini, F.}{7,25}
\orcidauthor{0000-0003-3959-2595}{Saracco, P.}{17}
\orcidauthor{0000-0003-2029-0626}{Scandariato, G.}{31}
\orcidauthor{0000-0002-6657-8054}{Schreiber, L.}{2}
\orcidauthor{0000-0001-5619-5896}{Severgnini, P.}{17}
\orcidauthor{0000-0002-7571-5217}{Silva, L.}{7}
\orcidauthor{0000-0002-7744-5804}{Simonetti, P. M.}{7}
\orcidauthor{0000-0002-7504-365X}{Sozzetti, A.}{5}
\orcidauthor{0000-0001-5252-5042}{Spinelli, R.}{13}
\orcidauthor{0000-0003-1859-3070}{Testi, L.}{14}
\orcidauthor{0000-0002-6725-3825}{Tozzi, A.}{3}
\orcidauthor{0000-0003-1665-6402}{Traficante, A.}{18}
\orcidauthor{0000-0002-0552-2313}{Tsantaki, M.}{3}
\orcidauthor{0000-0001-6926-1434}{Urso, R. G.}{31}
\orcidauthor{0000-0002-7585-8605}{Uslenghi, M.}{19}
\orcidauthor{0000-0002-5057-135X}{Vanzella, E.}{2}
\orcidauthor{0000-0002-3286-6349}{Vassallo, D.}{15}
\orcidauthor{0000-0002-8853-9611}{Vignali, C.}{14,2}
\orcidauthor{0000-0001-7604-8332}{Vladilo, G.}{7}
\orcidauthor{0000-0002-8864-506X}{Vogliardi, A.}{16}
\orcidauthor{0000-0001-8600-7008}{Zanella, A.}{2}
\orcidauthor{0000-0001-6880-5356}{Zingales, T.}{16, 15}
%\orcidauthor{0000-0000-0000-0000}{}{}
\affiliation{$^1$INAF Osservatorio Astronomico di Roma, Via Frascati 33, 00078 Monteporzio Catone, Rome, Italy}
\affiliation{$^2$INAF Osservatorio di Astrofisica e Scienza dello Spazio di Bologna, via Gobetti 93/3, Bologna, 40129, Italy}
\affiliation{$^3$INAF – Osservatorio Astrofisico di Arcetri, Largo E. Fermi 5, 50125, Firenze, Italy}
\affiliation{$^4$Dipartimento di Fisica, Università degli Studi di Torino,via Pietro Giuria 1, I-10125, Torino, Italy}
\affiliation{$^5$INAF Osservatorio Astrofisico di Torino, Via Osservatorio, 20, Pino Torinese, 10025, Italy}
\affiliation{$^6$INAF – Osservatorio Astronomico d’Abruzzo, Via Mentore Maggini, 64100 Teramo, Italy}
\affiliation{$^7$INAF - Osservatorio Astronomico di Trieste, Via Tiepolo, 11, Trieste, 34131, Italy}
\affiliation{$^8$INAF -  Osservatorio Astronomico di Capodimonte, Salita Moiariello 16, 80131 Naples, Italy}
\affiliation{$^9$Dipartmento di Biologia, Università di Padova, via U. Bassi 58b, 35131 Padova, Italy}
\affiliation{$^{10}$INAF-Osservatorio Astronomico di Brera, via Emilio Bianchi 46, 23807 Merate, Italy}
\affiliation{$^{11}$Universit\`a degli Studi di Palermo, Dipartimento di Fisica e Chimica, 
Via Archirafi 36, Palermo (Italy)}
\affiliation{$^{12}$Dipartimento di Fisica e Astronomia, Università degli Studi di Firenze, Via G. Sansone 1, Sesto
Fiorentino, 50019, Italy}
\affiliation{$^{13}$INAF – Osservatorio Astronomico di Palermo, Piazza del Parlamento, 1, 90134 Palermo, Italy.}
\affiliation{$^{14}$Dipartimento di Fisica e Astronomia “Augusto Righi”, Università di Bologna, Via Piero Gobetti 93/2, I-40129 Bologna, Italy}
\affiliation{$^{15}$INAF – Osservatorio Astronomico di Padova, Vicolo dell’Osservatorio 5, 35122 Padova, Italy}
\affiliation{$^{16}$Dipartimento di Fisica e Astronomia "Galileo Galilei", Universit\`a degli Studi di Padova, Vicolo dell’Osservatorio 3, 35122, Padova, Italy}
\affiliation{$^{17}$INAF-Osservatorio Astronomico di Brera, via Brera 28, 20121 Milano, Italy}
\affiliation{$^{18}$INAF - Istituto di Astrofisica e Planetologia Spaziali, Via del Fosso del Cavaliere 100, 00133 Rome, Italy}
\affiliation{$^{19}$INAF – Istituto di Astrofisica Spaziale e Fisica Cosmica, Milan, Italy}
\affiliation{$^{20}$Department of Physics, University of Rome ``Tor Vergata'', Via della Ricerca Scientifica 1, 00133 -- Rome, Italy}
\affiliation{$^{21}$Dipartimento di Fisica "E.R. Caianiello", Universit\`{a} di Salerno, Via Giovanni Paolo II 132, Fisciano, I-84084, Italy}
\affiliation{$^{22}$Istituto Nazionale di Fisica Nucleare, Sezione di Napoli, Via Cintia, Napoli, I-80126, Italy}
\affiliation{$^{23}$INAF - Istituto di Radioastronomia, Via P. Gobetti 101, I-40129 Bologna, Italy}
\affiliation{$^{24}$Politecnico di Milano, Dipartimento di Elettronica, Informazione e Bioingegneria, Piazza Leonardo da Vinci 32, 20133 Milano, Italy}
\affiliation{$^{25}$Institute for Fundamental Physics of the Universe, via Beirut 2, 34151 Trieste, Italy}
\affiliation{$^{26}$ CNR - Istituto di Fotonica e Nanotecnologie, Via Trasea 7, 35131 Padova , Italy}
\affiliation{$^{27}$ Università di Padova, Dipartimento di Ingegneria dell'Informazione, Via Gradenigo 6/B, 35131 Padova, Italy}
\affiliation{$^{28}$ Como Lake centre for AstroPhysics (CLAP), DiSAT, Università dell’Insubria, via Valleggio 11, 22100 Como, Italy}
\affiliation{$^{29}$ Dipartimento di Fisica, Sezione di Astronomia, Universit\`{a} di Trieste, Via G. B. Tiepolo 11, 34143 Trieste, Italy}
\affiliation{$^{30}$ INAF, Istituto Nazionale di Astrofisica,, Viale del Parco Mellini 84 00136 Roma, Italy}
\affiliation{$^{31}$ INAF -- Osservatorio Astrofisico di Catania, Via S. Sofia 78,  I-95123, Catania, Italy }
\affiliation{$^{32}$ Dipartimento di Fisica e Astronomia ``Ettore Majorana'', Università di Catania, Via S. Sofia 64, I-95123, Catania, Italy}
\affiliation{$^{33}$ Dipartimento di Matematica e Fisica, Università degli Studi Roma Tre, via della Vasca Navale 84, 00146 Roma, Italy}
\affiliation{$^{34}$INAF - Osservatorio Astronomico di Cagliari, Via Della Scienza 5, 09047 Selargius, Italy}
\affiliation{$^{35}$Dipartimento di Fisica e Astronomia ``Galileo Galilei'', Università degli Studi di Padova, Vicolo dell'Osservatorio 3, 35122 Padova, Italy }

\author{\footnotesize{\bf Endorsed by:}
Abbas Ummi (INAF-OATo), Affer Laura (INAF-OAPa), Alcala' Juan Manuel (INAF-OACN), Andrea Rossi (INAF-OAS), Arcidiacono Carmelo (INAF-OAPD), Azevedo Silva Tomás (INAF-OAA), Bacciotti Francesca (INAF-OAA), Balbi Amedeo (Un. Roma Tor Vergata), Belfiore Francesco (INAF-OAA), Bellazzini Michele (INAF - OAS), Benatti Serena (INAF - OAPA), Bergomi Maria (INAF-OAPD), Berni Leda (INAF-OAA), Bevacqua Davide (INAF - OAR), Biagiotti Francesco (INAF-IAPS), Biassoni Federico (INAF-OAB), Caito Letizia (INAF-HQ), Calabro' Antonello (INAF-OAR), Capasso Giulio (INAF-OACN), Caratti o Garatti Alessio (INAF-OACN), Castellano Marco (INAF-OAR), Claudi Riccardo (INAF-OAPD; Un. Roma Tre), Colombo Salvatore (Inaf-OAPa), Comastri Andrea (INAF-OAS), Correnti Matteo (INAF-OAR), Corsini Enrico Maria (Un. Padova), Covone Giovanni (Uni.Napoli), D'Amato Quirino (INAF-IAPS), Damasso Mario (INAF-OATo), Dazzi Francesco (INAF-OAR), De Caprio Vincenzo (INAF - OACN), Di Antonio Ivan (INAF-OAAb), di Paola Alessia (INAF-OATo), Dolci Mauro (INAF-OAAb), Farina Serena (IUSS Pavia, INAF-IASF), Farinato Jacopo (INAF-OAPD), Focardi Mauro (INAF - OAA), Fumagalli Michele (Un.Milano-Bicocca), Gai Mario (INAF-OATo), Gandolfi Giovanni (INAF-OAR), Gargiulo Adriana (INAF-IASF Mi), Genoni Matteo (INAF-OAB), Giovannini Ilaria (Un.Padova "Galileo Galilei" - Un. Santiago Cile), Gorius Nicolas (INAF-OACT), Grillo Claudio (Un.Milano), Guarcello Mario Giuseppe (INAF - OAPA), Guglielmino Salvatore Luigi (INAF-OACT), Inno Laura (INAF-OACN and University "Parthenope" of Naples), Landini Federico (INAF-OATo), Leto Giuseppe (INAF-OACT), Ligori Sebastiano (INAF-OATo), Magrini Laura (INAF-OAA), Maldonado Jesús (INAF - OAPA), Mallia Franco (Campo Catino Observatory), Medinaceli Eduardo (INAF-OAS), Musella Ilaria (INAF-OACN), Nancy Elias-Rosa (INAF - OAPd), Pace Emanuele (Un. Firenze), Pérez-Díaz Borja (INAF-OAR), Piacentini Francesco (Un.Roma "La Sapienza"), Pillitteri Ignazio Francesco (INAF-OAPA), Poggiali Giovanni (INAF-OAA), Poggianti Bianca Maria (INAF-OAPD), Ragusa Rossella (INAF-OACN), Rebrysh Oleksandra (INAF-OAPd), Rinaldi Giovanna (INAF-IAPS), Ripepi Vincenzo (INAF-OACN), Riva Alberto (INAF - OAT), Romano Donatella (INAF-OAS), Santini Paola (INAF - OAR), Schiappacasse Jose (INAF-OAA), Scialpi Martina (Un.Firenze, Un. Trento, INAF-OAA), Sciortino Luisa (INAF-OAPA), Tripodi Roberta (INAF-OAR), Turrini Diego (INAF-OATO), Vettolani Giampaolo (INAF-IRA), Vito Fabio (INAF-OAS)

}

%\affiliation{$^X$}

%\orcid{}
%\inst{1}
%\and  B. J. Weiner \orcid{0000-0001-6065-7483}
% \inst{1}
 
% \institute{\textit{INAF – Osservatorio Astronomico di Roma, via Frascati 33, 00078, Monteporzio Catone, Italy}\\ 
%}}

\thanks{\vspace{0.1cm}$^\dagger$E-mail: \href{mailto:adriano.fontana@inaf.it}{adriano.fontana@inaf.it}}

\begin{abstract}

This paper summarizes the interest of the Italian astronomical community in the possible participation in the development of the Habitable Worlds Observatory (HWO). It is largely based on the outcome of a dedicated meeting held in Rome in July 2025, and it includes further revisions and new suggestions that have arisen in the following months.
It includes a succinct description of the main scientific interests and of the key technologies that the Italian community can contribute to the development of the mission. It also includes a short section dedicated to the instrument concepts that we deem important to develop for potential adoption on HWO, and that the Italian community is interested in developing. Finally, it includes a concise presentation of the main instruments for ground-based telescopes (VLT and ELT) and space missions that will be operational before HWO and in which the Italian community has leading positions. They will be invaluable to settle the scientific landscape for the mission and to prepare the Italian scientists to make optimal use of HWO. This paper is primarily aimed at paving the way to discussions and collaborations with worldwide scientists and agencies interested in the development of the mission.
\end{abstract}

\maketitle

\date{\today}

\newpage

\tableofcontents

\newpage

\section{Introduction}
\label{sec:introduction}
The Habitable Worlds Observatory (HWO) is a mission, currently at the pre-design level, with the ambition to deliver the future major space observatory for UV--optical--nearIR astronomy.  At the time of writing, the mission concept is drawn around a primary mirror of diameter at least 6.5m, potentially reaching 8m or above, diffraction limited around 0.5$\mu$m and with an elevated efficiency in the range $0.2-2.4\mu$m. Equipped with four instruments, placed in L2 but with potential to be serviced by robotic missions for upgrades and maintenance, and the ambition to be launched in the 40’s, it promises to be one of the most powerful observatories available to explore the Universe at the midpoint of this century. As the name suggests, the main scientific goal of the mission is the capability to explore for the first time the number, nature and atmosphere of Earth-like planets around solar-like stars – i.e. accessing for the first time the real Earth analogs. This will be accomplished by an ambitious coronagraph that aims at reaching the contrast level of 10$^{-10}$ at 1AU around a few dozen F/G/K stars in our solar neighborhood, coupled with state-of-the-art imagers and spectrographs to detect and study the reflected light by these planets. 
HWO, however, is much more than a planet-hunter. The large size ($\geq 2.5 \times$~HST) of the primary mirror will deliver an unmatched sensitivity and spatial resolution in the whole range from the UV to the near IR, complementing JWST in both respects today and  the ELTs at the time the mission will operate. Coupled with a full suite of instruments, HWO will be ideal to address a vast number number of science cases across all areas of astronomy.

Motivated by the interest in this ambitious mission, several scientists in the Italian community have explored options to contribute to the scientific and technological development of the mission. To bring together these initiatives and engage the broader community, a nation-wide workshop\footnote{https://indico.ict.inaf.it/event/3258/} was organized at the Rome University La Sapienza on July 10-11, 2025. More than a hundred scientists contributed to a very lively discussion and presented scientific cases, related missions that will operate before or along HWO, key technologies and instrument concepts that might be applied to the HWO mission.  This paper stems from  the contributions presented at the Rome meeting and builds on it to present a more comprehensive and updated picture of the Italian interests in HWO. 

As the project is now entering  a more active phase, this paper summarizes the ambitions of the Italian community and the desire to contribute to the development of the mission. The ultimate goal of this paper is to trigger interactions with the other individual, groups and communities that are starting to shape their contribution toward HWO. 

The paper is organized as follows. Section 2 describes the key science cases that we are interested in pursuing with HWO, and which instrument design would address them. In a nutshell, this section tries to address the key question \textit {“What will be the key science questions that HWO can explore, and which instruments we need to address them?”}. Section 3 will summarize the key instrument developments and space missions with Italian leadership or strong participation that will be deployed before HWO is launched. Although sometimes not immediately connected to HWO, they will define the scientific landscape where HWO will operate and will place a generation of Italian scientists in the ideal position to exploit HWO to its full potential. As such, they are important in this context. Section 4 will describe the key technologies that Italian scientists, sometimes in tight connection with leading Italian industries, are developing and that can be used to optimize HWO’s performance. Finally, Section 5 presents concepts of possible instruments for the HWO focal plane, either served by the coronagraph or outside it. In most cases, they directly descend from the science priorities presented in Section 2. Given the currently coarse specification for the telescope design and the early phase of development of the entire project, none of these ideas are fully-fledged, mature instrument designs: this section is more a \textit{manifesto}  of intent than a set of detailed, mature proposals. 

We hope this paper is just the beginning of a long road that will eventually lead to a strong Italian participation to this transformative mission. 

\newpage

\section{Science Cases}
\label{sec:sciencecases}

\subsection{Exoplanets: state of the art, prospects and constraints}\label{sec:sci.exoplanets}
%\textit{Text by: V. Bozza, M. Brogi, V. D'Orazi, N. La Rocca, E. Liistro, M. Marcellino, D. Mesa, G. Micela, V. Nascimbeni, L. Pino, G. Piotto, D. Polychroni, A. Sozzetti, M. Tsantaki}

\begin{figure}[t]
  \centering
  \includegraphics[width=.9\linewidth]%{Figures/summary_plot_PCS_rocky_planet_science_case.png}
  {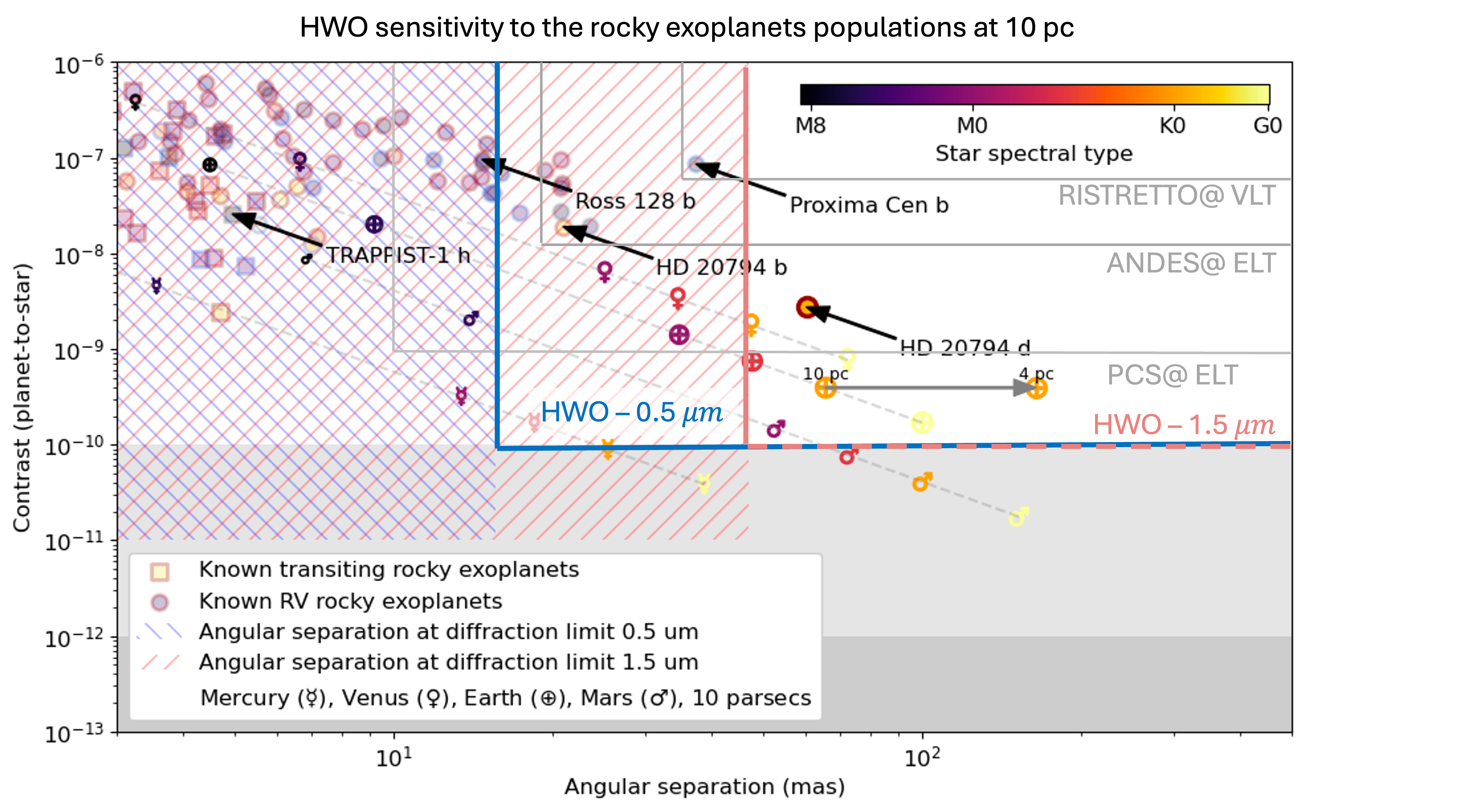}
  \caption{{\small HWO’s discovery and characterization parameter space (spatial separation versus contrast). Currently known rocky exoplanets (mass $<$ 5 Earth masses, radius $<$1.7 Earth radii) as a function of their expected planet--star contrast ratio and their angular separation. Filled squares/circles represent known transiting/non-transiting exoplanets, color-coded by spectral type of their host star. Symbols represent Venus, Earth Mars and Mercury.  The color the planet markers show the spectral types of the host star while the color of the circular outline show the planet insolation regime: blue for planets within the Habitable Zone (HZ), red for those interior to the HZ inner edge, and grey for those beyond the outer edge. The expected contrast ratios of Earth, Mars, Venus, and Mercury at a distance of 10 pc are also shown. The expected detection limits for the proposed RISTRETTO instrument on the VLT and the future ANDES instrument on the ELT are shown. The areas below diffraction limit of ELT and below the expected constrast limit for PCS are displayed.
  Even the most performing ELT/PCS observations will have a contrast 2 orders of magnitude worse than HWO. While they will access a sample of temperate planets
around M-dwarf stars and a few late K dwarfs, they will not be able to investigate a significant sample of temperate rocky exo-planets orbiting stars from early G to early K (i.e. solar-type stars). An 8-meter class HWO will be able to observe the habitable zone of 10 pc-away G0 stars out to about $1.5 \mu\mathrm{m}$, and push to later types or larger distances in the optical. Adapted by L. Pino from Fig. 2 in PCS Roadmap: Science Cases \& Top Level Requirements, prepared by G. Chauvin, B. Charnay and S. Desidera with the contributions of all the scientists involved. Credit for the original figure M. Turbet, P. Palma-Bifani and L. Pino.}}
  \label{fig:PCS_rocky_planet}
\end{figure}

Understanding the emergence and diversity of habitable worlds similar to our own is arguably the ultimate goal of exoplanet science. In this quest, the main exoplanet science case for HWO is the measurement of the atmospheric composition and surface properties of a representative sample of temperate exoplanets orbiting solar-type stars, i.e. planets with size, host star, and irradiation levels similar to the Earth. The aim of the mission is determining the likelihood of hosting life and recognizing false positives. The mirror size and hence collective power of HWO is tuned to access at least a couple dozen, ideally 40--50 temperate planets \citep{Stark2014, Stark2024} around 100--200 F-G-K stars, the exact final number depending on planet occurrence and final mirror diameter, still unknown.

HWO observations will capture the starlight reflected off the planet's surface to discriminate between different gases, estimate their volume mixing ratio to within an order of magnitude, and constrain surface temperature within a few tens of K (10--20 K, \citealt{Alei2024}). The expected variety of secondary atmospheres is high, and thus the spectroscopic capability of HWO instrumentation
needs to encompass as many key gasses as possible (Section~\ref{sec:core_hwo}). Such a measurement requires spatially separating planets from their host stars at a raw contrast ratio of $10^{-10}$ or better, and accumulating enough photons from targets which are extremely faint ($m_V \ge 30$) and inherently $\sim$25 magnitudes fainter than their parent stars. \textit{Without a dedicated coronagraphic mission such as HWO, it will not be possible to achieve such a fundamental scientific goal} (Fig.~\ref{fig:PCS_rocky_planet}).

HWO's exoplanet science will inevitably build on key deliverables of missions planned for the late 2020s and 2030s (Section~\ref{sec:future}). Most of them saw a strong involvement of the Italian community, laying the instrument groundwork for exploiting HWO. Among them, the ESA CHEOPS mission \citep{Benz2021}, combined with ground-based radial velocities, delivers bulk densities and first-order compositional classifications, thereby providing essential priors for interpreting future HWO atmospheric spectra \citep{Lacedelli2021, Leleu2021}. The final Gaia Data Release, as well as the yield of the PLATO mission, will revolutionize our understanding of the frequency and architecture of planetary systems like our own, complementing the astrometric capabilities of HWO itself as a discovery machine. In terms of precursor atmospheric surveys, the Ariel mission will survey a thousand exo-atmospheres, revealing population-level trends and even time variability for the brightest gas giants. From the ground, the ELT instrumentation (notably the ANDES and PCS spectrographs) will bring us closer to temperate rocky exoplanets, but only around M dwarfs and a few late K dwarfs, short of the contrast required for solar-type stars (Section~\ref{sec:fut.PCS}).

The ultimate question of habitability on Earth-like worlds is beyond the reach of the 2030s missions and requires an HWO-class observatory. Through its involvement in those missions, Italy is laying the groundwork to address this goal, for which possible contributions to HWO can now be evaluated. The following sub-sections describe the ideal case for HWO observations and how they build on previous facilities, starting from the most pressing challenge: the lack of prime targets, as terrestrial planets around solar-type stars have yet to be consistently found.

\subsubsection{Discovering the HWO targets: Demographics of Earth-Like Planets Around Sun-Like Stars in the HWO Era}\label{sec:demographics}

\begin{wrapfigure}{l}{0.5\textwidth}
%\begin{figure}[t]
  \centering
  \includegraphics[width=.9\linewidth]{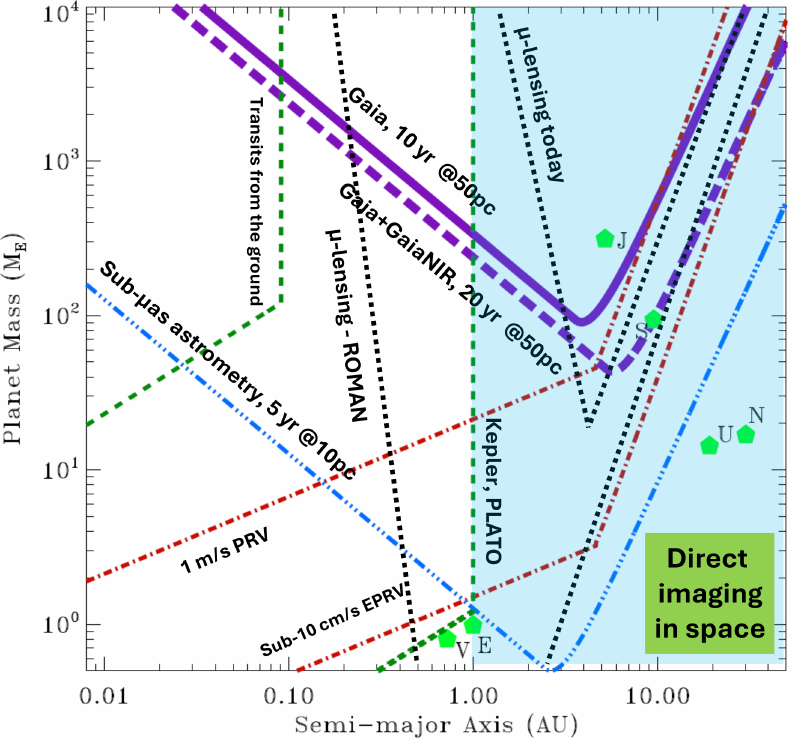}
  \caption{{\small Exoplanets discovery space in mass-separation space for astrometry (purple and blue lines), Doppler (red lines), transit (green lines), and microlensing (black lines) techniques. For all techniques, sensitivity curves are drawn for present-day programs and future projects (both approved and under study). The area of parameter space filled in light blue identifies the approximate region of interest for space-based direct imaging with HWO. Solar system planets are shown as large green pentagons.}}
  \label{fig:discovery_space}
\end{wrapfigure}

The prime sample of exoplanets to be targeted by HWO has yet to be found. The primary list of targets for the detailed atmospheric characterization of temperate terrestrial exoplanets directly detected by HWO (and possibly LIFE) will ultimately encompass a sample of $\sim$100--200 F-G-K-type dwarfs within 20 pc from the Sun\footnote{For example, the HWO reference sample contains 164 F-G-K dwarfs \citep{Mamajek2024, Tuchow2025}.}. Achieving this requires accurate and precise measurements of the full architectures (i.e., orbits and masses) of the planetary systems containing Earth-analogue exoplanets, in order to correctly interpret atmospheric spectra given the foreseeable lack of radius measurements, and to assess (long-term) habitability metrics when these planets will be directly imaged \citep{Plavchan2024, Stark2024, Kane2024, Painter2025, Damiano2025}. Today, the frequency $\eta_\oplus$ of true Earth twins\footnote{Planets with approximately Earth's mass and radius, sitting in the middle of the temperate zone of a solar-type primary star.} and the occurrence rate of Solar-System-like architectures\footnote{Systems with temperate terrestrial planets and gas giants beyond the snowline of a Sun-like star.} have yet to be measured directly \citep[for a review, see e.g.][]{Biazzo2022}, and the most successful detection techniques (radial velocity (RV) and transits, and soon absolute astrometry, direct imaging and microlensing) remain far from the sensitivity required for Earth-like planets around the nearest F-G-K dwarfs (Fig.~\ref{fig:discovery_space}).

This sensitivity will be approached only through convergent, multi-technique efforts over the coming two decades, in most of which the Italian community is strongly involved (see Section~\ref{sec:future}). In particular, the PLATO mission \citep{Rauer2025} will estimate $\eta_\oplus$ for solar-type stars from actual detections \citep{Heller2022,Matuszewski2023} rather than from extrapolations \citep{Bryson2021} of Kepler data, and will deliver accurate stellar parameters and ages through asteroseismology \citep{Goupil2024}, essential to date the systems and reconstruct their evolutionary context. Combined with Roman microlensing, Gaia astrometry, precision RVs and Kepler and TESS transit data, this will eventually enable statistically robust inferences on the frequency of true Solar-System analogues. Even so, Earth-like planets around the nearest solar-type stars may remain out of reach, and the best prospects for measuring the true mass function of temperate rocky planets ($\sim$1--5 M$_\oplus$) in the HWO reference sample, today essentially unknown, come from extreme-precision ground-based RVs and space-based ultra-high-precision differential astrometry (Fig.~\ref{fig:discovery_space}).

On the RV side, this is a major observational effort to be carried out from the ground: current high-resolution spectrographs reach $\sim$30 cm/s formal precision on the brightest nearby FGK dwarfs, in principle sufficient for a super-Earth with $M_p \sin i \ge 3$ M$_\oplus$ in the HZ. While ESPRESSO \citep{Pepe2021} was designed for the $<$10 cm/s level required for true Earth analogues, the limiting factor is the RV jitter at the $\gtrsim$1 m/s level from instrument systematics and stellar activity, granulation and super-granulation \citep{Meunier2024}, whose mitigation requires ever-improving corrections of instrument systematics (e.g., \citealt{Cretignier2021, Cretignier2023}), as well as the identification of robust, noise-resilient diagnostics and modeling approaches directly at the spectral level (e.g., \citealt{Cameron2021, Cretignier2022}), and/or on the derived RVs and stellar magnetic activity indicators (e.g., \citealt{Rajpaul2015}). High-precision RV monitoring of the future HWO's Tier-1 stars with stable spectrographs on 4m- (e.g., HARPS and HARPS-N) and 8m-class (ESPRESSO) telescopes remains crucial to characterize system architectures, to discover super-Earths revealable already from the ground (e.g., \citealt{Dalal2024}) that HWO should then observe with the highest priority, and to flag eccentric giant planets that would dynamically perturb the HZ. 

On the astrometric side, the signature $\alpha = 0.3$ $\mu$as induced by an Earth twin around a Sun-like star at 10 pc is much larger than the typical $\sim$0.05 $\mu$as positional jitter due to stellar variability \citep{Sozzetti2005, Makarov2009, Meunier2022}, so that differential astrometry with $<$1 $\mu$as precision would enable orientation-independent full orbit reconstruction and true mass estimation, provided instrument systematics (e.g., geometric stability, optical aberrations) are kept under control. This precision is unachievable from the ground due to the ultimate limitations
from Earth’s atmosphere,\footnote{Best-case estimates for the GRAVITY+ instrument on the VLT reach the 20--30 $\mu$as level \citep{GravityCollaboration2021}, comparable to the limiting per-measurement precision attainable by global astrometry with Gaia.} and possible paths to perform ultra-high-precision astrometry with precursor missions and with HWO itself are discussed in Section~\ref{sec:strum.astrometry}. Securing these orbits and masses is essential both to mitigate risk, through a thorough screening of planetary companions that enables robust dynamical viability assessments of the target HZs \citep{Kane2024, Painter2025, Sagynbayeva2025}, and to maximize yield, since accurate prior knowledge of orbital elements and masses helps optimize the scheduling of direct-imaging observations \citep{Spohn2022,Spohn2024,Li2025} and limits biases in the assessment of habitability and biosignatures \citep{Damiano2025}.

f\subsubsection{The core HWO science: the atmosphere of Earth-like planets}\label{sec:core_hwo}

To recognize false positives, standard biomarkers such as oxygen and ozone are insufficient, and recent studies, also in conjunction with the European mission concept LIFE, point to the need to access a full set of molecules, including methane, water vapor, carbon monoxide and carbon dioxide \citep{Abbott2017}. Beyond reflective surfaces, pinpointing the water content of temperate rocky planets is key to recognizing habitable conditions and discriminating against alternative scenarios, from Venus-like runaway-greenhouse states to abiotic oxygen build-up.

To characterize, as opposed to simply detect, these atmospheres, HWO should resolve key molecular bands of O$_2$, O$_3$, H$_2$O, CH$_4$, CO and CO$_2$ \citep[e.g.][]{Meadows2018, Schwieterman2018}. While O$_2$ and O$_3$ fall in the UV--optical, the diagnostic bands of the carbon- and water-bearing species lie in the near-infrared: the strongest water-vapour band is around 950 nm, with further features near 1150 and 1500 nm, CO$_2$ arises around 1600 nm and becomes dominant near 2000 nm, and CO does not absorb significantly shortward of 2300 nm. Fully exploiting HWO's coverage out to $\sim$2.5 $\mu$m with adequate spectroscopic capability, at a minimum resolving power $R\sim100$, is therefore essential to recognize species in strong chemical disequilibrium compatible with a biogenic origin.

\textit{Oxygen and Ozone as life tracers.} Two strong biosignatures on Earth are linked to oxygenic photosynthesis: the accumulation of atmospheric oxygen and the Vegetation Red-Edge in surface reflectance spectra. The feasibility of this metabolism on exoplanets, particularly in the worst-case low-visible-light environment of M-dwarf habitable zones, is supported both by theoretical models \citep{galewandel2016,ritchie2018,gray2025} and by recent laboratory experiments showing that cyanobacteria and microalgae can photosynthesize and release O$_2$ under simulated M-dwarf spectra and Archean atmospheres \citep{boccia2026, Liistro2024, Liistro2026, battistuzzi2023a, battistuzzi2023b}. Within the HWO bandpass, the most prominent O$_2$ features lie at 0.69 and 0.76 $\mu$m, while O$_3$ absorbs in the Chappuis bands (0.5--0.7 $\mu$m) in the optical. Caution is nonetheless required, since O$_3$ is created by the variable UV activity of M dwarfs and its vertical structure can be reshaped by atmospheric photochemistry \citep{Meadows2018a, Kaltenegger2021}. The interpretation of these tracers, and of the surface reflectance signatures of photosynthetic pigments, can rely on the climate--chemistry--biology modelling and laboratory expertise of the Italian community detailed in Section~\ref{sec:fut.biosignatures}.

\textit{Beyond Oxygen: Volatile Organic Compounds as life tracers.} Single-molecule detections, such as O$_2$, CO/CO$_2$ depletion, or phosphine, remain susceptible to abiotic false positives. The terrestrial model demonstrates that a biosphere instead drives systemic disequilibrium across an entire Chemical Reaction Network, generating interconnected suites of Volatile Organic Compounds (VOCs). If HWO detects a candidate biogenic source or sink, it should therefore concurrently search for the predicted short-list of trace VOCs completing that metabolic signature, for instance N$_2$O and CH$_4$ in cold ($\sim -20$ to $20\,^\circ$C) regimes, isoprene or carbonyl sulfide in temperate, Earth-like ones ($\sim 20$--$50\,^\circ$C), or reduced sulfur species such as H$_2$S in hot ($\sim 50$--$120\,^\circ$C) regimes. Furthermore, temporal dynamics offer a definitive discriminant: seasonal variations can trigger microbial blooms whose cyclic signatures, tracked through multi-epoch monitoring, would distinguish biological forcing from steady abiotic outgassing. Harnessing this systemic perspective requires high-resolution, broad spectral coverage for simultaneous multi-molecule fingerprinting, and multi-epoch observations to follow VOC fluctuations over time.

The required spectral resolution is not merely a matter of identifying bands: as shown by the simulation framework of \citet{2026arXiv260417554R}, biosignature detection in reflected light depends critically on resolving power, with a moderate-to-high resolution ($R>1000$) fundamental to disentangle narrow molecular features from correlated speckle noise, whereas at $R\sim140$ spectral features cannot be separated from residual starlight speckles, which can suppress the biosignature signal or produce false detections. These requirements, together with the importance of fully exploiting HWO's near-infrared coverage, directly motivate the spectroscopic instrument concepts that the Italian community proposes to explore and lead for HWO (Sections~\ref{sec:strum.coro_instruments} and \ref{sec:strum.NIRHRspec}).

%For future reference, we provide in Table~\ref{tab:atmolines} a full list of the most prominent lines that are good tracers of biotic elements in the atmosphere.

\textit{Surface Biosignatures.}While the exploration of exoplanetary life predominantly focuses on identifying atmospheric chemical disequilibria, surface biosignatures - specifically those originating from photosynthetic pigments - provide a direct link to a planet's surface habitability and biological evolution, and represent a complementary biosignature to atmospheric ones (Parenteau et al. 2026). On Earth, the presence of widespread vegetation introduces spectral features observable from space: a small chlorophyll bump around 500 nm and the "red edge," a sharp reflectance increase shortward of 700 nm (Arnold et al. 2002). These unique signatures represent light-harvesting architectures that are optimized for the specific radiative environment of terrestrial surface around a G-type star. 

\textit{Volcanism.} During explosive volcanic activity, energetic perturbations travel upward and create structured electron density variations in this plasma layer. These wave-like variations are called Co-Volcanic Ionospheric Disturbances (CVIDs) \citep{Heki2004}. Beyond the physical propagation of CVIDs, volcanic eruptions modify the chemistry of the troposphere, stratosphere, and mesosphere through direct volatile injection, wave-induced transport, and lightning-driven synthesis \citep{Cimarelli+2022}. While detecting transient, wave-like plasma oscillations on extrasolar worlds is currently (and for many years) impossible with existing technology, the physical, chemical, and electromagnetic processes intrinsically tied to CVIDs provide several highly promising indirect detection pathways to identify active volcanism on rocky exoplanets. Volcanic outgassing represents a key mechanism for injecting volatiles ($H_2O$, $SO_2$, $CO_2$) into planetary atmospheres \citep{vonGlasow2010}, helping secondary atmospheres survive stellar erosion over billions of years. Furthermore, characterizing these outgassing dynamics provides a critical metric to discriminate between planetary geodynamic stages: primitive worlds hosting global magma oceans sustained by intense stellar or tidal heating, and evolved worlds possessing a solid crust/lithosphere regulated by active volcanic and tectonic regimes (such as stagnant lid or mobile plate tectonics). While global magma oceans maintain a continuous, highly dense blanket of volatiles in equilibrium with global silicate melt  \citep{Wordsworth+2022}, geologically active worlds with solid crusts exhibit transient, localized, and episodic volatile pulses (such as SO2 spikes or transient sulfate hazes, \citep{Misra+2015} that undergo rapid chemical processing and atmospheric transport. By mapping how volcanic lightning, ionospheric depletions, and sulfate aerosol hazes modify Earth's spheres, next-generation space missions can remotely probe the interior states and surface processes of rocky worlds.

\subsubsection{The power of polarimetric observations}\label{sec:sci.polarimetry}

While the coronagraphic design is the key selling point of the HWO mission, extremely high raw contrast is only one of the elements for the success of the mission. \citet{Vaughan2023} showed that inner working angles (IWAs) at visible wavelengths between $2\lambda/D$ and $3\lambda/D$ would allow us to detect not only the planet, but also phase modulations due, for instance, to the scattering properties of water vapour clouds or oceans. With an 8-meter class HWO and an IWA of $2\lambda/D$, these features remain observable also at 1.5 $\mu$m (77 mas) for a few (ocean glint) to a few tens of planets (Rayleigh scattering and water vapour cloud scattering). At these wavelengths, the hierarchy of scattering processes fundamentally reorganizes, which is key to break degeneracies among scattering surfaces. For example, Rayleigh scattering is a dominant source of scattering at blue wavelengths, and partially occults surface signals from the ocean glint, which is more wavelength-independent \citep{Robinson2010}. In polarized light, the effect becomes even more dramatic, with near-infrared polarization now identified as the most unambiguous probe of surface oceans, being free of the cloud and dry-surface false positives that affect the total-intensity glint \citep{TreesStam2022OceanSigs}. Realistic 3D cloud modelling moreover confirms that polarization is a stronger diagnostic than total intensity for these targets, and that simplified models can overestimate the reflected-light contrast by up to a factor of two \citep{Roccetti2025}.

\subsection{Exoplanets}
\label{sec:exo_requirements}
%\textit{Text by: }

Building on the science cases of Section~\ref{sec:sci.exoplanets}, we set out here their requirements for HWO. Most of the reflected-light cases below require a coronagraphic contrast of order $10^{-10}$ at small inner working angles (of order $2\lambda/D$), the regime that isolates a temperate planet from its host star; astrometry and transiting-planet spectroscopy instead set requirements independent of the coronagraph. Items 1–3 address target and system architectures, Items 4–9 cover atmospheric characterization, and Items 10–13 focus on volcanism and biosignatures. Some of the requirements below are further tailored to possible instrument designs in Sections~\ref{sec:strum.coro_instruments} and \ref{sec:strum.NIRHRspec}, leveraging the leadership of the Italian community.

\subsubsection{Detection, orbits, true masses and system architecture.} 
\label{sec:2.2.1}As set out in Section~\ref{sec:demographics}, interpreting any atmospheric spectrum requires the orbit, true mass and dynamical context of the target: in the absence of a transit the planetary radius is unknown, so the reflected-light flux is degenerate between radius and albedo, and a habitable-zone candidate cannot be validated without screening the system for eccentric giants that would destabilise it. HWO addresses this by combining direct detection and the tracing of the planet's phase and orbit with a measurement of the host-star astrometric reflex, which yields the true mass and hence, through a mass-radius relation, the radius prior that breaks the degeneracy; unlike radial velocities, astrometry returns the true mass rather than a minimum mass. \\
\textbf{Instrument requirements:} a High-Contrast Imager working behind the coronagraph, with broad photometric bands across the coronagraph's operating range to detect the planet and follow its phase and orbit, and for differential astrometry of the host reaching $<$1 $\mu$as per-epoch precision, accumulated over a multi-year baseline, with a field of view wide enough to secure a sufficient grid of reference stars and the control of geometric and optical systematics at the sub-$\mu$as level.

\subsubsection{Reflected-light diversity: giant, Neptune- and sub-Neptune-class planets.} 
\label{sec:2.2.2}Beyond Earth analogues, HWO's reflected-light reach opens the systematic characterization of mature giant, Neptune-class and sub-Neptune planets at large orbits, a population inaccessible to current high-contrast imaging (confined to young, self-luminous giants), and potentially at the limit of contrast of ground based ELT instrumentation such as PCS \citep{Kasper2021}. Because their planet-to-star contrast is more favourable than that of an exo-Earth, these objects can be reached at intermediate to wide separations, where reflected-light spectra constrain cloud and haze structure, bulk atmospheric composition and metallicity across a mass range essentially unexplored in reflected light, including candidate water-rich sub-Neptunes. Probing this diversity connects formation history to present-day structure and provides the comparative planetology context for the terrestrial targets. 

\textbf{Instrument requirements:} this case requires a Coronagraph-fed Spectrograph spanning the optical to $\sim$2.5 $\mu$m at a resolving power of order 100, sufficient to resolve the H$_2$O and CH$_4$ bands and constrain the cloud and haze continuum; the near-infrared extension is essential, since the diagnostic H$_2$O bands and the strongest CH$_4$ bands (near 1.7 and 2.3 $\mu$m) lie in the near-infrared, beyond the reach of the optical channel alone.

\subsubsection{Exomoons and binary planetary companions.}
\label{sec:2.2.3}The detection and characterization of exomoons and binary planetary companions represent a unique opportunity to probe the formation and dynamical evolution of planetary systems beyond the Solar System. HWO will provide the first statistical exploration of rocky satellite systems around giant and terrestrial planets through its unmatched combination of angular resolution, high-contrast imaging, and PSF stability in the UV--optical--NIR regime \citep[see for example][]{Limbach2026}. By directly resolving portions of the Hill spheres of nearby companions \citep[see for example][]{Lazzoni2020} emitting at the aforementioned wavelengths, HWO will constrain the occurrence, orbital architectures, and photometric properties of Earth-like exomoons and circumplanetary material, revealing how satellite formation depends on planetary mass, age, and environment. Multi-epoch observations will enable the detection of orbital motion, mutual events, and photocenter variations induced by massive satellites, while moderate-resolution spectroscopy will provide the first constraints on the composition of circumplanetary disks and candidate exomoon atmospheres. In synergy with ELT/PCS, HWO will extend these studies toward shorter wavelengths and higher photometric stability, enabling population-level investigations inaccessible from the ground. 

\textbf{Instrument requirements:} this science case calls for a high-contrast coronagraphic imager operating at contrasts better than $10^{-9}$ at separations of a few tens of mas, coupled with a UV--optical--NIR integral field spectrograph ($R\sim100$--$1000$) and high-precision time-resolved photometric and astrometric capabilities.

%\subsubsection{High-dispersion coronagraphy.} Feeding the high-contrast image formed by the coronagraph into a high-resolution spectrograph adds spectral starlight suppression to the coronagraph's spatial suppression. Cross-correlating the planet's dense line forest against templates raises the effective contrast roughly as the square root of the number of resolved lines. This would let HWO not only characterise the molecular composition of directly imaged planets at high fidelity, but also measure their rotation through line broadening, on companions at wider separations and around the brightest hosts. 

%\textbf{Instrument requirements:} this case calls for a fibre-fed coupling of the coronagraph to a high-resolution spectrograph ($R$ of order $10^5$) covering the optical and near-infrared, a cross-instrument architecture to be proposed as a further capability of the telescope; it requires ultra-stable wavelength calibration (Fabry-Perot etalon or laser frequency comb) and the ultra-low-noise detectors suited to the photon-starved companion regime.

\subsubsection{Transiting-planet atmospheres.}
\label{sec:2.2.4}Independently of the coronagraph, high-resolution transmission and emission spectroscopy of transiting planets opens a further parameter space. Cross-correlating the planet's resolved line forest, here separated from the stationary stellar lines by the planet's Doppler shift, reaches temperate planets around host stars of all spectral types, including the habitable-zone planets of late-type stars such as TRAPPIST-1 that the reflected-light mode cannot reach. 
From the ground, this regime has an added challenge: telluric contamination appears as quasi-stationary lines for relatively long periods, and it becomes more challenging to disentangle in Doppler space, and more severe at wavelengths longer than $2.5~\mu\mathrm{m}$. While ground-based characterization of long period exoplanet atmospheres is a topic of active research, HWO provides the opportunity to fundamentally remove the challenge of telluric lines and therefore extend ground-based capabilities for long period planets (see Section \ref{sec:strum.coro_instruments}).

\textbf{Instrument requirements:} independent of the coronagraph, this case calls for a high-resolution spectrograph ($R>50{,}000$) covering the ultraviolet and optical features and lines of interest, and the near-infrared molecular bands, ideally extending to the thermal-infrared L and M bands; it requires a stable wavelength solution and detectors combining low read noise with the dynamic range for bright transit hosts.

\subsubsection{Mass loss from planetary atmospheres.}
\label{sec:2.2.5}
Hydrodynamic escape, driven by the host star high-energy X-ray and extreme ultraviolet (EUV) irradiation, represents a primary mechanism for the mass loss of planetary atmospheres \citep[e.g.][]{Watson1981}. Characterizing the efficiency of this mechanism is essential to establish the conditions for whether a planet is able to develop and hold an atmosphere \citep{Kubyshkina2018, Pezzotti2021a}, and more in general, an environment amenable to life. 
The JWST ``Rocky worlds DDT''\footnote{https://rockyworlds.stsci.edu/} will indirectly tackle this science question by using measurements of the planet brightness at 15$\mu$m to infer the presence of an atmosphere, for planets 
%seraching for atmospheres of planets 
orbiting M-dwarfs across the cosmic shoreline \citep{Connors2026}. ELT instrumentation will be able to directly detect surviving secondary atmospheres of M-dwarf planets and measure their composition. Spectroscopic observations in the UV could complement these efforts to search for and quantify atmospheric mass-loss rates for planets transiting M-dwarfs, thanks to the presence of resonant individual lines of neutral hydrogen and heavier elements, which are ideal for observing atmospheric species \citep{DosSantos2025}. 

\textbf{Instrument requirements:} Extending these atmospheric escape characterizations to a wide variety of stellar spectral types necessitates moderate-to-high resolution UV spectrograph, with spectral resolving power $R > 10,000$ and a core wavelength coverage ideally in the range $970\text{--}3000 \text{ \AA}$.

\subsubsection{Rotational variability.} 
\label{sec:2.2.6}
As the planet spins, the disk-integrated light curve and its colour variations map the longitudinal distribution of clouds, continents and oceans \citep{Cowan2009} and can yield the rotation period \citep{Palle2008} and the presence of dynamic weather systems. Following both regimes turns the planet from a single snapshot into a time-resolved system, a capability that distinguishes a dedicated direct-imaging observatory from a survey facility. 
We also note that the seasonal modulation of trace gases is potentially a key discriminant between biological forcing and steady abiotic outgassing, as introduced in Section~\ref{sec:core_hwo}. Measuring these variations would require observing molecules detectable at $\lambda >3\mu$m, which are currently out of the baseline design, but they would represent an important  extension to this science case.

\textbf{Instrument requirements:} this case calls for the ability, and the mission time, to revisit high-priority targets at multiple epochs, on timescales from the months and years of seasonal change down to the hours-long cadence that resolves a rotation period of order a day, with photometric and spectro--photometric stability across epochs so that intrinsic variability is not mistaken for instrument drift. Because each epoch is itself a long integration on a faint source, such time-resolved characterization is photon-limited and realistic mainly for the closest and brightest targets.

\subsubsection{Molecular characterization and robust biosignature detection.} 
\label{sec:2.2.7}The diagnostic bands that establish habitability and discriminate the main false positives are detailed in Section~\ref{sec:core_hwo}: O$_2$ and O$_3$ in the UV and optical, and the carbon- and water-bearing species (H$_2$O, CH$_4$, CO, CO$_2$) in the near-infrared. 
As argued in Section~\ref{sec:core_hwo}, single-molecule anomalies are vulnerable to abiotic false positives; the robust diagnostic is the systemic disequilibrium a biosphere drives across a chemical reaction network, expressed as interconnected suites of VOCs. The UV and optical provide essentially only O$_2$ and O$_3$, the tracers most exposed to false positives, whereas the species that complete the disequilibrium absorb in the infrared: H$_2$O, CH$_4$, CO and CO$_2$ retain near-infrared overtone bands accessible to HWO, placing the core disequilibrium of O$_2$ and CH$_4$ within reach. The trace VOCs, lacking detectable near-infrared overtones, remain the domain of a mid-infrared facility such as LIFE. 
The goal here is to characterize rather than merely detect: to measure the abundances of these molecules to within an order of magnitude and the surface temperature to within a few tens of K, and to separate genuinely habitable states from Venus-like or abiotically oxygenated ones. 
This argues both for pushing HWO's red coverage as far as the design allows. In principle, resolving the molecular bands into many individual lines also helps: their cross-correlated signal stays effective despite the low per-channel signal-to-noise of these faint targets, unlike the band-level $R\sim100$ abundance retrieval of the molecular-characterization case.
To define the instrument properties of the HWO required to detect surface vegetation, we have developed a preliminary set of simulations (Spinelli et al. in preparation)  adopting a comprehensive grid of benchmark Earth-sized planet characterized by a surface composition of 20\% forest, 60\% ocean, 10\% desert, and 10\% polar ice \citep{2009EPJWC...1...75M}. The target was modeled under two distinct stellar configurations: a solar analog ($T_{eff}=5700\text{ K}$) with the planet at 1 AU (projecting an angular separation $\theta=100\text{ mas}$ at 10 pc, and a cooler K-dwarf ($T_{eff}=4900\text{ K}$) with the planet at 0.7 AU ($\theta=70\text{ mas}$ at 10 pc. These simulations complement recent results like for instance \citet{2026arXiv260417554R} and \cite{Wang2026}, which have also raised concerns about the correlated, chromatic speckle noise that can entirely suppress the molecular signal. They both show that high resolution spectrograph ($R>1000$) can better detect the most crucial molecules compared to a low resolution spectroscopy mode, although the latter can be viable especially under the best coronagraphic performances. 

\textbf{Instrument requirements:} this case requires coronagraph-fed spectroscopy with continuous coverage from the UV and optical out to $\sim$2.5 $\mu$m, the near-infrared part being mandatory because the carbon- and water-bearing species absorb predominantly long--ward of $\sim$0.95 $\mu$m (the optical CH$_4$ bands being negligible at terrestrial abundances); a resolving power of $R\sim100$, set by and matched to the photon budget of these faint targets yet sufficient to retrieve abundances to within an order of magnitude; and, as the binding constraint, the throughput and integration capability to reach the required signal-to-noise on sources that are extremely faint ($m_V \ge 30$) and some 25 magnitudes fainter than their host stars.
We assume a baseline raw contrast noise floor of $\text{error} \le 1 \times 10^{-10}$ and an aggressive deeper contrast of $1 \times 10^{-11}$. Depending on the specific targets, HWO can leverage distinct instrument combinations across intermediate and high resolutions. For a target at a nearby distance of 10 pc under the baseline noise floor ($\text{error} = 1 \times 10^{-10}$), while low resolution spectroscopy of $R=100$ is sufficient for a marginal detection, an intermediate resolution of $R=1000$ provides a significant detection. In the hypothesis of a contrast of $1 \times 10^{-11}$, even with a lower resolution it is possible to obtain a significant detection. 
We also advocate for detailed simulations aimed to access whether  a resolving power above the $R\sim1000$ floor, provided either by coronagraph-fed integral-field spectroscopy at $R\sim1500$--$5000$, surveying the field for faint habitable-zone planets, or, for a single brighter or more widely separated companion, by fibre-fed high-dispersion coronagraphy at $R\sim10^5$, are a viable object for these faint targets.

\subsubsection{Surface, cloud and ocean diagnostics.} 
\label{sec:2.2.8}
Interpreting an atmospheric spectrum requires the surface and cloud context, reached through two complementary routes: in total intensity, surface reflectance signatures (the Vegetation Red Edge and photosynthetic pigments, Section~\ref{sec:core_hwo}; surface mineralogy and volcanic state, Section~\ref{sec:fut.biosignatures}); and in polarized light, the breaking of cloud, aerosol and ocean degeneracies, with near-infrared polarization as the most unambiguous probe of surface oceans (Section~\ref{sec:sci.polarimetry}; \citealt{TreesStam2022OceanSigs}). The requirement that follows is that these two observables, total-intensity reflectance and polarization, be available together rather than in isolation, since each lifts degeneracies the other cannot. 

\textbf{Instrument requirements:} this case requires the two observables together. On one side the coronagraph-fed reflectance spectroscopy already described, extended into the near-infrared and sampled at several orbital phases; on the other, imaging polarimetry of the same targets at small inner working angles ($2\lambda/D$ to $3\lambda/D$), measuring the linear Stokes parameters $Q$ and $U$ of the cloud- and ocean-induced phase modulations across the visible and near-infrared.

\subsubsection{Volcanism}
\label{sec:2.2.9}
Interior and surface characterization: map active surface volcanism and tidal/radiogenic heating rates by detecting short-lived species like $SO_2$ (whose lifetime is constrained by rapid atmospheric sinks). An $SO_2$-rich volcanic atmosphere would imply a highly oxidized mantle and require volcanic and tidal heating rates. Confirming an $SO_2$-dominated atmosphere would have significant implications for our understanding of atmospheric retention and secondary atmosphere formation \citep{Bello-Arufe+2025} , e.g., storage of a fraction of the primordial volatiles. 
Ionospheric biosignatures: discriminate between geologically outgassing, prebiotic worlds and active biospheres;   an upper ionosphere dominated by molecular ions ($O_2^2+$, $CO_2^+$, $H_2O^+$) indicates abiotic volcanic degassing, whereas an ionosphere dominated by atomic oxygen ($O^+$) serves as a robust biomarker for global photosynthetic activity.
Prebiotic synthesis pathways: trace volcanic lightning discharges and electrified plumes. Highly energetic volcanic lightning breaks $N_2$ triple bonds, synthesizing $NO$ and subsequent bio-available nitrates while generating diagnostic trace-gas anomalies ($CO$, $CH_4$). We have successfully proved in laboratory the existence of lightening induced generation of nitrates in $CO_2$ atmospheres contaminated by ammonia delivering  \citep{Jimenez+2025}.
Transient outgassing dynamics: detect episodic explosive eruptions (similar to Earth’s Pinatubo or Sarychev events) via high-altitude sulfate hazes, which cause temporary, observable expansions ("puffing") of the planetary transit radius.

\textbf{Instrument requirements:}
Observing  co-volcanic proxies requires multi-wavelength coverage across ultraviolet to low frequency radio. In the context of HWO bands would be of interest:
Ultraviolet (0.2–0.4 $\mu$m): strong $SO_2$ absorption bands (particularly below 320 nm), the $O_3$ Hartley (200–300 nm) and Huggins (310–360 nm) bands, and resonance lines of ionospheric/exospheric metal species such as $Mg II$ (280 nm) and $Ca II $H\&K (393–397 nm). Low spectral resolution is generally sufficient for detecting broad $O_3$ absorption, whereas high spectral resolution is advantageous for studying upper-atmosphere composition.\\
Visible/NIR (0.4–1.0 $\mu$m): $O_2$ absorption bands (especially the 0.76 $\mu$m A-band), $H_2O$ bands (including 0.94 $\mu$m), and continuum spectral slopes produced by Rayleigh scattering, aerosols, and clouds. Baseline exoplanet characterization studies typically adopt resolving powers of R about $\approx$ 50–200, which are sufficient to detect and characterize these broad molecular and continuum features while maintaining adequate signal-to-noise ratios for faint reflected-light observations of terrestrial planets.\\
(0.85–2.0 $\mu$m): useful for broad molecular characterization of $CO_2$, $CO$, and $CH_4$. For HWO, the relevant spectroscopy is low-to-moderate resolution. $SO_2$ is only weakly expressed in this band and is much more readily probed at longer wavelengths, particularly near 7.3 $\mu$m.

\subsubsection{Host-star high-energy environment and habitability.} 
\label{sec:2.2.10}The interpretation of biosignatures cannot be separated from the high-energy environment of the host star, a point already raised in Section~\ref{sec:core_hwo} in connection with ozone. Ozone is produced and continuously reshaped by stellar ultraviolet radiation, and the UV activity of young stars and M dwarfs is both strong and highly variable, so that the same atmospheric oxygen content can correspond to very different O$_3$ columns; strong XUV irradiation can moreover drive abiotic oxygen build-up that mimics a biosignature. Characterising the ultraviolet output, flare rate and high-energy spectral shape of each host star is therefore a prerequisite for the photochemical models used to interpret HWO spectra and for telling biological from abiotic oxygen apart. HWO's own ultraviolet sensitivity supplies the directly observable part of this picture, the far- and near-ultraviolet that drives the ozone photochemistry and anchors the reconstruction of the otherwise unobservable extreme-ultraviolet, for the very systems it images rather than through heterogeneous archival data. 

\textbf{Instrument requirements:} this case requires ultraviolet spectroscopy of the host stars, possibly resolved in time to identify flares, covering the chromospheric and transition-region diagnostics and the far-ultraviolet continuum relevant to atmospheric photochemistry, and exploiting HWO's UV-to-optical coverage.

\subsection{Star formation}
\label{sec:sci.starformation}

\begin{figure}[h]
  \centering
  \includegraphics[width=.9\linewidth]{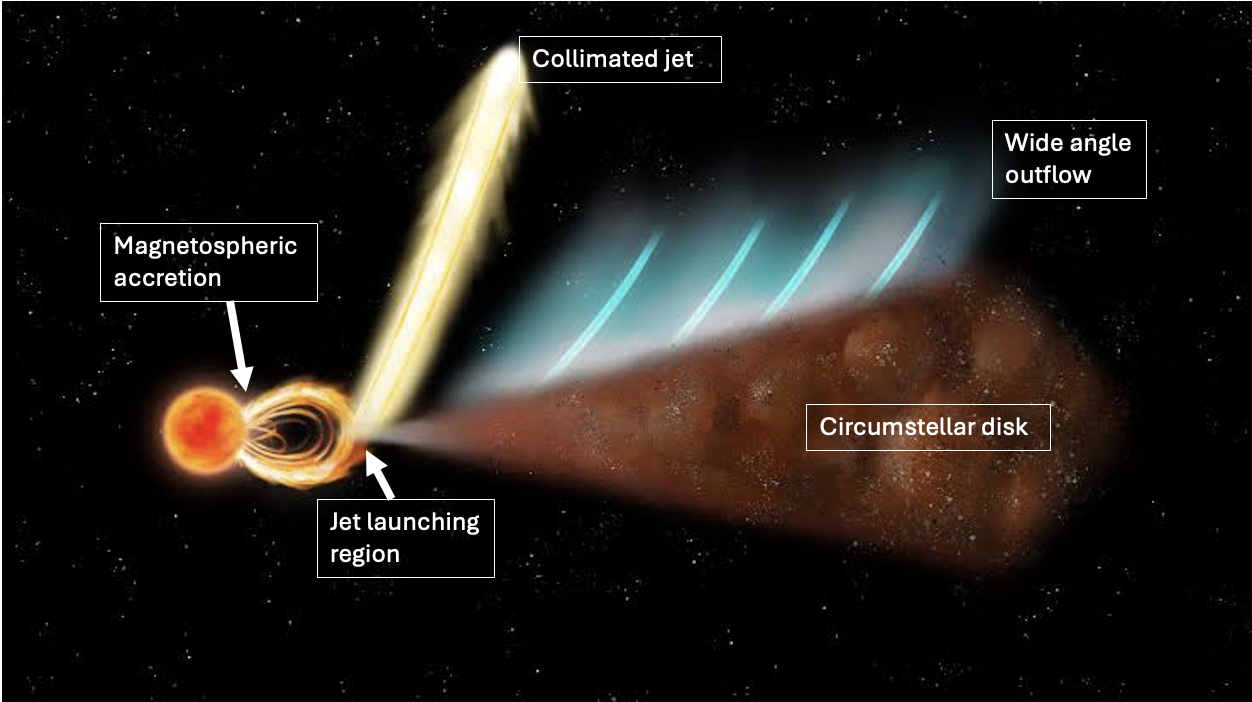}
  \caption{{\small Sketch of the different players in the star formation process that can receive a boost from HWO.  HWO will be able to observe the jet launching region as close as 2-3 au from the central star. Moreover, it will enable the measuring kpc distant sources. }}
  \label{fig:star_formation_launches}
\end{figure}

%\textit{Text by: Garufi, A., Majidi, F. Z., E. Rigliaco, G. Sacco, L. Testi, A. Traficante, K. Biazzo } \\
Star formation has received a great boost in the past years with the advent of observing facilities that were partially devoted to answer some of the most interesting questions in the field, such as: what is the role of disk dispersal in the stellar formation? Are all stars acting the same in their path of formation? 
However, there are some questions that remain unanswered. The HWO observatory will be able to address some of the challenges that are still open. 
In the following we report some of the science cases that are of huge interest for the Italian community working on star formation, and that can be performed thanks to the characteristics of HWO.  \\

\subsubsection{Accretion onto young stars.} 
\label{sec:2.3.1}The main open question that we will be able to answer thanks to HWO is how accretion onto young stars works in low-metallicity environments (\citealt{DeMarchietal2010, DeMarchietal2017, Biazzoetal2019, DeMarchietal2024, Ashrafetal2026}; Biazzo et al., subm., and references therein). There are indications that accretion operates differently in these environments. However, due to the large distances and the limited resolution of current instruments, we cannot use the UV Balmer Jump in the wavelength range 300-400 nm and the U-band photometry to measure accretion rates onto these stars (see, e.g., VLT-based studies in nearby regions; \citealt{alcala14, alcala17, manara21, manaraetal2023}, and references therein).
With HWO we will be able to explore the outer Galaxy, at kpc distances, investigating how accretion works in low-metallicity environments and how the disk evolution and planet formation are affected, overcoming the resolution and sensitivity constraints. 

{\bf Instrument requirements:} for the proposed science case we require a UV spectrographs, covering the wavelengths range between 2000-5000\AA\ with a spectral resolution of the order of a few thousand will probe the region around the balmer jump allowing the measurement of the accretion of gas onto the stars at Kpc scale distance. This science case clearly calls for a UV MOS instrument with high sensitivity and field-of-view of a few arcmin$^2$.

\subsubsection{UV Magnetic Activity and Disk Regulation Across the Stellar-Planetary Boundary.}\label{sec:2.3.2}

HWO will provide the first statistical view of how magnetic activity, accretion, and UV-driven disk evaporation regulate the formation of objects across the stellar-planetary mass spectrum in low-metallicity environments. Using UV MOS, HWO will simultaneously observe large samples of young stars, brown dwarfs, and ultra-cool objects in distant outer-Galaxy star-forming regions, combining Balmer continuum measurements, fluorescent H\({}_{2}\) emission features, with main UV tracers such as Mg II, C IV, Si IV, H2, and many more lines to probe accretion \citep{Ingleby_2013}, flare activity \citep{jackman2024opticallyquietfuvloud}, and photoevaporative winds \citep{France_2012}. In this context, UV spectroscopy will directly trace how high-energy irradiation from young objects controls disk mass-loss rates and dispersal timescales, effectively regulating the reservoir of material available for planet formation across the entire mass spectrum \citep{Best_2024, Gennaro_2020, Luhman_2025}. This transformational capability will reveal whether metallicity modifies the physical boundary between stars, sub-stellar objects, and giant planets through its impact on fragmentation, disk evolution, and magnetospheric accretion \citep{Walch_2011, Nakatani_2018}. By extending UV accretion and activity studies into the poorly explored low-mass regime (especially multiple stellar systems emerging collectively from LSST, Roman, and Euclid surveys), HWO will establish whether the pathways leading from star formation to planet formation are universal or fundamentally dependent on environment, providing unique insight into the origins of ultra-cool dwarfs and free-floating planetary-mass objects under conditions analogous to the early Milky Way.

{\bf Instrument requirements:}  This science case clearly calls for a UV MOS instrument with high sensitivity and field-of-view of a few arcmin$^2$.

\subsubsection{Jet launching regions around young stars.}
\label{sec:2.3.3}The determination of the jet launching region around young stars, i.e. the exact site where jets form and start to propagate, remains an open issue in star formation (see, e.g., studies based on VLT and TNG observations \citealt{Bacciottietal2011, nisini18, Gianninietal2019}). There are several theoretical models, assessing that the inner region from which the jet is launched should be diagnostic of the mechanism that is driving the jet (if MHD wind, or disk wind, or X-wind). With HWO we might be able to resolve the Alfven radius where the magnetic field's energy density equals the kinetic energy density of the surrounding plasma. The limiting factors at present are the spatial and spectral resolutions.  \\
{\bf Instrument requirements:}  for this science case, we  require an Integral Field Spectrograph with   NUV coverage and appropriate spatial and spectral resolution. A spectral resolution between 3000 and 5000 is required to resolve the lines. 
In a typical SFR-ing region at 150 pc we can reach a resolution of 2-3 au at the base of the jet. In the best case scenario, if we investigate the NUV line in the Lyman band at 150 nm, we could reach a diffraction limit of 7 mas for an object at 100 pc, which would translate into a resolution of 0.7 au. 
It is important to reach this kind of resolution for this proposed science case, because it is necessary to resolve the wind coming from the inner disk, below the Alfven radius limit. The observation of these innermost regions will allow us to disentangle the driving mechanism of the jets and angular momentum extraction. 
This is a unique science case for HWO since we do not have the UV coverage in the ELT instruments.\\

\subsubsection{Collimated jets and wide-angle outflows.}\label{sec:2.3.4}
%\noindent {\textit{New text added by A. Traficante}}\\
Thanks to its combination of sub-arcsecond UV–optical imaging and integral-field spectroscopic capabilities, HWO will unveil the full feedback cascade driven by highly collimated jets and wide-angle molecular outflows, extending the investigation far beyond the jet-launching zone in nearby protostars, revolutionizing our understanding of protostellar feedback mechanisms. 

At larger scales, HWO will capture the global impact of outflows on their parental cores and clusters, revealing how hundreds of protostellar jets collectively drive turbulence, regulate core lifetimes, and reshape cloud structure \citep{Lebreuilly24}. The UV imaging and IFU spectrograph will trace the excitation and ionization structure of the outflows through diagnostic lines such as CIV, SiII, FeII, and H$_{2}$, allowing precise determination of shock velocities, densities, and mass-loss rates. By mapping shock knots, bow fronts, and cavity walls in these lines, HWO will distinguish between shock-heated and UV-illuminated gas components, thereby quantifying how energy and momentum are transferred from the jet to the ambient medium. These diagnostics will directly constrain the interplay between accretion and ejection, testing theoretical models of magneto-centrifugal launching and the relative contribution of disk winds and stellar winds to jet collimation. The incredible angular resolution of $\simeq0.01^{"}$ corresponds to $\sim500$ AU at extragalactic distances, such as the Large Magellanic Cloud \citep[$\sim50$ kpc, ][]{Pietrzynski19}. HWO will probe for the first time direct imaging of jets and bow shocks in individual protostars in an external galaxy. Such capability will bridge the gap between Galactic and extragalactic star-formation studies, connecting the physics of jet collimation and feedback across metallicity regimes. \\
{\bf Instrument requirements:} IFU spectrograph in the UV with angular resolution of $\simeq0.01^{"}$ is requested to accomplish the proposed science case. \\

% Through synergy with ALMA and ngVLA observations of the cold entrained gas and magnetic fields, HWO will provide a multi-phase view of outflow feedback, linking atomic, molecular, and ionized components into a coherent physical picture.

\subsubsection{Planet-forming disks}
\label{sec:2.3.5}
%\textit{Contact author: A. Garufi}

The dust component of young circumstellar disks encodes the initial conditions and early evolution of planet formation. Of particular importance are the characterization of dust grain properties (their size, composition, and spatial segregation) and the emergence of disk sub-structures (e.g., rings, cavities, spirals) that are possible tracers of ongoing planet formation. Current observations from JWST and HCI of 8-m telescopes reveal a rich diversity of disk brightnesses and morphologies \citep{Avenhaus2018, Benisty2023, Garufi2024, Duchene2024}, but are limited to the outer disk (tens of au in star-forming regions at 150 pc) and to the near-IR regime, limiting multi-wavelength diagnostics of the scattered light and of the dust grain properties.

HWO will be capable of resolving disk sub-structures closer to terrestrial separations in nearby disks as well as extending such studies to a much broader range of targets at further distances (1 -- 2 kpc). In scattered light, HWO will directly trace the distribution of small dust grains in the disk surface layers, allowing the identification of sub-structures induced by forming terrestrial planets. The combination of spatial resolution and sensitivity will make it possible to detect low-contrast features and subtle radial or azimuthal variations that are currently beyond reach. Compared to current 8-m class facilities mostly operating in the near-IR, HWO will combine superior angular resolution in the optical with significantly improved stability and contrast, opening access to the innermost disk regions in a statistically meaningful sample of systems. ELT will achieve comparable angular resolution in the near-IR making HWO and ELT observations complementary, as they will probe different grain populations and scattering regimes.

A particularly powerful diagnostic would be provided by polarimetric imaging. The polarization fraction and angle of scattered light are highly sensitive to dust grain size, shape, and composition \citep{Perrin2015, Min2016, Tazaki2022}. A polarimeter on HWO would therefore allow a direct, spatially resolved characterization of dust properties in the inner disk regions, breaking key degeneracies of total-intensity imaging. Polarimetry would also be essential to enhance the contrast between the inner disk and the stellar halo, enabling the detection of faint structures at small angular separations. 

\textbf{Instrument requirements}: this science case requires optical 10--20 mas resolution, coronagraphic imaging mode with capability to access regions within a few $\lambda$/D from the star, and broad multi-filter coverage (300--1000 nm) to probe the wavelength dependence of scattering. A high-precision imaging polarimeter compatible with HCI is a key component, enabling measurements of polarized intensity and polarization angle with high spatial fidelity. A field of view of a few arcseconds is sufficient to encompass the relevant disk regions while maintaining optimal wavefront control.

\subsubsection{Accretion on forming planets}
\label{sec:2.3.6}
Beyond tracing the disk structures associated with planet formation, HWO will provide a unique opportunity to directly characterize actively accreting protoplanets. Gas giant planets are expected to undergo prolonged phases of runaway accretion, during which the accretion shock and circumplanetary environment produce strong emission in hydrogen recombination lines (e.g. H$\alpha$) together with the optical/near-UV accretion continuum \citep{Zhu2015, Aoyama2018, Marleau2019}. Over the past decade, H$\alpha$ high-contrast imaging has enabled the first detections of accreting protoplanets such as PDS~70~b and c \citep{Wagner2018, Haffert2019}, demonstrating the power of optical observations to directly probe planetary mass assembly.
The combination of HWO's stable space-based point spread function, high-contrast coronagraphy, and diffraction-limited optical resolution will extend these studies to fainter accretion luminosities and lower-mass forming planets than currently accessible. Simultaneous observations of disk substructures and accreting protoplanets will provide a direct link between disk morphology, planet-disk interactions, and planetary growth. Such observations will also enable statistical studies of planetary accretion rates across different stellar masses, disk evolutionary stages, and environments, placing strong constraints on planet formation timescales and accretion mechanisms.

\textbf{Instrument requirements}: This science case requires diffraction-limited optical imaging with angular resolution of 10–20 mas, together with high-contrast coronagraphy providing access to separations of only a few $\lambda$/D from the host star. Broad wavelength coverage from the near-UV to the optical (300–900 nm), including hydrogen recombination lines and the Balmer continuum, is required to simultaneously detect the accretion continuum and hydrogen recombination lines such as H$\alpha$. An optical/near-UV integral field spectrograph with spatial resolution of $\sim$10-20 mas and spectral resolution $R \simeq 5000$ to resolve hydrogen recombination lines and derive robust accretion diagnostics would enable measurements of accretion rates and characterization of the circumplanetary environment. High contrast performance at small angular separations is essential to detect the faint accretion signatures of low-mass forming planets.

\subsubsection{Photoionization Feedback from HII Regions.}
\label{sec:2.3.7}
Massive stars profoundly reshape their surroundings through intense photoionization, stellar winds, and radiation pressure, creating complex networks of HII regions, photon-dominated regions (PDRs), and photoevaporative flows \citep{Beuther25}. These processes govern the dispersal of molecular clouds, trigger or quench subsequent star formation, and ultimately regulate the structure and energetics of the ISM \citep{Suin24, Vazquez-Semadeni26}. Yet, despite their central importance, the coupling between radiation and gas, especially across the ionization front and into the adjacent molecular material, remains poorly resolved even in the nearest star-forming regions.

HWO will provide an unprecedented view of photoionization feedback by combining high-resolution UV imaging and spectroscopy. With angular resolutions down to $\sim 0.01$", HWO will directly resolve ionization fronts and PDRs down to $\sim10-50$ AU scales in the nearest massive star-forming regions, unveiling the transition layers where ionized, atomic, and molecular gas coexist. Such imaging will map the fine structure of photoevaporating pillars, globules, and disks, tracing how ionizing photons organize the surrounding medium and drive mass-loss from dense clumps and protostellar envelopes.

{\bf Instrument requirements:} The UV spectrographs, covering diagnostic transitions such as CIV, SiIV, and H$_{2}$, will probe the multi-phase gas within these regions, from the hot, shock-heated plasma inside expanding bubbles to the warm molecular layers of the adjacent PDRs. The spectral resolving power of R$\sim 10000-30000$ will deliver velocity-resolved line profiles, allowing us to get direct measurements of expansion velocities, turbulence, and photoevaporation flows at the interfaces between ionized and molecular gas. These diagnostics will quantify the rate at which stellar radiation erodes nearby material, thus constraining the efficiency of cloud dispersal and triggered star formation.\\

In essence, HWO will provide the first comprehensive, spatially and spectroscopically resolved view of jests, outflow and photoionization-driven feedback, unveiling the physics that determines whether stellar radiation disrupts, compresses, or triggers the next generation of star formation across the Milky Way and nearby galaxies. In synergy with ALMA, JWST, and future radio surveys of the Galactic Plane with MeerKAT and SKA, HWO will enable a multi-wavelength reconstruction of feedback processes, from the ionized plasma traced by HWO and SKA to the cold molecular gas seen in the sub-mm regime.\\

\subsubsection{Star-forming regions at Cosmic Noon}
\label{sec:2.3.8}
\begin{figure}
    \centering
    \includegraphics[width=\linewidth]{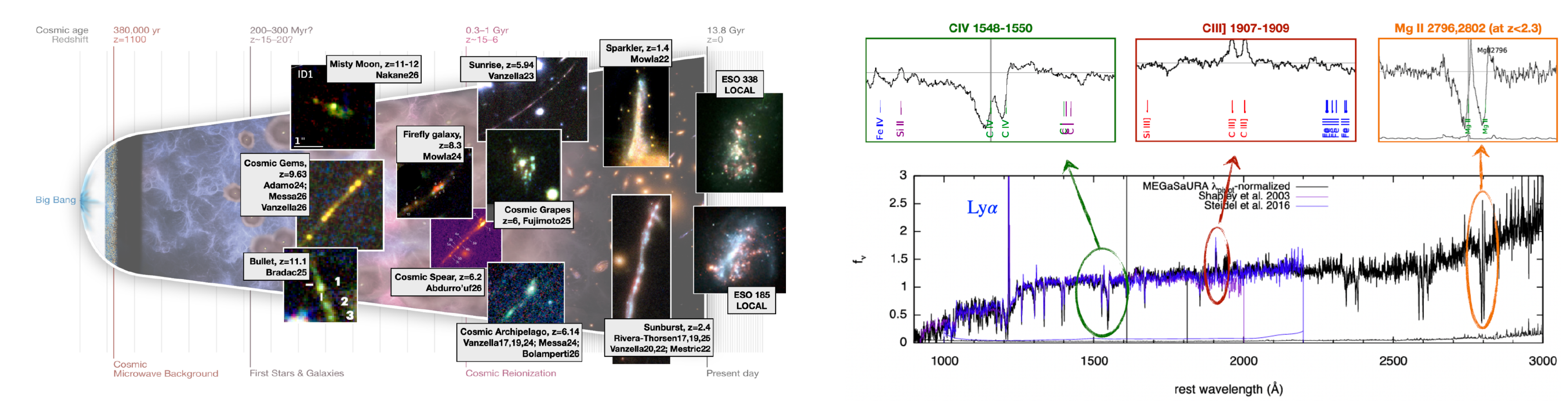}
    \caption{\textit{Left:} Schematic view of the cosmic history in the Universe and the rapid growth of detections of star clusters from cosmic noon up to reionization epoch, all of them based on strong gravitational lensing. On the right, two local starburst galaxies showing star clusters (ESO 338 and 185). \textit{Right:} A rest-UV stacked spectrum from the MEGaSaURA sample \citep{Rigby2018}, with some of the main lines highlighted. This serves as
an example of what is expected to be detected with HWO at cosmic noon.}
    \label{fig:star_clusters}
\end{figure}

%\textit{Text by: M.Messa, E.Vanzella, A.Zanella}\\
Star formation at cosmic noon ($\rm z\sim 1-3$), the epoch of peak star formation rate density \citep[][see also Fig.~\ref{cmd}]{Madau2014}, is not smoothly distributed across galaxies but concentrated in compact, dense clumps, orders of magnitude denser than typical HII regions and star-forming regions in the local Universe \citep[e.g.][]{Elmegreen2005,ForsterSchreiber2011}. 
The combination of gravitational lensing with HST and JWST has brought these structures into sharp focus: in strongly lensed fields, individual star-forming regions and star clusters can be resolved down to tens of parsecs at $\rm z\gtrsim1$ \citep[e.g.,][]{Vanzella2022,Messa2022,Claeyssens2023,Claeyssens2025}, revealing massive compact clusters that may represent the formation sites of present-day globular clusters (Fig.~\ref{fig:star_clusters}, left). 
%First ALMA observations of the cold molecular gas in a handful of lensed galaxies at cosmic noon are already suggesting that the denser, more turbulent ISM conditions at these redshifts may lead to the formation of denser GMCs and a more efficient conversion of gas into stars (ref to DZ+19,23), providing a physical explanation for the extreme properties of the star-forming clumps observed in the rest-UV.

The coming decade will witness a dramatic expansion of this field, with Euclid already discovering hundreds of new strongly lensed cosmic-noon arcs across the full sky \citep{Euclid2025d,Euclid2025a,Euclid2025b,Euclid2025c} and the upcoming ELT's extreme angular resolution in the near-infrared enabling the study of compact star-forming clumps at sub-100 pc scales even in unlensed fields \citep[e.g.,][]{Ciliegi2024}. Together, these facilities will transform the study of sub-galactic star formation from an exploration of individual iconic systems into a statistical science. 
A critical diagnostic window will remain out of reach without HWO. Rest-frame UV spectroscopy at sufficient sensitivity to characterise the nebular gas conditions (ionisation, chemical abundances, outflow kinematics, and $\rm Ly\alpha$ escape) at the scales of individual star-forming regions ($\rm \lesssim100$ pc) is entirely inaccessible from the ground at cosmic noon redshifts; HWO, with its UV-to-optical coverage, will provide access to the full suite of rest-frame UV diagnostics, e.g. $\rm Ly\alpha$, C~IV, He~II, C~III], and Mg~II, for galaxies at z$\sim$1–3 (Fig.~\ref{fig:star_clusters}, right). 
Operating in synergy with ELT and future improvements to millimeter interferometry such as the proposed ALMA2040 (which will for the first time enable systematic studies of cold molecular gas and GMC properties at cosmic noon), HWO will connect cold gas reservoirs to ionised nebular emission and stellar populations, providing the ionising budget, chemical enrichment, and feedback diagnostics needed to follow the full star formation cycle at sub-galactic scales, from GMC to UV-bright clumps.

\textbf{Instrument requirements:} HRI, wide-field UV/optical/NIR imager, fundamental to determine the rest-frame ultraviolet clump sizes at redshift $\simeq 0.5-1$, along with those up to redshift 6. UV-IFS (UV integral-field) spectrograph covering the wavelength range UV-to-optical spanning 
$\sim 1000-11000$\AA in the observed frame is needed
to blindly access the key rest-frame UV diagnostics (Ly$\alpha$, NV~1240, NIV~1483,1486, HeII~1640, OIII]1661,1666, CIII]1907,1909 and Mg~II) for sources at $\rm z\sim 1-8$ in a single spectral window. 
A spectral resolution of $\rm R\sim 2000-5000$ is required to resolve the profiles of key lines, in particular (1) the spatially diffuse Ly$\alpha$ profile which is a factor of $\times$10 more extended than the UV continuum \citep{wisotzki2016} and/or multi-peak structure in the wavelength domain (also key to identify candidate Lyman continuum emitters, see Sect.~\ref{sec:2.5.4}, and stellar feedback \citep{Bhagwat2025}), (2) the high-ionization doublets which provide key physical properties of the ISM, electron temperature and density and (3) the velocity structure of UV absorption features tracing outflows. Spatial resolution: diffraction limited. Field of view: as wide as possible, to enable observations of multiple targets within a single pointing. The current JWST/NIRSpec IFS field of view of $3''\times3''$ allows spatially resolved spectroscopy from space for only one galaxy at a time, significantly limiting the ability to build statistically meaningful samples in flux-limited surveys. \\ %Add also sensitivity requirements and/or spatial resolution?

\subsection{Stellar populations}
\label{sec:sci.stellarpopulations}
%\textit{Text by: F. Annibali, D. Massari, G. Cescutti}\\

\subsubsection{Tracing galaxy evolution with resolved stars} 
\label{sec:2.4.1}

Resolved-star color–magnitude diagrams (CMDs) provide one of the most powerful tools for reconstructing the star formation history (SFH) of nearby galaxies \citep{Weisz2014,Cignoni2019,Annibali2022}. By combining the observed luminosities and colors of individual stars across all evolutionary phases with state-of-the-art stellar evolution models \citep[e.g.,][]{Bressan2012,Marigo2017} and assumptions on the initial mass function (IMF), it is possible to directly infer the rate of star formation (SFR) as a function of cosmic time.
The main observational limitation of this approach is that, while the brightest regions of the CMD are populated by relatively young stars, old, low-mass stars occupy the faintest portions of the diagram. Probing these ancient populations therefore requires facilities with extremely high sensitivity and spatial resolution, capable of reaching fainter magnitudes and resolving stars at increasingly larger galaxy distances and in increasingly crowded environments. 

Current facilities such as HST and JWST can reach the horizontal branch (HB)--the brightest evolutionary phase that reliably traces populations older than 
$\sim$10 Gyr (roughly corresponding to the cosmic noon at $z \sim 2$)--in galaxies closer than about 3--4 Mpc. The ELT, offering JWST-like near-infrared sensitivity but $\sim$7 times better angular resolution, will extend these studies to the dense central regions of galaxies within this distance range, providing essential synergy with JWST.
However, due to the intrinsically blue colors of HB stars, only HWO, with resolving power in the UV comparable to the ELT’s resolution in the NIR, will be able to extend these studies to significantly larger distances, potentially out to the Virgo cluster. Below we outline representative science applications.

\textbf{Instrument requirements:} An imager with relatively large field of view (about 10 arcmin$2$ or more) and diffraction-limited spatial resolution sensitive at UV-to-NIR wavelengths is ideal for this science goal.

\subsubsection{Reaching the Oldest Stellar Populations in Giant Elliptical Galaxies}
\label{sec:2.4.2}
Massive spheroidal systems contain more than half of the stellar mass in the local Universe \citep[e.g.,][]{Fukugita1998}. 
Yet, when and how 
giant early-type galaxies (ETGs) assembled their stars is still highly debated. While the enhanced [$\alpha$/Fe] ratios observed in local ETGs 
have long been suggested to be due to an early, rapid formation, direct evidence for large number densities of massive quiescent galaxies at 
$3 < z < 5$ has emerged only recently from JWST observations \citep[e.g.,][and references therein]{Carnall2024}.

In this framework, the stellar content of present-day massive ETGs should be dominated by stars older than $\sim$10 Gyr. 
Unfortunately, no massive ``classical'' elliptical galaxies are close enough to be resolved into stars down to the HB level with current facilities. 
The nearest giant ETG, Cen A (at $\sim$3.8 Mpc), is a peculiar merger remnant, while the closest classical E galaxy 
is NGC\,3379 at $\sim$10 Mpc.
For this galaxy, deep HST and JWST imaging \citep[e.g., ]{Harris2007,Anand2025} have reached the brightest $\sim$2 mag of the red giant branch 
(RGB), which samples stars with ages between 2--13 Gyr but provides limited temporal resolution for disentangling early SF episodes. 
Inferring a reliable SFH back to the earliest epochs requires reaching the fainter HB. At a distance of 10 Mpc, the HB is expected at F090W$\sim$30 and F150W$\sim$29.5, demanding prohibitively long integrations with JWST (and even with the ELT).

\begin{figure}[!t]
%\vspace*{-2\baselineskip}
\centering
\includegraphics[width=\textwidth]{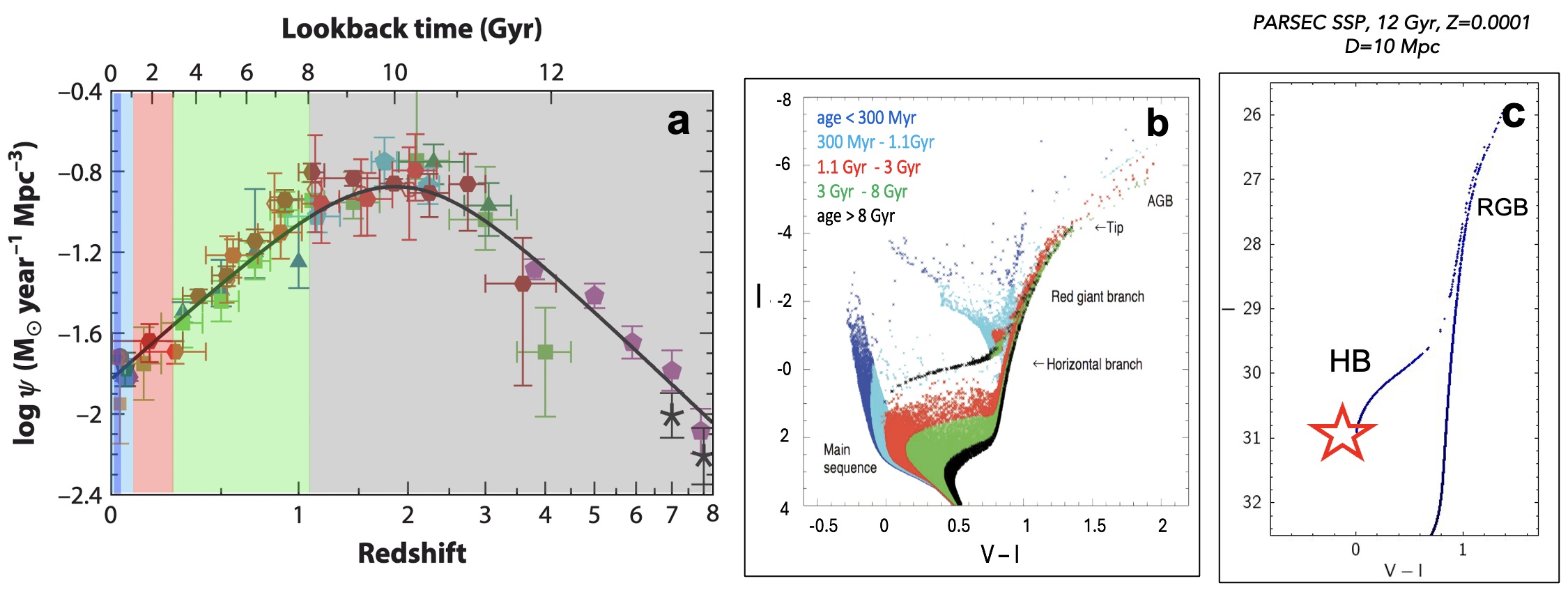}
\caption{Panel a: cosmic star formation from \cite{Madau2014}, with different look-back times color-coded according to the same color 
scheme as in the color-magnitude diagram of panel b: blue for age $<300$\,Myr, cyan for 300\,Myr$\leq$age$<$1.1\,Gyr, red for 1.1\,Gyr$\leq$age$<$3\,Gyr, green for 3\,Gyr$\leq$age$<$8\,Gyr, and black for age$>$8 Gyr. The horizontal branch (HB) is the brightest evolutionary phase that can act as 
a reliable clock for populations older than 8-10 Gyr. Panel c: stellar isochrone from the PARSEC models showing the magnitude limit needed to 
reach the HB at 10 Mpc distance; HWO will allow to measure those stars with just $\sim$1\,h exposure time in V and $\sim$1\,h in I. 
}
\label{cmd}
%\vspace{-0.3cm}
\end{figure}

By contrast, HWO can efficiently reach these magnitudes with modest exposure times. Based on the current HWO exposure time calculator \url{https://hwo.stsci.edu/camera_etc}, we estimate that detecting the HB in NGC\,3379 would require only $\sim$1 hr in V, and $\sim$1 hr in I (see Fig.~\ref{cmd}).

Even more compellingly, HWO could make it possible to detect ancient HB stars in Virgo, the nearest major galaxy cluster (distance $\sim$16.5 Mpc), which hosts numerous giant ellipticals as well as galaxies spanning all morphological types and masses. Reaching the HB in Virgo would require longer--but still reasonable--exposure times of a few hours in V and I.

\textbf{Instrument requirements:} An imager with relatively large field of view (about 10 arcmin$2$ or more) and diffraction-limited spatial resolution sensitive at UV-to-NIR wavelengths is ideal for this science goal.

\subsubsection{The Origin of the UV Upturn in ETGs and Spiral Bulges}
\label{sec:2.4.3}

\noindent ETGs and spiral bulges host a hot stellar component that produces an excess in their UV spectral energy distribution shortward of 
$\sim$2000 \AA \citep{Oconnell99}. Despite their remarkably similar optical and NIR spectra, this UV ``upturn'' varies by nearly three orders of magnitude, as quantified by the (1550$-$V) color. The UV-upturn strength also correlates with galaxy velocity dispersion, suggesting that it carries key information about how stellar population properties depend on galaxy mass.

However, the physical origin of the UV upturn remains elusive because the responsible stellar population has so far escaped direct detection. Indirect evidence points to extremely hot HB stars and their progeny, yet the nature of these sources—whether arising from helium-enriched stars, enhanced mass loss at high metallicity, or other channels—remains uncertain, partly due to limitations in current stellar evolution models.

HWO, with its high-resolution ultraviolet spectroscopic capabilities, could directly constrain the chemical abundance properties of the population responsible for the UV upturn, offering crucial insights into its origin and its connection to the formation histories of ETGs and bulges.

\textbf{Instrument requirements:} A UV-optical MOS with relatively large field of view (about 10 arcmin$2$ or more) and diffraction-limited spatial resolution sensitive is ideal for this science goal.

\subsubsection{Massive Stars in Extremely Metal-Poor Environments}
\label{sec:2.4.4}
Nearby galaxies with extremely low metallicity and ongoing star formation provide the best local laboratories for understanding the physical processes that governed star formation in the early Universe. The blue compact dwarf IZw\,18 (distance $\sim$18 Mpc), with a metallicity of only a few percent solar, is one of the most remarkable examples. Recent JWST observations have revealed high-ionization emission lines such as [Ne V] and [O IV] \citep{Hunt2025}, indicative of an exceptionally hard radiation field, comparable to that inferred in Lyman continuum leakers.
This intense ionizing radiation could arise from very metal-poor, super-massive stars, from metal-poor stellar populations hosting a self-consistent contribution from X-ray binaries, from X-ray emission powered by star-cluster winds, or from an AGN. Discriminating among these scenarios is crucial for understanding the nature of the ionizing sources that drove cosmic reionization.

HWO will be capable of resolving individual massive stars in these nearby metal-poor systems and characterizing the surrounding ionized gas through high-resolution UV spectroscopy. This will enable direct measurement of stellar temperatures, ionizing photon fluxes, and nebular conditions, providing unprecedented insight into the nature of the first generations of massive stars and their role in the reionization of the Universe.

{\bf Instrument requirements:} reaching HB stars in galaxies in the Virgo cluster (D$\sim$16.5 Mpc) translates into being able to measure a star with B$\sim$31.5, V$\sim$31.5, and I$\sim$31.5 (Johnson-Cousins) with a signal-to-noise of at least S/N$\sim$3. 
Spatial resolution is crucial to resolve stars in crowded regions, as those typical of ellipticals in Virgo, and of 
the central star forming regions in dwarf galaxies. For a mirror diameter  $\geq$6 m, the diffraction limit is  
$\leq$10 mas in the UV, adequate for our purposes; nonetheless, it is important that the pixel scale is sufficiently small ($\leq$5 mas) to avoid PSF undersampling. The implementation of an IFU instrument in the FUV-NUV range would be extremely valuable to study the spatial properties of the ionized gas in metal poor star forming systems with unprecedented details (spaxels $\leq$5 mas). UV spectroscopy with R$\gtrsim$10,000 in longlist, MOS, or IFU mode are all adequate to study chemical properties of hot HB stars in Virgo ellipticals. 

\subsubsection{Exploration of r-Process Nucleosynthesis}
\label{sec:2.4.5}

\noindent High-resolution UV spectroscopy with HWO will revolutionize our understanding of the origin of the heaviest elements produced by the r-process. While optical facilities and instruments such as CUBES (Evans et al. 2020) will identify and characterize large samples of metal-poor and r-process-enhanced stars through key tracers observable from the ground, many crucial heavy elements (e.g. Se, Te, Pt, Au, Pb, Bi) are accessible only in the far-UV. HWO will therefore provide the natural and necessary follow-up to CUBES discoveries, enabling the reconstruction of nearly complete r-process abundance patterns across a wide range of Galactic environments.

This synergy is particularly powerful because CUBES can efficiently discover and chemically pre-characterize faint metal-poor stars, while HWO can extend the analysis to the UV domain with higher spectral resolution and sensitivity than currently possible with HST/STIS \citep{Roederer22}. Such observations will constrain the physical conditions, nuclear physics, and astrophysical sites of the r-process, including neutron-star mergers and exotic supernovae \citep{Ji&Frebel18,Holmbeck23}.

{\bf Instrument requirements:} include high-resolution UV spectroscopy with $R \approx 40,000-100,000$ (ideally $\sim100,000$), wavelength coverage extending from at least 2000-3100\AA{} and preferably down to 1500-1700\AA{}, together with $S/N \ge 20-50$ per pixel to detect weak and blended absorption features of heavy neutron-capture elements.

\subsubsection{Tracing the First Stars}
\label{sec:2.4.6}

High-resolution UV spectroscopy with HWO will provide a unique window on the first stellar generations and the earliest phases of chemical enrichment in the Universe. Instruments such as CUBES \citep{Evansetal2020} will efficiently identify extremely metal-poor stars through near-UV molecular features such as NH and OH, building large samples of candidate second-generation stars. HWO will extend these studies into the far-UV, where many crucial transitions of elements such as C, Mg, Al, Si, P, S, Sc, V, Cr, Mn, Fe, Co, Ni, Cu, and Zn remain detectable even at metallicities below [Fe/H] $<-5$.

The combination of CUBES and HWO will therefore enable both the discovery and the detailed chemical characterisation of the most primitive stars known. UV abundance measurements will constrain the masses, explosion energies, and nucleosynthetic yields of Population III supernovae, while also testing the possible existence of surviving metal-free low-mass stars \citep[e.g.][]{FrebelNorris2015,Keller2014Natur}.

{\bf Instrument requirements:}  Key requirements include UV spectroscopy with $R \approx 40,000-100,000$ (ideally $\sim100,000)$, wavelength coverage from about 1800-3100\AA\ and preferably down to 1400\AA\ , together with S/N $\ge 50-100$ per pixel to detect extremely weak absorption features in faint metal-poor stars.

\subsubsection{Probing Cosmic-Ray Spallation with boron}
\label{sec:2.4.7}

Beryllium and boron are unique tracers of the earliest phases of Galactic chemical evolution, as they are produced almost exclusively by cosmic-ray spallation processes. CUBES \citep{Evansetal2020} will dramatically expand Be measurements through the Be\,II resonance lines at 313\,nm, enabling studies of stellar mixing, Galactic enrichment, and the evolution of cosmic rays in very metal-poor stars. Boron, instead, is accessible only in the ultraviolet, with the B\,I line at 2089.50\AA\  identified as the most reliable abundance indicator in metal-poor stars, and currently observable only with HST/STIS \citep{Spite2025}.

HWO will extend boron studies to much fainter metal-poor stars across the Milky Way and nearby systems, enabling a direct comparison between Be and B over a broad metallicity range. Together, CUBES and HWO will provide powerful constraints on early cosmic-ray nucleosynthesis and the chemical evolution of the Galaxy \citep{Smiljanic2014}.

{\bf Instrument requirements:} Key requirements include high-resolution UV spectroscopy ($R \approx 40,000-100,000$, ideally $\sim 100,000$) with wavelength coverage extending from $\sim$2000\AA\  to $\sim$3150\AA\ , in order to access both the B\,I 2089.50\AA\ line and the Be\,II resonance lines. High signal-to-noise ratios (S/N $\ge$ 50–100 per pixel) and sensitivity to stars as faint as V $\approx$ 14-18 are essential to move beyond the current HST/STIS limits.

%\subsection{Galaxy formation and evolution}
%\label{sec:drivershos}
%\textit{Text by : P. Saracco, S. Bisogni, R. Bonito, A. Caratti o Garatti,  F. D’Ammando, E. Dalla Bont\`a, P. Franzetti, A. Gargiulo, M. G. Guarcello, L. Izzo, F. La Barbera, A. Longobardo, C. Mancini, M. Mirabile, A. Pizzella, L. Prisinzano, G. Vietri, H.-F. Wang
%}

\subsection{Galaxy and AGN evolution}
\label{sec:sci.galaxiesAGN}
%\textit{Text by: V. D'Odorico, A. Marconi, L. Pentericci, P. Saracco}

Understanding the formation  of cosmic structures remains one of the most compelling frontiers in astrophysics. Despite significant progress, key physical processes that shaped the early Universe are still poorly constrained. The Epoch of Reionization marks the final major phase transition in cosmic history, when the first galaxies and active galactic nuclei (AGN) ionized the intergalactic medium (IGM). Yet, the mechanisms driving this process remain elusive: we do not know which sources dominated ionizing photon production or how Lyman-continuum radiation escaped from their interstellar and circumgalactic media into the surrounding gas.
The first stellar generation, the so-called Population III stars, formed from pristine gas and initiated both the re-ionization and the chemical enrichment of the Universe. While direct detection of these stars is unlikely, the uncovering of their nucleosynthetic signatures — preserved in extremely metal-poor stars today and in high-redshift gaseous environments — would provide critical constraints on their initial mass function and their contribution to early cosmic evolution. 

These early stellar and galactic processes ultimately connect to the co-evolution of galaxies and their central supermassive black holes. Despite major advances, the physical mechanisms regulating black hole growth and feedback remain unclear. Key questions persist: how did the first SMBHs form and grow so rapidly, and how does accretion interact with the interstellar medium to influence star formation? Answering these questions requires capabilities such as rest-frame ultraviolet spectroscopy, polarimetry, and sub-arcsecond imaging — capabilities that future observatories like  HWO are uniquely positioned to provide.

A deep understanding of the galaxy life-cycle requires understanding the mass assembly
history, the star formation processes, the mechanisms that regulate the shutdown of 
star formation, what keeps a 
galaxy in a quiescent state and the role played by the environment in these processes.
In this regard, the connection between the galaxy life-cycle and the intergalactic 
(IGM) and circumgalactic (CGM) medium is still rather unclear.

%Although the above topics focus on a different scale — ionizing photons escaping galaxies, the birth of the first stars, and SMBH–galaxy co-evolution — all share the same fundamental goal: to uncover the physical mechanisms that governed the birth and transformation of structures in the early Universe. 
Progress demands next-generation observational facilities capable of probing distant epochs with unprecedented resolution and sensitivity.
%Understanding how baryons assembled to form the first stars, galaxies, and cosmic structures, and how these systems subsequently evolved over cosmic time, is a central theme in contemporary astrophysics. 

\subsubsection{Massive Galaxies}
\label{sec:2.5.1}
Over the past decade, deep ground based observations and, more recently, JWST observations
have revealed numerous massive galaxies up to $z$$\sim$5 challenging our understanding 
of galaxy formation physics \citep[e.g.,][]{boylan23, chworowsky24}. 
These massive (log(M$_*$/M$_\odot$)$>$11) quiescent galaxies 
include both older systems and younger galaxies resembling post-starburst 
populations \citep[e.g.,][]{glazebrook17, tanaka19, saracco20apj, forrest22, antwi25, carnall24}. 
The oldest galaxies host stellar populations with an age comparable to the cosmic epoch, that is formed within about 500 million years of the Big Bang \citep[e.g.,][]{glazebrook24}.
The youngest massive galaxies, with strong Balmer absorption lines, formed within a 
few hundred Myr before observation \citep[e.g.,][]{deugenio20}. 
Given the high stellar mass of these galaxies, star formation rates 
$>$400 M$_\odot$/yr are required. 
Near-solar or supersolar metallicities detected in such galaxies further challenge theories of chemical enrichment, as such abundances require mechanisms inconsistent with short star formation times. 
Therefore, massive galaxies challenge our understanding due to the following main issues:
\\
{\it Extreme Star Formation} - The high star formation rates required to form such high stellar masses are highly challenging, especially in the early Universe because of the  
lower cooling efficiency of pristine hydrogen and the lack of sufficiently massive 
dark matter halos.
\\
{\it Rapid Enrichment} - Solar or super-solar chemical enrichment levels typically require timescales $>$1 Gyr inconsistent with the short star formation timescales imposed by 
the age of their stellar population. 
\\
{\it Abrupt Quenching} - Many of these massive galaxies are completely quiescent, despite the high star formation rates that were required to form their stars so rapidly.
Therefore, an extremely efficient quenching mechanism must have abruptly halted star formation. 
\\
{\it Mass growth and assembly} -
Both old and young massive galaxies challenge hierarchical models since their 
stellar mass must have formed in situ, as there is insufficient time for hierarchical assembly through the merging of individual subunits, especially for the young population.

{\bf Instrument requirements} 
Although HWO is primarily focused on its UV and blue sensitivity, it may be able to open a new observational window up to $\simeq 2.5~\mu$m if it will be equipped with a near--IR spectrograph fully matched to the diffraction-limit PSF. We stress indeed that the diffraction limit of a 6.5m class telescope is about 0.05$''$ at $1.5~\mu$m, and is not well sampled by the current JWST NIRSpec capabilities, that are limited by the relatively large slit aperture (0.2") and fixed arrangement of the MSA. A proper design of a near-IR spectrograph would allow achieving the following performances:\\
$\bullet$ {\it Near-IR Coverage 1.0--2.5 $\mu$m} - Essential to trace rest-frame optical and UV features 
(e.g., Balmer lines, Ca~H\&K, D4000, [OII], Mg, Fe) necessary to derive stellar population parameters (ages, metallicity) and properties of interstellar medium (excitation, kinematics, outflows, star formation rate, metallicity), redshifted into the infrared 
for redshifts $z$$>$1.5. This allows for a homogeneous study of star formation, 
chemical enrichment, and quenching.
\\
$\bullet$ {\it Angular Resolution 30--40 mas} - Needed to resolve sub-galactic regions with 
physical scales comparable to the Giant Molecular gas Clouds (GMC, 150–250 pc) 
across all cosmic epochs. GMCs contain up to 10$^6$--10$^7$ M$_\odot$, they are the 
main star-forming units, the main places of metal production and sources of enrichment, 
the cradle of Globular Clusters (GCs). 
\\
$\bullet$ {\it Multiplexing} - To efficiently map extended structures and environment of galaxies, multiple clumps, clusters and overdensities of galaxies, study galaxy properties homogeneously across different environments.

These characteristics define the spectrograph SHOS described in Sec. \ref{sec:strum.SHOS}.

\subsubsection{Scaling Relations and the stellar Initial Mass Function} 
\label{sec:2.5.2}
Galaxies follow key empirical scaling relations, such as the Fundamental Plane 
\citep[FP;][]{djorgovski87,dressler87},
connecting velocity dispersion, effective radius, and surface brightness, 
the age-mass, and the metallicity-mass relations \citep[e.g.,][]{gallazzi05}. 
The remarkably small intrinsic dispersion of the FP represents a significant 
challenge to the hierarchical paradigm, which, being a stochastic merger-driven 
process, should produce a much wider dispersion. 
Tracing these relations back through cosmic time is essential to determine whether 
they are the result of later evolution or direct formation imprints.

A related critical factor is the stellar Initial Mass Function (IMF). 
Although traditionally thought to be universal, recent evidence contradicts this assumption \citep[e.g.,][]{hopkins18,smith20}. 
Observations of IMF-sensitive absorption features 
(such as the gravity-dependent NaI doublet at 8183, 8195 \AA\ and the Wing-Ford 
band FeH at $\sim$9900 Å) reveal that the central regions of massive early-type galaxies 
host a bottom-heavy (dwarf-enriched) IMF  \citep[e.g.,][]{conroy12,labarbera13,spiniello14,labarbera19,parikh18}. 
Measuring IMF radial gradients is paramount: it can distinguish between in situ star formation (which depends heavily on local gas density and cloud fragmentation) and stellar mass assembly via mergers (which flattens gradients).

{\bf Instrument requirements} 
$\bullet$ {\it Angular Resolution} - To probe IMF radial gradients before merger 
events erase them, observations must target galaxies at $z$$\gtrsim$1.
Since the most compact massive galaxies at these epochs typically have an effective radius smaller than 1 kpc, sampling scales below 500 pc is mandatory. 
This unexplored regime is entirely beyond JWST's spatial resolution 
(limited by its 0.1 arcsec/pixel scale).
\\
$\bullet$ {\it Near-IR coverage} - Near-IR is required to capture the faint 
gravity-sensitive absorption features redshifted into the infrared even at 
low redshifts.

\subsubsection{Origin of Ultra-Diffuse Galaxies}
\label{sec:2.5.3}
Ultra-Diffuse Galaxies (UDGs) feature sizes comparable to the Milky Way but are approximately 100 times less massive \citep[e.g.,][]{vandokkum15}. 
Their origin challenges current models of galaxy formation, dividing into two 
main evolutionary scenarios:
\\
{\it Failed Massive Galaxies} - Systems whose star formation was prematurely 
quenched leaving them with dwarf-like stellar masses inside massive dark matter halos \citep[e.g.,][]{vandokkum15,toloba23}. 
These appear red, old, metal-poor, Dark Matter-dominated, and host a rich 
population of GCs.
\\
{\it Expanded Dwarfs} - Originally low-mass dwarf galaxies that expanded within 
low-mass halos due to intense internal feedback or environmental processes
\citep[e.g.,][]{carleton19}. 
These exhibit bluer colors, extended star formation histories, and fewer GCs.

To distinguish between the two possible scenarios, the following instrument requirements must be met.
\\
{\bf Instrument Requirements}
\\
$\bullet$ {\it High Angular Resolution} - Absolutely critical an angular resolution
of about 30-40 mas to disentangle these highly crowded systems and accurately isolate 
the spectra of individual GCs from the background field stars.
\\
$\bullet$ {\it Multi-IFU} - To simultaneously acquire kinematics and metallicity 
(age and Z) for multiple GCs and field stars, which is required to determine the 
dark matter content and deduce the formation history.
\\
$\bullet$ {\it Near-IR coverage} - This is essential for observing through the 
dust-shrouded central regions of GCs and also through the diffuse dust within UDGs.

\subsubsection{The escape of Lyman continuum radiation: a key ingredient to understand reionizaton}
\label{sec:2.5.4}
%The Epoch of Reionization represents the last major phase transition in the early Universe, during which neutral hydrogen in the intergalactic medium (IGM) was gradually ionized by the first galaxies and AGN. 
A key challenge in fully understanding the reionization process is the lack of a clear picture of how the necessary ionizing radiation (i.e. in the Lyman continuum range) was generated by these sources and how it escaped into the IGM.
At present, our understanding is limited both regarding which sources dominated the production of ionizing photons (galaxies, AGN, or a combination of both) and the physical mechanisms that allowed these photons to escape from the interstellar and circumgalactic media into the surrounding intergalactic medium. 

Direct measurements of Lyman continuum (LyC) radiation 
at high redshifts is impossible due to the IGM opacity \citep{inoue2014}.
Consequently, we  must rely on local analogs and on simulations to develop diagnostics of LyC  to infer the contributions of individual galaxies to the total ionizing budget.
\\
Currently, the only statistical sample of sources with solid measurement of LyC escape is available at very low redshift $<0.3$, based on HST observations \citep{flury22,jaskot25}. At higher redshifts, the number of confirmed LyC emitters remains too small to robustly characterize the properties of the population. However, a few rare detections have demonstrated that LyC leakage can be observed directly out to z $\simeq$ 4--6 \citep[e.g.,][]{Vanzella2018, Marques-Chaves2022, Saha2026}. A powerful UV-sensitive space observatory  would enable transformative science by addressing the following issues:

\textbf{SC1:  Digging deep into the LyC LF} Even in the low redshift universe our knowledge of LyC radiation is limited to a sample of 90 sources down to L*:  major  progress to   identifying the main contributors to cosmic reionization would be enabled  by pushing detections down to at least $M_{\rm UV} =13$ for  several thousand galaxies, to calibrate indirect indicators such as ionizing conditions, the fraction of the ISM that is covered by thick neutral hydrogen, dust attenuation and SFR surface  density \citep{mascia}.

\textbf{SC2: The evolution of the LyC emission to cosmic noon}
 Direct detection of ionizing radiation from a statistically representative sample of sources at redshift $\sim1$ (when the IGM is still relatively transparent) and at redshift $\sim3$ (the highest redshift where such detections are feasible and the closest observational window to the Epoch of Reionization) is crucial to understanding how the escape of ionizing radiation evolves over cosmic time. At present, we do not know whether the galaxy properties that enable Lyman-continuum escape change with redshift \citep{jaskot25,liu25}.

\textbf{SC3: The physics of LyC escape: }
Understanding the physical mechanisms that govern LyC escape is a multiscale problem, starting from the production of LyC photons in super star clusters, their propagation through the interstellar medium, and their eventual escape through the CGM into the IGM. Only a few exceptional cases, such as the strongly lensed Sunburst Arc \citep{Kim2023}, have provided spatially resolved insights; most other studies rely on integrated spectra, making it difficult to identify the small-scale processes that enable LyC leakage. 
\\
\\
\textbf{Instrument requirements} For SC1 and SC2  we require photometric filters in the far-UV (for the 0.1-0.3 redshift range) and UV (for cosmic noon)  filters and the ability to reach observed magnitudes of $AB \sim 32$ at wavelengths of interest  in order to directly detect Lyman-continuum (LyC) radiation. In addition, relatively high-resolution multi object UV and optical spectroscopy (R $\sim$ 2,000) is needed to identify and characterize indirect indicators of LyC escape. To map small scale  LyC emission (SC3), characterize the physical conditions of the surrounding interstellar medium, and directly observe feedback-driven outflows that facilitate LyC escape, an IFU with high spatial resolution (~10-100 pc) and UV wavelength coverage would be transformative. Such observations would enable us to establish direct links between LyC escape in local environments and the mechanisms operating in high-redshift, clumpy star-forming galaxies.

\subsubsection{Measuring chemical abundances in the Reionization epoch.}
\label{sec:2.5.5}

The first stellar generation (Population III) was fundamentally different from the following ones because it was born from chemically pristine gas. While direct detections of first stars are extremely challenging and maybe impossible \citep[e.g.][]{katz23}, we can observe their chemical fingerprints in very old, and very metal-poor stars in our galaxy \citep[see][for a recent review]{Bonifacio2025} or in galactic gaseous environments at high redshift \citep[e.g.][]{Saccardi2023}. The chemical elements synthesized and dispersed in the surrounding environment after the death of a star, are probes of its properties and, in particular, they allow to put constraints on the unknown initial mass function of the Population III generation \citep[e.g.][]{Koutsouridou2024}.   

The chemical abundances of high redshift gas can be studied in the absorption spectra of bright background quasars. Currently, only a a few tens of the known quasars in the Reionization epoch can be observed with this goal and only at intermediate resolution \citep[e.g.,][]{Dodorico2023}. The first statistical measurements of chemical abundances of gas at $z\sim 6$ suggest a possible contribution to the enrichment from Pop III stars, but no clear signature is observed \citep{Christensen2023, Vanni2024, Sodini2024}. Next-generation spectrographs on 30-40m telescopes (e.g. ANDES at the ELT, see Section~\ref{sec:fut.ANDES}) will significantly enhance the quality of the results by increasing spectral resolution. However, the foreseen very high resolving power (R $\sim$ 50,000-100,000) will limit the observations to the brightest targets (AB $< 21-22$) penalising the $z > 6.5$ redshift regime  \citep{Dodorico2024}. Furthermore, ground based observations will always be heavily impacted by telluric absorption in the near-infrared. 

In this context, a space telescope with the characteristics of HWO would play a fundamental role in the measurement of the chemical abundances of galactic gaseous environments for a large sample of background quasars at the epoch of Reionization \citep{Fan2023,Belladitta2025,Martinez2026}, increasing the probability to find peculiar signatures like those of Pair Instability Super Novae \citep[PISN;][]{Ferrara2026}.
\\
\\
\textbf{Instrument requirements} 
For this science case, we require a wavelength coverage extending to 2.5  $\mu$m, in order to cover the stronger transitions due to MgII and FeII for the highest redshift quasars ($z\sim7.5-8.0$) and a resolving power $R \simeq 20,000-40,000$ to allow resolving metal absorption lines and detect the weaker ones which could trace very metal poor environments.

\subsubsection{The connection among black-hole accretion, feedback, and host-galaxy evolution across all epochs}
\label{sec:2.5.6}
By the 2040s, wide-field surveys from Rubin, Roman, Euclid, SKA, and {\it Athena} will have completed a nearly complete census of AGN up to $z>10$, while LISA and the Einstein Telescope will detect black-hole mergers across cosmic time \citep{Ivezic2019,EUCLID2025,ATHENA2025,AmaroSeoane2017,Maggiore2020}. ELT-class IFUs will spatially resolve feedback in nearby galaxies down to tens of parsec scales \citep[e.g.,][]{Nguyen2025}. On the theoretical side, GRMHD and radiation-hydrodynamic simulations are already linking accretion and feedback processes \citep[e.g.,][]{Guo2025}, while AI-driven emulators will enable direct inference of physical parameters from IFU datasets. HWO will bridge these multi-scale datasets, connecting sub-parsec accretion physics to the cosmological evolution of galaxies.

The next frontier is to establish a physically consistent connection between black-hole accretion, feedback, and host-galaxy evolution across all epochs \citep{Fabian2012,Heckman2014,Kormendy2013}. We identify three key science cases spanning the SMBH–galaxy cycle. In addition to spatially resolved spectroscopy, a multiplexed MOS capability is essential to obtain statistically significant samples across cosmic time.

{\bf SC1: The Black Hole Engine.} 
HWO will resolve the inner accretion and outflow regions of nearby AGN ($D \lesssim 100$ Mpc), probing $\sim$1–10 pc scales with $\sim$0.01\arcsec\ resolution and linking event-horizon-scale physics \citep{EHT2024} to galaxy-scale feedback. High-ionization UV lines (e.g. CIV, NV, SiIV), combined with optical and NIR diagnostics, will constrain gas density, ionization, and kinematics. Spectroscopy at $R \sim (1-2)\times10^4$ (FWHM $\sim$ 15--30~km~s$^{-1}$) will separate disk, wind, and jet components and measure outflow energetics. A sample of $\sim$50–100 AGN will provide the first statistical, spatially resolved view of AGN feeding and feedback on parsec scales.

{\bf SC2: SMBH-galaxy Co-evolution at Cosmic Dawn.} 
HWO will detect SMBHs and their host galaxies at $z \sim 6$–10, probing the interplay between accretion, star formation, and metal enrichment. Recent {\it JWST} discoveries of massive black holes at $z>7$ \citep[e.g.,][]{Maiolino2024,Li2025} highlight the need for a facility combining high sensitivity, high spatial resolution, and multiplexing. Rest-frame UV diagnostics (Ly$\alpha$, CIV, HeII, CIII]) will constrain ionization, metallicity, and gas density. IFU and MOS modes will enable both resolved studies and surveys of $\sim10^3$–$10^4$ sources. HWO will probe AGN vs stellar ionization, detect obscured black holes, and measure occupation fractions, with $\sim$50–100 pc resolution at $z\sim6$. It will also enable rapid follow-up of LISA events.

{\bf SC3: Metal-enriching due to Black Holes.} 
HWO will quantify how AGN-driven outflows redistribute metals and energy from galactic nuclei to the ISM and CGM across $z \sim 0$–4. Spatially resolved metallicity diagnostics (UV line ratios and optical indicators), combined with IFU kinematics, will constrain metallicity, ionization, and outflow dynamics, requiring moderate-to-high spectral resolution. Mapping metallicity gradients and outflow velocities will constrain metal loading factors and test simulation predictions, establishing the role of AGN in galaxy evolution. A sample of $\sim$200–500 galaxies will enable statistical constraints. Spectropolarimetry and time-domain observations will probe outflow geometry and variability.

\textbf{Instrument requirements.} The current 6–10 m HWO baseline (0.1–2.5\,\micron) satisfies most sensitivity goals but requires four key capabilities:  
(1) a UV–optical IFU with $R \approx 5{,}000$–$20{,}000$ and 0.01\arcsec\ sampling, crucial to resolve parsec-scale AGN feedback;  
(2) a multiplexed MOS mode for statistical surveys;  
(3) wavelength coverage up to 2.5\,\micron\ to access rest-frame UV diagnostics at high redshift;  
(4) polarimetric and time-domain capabilities to probe geometry and variability.

These requirements converge toward a JWST/NIRSpec-like UV–optical–NIR spectrograph combining MOS and IFU modes \citep{Jakobsen2022}. HWO will detect key UV emission lines down to $\sim10^{-19}$ erg s$^{-1}$ cm$^{-2}$ with $S/N \gtrsim 10$ in $\sim$10-hour exposures, enabling both resolved studies and large surveys of up to $\sim10^4$ galaxies and AGN, and providing a comprehensive multi-scale view of SMBH growth and galaxy evolution.

\subsubsection{Dual AGN}
\label{sec:2.5.7}
Dual AGN represent a key phase in the hierarchical growth of galaxies, tracing systems in which two SMBHs are simultaneously fueled during galaxy interactions while still separated on kiloparsec scales. At later stages, these systems evolve into gravitationally bound binary SMBHs on parsec or sub‑parsec scales; these are the direct progenitors of gravitational‑wave sources detectable by future facilities such as LISA \citep[e.g.,][]{AstroLISA23}, pulsar timing arrays (PTAs)\citep[e.g.,][]{epta2023}, and LGWA \citep[e.g.,][]{lgwa2025}. Observationally, dual AGN and their more compact post-galaxy-merger counterparts (``duets'') provide a unique framework to connect galaxy mergers with SMBH coalescence. In recent years, significant progress in the search for dual AGN has been achieved through multi‑wavelength approaches combining high-spatial-resolution imaging, astrometry, X-ray surveys, and optical spectroscopy. In particular,
Gaia \citep[e.g.,][]{Chen2022,Mannucci2022},
Euclid \citep{Ulivi2025},
and JWST \citep[e.g.,][]{Perna2025,Ubler2025}
are enabling the identification of close candidates up to the highest redshifts, while XMM-Newton and Chandra surveys probe the accretion properties of large statistical samples \citep[e.g.,][]{Hou2020,Derosa2023}, and optical spectroscopy characterizes their emission‑line properties and host galaxies
\citep[e.g.,][]{Mannucci2023,Scialpi2025,Tang2026}.
However, these studies are fundamentally limited by angular resolution: while kpc‑scale dual AGN can be resolved with current instruments, probing the crucial transition to sub‑kpc and parsec‑scale binaries remains extremely challenging. This limitation 
leads to biases against the identification of the most compact dual systems. As a result, the census of dual AGN at small separations is still severely incomplete below projected separations of $\sim$2 kpc, and the link between galaxy interactions, AGN triggering, and SMBH merging timescales remains poorly constrained.

HWO would enable the direct identification and characterization of dual nuclei at sub-kpc separations, even in obscured or morphologically disturbed systems where current facilities remain incomplete. This would provide the first systematic view of compact dual AGN at cosmic noon, bridging the gap between kpc-scale galaxy pairs and the onset of bound SMBH binaries.

\textbf{Instrument requirements.} This science case requires a UV--optical--NIR IFU spectrograph with angular resolution of $\simeq 0.01''$-- $0.03''$, corresponding to physical separations of $\sim 80$--$250$ pc at $z\sim1$--3, thus enabling the direct spatial resolution of dual AGN systems during the peak epoch of galaxy assembly and black-hole growth. A spectral resolution $R\sim5{,}000$--$10{,}000$ is required to disentangle the kinematics and ionization structure of both nuclei and of the surrounding gas, tracing inflows, outflows, merger-driven feedback, and the narrow emission lines of the host galaxies. In unobscured systems, broad-line spectroscopy will additionally enable single-epoch virial black-hole mass estimates based on lines such as H$\beta$, Mg{\small II} and C{\small IV} at increasing redshifts. 
Wavelength coverage extending to $2.5\,\mu$m is essential to access rest-frame UV and optical diagnostics at $z>1$--2, while a multiplexing MOS capability would maximize the identification and physical characterization of AGN and galaxies at comparable redshifts in overdense regions and protoclusters. Instrument concepts such as SHOS, currently proposed for HWO (see Sect.~\ref{sec:strum.SHOS}), would provide the high-angular resolution near-IR spectroscopic capabilities required to probe the transition from kpc-scale dual AGN to compact duets, including obscured nuclei and merger-driven gas dynamics.

\subsection{Cosmology}
\label{sec:_sci.cosmo}
%\textit{Text by: F. Annibali, D. Massari, M. Marconi, M. Moresco, D.\ Milakovi{\'c}}

\subsubsection{Distance ladders}
\label{sec:2.6.1}
Despite tremendous advances over the last decades, cosmology is currently at a crossroad. On the one side, we developed several independent techniques to probe the Universe that helped us shape our current
$\Lambda$CDM framework, but on the other side, the more precise the measurements become, the more clearly inconsistencies begin to emerge. A clear example is the determination of the Hubble constant $H_0$, commonly referred to as the Hubble tension \citep{Verde2019}.
This discrepancy is both a significant challenge and an extraordinary opportunity: it may reveal previously unrecognized systematics in our measurements, or it may be signalling that our current cosmological model is incomplete, pointing toward new physics. To move forward, it is essential to develop new ways to explore the Universe, because independent and innovative probes will allow us to cross-check systematics and significantly strengthen our constraints on its fundamental properties \citep[e.g.][]{Moresco2022}.
Although not primarily designed as an observatory for cosmological analysis, HWO also has significant potential in this field, because an observatory with broad wavelength coverage and precise spectroscopic capabilities can uniquely contribute across multiple fronts. 

One of the primary methods for determining the expansion history of the Universe is the cosmic distance ladder, in which several distance indicators are linked together to constrain the luminosity–distance relation up to $z \sim 1.5$.
In its standard form, parallaxes, Cepheids, and Type Ia supernovae are combined in a three-rung approach to infer cosmological parameters.
While powerful, this method would benefit greatly from reducing the number of intermediate steps, thereby limiting the propagation of systematics.
With HWO, it will be possible to develop an improved, two-rung distance ladder \citep{Anand2025}.
Thanks to its sensitivity and angular resolution, HWO will be able to measure Cepheid distances out to $\sim 100$ Mpc, enabling a direct determination of $H_0$ to the 1\% level without relying on SNe Ia. 
This would bypass several sources of systematic uncertainty affecting secondary distance indicators (e.g. SNIa) and provide a substantially more robust calibration of the local expansion rate.
The 1\% level of precision will be possible also thanks to the improved geometric distances of Galactic Cepheids that at the time of HWO operations will have been provided by the next Gaia data releases and by the improved knowledge of the debated metallicity dependence \citep[][and references therein]{Breuval25,Ripepi26} of Cepheid PL relations expected from e.g. 4MOST surveys in combination with Roman, JWST and Rubin-LSST data, as well as from ongoing theoretical efforts \citep[see e.g.][and references therein]{DeSomma24,Marconi24cep}.

HWO will also enable major progress on several other promising distance indicators, such as TRGB, J-AGB stars, RR Lyrae, and surface-brightness fluctuations (SBF) by providing alternative calibrations of the cosmic distance scale to several hundred megaparsecs. This will deliver independent, cross-validated calibrations of the distance scale, offering complementary anchors and reinforcing the robustness of cosmological constraints \citep{Anand2025}.

Significant advances will also be possible in probing the dark sector.
HWO high-resolution imaging of lensed arcs and Einstein rings will enable the detection of low-mass dark-matter subhalos through perturbations in the lensing potential, allowing the dark-matter halo mass function to be constrained down to $\sim 10^7,M_\odot$.
An independent route to constrain dark matter comes from counting low-mass satellites around Milky Way–like galaxies, since different dark matter models predict substantially different satellite abundances.
With HWO, it will be possible to detect the faintest satellites of nearby large galaxies by resolving their individual stars, enabling cosmological constraints from samples of 500–1000 ultra-faint satellites.
This represents a major step forward, beyond the reach of current facilities, because Hubble lacks the required sensitivity, and JWST does not provide the necessary optical-band resolution.

{\bf Instrument requirements:} The primary instrument for this science goal is an efficient multi-object spectrograph with spectral resolution $R>3000$ over a large wavelength range, possibly extending into the near IR to follow--up interemediate redhift targets. The optimal field of view is of  at least 2 arcmin by size and with the capability to follow-up a minimum of 100 objects.

\subsubsection{Astrometry to shed light on the nature of dark matter}
\label{sec:2.6.2}
Despite the success that $\Lambda$-cold dark matter ($\Lambda$CDM) cosmology has in describing many of the observed global properties of the Universe, this model is subject to some inconsistencies when considering the properties of dark matter haloes on small cosmological scales, such as dwarf galaxies.
One example is the so-called cusp-core problem, according to which the observed central density profile of dwarf galaxies is less steep than that predicted by CDM simulations \citep[e.g.,][]{Moore1994}. While several solutions have been proposed to explain the evolution of cusps into cores due to the interaction with baryons \citep[e.g.,][]{Navarro1996, Read2005}, it remains critical to directly measure the dark matter density profile in small, dark matter dominated stellar systems like the dwarf spheroidal galaxies satellites of the Milky Way.
One of the best ways to do this is to measure their stellar kinematics, but to avoid degeneracies that prevent an unambiguous determination of the dark matter density \citep{Binney1982} the simultaneous measurements of line-of-sight velocity and proper motions is required \citep{Walker2013}.
So far, the measurement of the 3D kinematics has been possible only for a few dwarf spheroidal galaxies, namely Sculptor \citep{Massari2018} and Draco \citep{Massari2020}. Despite the recent improvements on the proper motion uncertainties due to the availability of several {\it HST} epochs well separated in time \citep{Vitral2024, Vitral2026}, the current precision on the proper motion dispersion for these stellar systems ($\sim 2$ km/s) is not sufficient to solve the cusp-core problem at high significance. In fact, a precision of 1 km/s at the centre of these galaxies is the requirement to achieve such an objective \citep{Lazar2020}.

HWO has the potential to be groundbreaking in this sense. With the spatial resolution of a space-based 6-m diameter telescope, provided $i$) that its PSF will not be undersampled \citep[to avoid the so-called pixel-phase error, see][]{Anderson2000} and $ii$) an HST-like astrometric stability, HWO will be able to achieve a precision in the proper motion measurements of $1-2~ \mu$as/yr over a baseline of 10 years. With such an outstanding performance, HWO will enable measuring the internal velocity dispersion of dwarf galaxies out to a distance of 1 Mpc, therefore including the entire system of M31 dwarf galaxy satellites.
This means not only that the dark matter density profiles will be measured with high precision for nearby dwarf spheroidals, but that this kind of information will be derived for an entire population of dark matter dominated systems. The combination of these features will ultimately shed light on the nature of dark matter.

{\bf Instrument requirements:} Diffraction limit $<10$ mas, which is achieved at $\sim300$ nm with a telescope diameter of 6.5 meters. Imaging camera with a pixel size $<5$ mas, in order to avoid PSF undersampling at least for $\lambda>300$ nm. Astrometric stability comparable to HST. Field of view of a few arcmin$^{2}$, to probe the core of nearby dwarf galaxies and at least the half-light radius of those around M31. With these ingredients, proper motions will be measured with $\mu$as precision over a temporal baseline of 10 years for a sufficiently large sample of stars in each target. \\

\subsubsection{Fundamental constants}
\label{sec:2.6.3}
The values of fundamental physical constants determine everything from the properties of elementary particles, atoms, and molecules to the conditions in stellar interiors, nucleosynthesis pathways, and the growth of structure in the Universe. Many extensions of $\Lambda$CDM introduce an additional degree of freedom which elevates these constants into a dynamical field that can or must evolve in time or ramble in space \citep[see the reviews by][]{Uzan2025LRR....28....6U, Martins2017}. Predictions on which constants are affected, together with mechanism of their variation (temporal, spatial, in the presence of strong gravitational fields) depends on the specifics of each theory. 
Interestingly, in a recent comparison of extensions of $\Lambda$CDM aimed at resolving the Hubble tension, the model with a varying electron mass at recombination took the gold medal, providing the largest relief while introducing only a single new parameter \citep{Schoneberg2022PhR...984....1S}. 

In scalar-tensor theories of gravity and Brans-Dicke-type frameworks, the scalar field that modifies the gravitational sector couples universally to all mass scales, including the QCD confinement scale that sets the proton mass. In this case, the modified gravitational dynamics introduces variations in the proton-to-electron mass ratio, $\mu$ \citep{Uzan2025LRR....28....6U, Mohamadnejad2019MPLA...3450277M}. In string and M-theory, all coupling constants are determined by the vacuum expectation values of moduli fields controlling the geometry of compactified extra dimensions. Cosmological evolution of these moduli simultaneously shifts $\Lambda_\mathrm{QCD}$, the Yukawa couplings, and hence $\mu$ \citep{Damour2002PhRvD..66d6007D, Calmet2002EPJC...24..639C}. In this scenario, correlated variations in the fine-structure constant ($\alpha$) are also expected, albeit variations in $\mu$ are expected to be much stronger \citep{Calmet2002EPJC...24..639C, Dine2003PhRvD..67a5009D}. Runaway dilaton models produce a similar hierarchy, with $\Delta\mu/\mu$ growing logarithmically with redshift through the matter-dominated era \citep{Damour2002PhRvL..89h1601D, Martins2015PhLB..743..377M}. Chameleon and symmetron screening models \citep{Khoury2004PhRvD..69d4026K, Olive2008} add a further dimension: they predict that constants depends not only on cosmic time but on the local gravitational potential and matter density, meaning that measurements in qualitatively different environments: the intergalactic medium, Galactic molecular clouds, and the surfaces of white dwarfs probe genuinely distinct sectors of the theory parameter space.

Vibro-rotational transitions of molecules such as H$_2$, HD, CH, H$_2$O, CH$_3$OH have been used to constrain $\Delta\mu/\mu$ in astronomical spectra \citep{Thompson1975ApL....16....3T, Rahmani2013, Dapra2015MNRAS.454..489D, Dapra2017MNRAS.465.4057D, Ubachs2018, Muller2026A&A...706A.365M}. The relatively larger number of observed transitions of H$_2$ and its higher abundance make it one of the best molecules to measure $\Delta\mu/\mu$ from astronomical spectra. Current constraints, however, are mostly limited to $z>2$, where Lyman and Werner transitions are redshifted into the optical wavelength range, but are more difficult to observe due to the faintness of background sources. Quasar absorption systems yield $\Delta\mu/\mu < 1 \times 10^{-5}$ at redshifts $z = 2-3.5$ \citep{Rahmani2013, Dapra2015MNRAS.454..489D, Dapra2017MNRAS.465.4057D}. This redshift range lies within the matter-dominated epoch observed along a handful of sightlines with ThAr-calibrated spectrographs whose systematic wavelength distortions are comparable to the signal being sought, and are hence the constraints are not discriminatory. 

Measurements of $\Delta\alpha/\alpha$ with VLT/ESPRESSO reach statistical uncertainties of $\sigma_{\mathrm{stat}}(\Delta\alpha/\alpha) \sim 1\times10^{-6}$ \citep{Murphy2022HE0515}, albeit astrophysical uncertainties, due to the isotopic splittings of Mg and unknown isotopic abundances at high-redshift, are larger at $\sigma_{\mathrm{sys}}(\Delta\alpha/\alpha) \sim 5\times10^{-6}$ \citep{Webb2025MNRAS.539L...1W}. In the future, Mg transitions (used in almost all previous analyses) will be replaced by more suitable transitions; such as Al\,{\sc ii} $\lambda 1670$, Al\,{\sc iii} $\lambda\lambda\, 1854, 1862$, (Al has one stable isotope) and Ni\,{\sc ii} $\lambda\lambda\lambda\, 1317, 1741, 1752$ (Ni is less affected by isotopic splittings). To obtain the best constraints on $\Delta\alpha/\alpha$, these transitions must be combined with, e.g.\ Fe\,{\sc ii} $\lambda\lambda\lambda\, 2344, 2383, 2600$, which constrain the kinematic structure of the absorbing gas better than Al or Ni. The wavelength coverage of existing high-resolution terrestrial spectrographs restricts the combination of these transitions to absorption systems lying in a narrow redshift range $z\approx2.3\pm0.2$. An UV/optical spectrograph covering 1000\,Å to 1.8{\micron} would extend this range to $0<z<5$, covering 90\% of the universe's history.   

{\bf Instrument requirements:} A high-resolution spectrograph with a broad simultaneous wavelength coverage, starting from approximately 912\,Å, allowing access to the full Lyman and Werner bands starting from $z=0$ (useful for measurements in white dwarf photospheres and Galactic molecular clouds). Because molecular clouds are cold, a resolving power of $R \gtrsim 100{,}000$ is needed to fully resolve narrow velocity components and separate blended transitions whose sign-alternating sensitivity coefficients underpin the measurement. Wavelength calibration should have no distortions larger than $\lesssim 10$\,m\,s$^{-1}$ across the full bandpass. Accurate centroiding requires that the shape of the line-spread function is characterised at scales $\sim$1/10th of a pixel. 

\subsubsection{Redshift drift}
\label{sec:2.6.4}
A small temporal change -- a drift -- of the redshift of cosmological objects is a general consequence of an expanding universe \citep{Sandage1962ApJ...136..319S, McVittie1962ApJ...136..334M}. This makes it a powerful cosmological probe, and the only one that can directly measure the acceleration of the universal expansion without making any assumptions on the physics of supernovae explosions or about the correct theory of gravity, in contrast with more traditional probes like standard candles, standard rulers, and the CMB. 

In general theory of relativity, the drift is given by the difference between the expansion rate of the universe at the times of light emission and light observation, $\dot{z} =  \frac{\mathrm{d}z}{\mathrm{d}t}|_{t_\mathrm{obs}} = H(z_\mathrm{em}) - (1+z_{\mathrm{obs}})H(z_\mathrm{obs})$. In $\Lambda$CDM, the expected amplitude of $\dot{z}$ is of the order of $\sim10^{-11}$ per year, with a small positive drift in the redshift range $0<z<2$, a peak at $z\sim1$ and a turnover to larger negative drifts beyond $z>2$ \citep[c.f.\ figure 2 of ][]{Liske2008MNRAS.386.1192L}. In $\Lambda$CDM, the positive drift is a kinematic signature of an accelerated expansion, making $0<z<2$ the crucial redshift range to probe for dark energy studies. Alternative cosmological models make quantitatively different predictions. For example, the scale invariant cosmology of \cite{Maeder2017ApJ...834..194M} predicts a larger, positive $\dot{z}$ at all redshifts. 

The Lyman-$\alpha$ forest is the most unbiased tracer of $\dot{z}$ \citep{Loeb1998ApJ...499L.111L, Cooke2020MNRAS.492.2044C}. Forest absorbers reside in the diffuse, underdense intergalactic medium, where peculiar accelerations are more than an order of magnitude below the expected cosmological signal \citep{Cooke2020MNRAS.492.2044C}. By contrast, H\,{\sc i} 21-cm emission from galaxies and absorption from Damped Lyman-$\alpha$ (DLA) systems trace overdense, gravitationally bound environments in which gas dynamics produces spurious drifts that can exceed the cosmological signal \citep{Cooke2020MNRAS.492.2044C}. ELT/ANDES will exploit the forest at $2 \lesssim z \lesssim 5$, while SKA and FAST will probe H\,{\sc i} 21-cm at $0 \lesssim z \lesssim 2$ and $z \lesssim 0.35$, respectively \citep{Liske2008MNRAS.386.1192L, Marconi2024SPIE13096E..13M, Kang2024RAA....24g5002K, Kang2025ApJ...982..177K}. The atmospheric cut-off at $\sim3100$\,Å prevents any ground-based facility from observing the forest below $z \sim 1.5$, leaving the dark-energy dominated epoch entirely uncovered by the optimal tracer. A space-based UV–optical spectrograph is the only facility that can apply the Lyman-$\alpha$ forest method at $0 < z \lesssim 1.5$, and would simultaneously overlap with ANDES at $z > 2$ to enable important cross-checks.

{\bf Instrument requirements:} Wavelength coverage between 1200\,Å and 8500\,Å would provide access to the Lyman-$\alpha$ forest in the redshift range $0 \leq z \leq 6$. Extension to redder wavelengths enables the identification of intervening metal absorption systems that blend with the forest, contaminating the cosmological signal. While a resolving power of $R \sim 20{,}000$ suffices to resolve the forest lines (typical Doppler b-parameter is $\sim20$\,km\,s$^{-1}$), accurate modelling of contaminating metal lines ($b\sim1$ to 5\,km\,s$^{-1}$) requires $R \gtrsim 100{,}000$. Detecting $\dot{z} \sim 10^{-11}$\,yr$^{-1}$ over a $\sim$10\,yr baseline requires measuring velocity shifts of $\sim$1\,cm\,s$^{-1}$, demanding long-term wavelength zero-point stability at the same level, achievable with a laser frequency comb or a similarly stable calibration source.

\subsubsection{Reionization of He II}
\label{sec:2.6.5}
The epoch of helium reionization represents the ultimate major phase transition of cosmic baryons, fundamentally shaping the thermal and ionization state of the intergalactic medium (IGM). While hydrogen reionization is largely complete by $z \simeq 5.3$ \citep[e.g.][]{Bosman22}, the final reionization of He\,{\sc ii} is delayed until $z \sim 3$. At this epoch, the cosmic quasar population becomes sufficiently numerous to provide the necessary budget of hard ($E = h_{\rm P} \nu > 54.4$ eV) UV photons \citep[e.g.][]{Compostella13,Compostella14}.

Substantial uncertainties remain regarding the precise timeline and spatial morphology of He\,{\sc ii} reionization. The detailed properties of the IGM during this transition depend on several poorly constrained parameters of high-redshift AGN (such as their duty cycles, spectral energy distributions, opening angles, and ionizing photon escape fractions), as well as the distribution and geometry of self-shielding absorbers (including Lyman limit systems and higher column density absorption lines). Because a given patch of the IGM typically requires illumination from successive generations of quasars to become fully ionized, the process imprints a highly complex thermal and ionization structure onto the gas \citep[e.g.][]{Compostella13,Compostella14,Garaldi19}.

Unraveling the details of He\,{\sc ii} reionization has gained renewed urgency given the abundance of high-$z$ AGNs recently uncovered by JWST \citep[e.g.][]{Maiolino24a}, sparking intense debate over their potential role in driving H\,{\sc i} reionization \citep[e.g.][]{Madau24}. A dominant contribution from early quasars/AGNs could shift the timeline of He\,{\sc ii} reionization too early into cosmic history, introducing a tension with current empirical constraints.

Direct insights into this epoch are achieved through transmission spectroscopy of the intergalactic He\,{\sc ii} Lyman-$\alpha$ forest ($\lambda_{\rm rest} = 303.78$ \AA) along the lines of sight to far-UV (FUV)-bright quasars at $z > 2$. However, this observational technique is severely limited by the necessity of space-based facilities and the scarcity of suitable background quasars whose intrinsic FUV flux remains unabsorbed by intervening, optically thick H\,{\sc i} systems.

The current state of the art on this topic \citep{Worseck19} relies on a heterogeneous sample of 25 quasar spectra gathered with HST/COS, characterized by widely varying spectral resolutions and signal-to-noise ratios (S/N). Spanning the redshift interval $2.3 \le z \le 3.85$, these sightlines display a pronounced dispersion in measured optical depths at $z \gtrsim 2.6$, providing robust evidence for a highly inhomogeneous and patchy reionization topology.

While further FUV-selected quasar candidates have been identified \citep[e.g.][]{Worseck11,Syphers12} --and are critically needed to expand sample sizes and minimize cosmic variance-- they are generally too faint for high-quality follow-up with HST/COS. Breaking through this observational bottleneck and driving significant progress in the field will ultimately necessitate a next-generation space telescope with a larger aperture than HST, equipped with a high-resolution, UV-sensitive spectrograph.

{\bf Instrument requirements:} The wavelength range $900-2000$ \AA\ would allow to study the  He\,{\sc ii} Lyman-$\alpha$ forest in the redshift range $z\sim2-4$, critical to investigate the final phases of percolation of the He~III bubbles. An extension to the optical wavelength range would allow to measure the H\,{\sc i} Lyman-$\alpha$ forest corresponding to the He\,{\sc ii} one: the ratio of the two optical depths and its evolution with redshift are a probe of the contribution of quasars to the H\,{\sc i} reionization \citep{Garaldi19}. The minimum resolving power required to resolve the He\,{\sc ii} Lyman-$\alpha$ forest is $R\sim20{,}000$ but $R\sim 40{,}000-50{,}000$ could be desirable.

%%%%%%%%%%%%%%%%%%%%%%%%%%%%%%%%%%%%%%%%%%%%%%%%%%%%%%%%%%%%%%%%%%
\subsection{The transient sky}
\label{sec:sci.transient}
%\textit{Text by: L. Izzo, S. Piranomonte, F. D'Ammando}

\subsubsection{Supernova science with HWO}
\label{sec:2.7.1}
The HWO facility offers a unique opportunity for breakthrough discoveries in supernova (SN) science. Despite most SN science being performed through target-of-opportunity (ToO) observations, there is still great potential to perform SN science with HWO and its instrumentation without considering time-critical observations, and disrupting the planned observations schedule. This section explores the potential of HWO for SN research, focusing on the unprecedented sensitivity and spatial resolution provided by the High-Resolution Imager (HRI).

HRI will allow for the direct detection of SN progenitor stars before they explode. This observational strategy has been successfully applied using Hubble Space Telescope imaging with the Wide Field Camera and the Advanced Camera for Survey on data obtained for very nearby galaxies. For example, the direct progenitors of SNe like SN 2003gd \citep{Smartt2004} and SN 2005gl \citep{Gal-Yam2009} 
were reported after their detection. A similar strategy can be employed using HWO to identify cases of direct collapse, where massive progenitor stars, previously observed in pre-imaging, collapse directly into a black hole without producing a visible explosion typical of core-collapse SNe. This would provide valuable insights into the problem of high-mass progenitors of Type II SNe \citep{Smartt2009}, and the direct formation of stellar black holes. Existing archival images of nearby galaxies from space telescopes such as the HST and JWST can be used as reference images, with HWO obtaining new observations years later. This approach led to the discovery of the first failed SN events in NGC 6946 \citep{Adams2017}. 
Collaborative monitoring efforts with JWST and HWO can help eliminate alternative scenarios for specific events, further refining our understanding of SNe.

Another area of interest for HWO is the implementation of dedicated deep survey programs for cosmological fields, following the precedent set by JADES and NEXUS at the JWST \citep{Eisenstein2026,Shen2024} and upcoming space-based deep surveys from Euclid and Roman. By observing a single field with a specific temporal cadence, this approach would enable systematic searches for distant stellar explosions ($z > 2$) and increase the likelihood of detecting unique events such as super-luminous supernovae (SLSNe) and Population III analog explosions. These explosions involve very massive stars formed in extremely low metallicity environments \citep{Moriya2023}. 
Such a program would also facilitate investigations into the rate of core-collapse supernovae at greater distances, offering insights into the behavior of the initial mass function at various redshifts. Furthermore, it would allow for the construction of a sample of Type-Ia supernovae across a broad redshift interval for cosmological studies. 

{\bf Instrument requirements:} This specific program would require the use of near-IR filters, particularly those with wavelengths greater than 1.5$\mu$m, in conjunction with the HRI instrument. 

However, the opportunity to utilize the UV MOS/IFU detector, combined with a relatively large field of view, will provide an unprecedented perspective on the immediate environment of transient host galaxies. This will be particularly valuable for very nearby targets in a wavelength range that has not yet been extensively studied. The spatial resolution of HWO will enable the examination of spatially-resolved physical properties, such as metallicity, through absorption spectroscopy. Additionally, it will facilitate accurate estimates of intrinsic extinction measurements and provide a detailed analysis of the young stellar population components. For regions where core-collapse events have previously been detected, this information will be crucial to obtain precise measurements of their progenitor ages.

\subsubsection{HWO and Multi-Messenger Astronomy}
\label{sec:2.7.2}
The era of multi-messenger astronomy, started with the joint gravitational wave (GW) and electromagnetic (EM) detection of GW170817 \citep{Abbott2017}, will reach full development in the 2030s-2040s with next-generation GW detectors (e.g. Einstein Telescope, Cosmic Explorer, LISA, LGWA) capable of detecting several thousands to millions of sources per year spanning the entire mass spectrum: from stellar-mass binary black holes (BBH) and binary neutron stars (BNS) at cosmological distances ($z \sim 10$ and beyond), to supermassive black hole mergers ($10^3$--$10^7$ M$_{\odot}$), intermediate-mass black hole binaries (IMBHs), extreme and intermediate mass-ratio inspirals (EMRIs and IMRIs), galactic compact binaries, and potentially core-collapse supernovae and stochastic backgrounds from the early Universe. For a review see \citet{Bailes2021}.
%binary neutron star (BNS) mergers up to cosmological distances and tens of thousands of compact binary coalescences per year. 
In this context, HWO, with its unprecedented UV/optical/NIR sensitivity and capabilities, will play a unique and transformational role in the electromagnetic follow-up and characterization of multi-messenger transients.

\paragraph{Kilonovae in the HWO Era}

Kilonovae are powered by the decay of heavy radioactive species produced by rapid neutron capture (r-process) and ejected during the merger \citep{LiPaczynski1998, Metzger2010}.
The bulk of KN emission is expected to peak within a few days in the UV/optical band (blue KN, arising from lanthanide-free material of the polar ejecta) and in the NIR band on a timescale of a week (red KN, from lanthanide-rich equatorial ejecta; for a review see \citep{Metzger2019}). These sources are of great interest because 1) they are major sites of heavy-element production through efficient r-process nucleosynthesis, and 2) while emission from Short Gamma Ray Bursts (SGRBs) is confined within narrow jets \citep{Berger2014, DAvanzo2015}, KNe isotropic emission makes them detectable from any orientation.

This scenario was confirmed by the first detection of a GW event from a BNS merger (GW170817) associated with the weak SGRB 170817A \citep{Goldstein2017, Savchenko2017} and the bright KN AT2017gfo \citep{Pian2017}.
The emergence of delayed X-ray and radio emission was suggestive of off-axis afterglow emission \citep{Hallinan2017, Troja2017}.
Beyond providing the "smoking gun" of SGRB progenitors, it showed that the GRB geometry differs from a simple top-hat jet \citep{DAvanzo2018, Ghirlanda2019}.

KNe have also been identified in several SGRB light curves \citep{Rastinejad2025}, suggesting that KNe are ubiquitous and can probe NS mergers beyond the GW horizon.
Additional events will test new models \citep{Gillanders2022, Bulla2023}, helping disentangle micro-physics, geometry, and energetics.
By the 2040s, third-generation GW detectors will detect $\sim 10^5$ BNS mergers per year, creating an unprecedented demand for rapid EM follow-up capabilities.

HWO's far-UV coverage (down to $\sim$0.2–0.3 $\mu m$) will provide unique capabilities to study the early blue phase of kilonovae. The early blue emission, which fades within 1–2 days, may originate from shock-heated interface material or neutron precursor decay, but its physical origin remains debated.

HWO’s UV spectroscopic capabilities will enable early-phase diagnostics, constraining the velocity, composition, and lanthanide-free material of the ejecta. Multi-epoch UV spectroscopy will track rapid spectral evolution as the photosphere recedes and heavier elements emerge, testing models of stratified ejecta.
Additionally, UV spectroscopy of host environments will constrain metallicity, star formation history, and progenitor populations.
Combined with the heritage provided by ground-based CUBES UV spectroscopy and from space by JWST observations, HWO will enable unprecedented multi-wavelength studies of KN emission and r-process nucleosynthesis.

\paragraph{Black Hole Mergers in AGN Disks}

Recent theoretical models predict that BBH mergers embedded in AGN accretion disks can produce prompt, luminous UV flares \citep{McKernan2019, Antoni2019, Tagawa2024, Wang2022}.
These UV flares may outshine the host AGN and provide unique EM signatures of BBH mergers.

HWO’s UV sensitivity, resolution, and rapid response make it ideal for detecting these events. Its angular resolution will localize flares within AGN environments and distinguish them from AGN variability.
Multi-epoch UV spectroscopy will distinguish BBH-disk signatures from TDEs and AGN flares.

\paragraph{Investigating neutrino-emitting AGN with HWO}

High-energy neutrinos, in conjunction with gravitational waves, have indeed initiated a new era in astronomy, shifting from single-messenger observations to multi-messenger investigations of the cosmos. Neutrinos will provide a direct window into the Universe's most violent phenomena, enabling researchers to probe regions previously hidden from traditional electromagnetic astronomy. Among the few classes identified as promising high-energy neutrino sources are AGN, including blazars and Seyfert galaxies \citep[see e.g.,][for a review]{dammando26}. In particular, for two sources a statistically significant correlation between high-energy neutrinos and known extragalactic sources has been established so far: the blazar TXS\,0506$+$056 \citep{IceCube2018} and the Seyfert II galaxy NGC\,1068 \citep{Abbasi2022}. 

Recent findings suggest that neutrino-emitting AGN, such as the Seyfert galaxy NGC\,1068, are often "gamma-ray obscured". These sources have massive amounts of surrounding gas and dust that block high-energy gamma-rays but allow neutrinos and lower-energy electromagnetic radiation (X-ray, UV, and optical) to escape \citep[e.g.][]{inoue2020,murase2020}. HWO’s sensitivity will allow researchers to study these dense environments directly. Theoretical models suggest that neutrinos are produced in the immediate vicinity of the SMBH, particularly in the corona or the base of jets. The unprecedented precision of the HWO will be instrumental in imaging the complex structures surrounding SMBHs, allowing us to distinguish between coronal emission, jet emission, and interactions within the BLR. Specifically, HWO can characterize a neutrino-emitting candidate AGN by observing the accretion disk, jet-driven outflows, and host galaxy properties. Its superior angular resolution will help pinpoint exact particle acceleration locations, often situated within the inner few parsecs of the central SMBH, and characterize the AGN's structure, confirming if the properties of the core (such as dense winds or jets) are consistent with neutrino production \citep[see e.g.,][]{murase2023}. In addition, HWO will study AGN feedback by analyzing interstellar gas and star-forming regions with unprecedented precision. Correlating neutrino flux with AGN feedback activity observed by HWO will help us to understand how much energy is effectively transferred from the SMBH to the galaxy. In particular, HWO can study AGN feedback by using its unprecedented spatial resolution and advanced UV spectroscopy to map how SMBH eject energy into their host galaxies. By observing interstellar gas and star-forming regions at the milliarcsecond scale, HWO will trace the velocity, temperature, and composition of cosmic gas. This allows scientists to determine exactly how black holes suppress or trigger the birth of new stars. In this context, high-energy neutrinos act as direct tracers of the physical processes that drive AGN feedback, such as cosmic ray interactions within accretion disks, turbulent coronae, and massive outflows.
Finally, considering that HWO is designed to observe galaxy evolution over cosmic timescales, studying how neutrino production in AGN and their physical properties measured by HWO change throughout the history of the Universe can provide important information about the link between the growth of SMBH and their host galaxies.

In conclusion, by providing high-angular-resolution imaging and sensitive spectroscopy of the engines of AGN, HWO will help transform current correlation studies into a detailed, physical understanding of how AGN accelerates particles and produces high-energy neutrinos.

\paragraph{Synergy with future EM, GW and neutrinos facilities}

HWO's multi-messenger capabilities will be maximized through coordination with a diverse array of facilities operating across the electromagnetic spectrum and beyond. Third-generation ground-based GW detectors (e.g Einstein Telescope, Cosmic Explorer) and space-based detectors (e.g LISA, LGWA) will provide GW triggers and precise sky localizations, enabling targeted follow-up observations. At the same time, neutrino facilities such as IceCube-Gen2, KM3NeT, and Baikal-GVD will be fully operational, providing real-time alert for EM follow-up of high-energy neutrino events. These three major detectors, covering both hemispheres, will create a "global neutrino network" capable of ensuring constant monitoring  of the full sky. The Rubin Observatory's Legacy Survey of Space and Time (LSST) will discover approximately $10^5$ transients per day, which require rapid UV follow-up for classification and detailed characterization. High-energy facilities such as 
%THESEUS (Transient High Energy Sky and Early Universe Surveyor) and 
NewAthena (Advanced Telescope for High-ENergy Astrophysics) and CTAO (Cherenkov Telescope Array Observatory) will provide crucial X-ray and $\gamma$-ray counterpart detection, while ground-based facilities including the Extremely Large Telescope (ELT), Very Large Telescope (VLT) equipped with CUBES (Cassegrain U-Band Efficient Spectrograph), and Wide-field Spectroscopic Telescope (WST), will offer complementary UV, optical and near-infrared spectroscopy. This coordinated approach across facilities will enable comprehensive characterization of multi-messenger transients from radio to gamma-ray wavelengths.
\\

In this broader time-domain ecosystem, the recently announced Eric and Wendy 
Schmidt Observatory System deserves explicit mention as a highly complementary 
facility suite. The Argus Array \citep{Law2022} will provide wide-field, 
real-time optical survey data with an open-data philosophy, making it an 
excellent trigger source for rapidly evolving transients. The Large Fiber Array 
Spectroscopic Telescope \citep{Angel2022} will enable high-resolution 
optical spectroscopy for rapid follow-up. The Lazuli Space Observatory 
\citep{Roy2026}, optimized for optical and near-infrared follow-up on timescales 
of hours, will support the characterization of gravitational-wave counterparts. 
In parallel, the Deep Synoptic Array \citep{Law2024} will provide radio 
localizations of radio transients, enabling high-energy and multiwavelength 
follow-up. Together, these facilities strengthen the discovery and 
trigger-generation layer, while HWO will uniquely contribute to rapid UV/optical 
characterization and deeper physical interpretation of selected counterparts.

\paragraph{Operational Considerations}

To maximize HWO's impact on multi-messenger astronomy, several operational aspects require careful consideration. The observatory could implement robust protocols with rapid response capability for GW and neutrino triggers, ideally with decision timescales of hours to days. Multi-object spectroscopy capabilities will be essential to observe multiple candidates within a GW  and neutrino localization region simultaneously, significantly improving the efficiency of follow-up campaigns. In addition, continuous monitoring capabilities will be crucial for events with poorly constrained trigger times or for studying the evolution of long-duration transients over extended periods.

\paragraph{Scientific Impact}

HWO will uniquely address fundamental questions in multi-messenger astronomy that remain unanswered despite decades of theoretical and observational efforts. Understanding the origin of the blue kilonova component and what it reveals about merger dynamics remains a key challenge that HWO's capabilities are uniquely positioned to address. The detection and characterization of electromagnetic counterparts from BBH mergers in AGN disks would open an entirely new window on stellar-mass black hole physics in dense environments. Finally, measuring the rate and nature of r-process element production across cosmic time will fundamentally improve our understanding of how the heaviest elements in the universe are synthesized and distributed.

The combination of HWO's UV coverage, high sensitivity, multi-object spectroscopy, and space-based stability free from atmospheric absorption and seeing positions it as the definitive platform for multi-messenger follow-up in the 2040s. HWO will complement and extend the discoveries of Einstein Telescope, Cosmic Explorer, LGWA and LISA, enabling a comprehensive understanding of the transient universe across all cosmic messengers. 
In addition, studying neutrino-emitting AGN with HWO, in conjunction with neutrino facilities such as IceCubeGen2, KM3NeT, and Baikal-GVD, presents a significant opportunity for the multi-messenger study of the Universe. This research will enhance our understanding of cosmic ray acceleration, SMBH physics, and galactic evolution, while providing new insights into the extreme environments of our Universe.

{\bf Instrument requirements:} To satisfy the scientific requirements for studying neutrino-emitting AGNs, HWO must combine extreme spatial resolution with high-sensitivity spectroscopy across the ultraviolet, optical, and near-infrared wavelengths. In particular, 
Specifically, these studies require diffraction-limited resolution in the near-UV and optical bands to isolate the inner parsecs around SMBH. A primary mirror diameter of at least 6 meters is indispensable for achieving the angular resolution power necessary to separate coronal emission from jet structures. Furthermore, wavefront stability is critical to prevent artifacts from masking faint emissions near the nucleus, and far-UV extension is required to observe high-temperature gas in the accretion disk and corona. The primary instruments identified for these tasks include a UV IFU, a UV Multi-Object Spectrograph, and a High-Resolution UV/Vis Spectropolarimeter.

Regarding GW and kilonova astrophysics, the mission requires rapid response times and simultaneous broad spectral coverage from 100 to 2500 nm to capture rapid thermal and elemental evolution. Spectrographs must support low-resolution modes to accurately map the broad spectral continuum and determine global chemical abundances. Additionally, HWO’s scheduling and ground control systems must be capable of interrupting routine observations within 12 to 24 hours of receiving a trigger from next-generation detectors or neutrino observatories.

\newpage

\section{Future facilities and preparatory work}\label{sec:future}
\label{sec:futurefacilities}
\subsection{PLATO}
\label{sec:fut.PLATO}
%\textit{Text by:V. Nascimbeni, G. Piotto}

\begin{wrapfigure}{l}{0.55\textwidth}
%\begin{figure}
    \centering
    \includegraphics[width=0.75\linewidth]{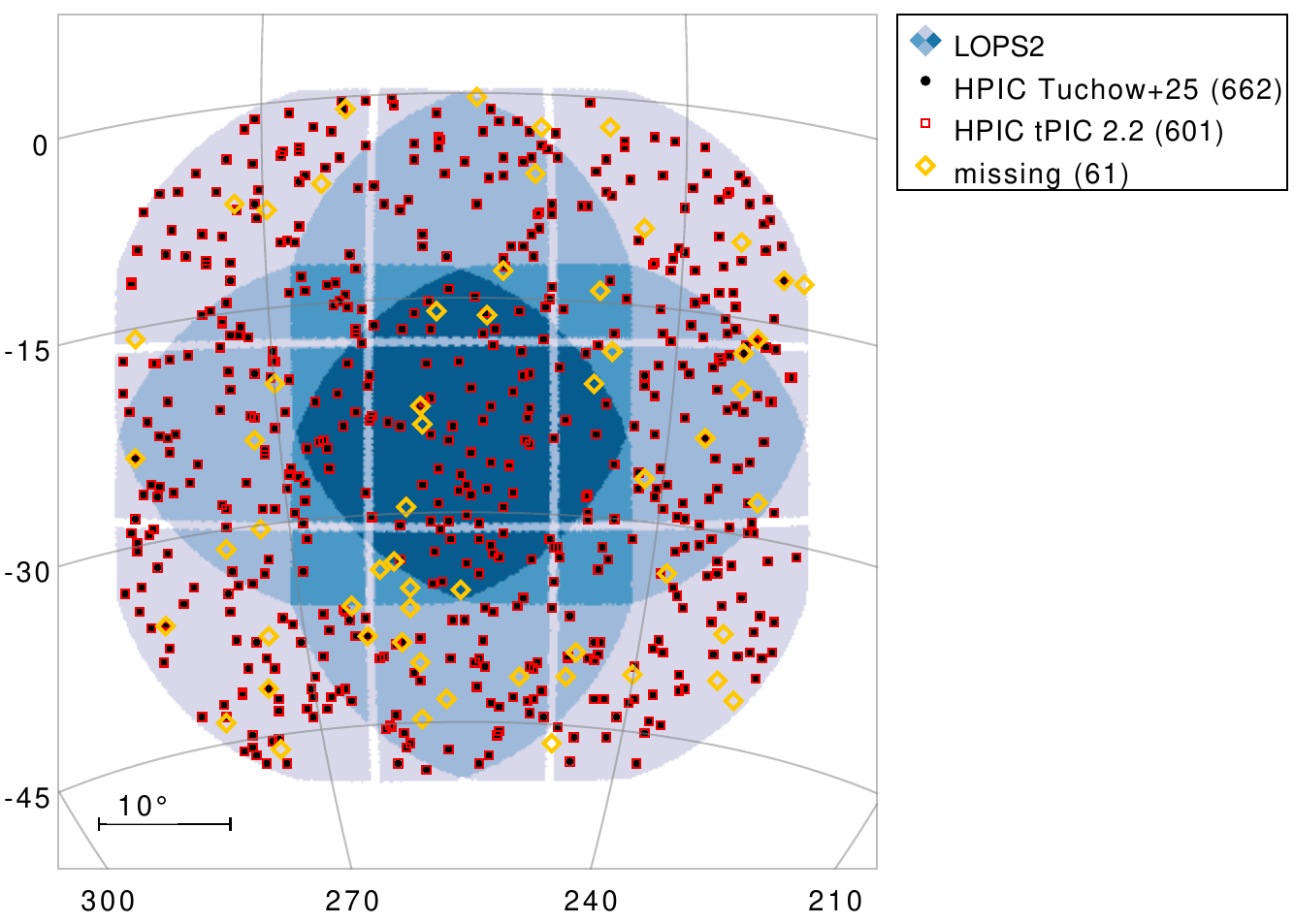}
    \caption{PLATO LOPS2 field of view (blue shades according to the number of cameras pointing in that direction). Black dots are HPIC targets over the LOPS2 area; red squares mark the HPIC entries that are in the tPIC; yellow diamonds are HPIC targets missing in the tPIC.}
    \label{fig:lops2}
\end{wrapfigure}

PLATO (PLAnetary Transits and Oscillations of stars) is ESA's M3 mission, ready for launch at the beginning of 2027, designed to detect and characterize extrasolar planets and to perform asteroseismic monitoring of a large number of stars. Its main program is the search for and determination of the bulk properties (radius, mass, mean density) of planets in a wide range of systems, including terrestrial planets in the habitable zone (HZ) of solar-like stars: with the complement of ground-based radial-velocity follow-up, these planets will be characterized for radius, mass and age with high accuracy (5\%, 10\%, 10\% respectively for an Earth-Sun analog). PLATO will detect and confirm small planets down to $R_\oplus$ around bright stars ($V<11$), and will put strong constraints on $\eta_\oplus$, important for the development of the final HWO design and observation strategy.

This makes PLATO uniquely placed in the HWO preparatory landscape. As anticipated in Section~\ref{sec:demographics}, the prime targets for HWO must first be found and characterized, and PLATO will be the only mission before HWO launch able to detect small planets (down to $R_\oplus$) in the habitable zone of bright nearby stars, including Earth twins, some of which may be interesting targets for HWO direct imaging. Furthermore, the asteroseismic characterization of nearby bright solar-type stars will deliver the accurate stellar parameters, including the ages, that are essential to identify the most promising HWO targets and to place the discovered systems in an evolutionary timeline.

\begin{wrapfigure}{r}{0.55\textwidth}
%\begin{figure}
    \centering
    \includegraphics[width=0.75\linewidth]{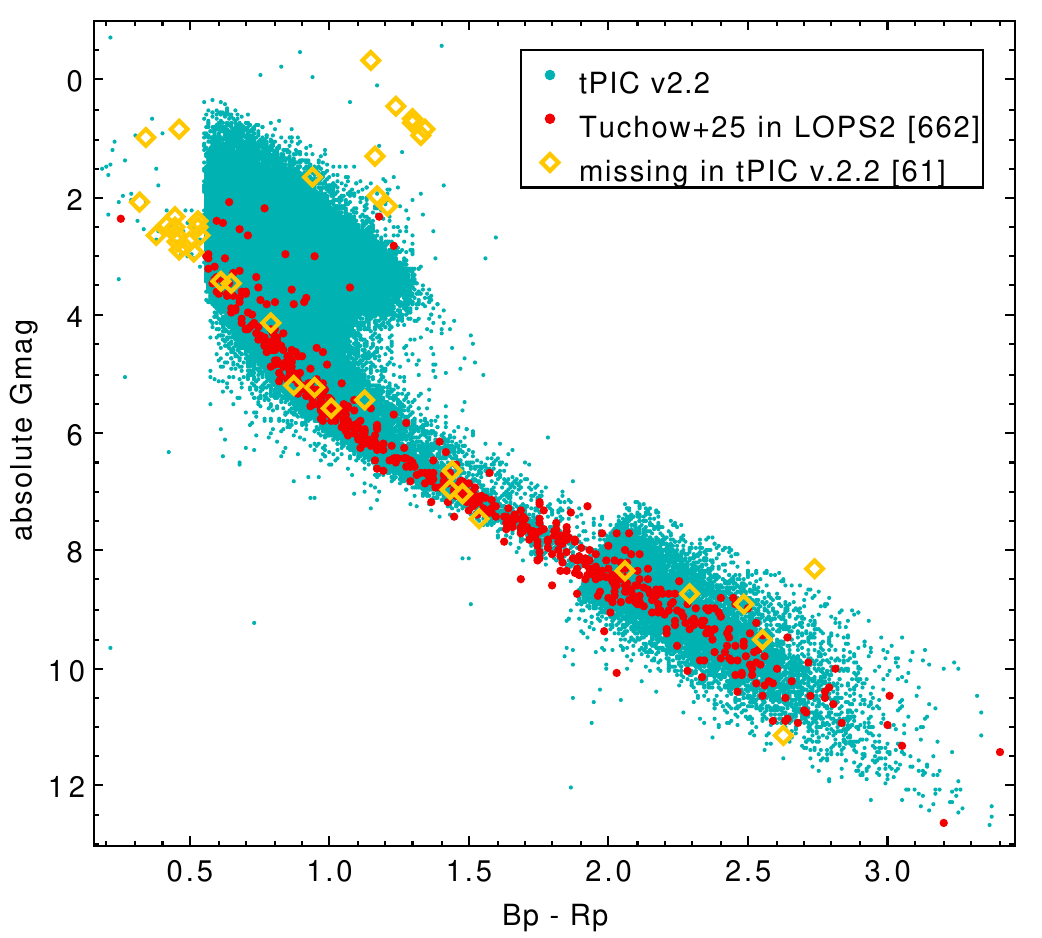}
    \caption{Color-magnitude diagram in Gaia DR3 bands, with all tPIC targets (teal dots), HPIC targets over the LOPS2 field (red dots) and the HPIC targets in LOPS2 not included in the tPIC (yellow diamonds)}
    \label{fig:PICCMD}
\end{wrapfigure}

The Italian community holds leading positions in PLATO, including the selection and definition of the first long-duration pointing field, LOPS2\footnote{See also \url{https://platomission.com/2023/07/11/first-plato-long-duration-observation-phase-lop-field-selected/}} (\citealt{Nascimbeni2025,Nascimbeni2026}), on which PLATO will stare for at least the first two years of operations. Given the small number of bright nearby stars suitable for the exoEarth survey, verifying their coverage by PLATO is a concrete and necessary step toward the HWO target list. We have cross-matched the latest HWO Preliminary Input Catalog (HPIC; \citealt{Tuchow2024}) with the latest PLATO target catalog (tPIC, in PIC 2.2.0)\footnote{For the PLATO Input Catalogue (PIC), see \citet{Montalto2021} and \citet{Nascimbeni2022}.} over the LOPS2 field: of the 662 HPIC targets falling within the LOPS2 footprint, $\sim$91\% (601) are already in the tPIC and will almost certainly be monitored by PLATO (Fig.~\ref{fig:lops2},~\ref{fig:PICCMD}). The small remaining fraction consists mostly of early-type dwarfs or evolved subgiants excluded from the tPIC by design, for which detecting or confirming a transiting habitable planet would be prohibitively expensive in terms of observing time (long orbital periods, large mass/radius, high $v\sin i$, pulsations, granulation, etc.).

In summary, PLATO will deliver the accurate stellar parameters and the candidate transiting Earth analogues needed to build the HWO target list and input catalog. Developing the PLATO--HWO synergy further is therefore a natural, high-priority avenue for the Italian community.

\subsection{Ariel}
\label{sec:fut.Ariel}
\textit{G. Micela, M. Tsantaki} %+K. Biazzo
The ESA Ariel (Atmospheric Remote-sensing Infrared Exoplanet Large-survey) mission \citep{Tinetti2018}, scheduled for launch in 2031, utilizes a 1-meter class telescope and transit spectroscopy across the visible to mid-infrared (0.5--7.8 $\mu$m), a range well suited to detecting abundant atmospheric molecules such as H$_2$O, CO$_2$, CH$_4$ and CO, to characterize the atmospheres of approximately 1,000 transiting exoplanets. It follows a three-tiered observing strategy \citep{Tinetti2022}, from a large-scale population survey (Tier 1) to increasingly detailed atmospheric characterizations of selected high-interest targets (Tier 3). Its primary objective is to investigate the chemical composition, atmospheric circulation and cloud patterns of individual planets, while also studying large population trends such as chemical diversity and the transition between terrestrial planets and sub-Neptunes. 
By delivering the first large-scale survey of exoplanet atmospheres, Ariel provides a statistical foundation, identifying global atmospheric properties, trends and outliers, which is crucial for HWO target selection. Working mainly with transiting planets, Ariel's observational focus is on the inner regions of planetary systems, characterized by shorter orbital periods, and its data will serve as the context for HWO's subsequent targeted observations. Ariel's data are expected in the early 2030s, while HWO data are planned for after 2040.

This sequential approach is critical for Comparative Planetology, allowing researchers to bridge the inner and outer regions of planetary systems. By combining Ariel's large-scale survey of inner, short-period planets with HWO's direct imaging of outer, long-period systems, we study planetary properties across an unprecedented range of orbital distances, stellar fluxes, and environmental conditions. This combined approach enables the study of system architecture, allowing investigation into how atmospheric composition and habitability potential vary with distance from the central star (chemical gradients).

Ariel is also a methodological and scientific precursor on the side of the host stars. The accurate characterization of host stars is essential for interpreting the subtle atmospheric signals that HWO aims to detect: stellar elemental abundances directly impact planetary interior structure, mineralogy and composition, with the Fe/Mg ratio tracing the relative size of a planet's core and Mg/Si constraining mantle mineralogy (see, e.g., the
leading efforts of the Italian community in this field within the GAPS - Global Architecture of Planetary Systems - project; \citealt{Biazzoetal2022, Filomenoetal2024}; Baratella et al., subm., and Sect.\,\ref{sec:fut.ANDES}). By observing planets orbiting stars spanning a wide range of metallicities, masses and evolutionary states, Ariel can also establish the link between stellar properties and planetary atmospheric composition, a goal that critically also relies on the synergy with high-precision stellar spectroscopy ($R>50{,}000$) for homogeneous parameter determination.

This is where the Italian community brings directly transferable expertise. Within the Ariel Consortium, the Italian community plays a central role in the stellar characterization of the planet-host sample: the Stellar Characterization Working Group is responsible for the homogeneous characterization of all stellar hosts included in the Ariel Mission Candidate Sample, producing a self-consistent catalogue\footnote{\url{https://sites.google.com/inaf.it/arielstellarcatalogue}} of stellar atmospheric parameters, elemental abundances (CNO, Li, refractory elements), activity indicators, ages, masses, and radii across a broad range of spectral types \citep{Magrini2022, daSilva2024, Tsantaki2025}.

\begin{figure}
\centering
\includegraphics[width=0.5\linewidth]{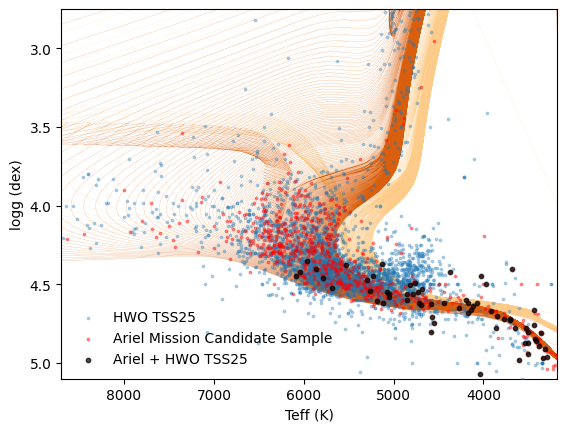}
\caption{\label{fig:hr_hwo} Kiel diagram of stars in common between the Ariel space mission Candidate Sample (version 2025) and the HWO Target Stars and Systems 2025 (TSS25) in black. The stellar parameters for the TSS25 are obtained from the high-resolution spectroscopic PASTEL catalog \citep{Soubiran2016} for the stars with available measurements.}
\end{figure}

Applying this homogeneous approach to the HWO TSS25 sample \citep{Peacock2025, Tuchow2025} will deliver the high-quality stellar parameters required for robust target prioritization and mission planning (see Fig.~\ref{fig:hr_hwo}). Looking ahead, further advances in stellar spectroscopic analysis are expected by the time of HWO’s launch, including improved 3D stellar model atmospheres, atomic data, and analysis techniques to provide increased accuracy in the stellar atmospheric parameters. At the same time, important gaps that remain in spectroscopic surveys are required to be filled, most notably
the lack of high-quality near-infrared spectra, which are crucial for measuring bio-essential elements such as phosphorus. Addressing these limitations will be critical for maximizing HWO’s scientific return. The experience and methodologies developed
within the Ariel stellar characterization program constitute essential precursor science, directly informing the target selection and
prioritization of HWO in the coming decade.

\subsection{Laboratory simulations of hazes and condensates in planetary atmospheres. }
\textit{Text from Cecchi-Pestellini C.Ciaravella A., Jiménez-Escobar A., Mangione A., Piazzese F. }

\label{life}

The characterization of habitable atmospheres and the unambiguous identification of biosignatures with the future HWO require an unprecedented level of support from laboratory astrophysics. To robustly decode high-contrast direct imaging and spectroscopic data, it is mandatory to constrain non-equilibrium chemistry, photolysis, and energy-driven processes that shape the lifetime of chemical species and the production of airborne condensed particles and hazes.

We can contribute to this effort through the LIFE+ (Light Irradiation Facility for Exochemistry in PLanetary Atmosphere Systems)laboratory. 
LIFE+ features a custom-made high-vacuum simulation chamber capable of processing gas mixtures with arbitrary molecular concentrations across wide temperature ($-20^\circ$ to $300^\circ$ C) 
and pressure ($10^{-8}$ to $10^{2}$ mbar) regimes. 
Chemical evolution is monitored in situ via a Quadrupole Mass Spectrometer and an ultra-sensitive FT-IR spectrometer equipped with a 5 m multi-path reflective mirror gas cell, delivering exceptional detection thresholds for trace volatile carriers. Unlike existing platforms, typically optimized for specific planetary environments, the enhanced performance of LIFE+ allows the simultaneous or alternative coupling of diverse energy inputs: a soft X-ray source (1487 eV/8055 eV), dedicated UV sources, a microwave generator linked to an ECR plasma, and an arc-discharge generator to simulate planetary lightning. The system includes a cooled substrate support that allows the collection and stabilization of condensation products and photochemical hazes, enabling subsequent detailed ex-situ spectroscopic analyses. Recent experiments within this facility have already demonstrated its capability by synthesizing ammonium nitrate ($NH_4NO_3$) aerosols from $CO_2$-rich atmosphere contaminated with $NH_3$ traces \cite{Jimenez+2025}.

Atmospheric hazes, clouds, and albedo modeling. Map the synthesis pathways, production rates, and optical properties of organic and inorganic aerosols generated under disequilibrium conditions, providing HWO's retrieval algorithms with crucial reference spectra to disentangle haze opacity from bulk molecular absorption.

Decoding photochemical biosignatures and false positives. Discriminate between biological and abiotic origins of potential planetary biomarkers, including volatile nitrogen/oxygen compounds (e.g., $N_2O$) and sulfur-bearing metabolic indicators (e.g., DMS). By exposing customized gas mixtures to energetic stellar fluxes, and atmospheric electric phenomena, LIFE+ will determine the exact non-biological generation rates and destruction lifetimes of these gases. 

Characterization of solid condensates and surface deposits. Systematically isolate and analyze the infrared and ultraviolet signatures of exotic solid condensates collected on the cooled window, providing experimental analogs for transient planetary surfaces, cold traps, or cloud-deck particles detectable by HWO.

Prebiotic synthesis on icy and ocean worlds. Investigate the radiation-driven chemistry of planetary ice shells through the LIFE companion facility. Capable of replicating the extreme vacuum and cryogenic conditions of icy moons and frozen exoplanets, LIFE provides a versatile platform to study solid-phase astrochemical evolution. A prominent application, highly relevant to HWO, is the prebiotic synthesis of organo-phosphorus compounds on Enceladus analogs. By mapping P–O bond fragmentation, the formation of reactive defect centers, condensation of pyrophosphates, and the phosphorus incorporation into organic compounds, LIFE explores the transition of reactive phosphites into the fundamental structural and energetic building blocks of biochemical compounds.

\subsection{The contribution of ion irradiation experiments to the HWO science}
\label{sec:ion irradiation exps}
%\textit{Text by: D. Fulvio, M. E. Palumbo, R. G. Urso}

Planetary surfaces and atmospheric hazes are exposed to extreme conditions, such as pressure and temperature variations and continuous bombardment by ionizing radiation, including energetic charged particles from stellar winds and cosmic rays. This ``space weathering'' can profoundly alter the physical, structural and chemical properties of surface and haze materials, with direct consequences on their spectral properties and hence on astronomical observations. Laboratory experiments on materials of astrophysical interest, such as ices, minerals, organic matter and meteorites, are therefore essential to correctly interpret observations from ground- and space-telescopes such as HWO.

In this context, the Laboratorio di Astrofisica Sperimentale (LASp) at INAF-OACT (Catania, Italy) is a facility dedicated to the study of solid materials under space-like conditions (ultra-high vacuum, $P<10^{-9}$ mbar, and temperatures in the range of 17--300 K). Its ion accelerator, which makes it a unique facility in Italy, can bombard selected samples of interest for planetary sciences, astrochemistry and astrobiology with energetic ions (up to several hundred keV), simulating the alteration induced by energetic charged particles. Changes are monitored in-situ through near-/mid-IR spectroscopy (in transmission or reflection mode) and Raman spectroscopy, complemented by ex-situ UV-Vis-NIR, IR and Raman characterization. The combination of in-situ ion irradiation and Raman spectroscopy further makes the LASp unique at the international level. Ion irradiation experiments can contribute to several of HWO's scientific objectives, which we outline here as starting points for preliminary laboratory work in view of the mission.

\textit{Detection of biosignatures.} The firm identification of a biosignature is meaningless unless abiotic sources can be ruled out. Ozone (O$_3$) is among the most promising observable biosignatures, yet ion irradiation experiments have shown that the bombardment of O-bearing ices (such as CO$_2$, H$_2$O:CO$_2$, CO:SO$_2$, N$_2$O, NO$_2$:N$_2$O$_4$ or pure O$_2$) can abiotically produce significant amounts of it \citep{Fulvio2025}, so that any HWO detection of ozone or other candidate biosignatures must account for such space-weathering sources.

\textit{Characterization of exoplanet surfaces.} Reflectance spectra in the FUV-VIS-NIR range are shaped by space weathering, which alters the spectral features of minerals and meteorites, causing variations in band shape and peak position and the appearance or reduction of spectral slopes, and must therefore be accounted for to correctly interpret surface observations. This can be done by studying the spectral changes induced by ion irradiation on meteorites and minerals considered possible analogues of exoplanetary surfaces \citep[e.g.,][]{Fulvio2018, Caminiti2024, Galiano2025}. Systematic experiments on such analogues containing traces of complex organic molecules or biological samples are particularly valuable, since ionizing radiation both destroys potential biosignatures \citep[e.g.,][]{Kobayashi2017} and forms new compounds of increasing complexity \citep[e.g.,][]{Urso2022, Capuano2026}, providing unique constraints on the expected detection limits of these species.

\textit{Search for technosignatures.} Among the atmospheric species that could reveal a technological civilization, NO$_2$ is one of the main candidate technosignatures, since its detection well above natural levels would indicate atmospheric pollution. Recent experiments on N$_2$O and NO$_2$:N$_2$O$_4$ ices show that these species are efficiently destroyed by ion bombardment and converted into other nitrogen-bearing molecules such as NO \citep[e.g.,][]{Fulvio2019, Oliveira2021}. Measuring their destruction cross-sections and reaction pathways allows us to estimate their expected lifetimes in solid phase on the surface before being injected into the atmosphere, and thus their viability as technosignatures detectable by HWO.

Beyond the experiments outlined above, the facility is being expanded: a new setup for Laser-induced ablation followed by Ionization and Time of Flight Mass Spectrometry (LI-ToF-MS), coupled to the ion accelerator, is currently under construction at LASp. Able to detect complex molecular species down to abundances of $10^{-7}$--$10^{-10}$ relative to the most abundant species in the irradiated sample, against the $\sim10^{-3}$ limit of IR spectroscopy, it will reach species beyond those accessible to the conventional techniques used in most laboratory astrophysics groups, with high impact on the robust interpretation of HWO observations.

\subsection{Formation, interior, climate and biosphere modelling: the interpretation framework}
\label{sec:fut.biosignatures}
%\textit{Text by: P. D’Incecco, S. Ivanovski, E. Liistro, M. Marcellino, D. Polychroni, P. M. Simonetti}

The existence of a rocky planet in or around the Circumstellar Habitable Zone \citep[CHZ;][]{kasting93} does not warrant its capability to support life, as both Venus and Mars teach us. To maximise HWO's return, in the years before launch we need to move beyond the classical, binary interpretation of the CHZ toward an integrated framework coupling formation, interior, climate, atmospheric chemistry and biology, capable of (i) prioritizing the best stellar and planetary targets, (ii) providing physically motivated priors to atmospheric retrieval pipelines, and (iii) establishing when a biosphere is detectable and when climate, atmospheric sinks or clouds may instead yield a false negative. The Italian community is already involved into several projects in this broad field, that could be coordinated into a coherent national chain of models and laboratory facilities that spans this entire sequence, summarized below.
\begin{figure}[t]
  \centering
  \includegraphics[width=\linewidth]{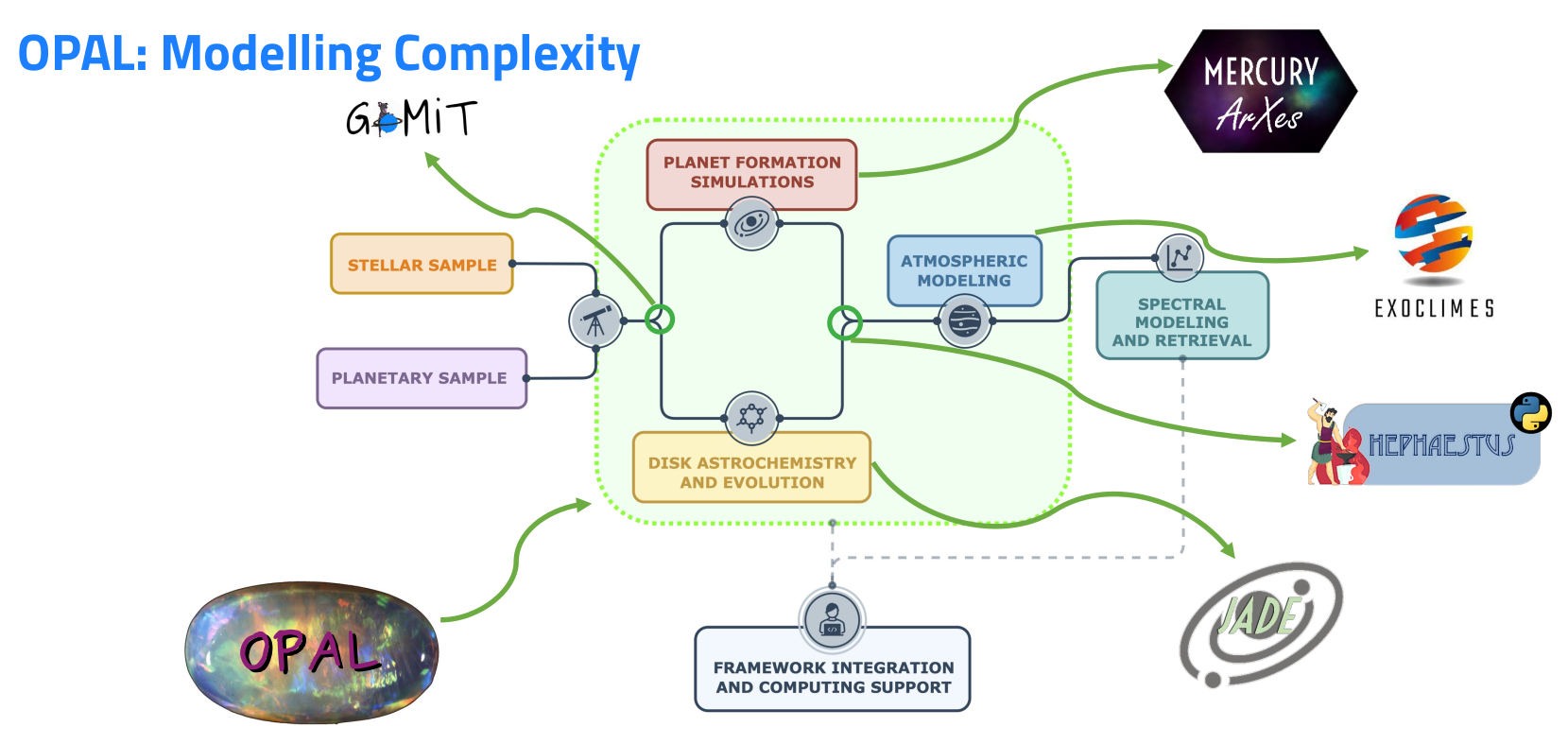}
  \caption{Schematic illustration of how the OPAL project maps the initial stellar and planetary data into the range of possible atmospheric compositions they can generate that are then fed to the spectral synthesis and analysis task.}
  \label{fig:opal}
\end{figure}

\textbf{\textit{Formation and interior: building terrestrial planets that can support life.}} Life as we know it requires astrobiologically important elements (mainly C, O, N, P, S), and a planet's capacity to host them is determined largely by its formation and early evolution: where it forms, the material it accretes, and the volatiles delivered throughout its history. In our Solar System, the predominantly chondritic\footnote{For the terrestrial planets, this chondritic composition corresponds to 93\% of their mass being oxygen bound to iron, magnesium and silicon \citep{Wasson98}.} building blocks of the terrestrial planets point to formation inwards of the water snow line, in disk regions devoid of water and organic material, implying a process that delivers volatiles from outer disk regions. This process is tied to system architecture, with giant planets, and Jupiter in particular, controlling the dynamical transport of such elements through asteroids and comets \citep{Morbidelli00, Turrini14, Turrini18}: the formation of one or more giants naturally excites the surrounding planetesimal disk, both in the Solar System \citep{Turrini11, Turrini14, Ronnet18, Pirani19} and in other systems \citep{Turrini18, Turrini19, Bernabo22, Polychroni25}. In particular, \citet{Polychroni25} showed that this process operates also on very large scales, up to an order of magnitude larger than the Solar System, and can effectively enrich planets orbiting tens of au from the giants in highly volatile elements such as nitrogen. The natural starting point to search for terrestrial planets capable of supporting life is therefore planetary systems with one or more giants beyond the water snow line \citep{Levison03, Turrini18}, with ice giants around M dwarfs playing the same role as gas giants around solar-type stars \citep{Bernabo22}. This understanding is being turned into a predictive tool by OPAL (Origins of Planets for ArieL), an end-to-end project that tracks the astrochemical paths of forming planets \citep{Polychroni2026}: taking advantage of the Ar$\chi$es suite of planet formation codes developed at INAF, OPAL simulates the chemically time-evolving disk and the multiple possible migration paths of growing planets to produce a library of different bulk elemental compositions of the simulated planet, sampling the multitude of possible initial conditions. These compositions are then combined with our expanded versions of the atmospheric tools in the {\sc Exoclimes} suite\footnote{\url{https://github.com/exoclime}} to produce realistic atmospheric compositions in equilibrium and disequilibrium conditions \citep[][]{Fonte2023,simonetti26}, from which the Ariel consortium creates the synthetic spectra on which to test its codes and pipelines (Fig.~\ref{fig:opal}). While OPAL currently focuses on the hot gas giants targeted by Ariel, the same methodology and codes can be extended to the creation of highly detailed terrestrial planetary compositions, identifying under which conditions the spectral features of interest can be distinguished and constraining the frequency of false positives, thereby helping to fine-tune which molecules are the best tracers to observe.

\textbf{\textit{Climate modelling: from 1D habitability metrics to 3D dynamics.}} A preliminary selection criterion is whether a planet lies within the CHZ, most widely defined through one-dimensional radiative-convective calculations for an Earth analogue \citep{Kopparapu13}. The CHZ edges, however, depend not only on the star and the orbit but also on the planet and atmosphere composition, so that water-poor planets may remain habitable over a much broader range of orbital distances \citep{abe11, kodama19}; and since climate feedbacks can induce hysteresis and multi-stability, even a low measured surface temperature does not necessarily imply that a planet is uninhabitable \citep{checlair17,murante20}. In binary star systems, the structure of the CHZ is further complicated by the overlap with regions of gravitational instability \citep{simonetti20}. Capturing this complexity requires a full hierarchy of models, from 1D frameworks for efficient parameter-space exploration to higher-dimensional models resolving the intrinsically 3D circulation of planetary atmospheres.

The Italian community is actively working on this full hierarchy. For rapidly rotating planets, the 1D Earth-like planets Surface Temperature Model (ESTM; \citealt{vladilo13, vladilo15}), and its recent versions EOS-ESTM \citep{simonetti22, biasiotti22} and pRT-ESTM \citep[][]{bisesi26a}, have been applied to detailed habitability studies of Kepler-452b \citep{silva17b}, Gl 514 b \citep{biasiotti24} and HD 20794 d \citep{biasiotti26b}, as well as to the past climate states of Mars \citep{simonetti24} and the Earth \citep{silva17a, biasiotti26a, bisesi26a}, including multi-stability and snowball transitions \citep{murante20}. EOS-ESTM also actively participates in the NASA Nexus for Exoplanet System Science CUISINES \citep{sohl24} FILLET \citep{deitrick23, barnes25}, CREME (Tsigaridis et al., subm.) and COD ACCRA (Chaverot et al., subm.) Projects, placing it at the forefront of international efforts to establish standardized benchmarks across the exoplanetary science field. For tidally-locked M-dwarf planets, both the 1D SYRO (SYnchronous ROtator) EBM \citep{simonetti25} and the 3D intermediate-complexity PLASIM-LSG code\footnote{\url{https://zenodo.org/records/4041462}} \citep{fraedrich05} have been adopted. PLASIM, in line with ExoPLASIM \citep{paradise22}, enable multi-parameter studies of large samples of such systems \citep[][accepted]{bisesi26b}, accounting for different dynamical regimes and their resulting effects on the habitability. These models allow planets to be ranked according to different habitability indices, beyond the simple presence of surface liquid water \citep{spinelli23,spinelli24}. Part of the codes and outputs are already publicly available through the ARTECS database\footnote{\url{https://wwwuser.oats.inaf.it/exobio/climates/}}. Crucially, when coupled with state-of-the-art radiative-convective codes \citep[e.g.][]{molliere19, Villanueva2022, MacDonald2024}, including those developed in-house \citep[e.g.][]{simonetti22}, these climate models can provide physically motivated constraints for retrieval pipelines, helping to reduce uncertainties in inferred atmospheric properties.

\begin{wrapfigure}{l}{0.55\textwidth}
\centering
  \includegraphics[width=\linewidth]{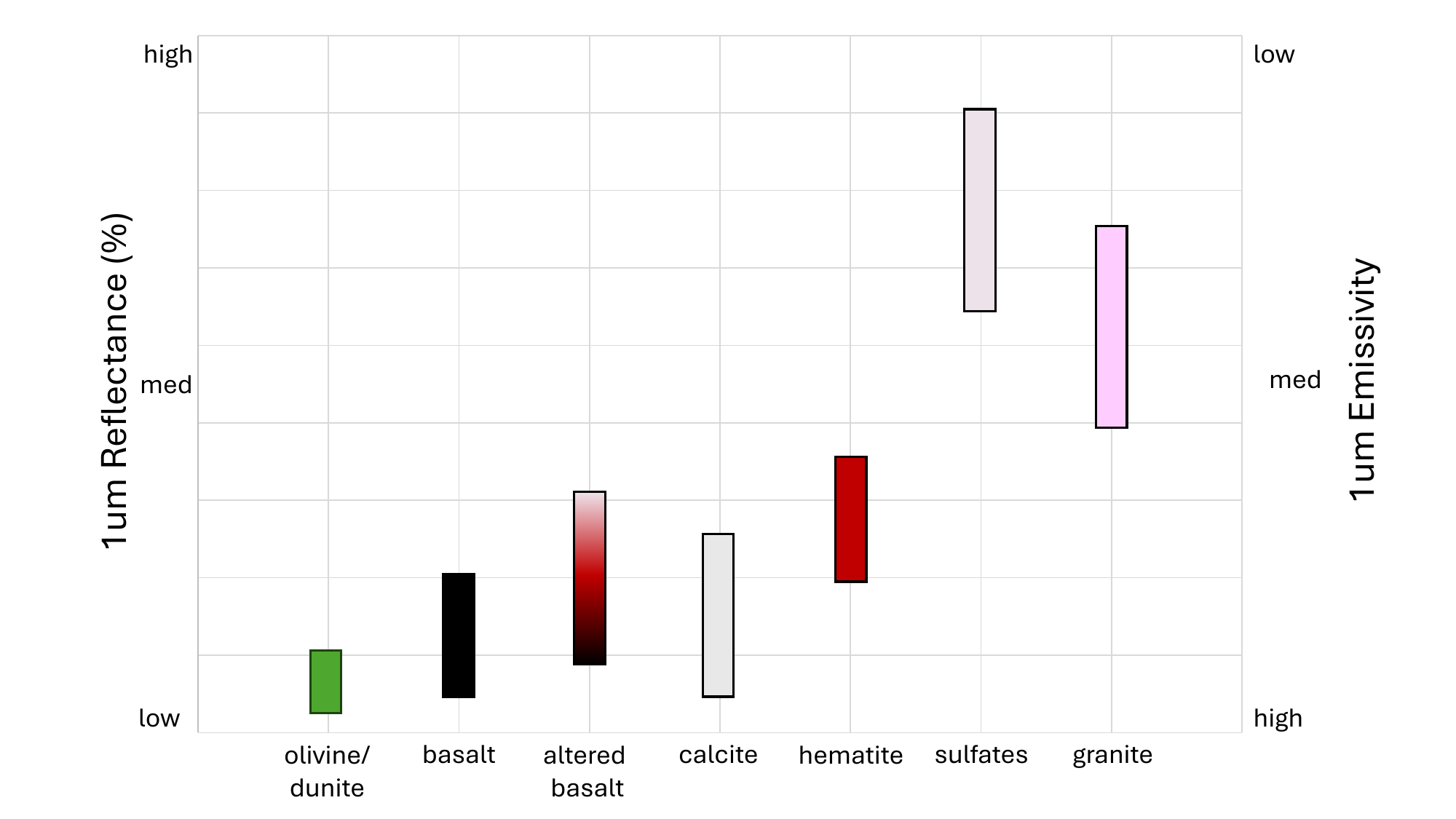}
  \caption{\textbf{1\,$\mu$m lookup for AVENGERS.} Approximate ranges of 1\,$\mu$m reflectance (left axis) and 1\,$\mu$m emissivity (right axis) for Venus-relevant materials (olivine/dunite, fresh basalt, altered basalt, hematite, sulfates, calcite, granite). Fresh basalts show low reflectance/high emissivity at 1\,$\mu$m; rapid oxidation and sulfate/Fe-oxide coatings under Venus-like conditions damp this signature.}
  \label{fig:one_micron_lookup}
\end{wrapfigure}

\begin{wrapfigure}{l}{0.45\textwidth}
\centering
  \includegraphics[width=\linewidth]{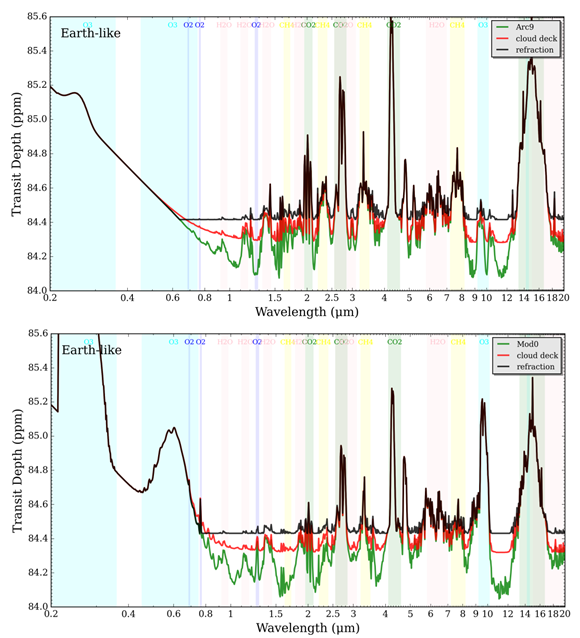}
  \caption{UV-to-mid-IR transit spectra for a low-O$_2$, post-GOE atmospheric 
  composition ($\sim 100$ ppm O$_2$; top panel), compared with a modern-Earth 
  atmospheric composition (bottom panel). The spectra show the clear-sky case 
  without refraction (green), the corresponding case including atmospheric 
  refraction (black; ExoRefract model, Maris et al., in prep.), and the 
  case without refraction but with a cloud deck at the tropopause (red).}
  \label{fig:transit_spectra}
\end{wrapfigure}

\textbf{\textit{Biosphere--climate feedbacks and the detectability of biosignatures.}} A further level of complexity arises once biological activity is included: life is not only a possible outcome of habitability but can modify the boundary conditions that regulate planetary climate and, therefore, the production and detectability of biosignatures. Through feedbacks such as the vegetation--albedo (Charney) mechanism \citep{charney75, aleina13}, biological activity can shift a planet from a snowball toward a partially habitable state near the outer edge of the HZ \citep{bisesi24, bisesi26a}; the sign of this feedback is not universal, however, since cyanobacterial blooms can instead raise the ocean albedo and favour glaciation \citep{battistuzzi2023a}. These feedbacks are directly relevant to biosignature interpretation, since the atmospheric abundance of O$_2$, O$_3$, CH$_4$, CO$_2$ and H$_2$O is controlled by the coupled climate--chemistry--surface system, and a planet with an active biosphere may still yield a false negative if climate, sinks or clouds suppress the observable signal. A key step is therefore to quantify whether a planet can sustain a biosphere capable of producing detectable biosignatures (e.g. \citealt{silva17a}). One example of this approach is the photosynthetic habitable zone (PHZ; e.g. \citealt{Hall:etal:2023}), namely the region, generally within the classical habitable zone, where oxygenic photosynthesis can occur at levels potentially relevant for atmospheric biosignature production; its reliability, however, depends critically on the adopted biological response functions, calling for a close coupling between planetary climate and atmospheric  models and experimentally constrained biological data.
Ultimately, biological production must be translated into atmospheric abundances and observable spectral signatures. Figure~\ref{fig:transit_spectra} illustrates this final step using spectra computed 
with pRT-ESTM \citep{bisesi26a}\footnote{Our climate model ESTM coupled 
with the petitRADTRANS radiative-transfer code \citep{molliere19}.} for low-O$_2$ and modern-Earth atmospheric compositions. The comparison 
highlights how biosignature detectability depends not only on atmospheric composition, but also on clouds and atmospheric refraction 
(Silva et al., in prep.).

On this front, the Italian community contributes through unique interdisciplinary collaborations and laboratory expertise.
A collaboration between the Department of Biology of the University of Padova, the CNR Institute of Photonics and Nanotechnology, and the INAF Observatory of Padova has developed, in the past decade, a setup to grow microorganisms under simulated planetary conditions, in terms of irradiance and atmosphere. Stellar spectra are reproduced with a custom Stellar Light Simulator, and O$_2$ production and CO$_2$ consumption are monitored inside a growth chamber with controllable atmospheres. Experiments with photosynthetic organisms exposed to simulated M-dwarf light and anoxic atmospheres have shown that oxygenic photosynthesis, and thus O$_2$ release in anoxic conditions, is feasible, with notable biodiversity and adaptive mechanisms across organisms with different levels of complexity \citep{battistuzzi2023a, battistuzzi2023b, Liistro2024, Liistro2026}. Along this line is the ongoing ASTERIA project (e.g. \citealt{Bellucci2024, Barbisan2026, bisesi26b}). This demonstrates that the limits of photosynthetic habitability cannot be inferred from incident stellar flux alone, but require a joint treatment of stellar spectra, atmospheric filtering, surface climate, water availability and biological adaptation. Building on this, the Exobioma project derives a continuous Multiparametric Life Score (MLS) by matching a database of tolerance limits for 300 terrestrial extremophile species against hypothesized atmospheric profiles, and integrates the metabolic pathways of the most compatible species into biochemical models to hypothesize atmospheric Chemical Reaction Networks (CRNs). The MLS provides a robust target-prioritization tool, while the CRNs provide a predictive ``dictionary'' of VOC biosignatures \citep{SousaSilva2020}, allowing HWO to move beyond single-gas anomalies toward the complex suites of gases that signify a functioning biosphere.

\textbf{\textit{Surface diagnostics: is it an Earth or a Venus?}} Understanding how volcanism regulates surface renewal, atmospheric composition, and long-term climate stability is central to the search for habitable planets, with Venus a prime example of volcanism going wrong \citep[][and references therein]{Gillmann2022}, showing evidence of volcanism as recent as the last 30--50 years \citep{Sulcanese2024}. The AVENGERS initiative (Analogs for VENus' GEologically Recent Surfaces; \citealt{DIncecco2025}) builds a calibrated bridge between Venus, Earth, and rocky exoplanets, combining field spectroscopy on young lavas (Etna, Cumbre Vieja), high-temperature laboratory emissivity measurements, and radar/spectral interpretation under Venus-like conditions. A key diagnostic emerging from this framework is the 1 $\mu$m mafic absorption and emissivity band: fresh basalts display low reflectance and high emissivity, while oxidation and the formation of sulfate and Fe-oxide coatings rapidly damp this feature under Venusian conditions \citep{Filiberto2024}, so that detecting an enhanced 1 $\mu$m mafic signature on a rocky exoplanet would indicate recent or ongoing resurfacing and active interior--atmosphere coupling (Fig.~\ref{fig:one_micron_lookup}). By further incorporating radar analyses that link surface roughness and dielectric properties to degrees of alteration, this combined spectral--radar approach can be applied to unresolved exoplanet surfaces, providing HWO with prior constraints on volcanic style, alteration state, and crustal differentiation, and offering a practical strategy to distinguish volcanically active, potentially habitable planets from evolved Venus analogs within the Venus Zone \citep{Kane2014}, where CO$_2$-rich and runaway-greenhouse conditions are expected \citep{Lincowski2018}.

Taken together, these formation, climate, biosphere and surface models can constitute a coherent interpretation framework that directly feeds the target selection and prioritization discussed in Section~\ref{sec:demographics}, and provides the physically motivated priors needed to interpret the atmospheric and surface spectra that HWO will deliver.

\subsection{The ELT Planetary Camera and Spectrograph}
\label{sec:fut.PCS}
%\textit{Text by: S. Desidera, L. Schreiber}

The Planetary Camera and Spectrograph (PCS) is a proposed instrument for the Extremely Large Telescope (ELT), whose main science goal is the detection and characterization of nearby exoplanets with sizes from sub-Neptune to Earth-size in the neighborhood of the Sun \citep{Kasper2021}, through a combination of Extreme Adaptive Optics, coronagraphy, high-dispersion spectroscopy, and possibly polarimetry. A three-year R\&D phase, building on the legacy of the ELT/EPICS Phase A studies \citep{kasper2010}, is currently consolidating the science case and instrument requirements before the formal start of the project.

PCS will significantly expand the detection space for rocky exoplanets, moving from the early detection of a very limited number of targets to a more comprehensive view (Fig.~\ref{fig:PCS_rocky_planet}). The currently known potential targets are mostly around very nearby M dwarfs, although the system around the G dwarf HD20794 \citep{Nari2025} looks promising for an extension to other spectral types. As anticipated in Section~\ref{sec:sci.exoplanets}, PCS will be a key and direct precursor for HWO (and LIFE): it will extend the direct detection and characterization of rocky planets from the handful of cases feasible with ANDES at ELT, while remaining about two orders of magnitude short of HWO's contrast and thus accessing temperate rocky planets around M dwarfs rather than around solar-type stars. This consolidates the preparation of HWO from a scientific, operational, and technological point of view, in particular through the optimization of the target list, the development of detection methods and data-analysis techniques tailored for planets in reflected light at close projected separation, and the tools for the interpretation of atmospheric signals.

On the technological front, PCS builds upon ongoing developments in extreme adaptive optics, including the realization of a new-generation deformable mirror with very high actuator density and fast response, and the implementation of predictive, multi-stage, and advanced wavefront-control schemes required to achieve sub-millisecond stability. These technologies, encompassing both wavefront sensing and real-time control, will be experimentally validated within the SAXO+ upgrade of SPHERE at the VLT \citep{sphere, Boccaletti2022}, which serves as a key on-sky demonstrator for PCS-related Adaptive Optics concepts, with first light expected in early 2028. INAF is actively participating in the PCS R\&D study, with responsibilities in the definition of the science cases and requirements as well as in the development and integration of the system and its scientific instruments, and contributes with roles of responsibility to the SAXO+ upgrade. Pursuing and expanding these activities will naturally place the Italian community in a strong position to make a significant contribution to HWO in its various phases.

\subsection{ANDES \& CUBES}
\label{sec:fut.ANDES}
%\textit{Text by: P. Di Marcantonio, G. Guilluy}

High-resolution spectroscopy (HRS, $R\gtrsim20{,}000$), pioneered by \citet{snellen2010}, is a powerful tool for exoplanet atmospheric characterization: molecular features are resolved into a dense forest of individual lines that can be detected even when bands overlap, via cross-correlation with model templates \citep{birkby2018}, while the planet's orbital motion during transit Doppler-shifts the planetary signal away from the stationary stellar and telluric lines. HRS and space-based low-resolution spectroscopy (LRS) are complementary: LRS retains the continuum information important for constraining global properties such as temperature and cloud coverage, but suffers from degeneracies, since different molecular species can produce overlapping bands, whereas HRS resolves these into individual lines but loses the continuum information during the analysis. Furthermore, ground-based HRS data can still suffer from residual telluric contamination, potentially biasing the inferred atmospheric properties, whereas space-based LRS observations are free from telluric effects. Their combined analysis therefore provides a more comprehensive understanding of exoplanetary atmospheres \citep[e.g.,][]{brogi2017}. The Italian community has built extensive expertise on this front: within the GAPS consortium, 180 nights have been collected through the Large Programs GAPS2 (P.I. Micela) and BRIDGES (P.I. Borsa) using GIARPS@TNG, which combines the two high-resolution spectrographs GIANO-B in the near-infrared (0.92--2.45 $\mu$m, $R\sim50{,}000$) and HARPS-N in the visible (0.38--0.69 $\mu$m, $R\sim115{,}000$). This decade-long GAPS effort, on mostly Jupiter-like planets, is a direct analogue of what ANDES at the ELT will achieve in the near future for rocky planets, and provides valuable lessons for upcoming ELT facilities: (i) the need for a robust treatment of stellar contamination; (ii) the crucial role of accurate orbital parameters, such as systemic velocity and orbital eccentricity, for optimal observation planning; (iii) the increasing difficulty of disentangling planetary signals from telluric contamination, particularly for the lower-mass planets that will be prime ELT targets; and (iv) the importance of exploiting synergies with space-based facilities such as HST and JWST. While ANDES will enable atmospheric studies of planets like Proxima Centauri b in the habitable zones of M dwarfs, accessing Earth-like planets around solar-type stars will require HWO. In the HWO era, combining space-based observations with high-resolution ground-based spectroscopy will also be highly beneficial. In particular, building on the approach of \citet{Snellen2014}, who joined high-resolution CRIRES/VLT spectroscopy with adaptive optics to characterise the non-transiting planet $\beta$ Pictoris b, ANDES+PCS will couple HRS with high-contrast imaging, enabling cross-correlation analyses of non-transiting, more massive and more distant planets than the Earth analogues accessible to HWO.

On the instrument side, INAF leads the ongoing construction of two advanced ground-based spectrographs, ESO/VLT CUBES and ESO/ELT ANDES. CUBES (Cassegrain U-Band Efficient Spectrograph; \citealt{Genoni2024}) is designed as the ultimate near-ultraviolet spectrograph for the VLT, providing continuous coverage from 305--405 nm in a single exposure at $R>19{,}000$ with an efficiency exceeding 40\% in the near-UV. ANDES (ArmazoNes high Dispersion Echelle Spectrograph; \citealt{marconi24}) is the high-resolution, high-precision, ultra-stable, fibre-fed echelle spectrograph for the ELT. ANDES baseline consists of three fibre-fed spectrographs (BV, RIZ, YJH) providing a spectral resolution of $\sim$100,000 with a simultaneous wavelength coverage of 0.4--1.8 $\mu$m and with the goal of extending it to 0.35--2.4 $\mu$m with the addition of a U arm to the BV spectrograph and a separate K-band spectrograph. ANDES operates both in seeing- and diffraction-limited conditions, the latter by including a single-conjugated adaptive optics module (baseline) and a coronagraph (goal) feeding an integral field unit in the NIR.

Although developed for ground-based telescopes, CUBES and ANDES tackle technological challenges directly relevant to HWO's technology roadmap, fostering synergies in three main areas. First, high sensitivity and efficiency in the UV/VIS range through optimized optics and coatings: CUBES demonstrates that high-reflectivity optics combined with optimized diffraction gratings and compact precision image slicers achieve both the high throughput and the spectral stability that HWO requires for photon-limited UV/VIS performance. Second, long-term spectral stability and thermal/mechanical precision, where the ultra-stable thermal control and fibre-fed modularity of ANDES offer lessons directly relevant to maintaining spectral stability in a space environment. Third, the integration of advanced calibration and wavefront-control systems to support high-contrast observations: ANDES couples high-dispersion spectroscopy with high-contrast imaging to reach contrasts of order $10^{-7}$, and its ultra-precise wavelength calibration, which combines traditional calibration sources with innovative technologies such as Fabry--Perot etalons and laser frequency combs (including in the UV), provides integration, operation and data-handling lessons of direct relevance for HWO's calibration plan, for guiding reference-source design and automated monitoring systems, with the potential to translate into space-qualified solutions. The data-reduction pipelines and complex control systems developed for both instruments provide a tested methodology adaptable to the continuous, unattended operations required in space-based platforms.

In parallel, INAF brings consolidated experience in instrument control software, electronics and data processing, coordinated through initiatives such as TETIS (Technologies for Telescopes and InStrument control software). While direct technology transfer to a space platform requires careful adaptation to the constraints of the extraterrestrial environment, the design philosophies, methodologies and R\&D outcomes from CUBES and ANDES provide a robust roadmap for HWO, positioning Italy as a strong partner for future high-precision spectroscopy collaborations both on the ground and in space.

\subsection{MAVIS}
\label{sec:fut.MAVIS}
%\textit{Text by: G. Cresci}

The MAVIS (Multi-conjugate Adaptive Optics Visible Imager and Spectrograph) instrument, currently under development for the Very Large Telescope (VLT) at ESO’s Paranal Observatory, will deliver diffraction-limited imaging and IFU spectroscopy over a wide field of view in the visible spectrum. By achieving unprecedented spatial resolution in the optical from the ground, MAVIS will be complementary to the IR-optimised instruments on the ELT, and will bridge the gap between current 8–10m class facilities and the next generation of space telescopes, including NASA’s HWO. 

MAVIS, operating on the Adaptive Optics Facility (AOF) of the Very Large Telescope (VLT), is a general-purpose instrument for exploiting the highest possible angular resolution of any single optical telescope available in the next decade, either on Earth or in space, and with sensitivity comparable to (or better than) larger aperture facilities. 

MAVIS comprises three principal modules \citep[see][]{Rigaut2021}: 

\begin{itemize}
\item An Adaptive Optics Module (AOM), responsible for wavefront sensing and correction. This features an innovative transmissive design that includes two post-focal deformable mirrors, totalling about 4,250 actuators (5,420, including the deformable secondary mirror of the AOF), eight 40x40 Shack-Hartmann LGS wavefront sensors, and three near-infrared natural guide star wavefront sensors, which also provide low-order truth sensing and slow-focus corrections. 
\item An imager module, Nyquist sampling the near-diffraction-limited optical beam over a 30”x30” field of view. The imager is equipped with a range of broad- and narrow-band filters, with throughput maximised via minimal re-imaging optics following the AOM. 
\item An image-slicing integral field spectrograph (IFU) module, with flexible spatial and spectral configurations. Exchangeable fore-optics and dispersing elements provide two spatial modes (0.025” and 0.050” square spaxels, covering respectively 3.6”x2.5” and 7.2”x5” fields of view), each of which has four spectral modes, covering 370nm-1μm at resolutions from 5,000-15,000. 
\end{itemize}

By probing the frontiers of angular resolution and sensitivity across a large portion of the observable sky (\textasciitilde 50\% at the galactic pole) at visible wavelengths, MAVIS will enable progress on an array of scientific topics, from studies of the Solar System to planetary systems around other stars, and from the physics of star formation in the Milky Way to the first star clusters in the Universe. By the early 2030s, MAVIS will be the instrument capable of delivering high spatial resolution visible diffraction-limited performance from the ground on timescales relevant to HWO’s science planning and commissioning (see Table~\ref{mavistab}).\\

\begin{table}
\centering
\begin{tabular}{llll}
\hline
 & MAVIS  & HST / WFC3  & HWO Target  \\
\hline
Wavelength range & 370--1000 nm & 200--1600 nm & 100 nm--2.5 $\mu$m \\
Angular resolution (V band) & $\sim$14 mas (diffraction limit) & $\sim$60 mas (F555W) & $\sim$20--30 mas (6-m class) \\
Field of view (imager) & 30 $\times$ 30 arcsec & 162 $\times$ 162 arcsec (WFC) & $\sim$few arcmin (TBD) \\
Spectral resolution & R = 4,000--15,000 & R = 200--10,000 (grisms) & R $\sim$140--70,000 (TBD) \\
Astrometric precision & $\sim$50--150 $\mu$as / epoch & $\sim$0.3 mas / epoch & $\sim$1 $\mu$as goal \\
Collecting area & 52.8 m$^2$ & 4.5 m$^2$ & $\sim$28--50 m$^2$ (6--8 m) \\
Sky coverage (MCAO) & $\sim$50\% at Paranal (4 LGS) & 100\% (space) & 100\% (space) \\
\hline
\end{tabular}
\caption{\small{The table illustrates that MAVIS occupies a compelling intermediate regime: superior collecting area to HST at comparable or better angular resolution, and wavelength coverage that substantially overlaps with HWO's optical prime focus. While HWO will reach far fainter limiting magnitudes in some regimes --- particularly in ultraviolet and in long space-based integrations --- MAVIS's ability to schedule large programs efficiently, revisit targets on short timescales, and coordinate multi-object spectroscopy over a wide field makes it an irreplaceable preparation facility.}}
\label{mavistab}
\end{table}

The science synergies between MAVIS and HWO are extensive, spanning multiple sub-fields of astrophysics. In the following, we highlight a few examples. \\

One of the most immediate areas of overlap is in the study of stellar populations in crowded environments — star clusters, galactic nuclei, and the resolved halos of nearby galaxies. In such fields, angular resolution is the critical limiting factor: two stars separated by less than a seeing disk blend into an unresolved source, corrupting photometry, color-magnitude diagrams, and spectral classifications. MAVIS's \textasciitilde 14 mas FWHM in V band corresponds to a physical resolution of roughly 0.07 parsecs at the distance of the Large Magellanic Cloud (50 kpc), 0.7 parsecs at the Andromeda galaxy (M31, \textasciitilde 785 kpc), and \textasciitilde 7 parsecs at the distance of the Virgo Cluster (\textasciitilde 16 Mpc). This capability will enable color-magnitude diagram (CMD) science in environments currently unaccessible even to HST, and will push stellar population studies to distances of several Mpc — resolving individual AGB stars, red giant branch stars, and luminous blue variables in galaxies beyond the Local Group. HWO's UV sensitivity will complement MAVIS's optical CMD work by revealing the hot stellar populations — white dwarfs, extreme horizontal branch stars, post-AGB objects, and young OB associations — that are luminous in the ultraviolet but difficult to access from the ground. MAVIS ground-based optical CMDs, combined with HWO UV photometry, will enable age and metallicity determinations for stellar populations in galaxies throughout the Local Volume and beyond.\\

The central few hundred parsecs of nearby galaxies host a rich array of structural components: nuclear star clusters (NSCs), nuclear stellar disks, compact stellar bars, and in many cases low-luminosity active galactic nuclei (AGN). Disentangling these components requires sub-arcsecond resolution photometry and spectroscopy over fields of tens of arcseconds — exactly the regime MAVIS is designed to cover. MAVIS IFU spectroscopy of nearby galactic nuclei will characterize stellar kinematics, stellar population gradients, and ionised gas distributions at physical scales of a few parsecs. These datasets will provide essential context for HWO UV spectroscopy of the same systems, which will probe the far-UV flux from young stellar populations and the UV signatures of AGN photoionisation.\\

Similar complementarity is foreseen also in the spectroscopic studies of the stellar populations in the core of nearby clusters, the immediate environments around individual stars, revealing the processes driving angular momentum loss that ultimately shapes star and planet formation, probing the later stages of stellar evolution, up to sensitive observations for minor bodies within (or passing through) the Solar System, and regular monitoring of major bodies at 10s-100s km resolution.\\

MAVIS thus represents a uniquely powerful ground-based instrument for advancing the science objectives of HWO across multiple fronts simultaneously. The convergence of MAVIS's operational timeline (early 2030s onward) with HWO's pre-launch preparatory period creates a strategic opportunity, offering critical science, technology validation, and strategic insights that inform and enhance HWO’s mission.

\subsection{Pioneering a Multi-Wavelength Technosignature Hunt: The HWO and Next-Generation Radio SETI Synergy }
%\textit{Text by: A. Melis, M. Pilia, A. Cabras}

The quest to discover extraterrestrial intelligence demands a fundamental paradigm shift. While HWO will revolutionize our understanding of Earth-like exoplanets through advanced optical and infrared characterization, atmospheric biosignatures alone may not definitively prove the existence of advanced civilizations. To establish a truly comprehensive search framework, radio SETI must be aligned with HWO's targeting strategies.

Our goal is to forge a multi-wavelength synergy, combining HWO’s precise exoplanetary data with the processing of radio signals observed by next-generation instruments to create a robust and ambitious technosignature detection pipeline. To match the magnitude of the HWO mission, radio SETI requires a transformative leap in data analysis. We will achieve this by introducing a sophisticated mathematical powerhouse to the radio domain: the Karhunen-Loève Transform (KLT). The KLT is already a proven, cutting-edge tool utilized for high-contrast exoplanet imaging in flagship space observatories like the James Webb Space Telescope (JWST), where it strips away blinding starlight to reveal faint, hidden worlds.

We are now pioneering a "Double KLT" architecture. The general idea is to utilize the beamforming technique with two distinct applications of the KLT: in the first, the on-target beam is cleaned of all instrument or terrestrial artificial components (Radio Frequency Interference, RFI), so that this ON beam contains solely the contribution of the star, any potential stellar aurora (generated by the interaction between the red dwarf and the planet), and the hypothetical technosignature. In the second application, auroras and SETI signals are isolated from the strong (but continuous, i.e., uncorrelated) signal of the star.

The Double KLT pipeline is currently undergoing validation using the formidable observational power of the MeerKAT radio telescope. By stress-testing our algorithms on real-world data today, we are hardening the system for the future era of radio astronomy. Looking forward, the true potential of this radio-HWO synergy will be unleashed with the advent of the Square Kilometre Array Observatory (SKAO). Combining the unprecedented sensitivity and vast field of view of the SKAO with the analytical impact of the Double KLT will exponentially multiply our search volume. This synergy will definitively transform radio SETI and perfectly synchronize it with HWO's exploration of the galaxy's most promising exoplanets.

\subsection{Technosignature search of laser emissions using HWO spectrographs}
%\textit{Text by: N. Antonietti, P. Pari}

An extraterrestrial intelligence may be thought of communicating or leaking communications using lasers. Lasers are non natural sources of coherent lights whose bandwidth is very narrow. Archive searches have been done using data available from HARPS spectrograph (at ESO's VLT) and currently archive searches are being done using data available from HARPSN spectrograph (at TNG) and ESPRESSO spectrograph (at ESO's VLT). This is an example of how SETI hunts for technosignatures at optical wavelengths.

The improved resolution of spectrographs onboard HWO, ranging from UV through optical bands to near infrared, will allow for a thorough search of extraterrestrial intelligence without being affected by airglow.

The pipeline does not require any additional instrumentation with respect to what is already in HWO for habitable world observations and is thus fully integrable.

\newpage

\section{Innovative Technologies}
\label{sec:innovativetechnologies}
HWO will be the most advanced optical telescope ever built. To meet its incredibly challenging scientific  and technological requirements, it will implement technologies that, to a large extent, still need to be developed.
In this section we briefly describes the key technologies that Italian researchers are developing and that can be crucial to achieve the requesteed performances.

\subsection{Mirror Technology$^*$}
\label{sec:tec.mirrors}
%\textit{Text by: R. Briguglio, G. Pareschi}
\blfootnote{$^*$For further information feel free to contact  Runa Briguglio \href{mailto:runa.briguglio@inaf.it}{\nolinkurl{runa.briguglio@inaf.it}{\nolinkurl{}} and/or \href{mailto:giovanni.pareschi@inaf}} and/or Giovanni Pareschi \href{mailto:giovanni.pareschi@inaf.it}{\nolinkurl{giovanni.pareschi@inaf.it}}  }
In the last few years some attention was devoted at INAF to the development and characterization of lightweight  ($< 20 kg/m^2$) and deformable mirrors for space telescopes. Two research teams are involved in such activity: INAF Osservatorio Astronomico di Brera (Milano/Merate) is focused on sandwiched mirrors technology and INAF Osservatorio Astrofisico di Arcetri (Firenze) on high orders active optics. The techniques have been already presented in several SPIE and topical workshop organized by ESA, NASA and ASI.

\subsubsection{Active Mirrors}
The starting point is the vast technological heritage from the field of electro-magnetic actuated, large format deformable mirrors (e.g. the secondaries of the LBT and VLT, M4 for ELT), and the associated network between INAF and industrial companies mastering the technology. As a reference, the reader may find some details in \cite{Esposito_2011}, \cite{2010SPIE.7736E..2CR}, \cite{2014SPIE.9148E..45B}, \cite{2018NatSR...810835B}.
The key element of such systems is the contactless actuation mechanism, with a thin Zerodur glass shell as optical surface controlled by non-contact voice coil actuator. Such control scheme implements the architectural decoupling between the mirror mechanics and the optical surface so that manufacturing errors and deformations of the support don't affect directly the optical quality, while high frequency vibrations on the mechanics are naturally low-passed by the actuators.
The concept of such contactless active mirror with a large actuator count has been explored in the 2010, in the context of a TRP funded by ESA. The LATT project demonstrated the manufacturability, controllability and optical quality of a 40 cm diameter active mirror controlled by 19 actuators, intended as a segment of an active primary. The prototype is significantly low-mass, with an areal density as low as $18 kg/m^2$ (including both mirror opto-mechanics and actuation system).
In 2021 a follow-up has been funded by INAF to further assess the concept and in particular to demonstrate in the optical laboratory the rejection of external disturbances and to evaluate a strategy to optimize the system mass budget.
Currently, the team is involved in a research program funded mostly by ASI and INAF (for a total budget close to 1M€), with a significant participation of the industrial partners.  More informations on the projects may be found in \cite{Briguglio_2023}, \cite{scalera2023lateralconstraintglassshell}, \cite{ScandagliaE2E}, \cite{TheLattWay},\cite{LATT_SPIE2016}, \cite{Menessini2026}.

\subsubsection{Sandwiched mirrors}
In this case a thin and floppy  glass shell is produced using the slumping of a glass sheet onto a ceramic mold that has a surface with a high optical quality. After this step, this curved shell is assembled and glued to a stiff substrate made in foamed and pre-shaped material. On the back of the substrate it is also glued a flat sheet of the same glass. This procedure combines the good optical performances achievable on optics produced by means of the hot slumping technique with the lightweight and stiffness of the foamed material and, finally, the good structural properties achievable in sandwich-like structures. The surface of the mirror can be further on corrected via bonnet and ion-figuring. The technique has been presented to several SPIE conferences and also to ASI during a recent workshop.

\subsubsection{Optical coatings}
%\textit{Text by:M.G. Pelizzo, A. J. Corso}

One of the most challenging aspects of the HWO is its extremely broad operational wavelength range, extending from the near infrared (NIR) down to the ultraviolet (UV), reaching wavelengths below 200 nm and potentially approaching 100 nm. Such demanding spectral coverage requires an extensive investigation of the optical coatings for the telescope mirrors, which must guarantee high performance across this wide wavelength range.
Currently, the baseline solution adopted for far-ultraviolet (FUV) telescopes consists of aluminum mirrors protected by fluoride coatings, which prevent oxidation while maintaining good transparency from the FUV to the NIR. MgF$_2$ is the most established and reliable protective material, although it introduces a reflectivity cut-off around 115 nm, typically providing more than 65–70\% reflectance near 120 nm while preserving reflectivities above 90\% at longer wavelengths \citep{deMarcos2018}. LiF coatings can extend reflectivities above 40\% down to approximately 100 nm, but their strong hygroscopicity makes it challenging to manage in ground operations \citep{hennessy2016}. More recently, alternative coating technologies based on materials such as AlF$_3$, SiC, and diamond-like carbon (DLC) thin films have been investigated. In particular, AlF$_3$ may extend the high-reflectance range of aluminum mirrors down to approximately 108–110 nm \citep{hennessy2016, Sales2025}. Carbon-based coatings, especially SiC and amorphous tetrahedrally coordinated carbon ultrathin layers, represent a promising new frontier, as they can be used for protecting aluminum from oxidation while preserving high transparency down to 100 nm and potentially below \citep{larruquert2013}.

The joint research group carried out by the National Research Council of Italy - Institute of Photonics and Nanotechnologies in Padova (CNR-IFN) and the University of Padua - Department of Information Engineering (UniPD-DEI) focuses on the development and space qualification of innovative optical coatings, with particular expertise in EUV and FUV normal-incidence mirrors. The group has extensive experience in this field through participation in several past and ongoing space missions, including Solar Orbiter, BepiColombo, and the MUSE mission, complemented by a broad network of collaborations with national and European research institutes and companies specialized in materials science and thin-film deposition. An additional research activity is dedicated to the study, modeling, and mitigation of coating degradation induced by the harsh space environment. In particular, primary mirrors are typically directly exposed to space radiation and continuous irradiation by low-energy particles, such as protons, alpha particles, and electrons, which can progressively degrade optical performance as particle fluence increases \citep{garoli2020}. The CNR-IFN and UniPD-DEI research group has developed extensive expertise in investigating these effects, including through ground-based irradiation experiments using particle accelerators. The main objective is to identify the physical mechanisms responsible for performance degradation and, whenever possible, develop mitigation strategies through optimized coating engineering \citep{corso2024}. These activities are supported by a set of facilities dedicated to the optical and structural characterization of advanced coatings. Available instrumentation includes X-ray diffractometry for crystalline-state studies, spectroscopic ellipsometry for optical constant measurements from the UV to the NIR range, surface morphology characterization, UV-VIS-NIR spectrophotometry, and an in-house reflectometry facility equipped with polarization control for reflectance and transmittance measurements at wavelengths down to 50 nm.

\subsection{High Contrast$^*$}
\label{sec:tec.contrast}
\blfootnote{$^*$For further information feel free to contact Fernando Pedichini \href{mailto:fernando.pedichini@inaf.it}{\nolinkurl{fernando.pedichini@inaf.it}} and/or  Daniele Vassallo 
\href{mailto:daniele.vassallo@inaf.it}{\nolinkurl{daniele.vassallo@inaf.it}}  }%\textit{Text by:F. Pedichini, D. Vassallo}

Italian scientists have a long tradition in the development of the most advanced optical systems for large optical telescopes, in particular those aimed at obtaining high contrast observations. In particular, they have developed key innovations, like adaptive secondary mirrors, \citep{Esposito2010}, large field adaptive optic systems \citep{Ragazzoni2000} or advanced sensors for wavefront sensors \citep{Ragazzoni1996}.

We shall describe here the key technologies developed by Italian teams that can be evaluated for implementation in the HWO context.

\subsubsection{Adaptive optical technologies for a space based telescope}
In a nutshell, these techniques aim at correcting in real time the inevitable optical distortions and deliver an unperturbed wavefront to the optical instruments. While in ground--based telescopes in due  these techniques have been mainly applied to correct for distortions due to the presence of the  atmosphere, requiring corrections of at kHz frequency (so called "adaptive optics"), for HWO the challenge is to improve the - already remarkable - optical quality that is ensured by the ultra--stable space environment and correct the residual aberrations, to an unprecedented level of 3nm.

The possibility to control the wavefront at such level of detail and within the framework of a stable space mission would allow an unprecedent realm of possibilities via wavefront engineering in order to deliberately model the wavefront to perform tasks aimed to the detection and characterization of an exoplanet.

Most of these techniques rely on the concept of engineering the wavefront in such a way to produce a limited region of the PSF where the halo is particularly faint, in order to detect or to examine spectroscopically with a better SNR the light form a close by exoplanet. These ranges from the use of the different statistical behaviour of photons in the core and in the halo of the PSF (also called “dark speckle”, see for instance Boccaletti et al. 1998, 2000) to the creation of dark holes (Give’ion et al., 2007; Potier et al., 2022) also shown to be engineered through the use of a Pyramid Wavefront Sensor (Goulas et al., 2026).

It is interesting to point out that most of these techniques are monochromatic or narrow band by nature. Modifications in order to make them white light, or at least with a significant large wavelength coverage (suitable for instance for spectroscopic examination an not just for exoplanet detection) have been explored so far (see for instance  Soummer et al. 2003). One can, in fact, conceive instrumentation where chromatism is deliberately introduced in order to achieve dark speckles or dark holes in principle in white light, in practice with a wavelength range large enough to be of practical purposes. Further engineering of the wavefront can allow for other approaches (see for instances Herscovici-Schiller et al., 2018)

\subsubsection{Pyramid Wavefront Sensor}

%PSF reconstruction, including for space-based astrophysics, may become a powerful tool to sharpen high-contrast images and reveal faint targets previously embedded in the background. It can work synergistically with ADI, PCA, and other post-processing techniques, most of which are based more on statistics than on linear operators.

The PWFS \citep[Pyramid Wavefront Sensor][]{Ragazzoni1996}), widely adopted on ground-based telescopes, is a pupil-conjugated wavefront sensor that is extremely sensitive to low-mid spatial scales. It provides not only the correction signal needed to keep the optics "on-shape" but also a time series of the residual errors in its telemetry files \citep{Pinna2015}.
In addition to provide feedback needed to correct for the optical distortions, it allows to use telemetry to reconstruct the instantaneous PSF of each image in a series\citep{Simioni2024}. These reconstructed PSFs are then useful for additional sharpening the data during the post-processing deconvolution phase.
An ad-hoc, space-based PWFS could feed an advanced PSF reconstruction algorithm capable of removing background noise even in coronagraphic instruments. This would push their contrast close to nominal specifications ($10^{-9}$–$10^{-10}$) even if the optical quality of the main telescope worsens by a few nanometers. 

\begin{figure}
    \centering
    \includegraphics[width=0.85\linewidth]{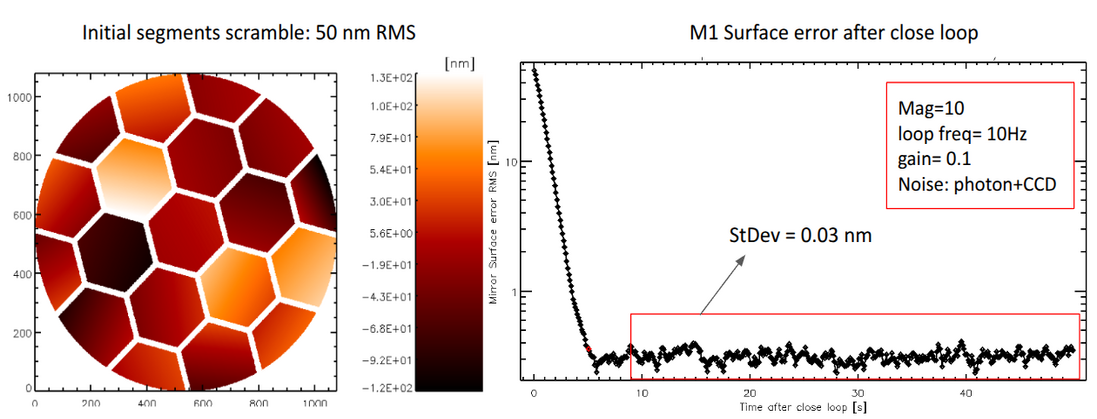}
    \caption{Simulation run of a Pyr WFS driving a 19 segments primary mirror. Left panel: the initial piston-tip-tilt Surface offset of the segments; right panel: residual Surface Error during the closed loop operations, with the indicated parameters.}
    \label{fig:PyrRes}
\end{figure}

A key point here is to speed up as much as possible the WF measurement rate, in order to feed at the fastest frame rate the metrology for the PSF reconstruction. The Pyr WFS, in this regard, offers a very high sensitivity so that a sub-nanometer residual noise is achieved with typical sub-second integration time even of relatively faint (V=10) guide star.
In a preliminary simulation run\cite{Briguglio_2023}, we created a 19 segments mirror with an initial WF offset (differential piston and tip-tilt amongst segments) and closed the loop at 10 Hz on a V=10 guide star; the simulation, which included photon and detector noise, demonstrated that the Pyr WFS recovered the initial WF offset in a bunch of seconds while the residual WF noise was as low as 30 pm STD, as can be seen in Fig. \ref{fig:PyrRes}.

Physical optics simulations  show that if the wavefront (WF) error is known with picometer-level precision, the contrast performance can be recovered to nearly its nominal value. For this reason, we investigated whether a PWFS could achieve the required sensitivity of less than 1 pm.
We have performed a number of  preliminary simulations of a potential "non-modulated PWFS" onboard a space telescope (with a 6–8 m unobstructed pupil) which revealed promising performance. 
\begin{figure}
    \centering
    \includegraphics[width=0.7\linewidth]{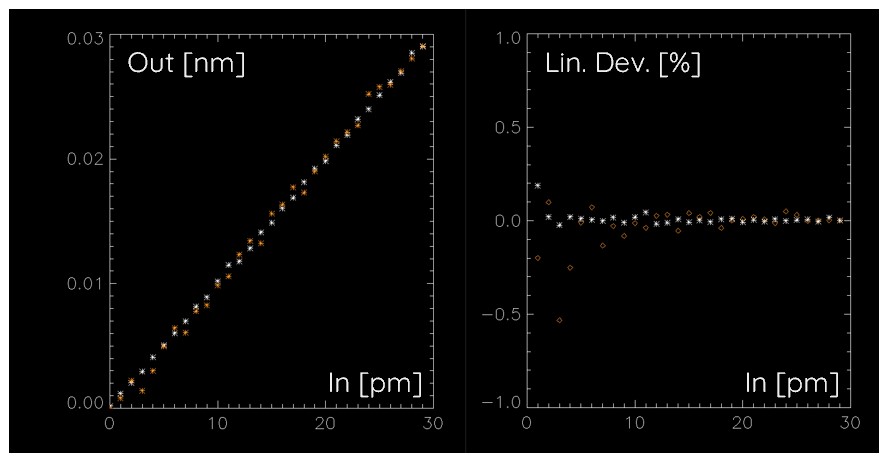}
    \caption{Linearity plot and linearity deviation for the simulated response of a non modulated Pyramid Wavefront Sensor in the picometer range on Zernike Z6. White dots is 60s exposure while Red dots are for 10s only on a magnitude R6 star. The error is always below the 1\%.}
    \label{fig:PYlin}
\end{figure}
The lack of turbulence allows this PWFS to operate in a high-Strehl non modulated regime and with high linearity (see Fig.~\ref{fig:PYlin}). Furthermore, the intrinsic brightness of typical HWO mission targets (stars in the R5 to R8 magnitude range) provides an excellent S/N ratio with reasonable integration times (see Fig.~\ref{fig:PYZ6}). Under these conditions, an integration time of only a few tens of seconds—well below the specified drift time for the opto-mechanics of the telescope and its optical train—yields a WFS S/N ratio so high that the residual WFS RMS error is only about 0.2 pm/mode. Moreover, the system shows exceptional linearity and negligible cross-talk between different Zernike modes.
\begin{figure}
    \centering
    \includegraphics[width=0.7\linewidth]{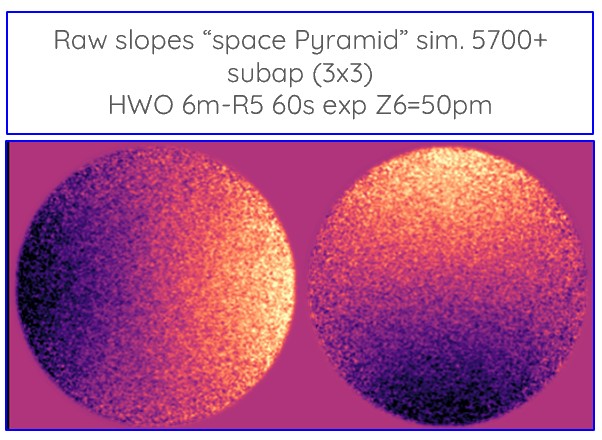}
    \caption{Wavefront slopes from the simulation of a super-sampled, not modulated Pyramid Wave Front Sensor sensing 50pm of Z6 astigmatism. Typical exposure time of 60s for an HWO Rmag 6 target.}
    \label{fig:PYZ6}
\end{figure}
The simulated hardware (5700 3x3 subapertures) is not significantly different from that already used for PYWFS in the most advanced AO systems on large, 8m-class ground-based telescopes \citep{Pinna2016} . Indeed, this technology, already at TRL9 for ground applications, can be smoothly upgraded for space applications. It is an almost static optoelectronic device with no moving parts and can be assembled from already space-qualified components like detectors, lenses, and mounts.

Such a sensitive PWFS may become a key element within a coronagraphic space telescope, such as HWO, for two main reasons: it keeps the telescope "on-shape" with sub-nanometer accuracy (which is mandatory for this kind of mission) and yields precious information for increasing contrast during data post-processing.

\textit{A spaced-qualified AO system }To achieve these goals, a group of INAF  scientist from the INAF Osservatorio Astronomico di  Arcetri and from the INAF Osservatorio Astronomico di Roma has been recently funded within a Italian Space Agency R\&D funding campaign. The goal is to develop a space-qualified AO computer: the missing link between the PWFS and the active/adaptive optics control system of a potential HWO telescope. This unit will be developed using a fast-track approach to design and create a fully reconfigurable layout for the main core processing units. It will be based on space-qualified FPGAs and programmable I/O interfaces to easily integrate different kinds of PWFS, including the advanced Pyramid sensor described above, at a quality level very close to that required for deep space missions.

%In summary, Italian researchers are active on  high-contrast imaging and exoplanet characterization from space, from both scientific and technological perspectives, including the investigation of advanced PSF reconstruction algorithms and the conceptual validation of a space-based high-order PYWFS through both numerical simulations and laboratory prototyping.
\newpage

\subsection{Detectors$^*$}
\label{sec:tec.detectors}
%\textit{Text by: M. Uslenghi}
\blfootnote{$^*$For further information feel free to contact Michela Uslenghi \href{mailto:michela.uslenghi@inaf.it}{\nolinkurl{michela.uslenghi@inaf.it} }}
HWO Wavelength Coverage spans the range 90–2500nm, including the FUV 90–200nm band: in particular, among the proposed instruments there is an UV multi-object spectrograph with FUV imaging. The FUV capability is essential for many HWO science objectives, but from a technological point of view this range is challenging, in particular for detectors. 
Moreover, the development of UV detectors has lagged behind that of IR and visible detectors, partly due to limited commercial and military interest, which in turn leads to reduced funding for R\&D. Astronomical applications also present unique requirements, as the flux of FUV photons from typical astronomical sources is 6 to 9 orders of magnitude lower than that of visible photons, thus, using detectors sensitive to visible photons (such as silicon-based detectors) shifts the problem toward the development of filters with excellent out-of-band rejection while maintaining high in-band transmission—a requirement that is equally challenging for the FUV regime.
MCP detectors have been for a long time the workhorse detectors for UV astronomy, allowing good efficiency in the UV along with excellent solar blindness, with the possibility of operating in photon counting with virtually zero readout noise. Furthermore, if coupled with a suitable readout system, they allow time resolutions down to 10ps, spatial resolutions down to 10 μm, and large sensitive areas, up to 20$\times$20cm. However, these detectors also have well-known drawbacks, including limited dynamic range and lifetime, crucial issues for space missions. Both parameters can benefit from the possibility of operating MCPs at low gains. By combining new production techniques with borosilicate glasses and ALD-Atomic Layer Deposition \citep{Siegmund2013,Ertley2017,Cremer2021}, significant improvements (2-3 orders of magnitude \citep{Lehmann2018}) have been made in terms of dynamic range and lifetime \citep{Cremer2020}). To fully exploit the potential of these new MCPs, however, it is essential to develop a new readout system capable of operating at low gains and high counts.
R\&D is ongoing in INAF and Politecnico di Milano \citep{Pelizzo2021, Uslenghi2022}, aimed to develop a detector based on MCPs with a readout system integrated into an ASIC (Application-Specific Integrated Circuit), MIRA -MIcrochannel plate Readout ASIC \citep{Fabbrica2022}, designed specifically for this application by Politecnico di Milano. 
The MIRA pixel readout chain is composed of the collecting anode, a low-noise charge sensitive amplifier (CSA) with a selectable analog processing time, a filtering stage, a discriminator with a 5-bit selectable threshold, a charge-sharing compensation logic (CSCL) granting a pixel-limited spatial resolution. Finally, two 17-bit counters alternating in parallel granting zero dead time in the serial digital readout.
\begin{figure}[t]
  \centering
  \includegraphics[width=\linewidth]{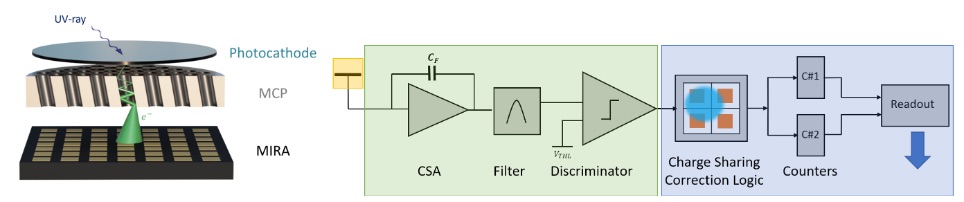}
  \caption{Schematic representation of the PLUS MCP detector with MIRA ASIC readout. \label{fig:Detectors}}
\end{figure}

A first version, 32 x 32 pixels, of MIRA has been realized using a scaled 65 nm CMOS technology in order to achieve the compact pixel size requirement with a 32\% Fill Factor \citep{Uslenghi2024}. After functional verification, the device has been integrated with the MCP and the prototype has been tested \citep{Farina2025}.
The main advantages over traditional MCP detectors include:
\begin{itemize}
    \item High detector lifetime and gain stability: the low noise (ENC$\sim$25e-) of the readout electronics allows the use of 3 to 6 orders of magnitude lower gain than traditional detectors
    \item High dynamic range, up to $10^5$ photons/s/pixel
    \item Use of lower voltage levels
    \item Compactness: Most part of the electronics is on--chip, integrated into the readout ASIC.
\end{itemize}
A new version of MIRA has been designed to enable the future upscaling of readout formats \citep{Nassi2024,Farina2026} and will be tested shortly.

\subsection{Artificial Intelligence$^*$}
\label{sec:tec.AI}
\blfootnote{$^*$For further information feel free to contact  Marco Landoni \href{mailto:marco.landoni@inaf.it}{\nolinkurl{marco.landoni@inaf.it}} and/or Tiziano Zingales \href{mailto:tiziano.zingales@unipd.it}{\nolinkurl{tiziano.zingales@unipd.it}}  }
%\textit{Text by: M. Landoni, T. Zingales}

Unlike transmission spectroscopy, direct-imaging spectroscopy couples the planet–star illumination geometry (phase), surface/cloud albedo structure, thermal emission, and the planet radius into the measured spectrum. These effects must be modeled within the retrieval to avoid biased parameter estimates \citep{Whiteford2023}, with particular care for wavelength–correlated residuals from high-contrast Integral Field Spectroscopy (IFS) extraction \citep{GrecoBrandt2016}. The methods below are tailored to the reflected-light, coronagraphic regime anticipated for HWO.
We propose the use of Transformers and Generative networks to improve atmospheric analysis on HWO spectra. In order to train the machine learning methods, we make use of a syntethic spectral dataset on large grids of atmospheric parameters. The synthetic data set can be built with the TauREx code \citep{TauREx2021} using atmospheric direct imaging models. 

\subsubsection{Atmospheric retrieval with informative priors and Transformers}
\label{subsec:transformer_hwo} 

Bayesian atmospheric retrievals for direct-imaging spectra are computationally intensive: each likelihood evaluation requires radiative-transfer forward modeling, and robust samplers (e.g., nested sampling) must explore high-dimensional, degenerate spaces—often necessitating $10^5$–$10^6$ model calls. Machine-learning accelerators—such as Transformer encoders used as learned likelihood/posterior surrogates and enable fast estimation of best-fit parameters and can provide even order of magnitude speedups, while preserving calibrated uncertainties in a hybrid ML+Bayesian framework \citep{Pagliaro2026}. In direct imaging techniques, inputs are coronagraphic IFS spectra $\mathbf{y}$ in reflected light, sampled on wavelengths $\lambda_{1:M}$ from the instrument reduction pipeline, similarly to \citet{GrecoBrandt2016}. Transformer encoder can couple input direct imaging spectra to an ensemble of reflected-light atmospheric parameters (molecular mixing ratios, surface/albedo classes, T--P controls, multi-layer clouds/hazes, reference radius, phase angle).
We propose the introduction of informative prior distributions to the Bayesian sampler using transformers \citep{Vaswani2017}
The target posterior is
\[
p(\boldsymbol{\theta}\mid \mathbf{y}) \propto \mathcal{N}\!\big(\mathbf{y}\,;\, \boldsymbol{\mu}_{\theta}\big)\; p(\boldsymbol{\theta}\mid \boldsymbol{\eta}),
\]
where $\boldsymbol{\mu}_{\theta}$ is the mean atmospheric parameter predicted by the Transformer and $p(\boldsymbol{\theta}\mid \boldsymbol{\eta})$ encodes informative priors. We adopt priors on: (i) radius/mass (from RV/astrometry when available) to break radius–albedo degeneracies; (ii) metallicity and C/O informed by stellar abundances/population synthesis; (iii) cloud vertical/optical properties and surface types consistent with reflected-light forward models. With the help of informative priors thanks to transformers techniques we may have significant speedups in atmospheric retrievals. On HWO-like simulations, informative priors mitigate credible-interval inflation at low S/N and help disentangle phase–radius–albedo degeneracies.

\subsubsection{Atmospheric detection and retrieval with Generative Networks}
\label{subsec:exogan_hwo}

We adopt the ExoGAN paradigm—training a deep generative model on paired spectra/parameters and casting retrieval as semantic inpainting on a joint array representation and train it on direct imaging forward models (geometric albedo spectra), conditioning on phase angle, radius priors, cloud/haze morphology. The generator acts as a fast surrogate for the spectral manifold; the discriminator regularizes toward physically plausible spectral solutions \citep{Zingales2018}. Post-training, we perform retrievals by optimizing latent variables to match the observed spectrum while remaining on-manifold; and (iii) use the generative posterior as a proposal/prior to accelerate Bayesian samplers, yielding large speedups for scan/triage while preserving traceability via hybrid hand-offs to full Bayesian inference \citep{Zingales2018}. All models are trained with HWO-like contrast, bandpass, and spectral resolution assumptions and incorporate realistic coronagraph noise terms \citep{Robinson2015PASP,CoronagraphPkg2019}.

\subsubsection{Quantum Machine Learning techniques}
Quantum Extreme Learning Machines (QELM) use a fixed, randomly initialized quantum reservoir (e.g., a short-time evolution of coupled qubits or continuous–variable modes) to generate a rich nonlinear feature map from inputs; only a linear readout is trained, yielding fast, stable fitting and making QELM attractive as for atmospheric retrieval components or proposal distributions in Bayesian samplers \citep{Vetrano2026,Mujal2021,FujiiNakajima2017}. Atmospheric retrievals using QELM are a game changer in atmospheric retrievals and may lead, especially in the 2040s when the technology will be mature enough, to a new class of fast retrievals with complex atmospheric models \citep{Vetrano2026}. In parallel, quantum generative adversarial networks (qGANs) parameterize a quantum generator (and, in some designs, a quantum discriminator) to learn the joint distribution of spectra and atmospheric parameters; this supports likelihood-free inference and rapid posterior proposal generation \citep{LloydWeedbrook2018,DallaireKilloran2018,Hu2018qGAN}. Training and evaluation of such quantum models can further leverage core quantum subroutines—for instance, Quantum Amplitude Estimation to quadratically speed up expectation estimation that appears in loss functions and detection metrics—providing a principled path to end-to-end acceleration once error-corrected hardware is available \citep{Brassard2002QAE}. This kind of application consist of a generatization of what demonstrated in \citep{Zingales2018} using quantum computing.

\subsubsection{Machine Learning techniques for direct imaging}
Direct imaging of exoplanets is a fundamental technique to enhance our understanding on the formation, composition and Physics of  exo-planetary systems. However, the faintness of the exoplanet signal with respect to the brightness of its host stars poses a significant challenges. In the past, many authors proposed a significant number of algorithms \citep[e.g.][]{flasseur24} to overcome the problem by combining statistical methods with deep learning and machine learning techniques. Briefly, those methods rely on the capabilities of machine learning and deep learning methods to detect the faint signal and structures imprinted by the presence of the exoplanet in the noise of the residuals of the subtracted speckle of coronagraph images. Many of these approaches combine both images from scientific detectors and data coming from wavefront sensors, telemetry, etc and frequently, to overcame the scarcity of the data to train and build models, synthetic dataset are generated by injecting on the observed data the expected PSF of the exoplanet at different contrast ratio. 
\\
Within this framework, we propose to apply and enhance deep learning techniques in the context of HWO by developing then along with the state of the art concept of the payload mission for the coronagraph to better assess the capability of the instrument to reach the 10$^{-9-10}$. Data augmentation for the training dataset, which currently rely mostly on classical method, will be based on both laboratory data and enhanced with the most advanced generative AI algorithms.

\newpage

\section{Instruments}
\label{sec:instrumentconcepts}

This section presents \textit{concepts} of possible instruments for the HWO focal plane, either served by the coronagraph or outside it. In most cases, they  derive directly from the science priorities presented in Section 2. We emphasize that, given the early stage of development of the mission,  none of these ideas can be defined as mature instrument design: rather than a collection of detailed proposals, this section is more a \textit{manifesto} of the ambition to participate in or to lead a specific concept. 

Before going into the details, it might be useful to summarize the connection between the science cases presented in Sect.~\ref{sec:sciencecases} and the various instrument categories, and  to identify \enquote{Contact Persons} for each instrument class. We encourage anyone interested in further collaborations about the development of these instrument cases to contact these persons directly.

\begin{tcolorbox}[
    colback=violet!10,     % background color of the box interior
    colframe=violet!50,      % color of the frame/border
    title={Post--coronagraphic instrumentation},      % optional title
    fonttitle=\scshape\bfseries    % optional: bold title
]
This is clearly the most interesting option for many science cases related to exo--planet science, but also for the study of star--forming regions:

$\bullet$ \enquote{Reflected-light diversity: giant, Neptune- and sub-Neptune-class planets.} (Sect.\ref{sec:2.2.2}) \\
$\bullet$ \enquote{Exomoons and binary planetary companions.} (Sect.\ref{sec:2.2.3}) \\
$\bullet$ \enquote{Molecular characterization and robust biosignature detection.} (Sect.\ref{sec:2.2.7}) \\
$\bullet$ \enquote{Surface, cloud and ocean diagnostics.} (Sect.\ref{sec:2.2.8}) \\
$\bullet$ \enquote{Planet-forming disks} (Sect.\ref{sec:2.3.5}) \\
$\bullet$ \enquote{Accretion on forming planets} (Sect.\ref{sec:2.3.6})

In particular, there is wide interest in developing imagers and especially low-resolution (R=100-1000) spectrographs for faint exoplanets, as further discussed in \ref{sec:strum.coro_instruments}. 

\smallskip
\textcolor{blue}{\bf Contact persons}:Matteo Brogi \href{mailto:matteo.brogi@unito.it}{\nolinkurl{matteo.brogi@unito.it}} and/or Lorenzo Pino  \href{mailto:lorenzo.pino@inaf.it}{\nolinkurl{lorenzo.pino@inaf.it}} and/or Fernando Pedichini  \href{mailto:fernando.pedichini@inaf.it}{\nolinkurl{fernando.pedichini@inaf.it}}
\end{tcolorbox}

\begin{tcolorbox}[
    colback=violet!10,     % background color of the box interior
    colframe=violet!50,      % color of the frame/border
    title={MOS and IFU},      % optional title
    fonttitle=\scshape\bfseries    % optional: bold title
]
These are  two \enquote{workhorse} capabilities that are demanded by a high number of science cases in most of the scientific areas listed in this document. We group them together as they are often conceived as different capabilities of a single, multi-purpose instrument. The science cases that would benefit of these facilities are:\\ 
$\bullet$ \enquote{Accretion onto young stars.} (Sect.\ref{sec:2.3.1})\\
$\bullet$ \enquote{UV Magnetic Activity and Disk Regulation Across the Stellar-Planetary Boundary.} (Sect.\ref{sec:2.3.2})\\
$\bullet$ \enquote{Jet launching regions around young stars.} (Sect.\ref{sec:2.3.3})\\
$\bullet$ \enquote{Collimated jets and wide-angle outflows.} (Sect.\ref{sec:2.3.4})\\
$\bullet$ \enquote{Photoionization Feedback from HII Regions.} (Sect.\ref{sec:2.3.7})\\
$\bullet$ \enquote{The Origin of the UV Upturn in ETGs and Spiral Bulges} (Sect.\ref{sec:2.4.3})\\
$\bullet$ \enquote{Massive Stars in Extremely Metal-Poor Environments} (Sect.\ref{sec:2.4.4})\\
$\bullet$ \enquote{Massive Galaxies} (Sect.\ref{sec:2.5.1})\\
$\bullet$ \enquote{Scaling Relations and the stellar Initial Mass Function} (Sect.\ref{sec:2.5.2})\\
$\bullet$ \enquote{Origin of Ultra-Diffuse Galaxies} (Sect.\ref{sec:2.5.3})\\
$\bullet$ \enquote{The escape of Lyman continuum radiation: a key ingredient to understand reionizaton} (Sect.\ref{sec:2.5.4})\\
$\bullet$ \enquote{The connection among black-hole accretion, feedback, and host-galaxy evolution across all epochs} (Sect.\ref{sec:2.5.6})\\
$\bullet$ \enquote{Dual AGN} (Sect.\ref{sec:2.5.7})\\
$\bullet$ \enquote{Distance ladders} (Sect.\ref{sec:2.6.1})\\
$\bullet$ \enquote{Supernova science with HWO} (Sect.\ref{sec:2.7.1})\\
$\bullet$ \enquote{HWO and Multi-Messenger Astronomy} (Sect.\ref{sec:2.7.2})

\smallskip
We remark that several of these science cases rely on the extension of the wavelength range as far as possible into the near--IR regime, at least up to $\lambda \sim 1.7\mu$m and possibly to $\lambda \sim 12.5\mu$m. In all cases, they request a spatial sampling adequate to exploit the diffraction limit of HWO at most wavelength. See Sect.\ref{sec:strum.SHOS} for more details.

\smallskip

\textcolor{blue}{\bf Contact person}: Paolo Saracco \href{mailto:paolo.saracco@inaf.it}{\nolinkurl{paolo.saracco@inaf.it}}
\end{tcolorbox}

\begin{tcolorbox}[
    colback=violet!10,     % background color of the box interior
    colframe=violet!50,      % color of the frame/border
    title={High resolution, wide field imager with astrometric capabilities},      % optional title
    fonttitle=\scshape\bfseries    % optional: bold title
]
This \enquote{workhorse} capability is clearly highly demanded across the whole range  of science cases, many of which request UV sensitivity and/or tight astrometric capabilities:\\
$\bullet$ \enquote{Detection, orbits, true masses and system architecture.} (Sect.\ref{sec:2.2.1}) \\
$\bullet$ \enquote{Photoionization Feedback from HII Regions.} (Sect.\ref{sec:2.3.7}) \\
$\bullet$ \enquote{Tracing galaxy evolution with resolved stars} (Sect.\ref{sec:2.4.1}) \\
$\bullet$ \enquote{Reaching the Oldest Stellar Populations in Giant Elliptical Galaxies} (Sect.\ref{sec:2.4.2}) \\
$\bullet$ \enquote{The escape of Lyman continuum radiation: a key ingredient to understand reionizaton} (Sect.\ref{sec:2.5.4}) \\
$\bullet$ \enquote{Astrometry to shed light on the nature of dark matter} (Sect.\ref{sec:2.6.2}) \\
$\bullet$ \enquote{Supernova science with HWO} (Sect.\ref{sec:2.7.1})

\smallskip
More details and background information are provided in Sect.~\ref{sec:instr_astrometry}
\textcolor{blue}{\bf Contact person}: Alessandro Sozzetti \href{mailto:alessandro.sozzetti@inaf.it}{\nolinkurl{alessandro.sozzetti@inaf.it}}
\end{tcolorbox}

\bigskip

\begin{tcolorbox}[
    colback=violet!10,     % background color of the box interior
    colframe=violet!50,      % color of the frame/border
    title={High Resolution Spectrograph},      % optional title
    fonttitle=\scshape\bfseries    % optional: bold title
]
This instrument is an obvious complement to the suite of already planned HWO instruments, and is unsurprisingly required by many science cases that are listed below:\\ 
$\bullet$ \enquote{Transiting-planet atmospheres.} (Sect.\ref{sec:2.2.4}) \\
$\bullet$ \enquote{Mass loss from planetary atmospheres.} (Sect.\ref{sec:2.2.5}) \\
$\bullet$ \enquote{Volcanism} (Sect.\ref{sec:2.2.9}) \\
$\bullet$ \enquote{Host-star high-energy environment and habitability.} (Sect.\ref{sec:2.2.10}) \\
$\bullet$ \enquote{Accretion on forming planets} (Sect.\ref{sec:2.3.6}) \\
$\bullet$ \enquote{Exploration of r-Process Nucleosynthesis} (Sect.\ref{sec:2.4.5}) \\
$\bullet$ \enquote{Tracing the First Stars} (Sect.\ref{sec:2.4.6}) \\
$\bullet$ \enquote{Probing Cosmic-Ray Spallation with boron} (Sect.\ref{sec:2.4.7}) \\
$\bullet$ \enquote{Measuring chemical abundances in the Reionization epoch.} (Sect.\ref{sec:2.5.5}) \\
$\bullet$ \enquote{Fundamental constants} (Sect.\ref{sec:2.6.3}) \\
$\bullet$ \enquote{Redshift drift} (Sect.\ref{sec:2.6.4}) \\
$\bullet$ \enquote{Reionization of He II} (Sect.\ref{sec:2.6.5}) \\
$\bullet$ \enquote{HWO and Multi-Messenger Astronomy} (Sect.\ref{sec:2.7.2})
\smallskip
We note that in several cases a resolution as high as $R\sim 100.000$ or above is not needed, being an intermediate $R\sim20.000$ adequate for some of this cases. We also remark that several of these science cases rely on the extension of the wavelength range as far as possible into the near--IR regime, at least up to $\lambda \sim 1.7\mu$m and possibly to $\lambda \sim 12.5\mu$m. 

\smallskip

\textcolor{blue}{\bf Contact person}: Valentina D'Odorico  \href{mailto:valentina.dodorico@inaf.it}{\nolinkurl{valentina.dodorico@inaf.it}} 
\end{tcolorbox}

\bigskip

\begin{tcolorbox}[
    colback=violet!10,     % background color of the box interior
    colframe=violet!50,      % color of the frame/border
    title={Polarimetric Capabilities},      % optional title
    fonttitle=\scshape\bfseries    % optional: bold title
]
Finally, a few science cases require polarimetric capabilities installed in the HWO instrumentation, as listed below:\\ 
$\bullet$ \enquote{Rotational variability.} (Sect.\ref{sec:2.2.6}) \\
$\bullet$ \enquote{Planet-forming disks} (Sect.\ref{sec:2.3.5}) \\
$\bullet$ \enquote{The connection among black-hole accretion, feedback, and host-galaxy evolution across all epochs} (Sect.\ref{sec:2.5.6}) \\
$\bullet$ \enquote{HWO and Multi-Messenger Astronomy} (Sect.\ref{sec:2.7.2})
\smallskip

More detail about innovative techniques to obtain polarimetric measurements are provided in Sect.~\ref{sec:strum.polarimetry}

\textcolor{blue}{\bf Contact person}: Andrea Vogliardi \href{mailto:andrea.vogliardi@phd.unipd.it}{\nolinkurl{andrea.vogliardi@phd.unipd.it}} and/or Gianluca Ruffato \href{mailto:gianluca.ruffato@unipd.it}{\nolinkurl{gianluca.ruffato@unipd.it}} and/or Filippo Romanato \href{mailto:filippo.romanato@unipd.it }{\nolinkurl{filippo.romanato@unipd.it }} 
\end{tcolorbox}

\bigskip
We defer to the next sections for a more detailed description of most of these instrument proposals. 

\newpage

\subsection{An infrared spectrograph for the coronagraph Instrument (CI)$^*$}
\label{sec:strum.coro_instruments}
\blfootnote{$^*$For further information feel free to contact  Matteo Brogi \href{mailto:matteo.brogi@unito.it}{\nolinkurl{matteo.brogi@unito.it}} and/or Lorenzo Pino  \href{mailto:lorenzo.pino@inaf.it}{\nolinkurl{lorenzo.pino@inaf.it}} and/or Fernando Pedichini  \href{mailto:fernando.pedichini@inaf.it}{\nolinkurl{fernando.pedichini@inaf.it}}}

As explained in section \ref{sec:sci.exoplanets}, the main goal of HWO for exoplanet science is to characterize the atmospheres of a sizeable sample of terrestrial temperate exoplanets, preferably around solar-type stars (see Fig. \ref{fig:PCS_rocky_planet}). To fulfill this mission, the main coronagraph of HWO is required to provide a raw contrast of 10$^{-10}$ relative to the star, and through post-processing a 1$\sigma$ photometric contrast of 10$^{-12}$.
To provide proper context, this requirement almost aligns with the performance goals of the Roman Space Telescope/CGI (ranging from $10^{-7}$ to $10^{-9}$), where Hybrid Lyot and Shaped Pupil coronagraphs are integrated with highly sophisticated wavefront control strategies.

It seems inevitable that for a {\sl detection} of the exoplanet itself and for tracing its orbit accurately, a photometric instrument will be needed. However, photometric capabilities are insufficient for atmospheric characterization, and thus an imager alone cannot obtain the main science goal of the mission. Spectroscopic capabilities are the only route to determine the nature of the temperate exoplanet, including the likelihood of hosting life.

Here we propose to capitalize on the significant experience within the Italian community, and have the Italian community explore and lead the design of a spectrograph directly fed by the main HWO coronagraph. While absolutely open to any international collaboration, we believe the Italian community can build  a coherent, national leadership thanks to the unique experience in both scientific and technological aspects. 

%However, to provide proper context, this requirement almost aligns with the performance goals of the Roman Space Telescope/CGI (ranging from $10^{-7}$ to $10^{-9}$), where Hybrid Lyot and Shaped Pupil coronagraphs are integrated with highly sophisticated wavefront control strategies.
%At the time of writing, novel architectural solutions are being explored, with the concrete possibility that the design and development of a Near-Infrared (NIR) coronagraphic instrument may be opened to the international community through a competitive call.

In the next two subsections we detail our path to the development of the case for a spectrograph directly fed by the main HWO coronagraph, that includes two synergic activities. The first subsection describes the activity ongoing to form a team willing and able to explore the science case and the design requirements for the instrument. 
In the second, we describe an innovative technological solution to avoid the usual drawbacks of standard coronagraphs (namely the chromatic dependence) that we are actively pursuing.

\subsubsection{The path to a coronagraph-fed spectrograph}
The Italian community has about a decade of experience in spectroscopic observations of exoplanets, especially with ground based instrumentation. This expertise stretches from optical to near-infrared wavelength, and includes both identification of species and measurement of their abundances through inversion techniques (retrievals) based on Bayesian Analysis. Furthermore, our community has a proven track record of leading design of spectroscopic instruments. The exoplanet community is also fairly united and compact, in particular thanks to the experience of the Global Architectures of Planetary Systems (GAPS) team. In more recent times, involvement into the design of the MORFEO, ANDES and PCS instruments at the Extreme Large Telescope has further reinforced the leading role of our community in shaping the science of the 2030s.

The main points to be explored would be the spectral resolving power and spectral range needed to produce unambiguous detection of species and measurement of their abundances. While arguably limited to $R=50-100$ by the small photon budget of the targets, spectral resolving power has nevertheless been shown to be a key discriminant in current simulations of possible HWO designs, and there is enough literature to start an informed study and thus choose an equally informed design.

Directly porting from ground-based studies, molecular mapping at intermediate resolving powers $R=1500-5000$ should also be explored, because cross correlation techniques can be used to filter and isolate the exoplanet signature from the underlying noise, generally providing an amplification in contrast proportional to the square root of the available spectral lines. This is not a mainstream technique worldwide, but Italy and in general the EU are leading on this front, thanks to the development of instruments for the ESO Extremely Large Telescope which will function on the same principles.

The spectral range should also be carefully evaluated. Limiting ourselves to optical wavelengths seems reasonable given the  coronagraphic designs typically assume optimal performances at $\lambda \sim 550$ nm and the corresponding spatial resolution scales linearly with the wavelength of the observations. However, key spectral information necessary to unambiguously claim biomarkers can be found at near-infrared wavelengths, particularly the signatures of H$_2$O, CH$_4$, and possibly CO and CO$_2$. Oxygen alone is not considered a robust biomarker, and even considering the time evolution of the Earth's atmosphere it would be desirable to measure and recognize other biotic scenarios where O$_2$ is not dominant. Thus, the benefit or hindrance of adopting a larger spectral range {\it even if the telescope is not optimized for the IR} should be carefully evaluated.

We are assembling a team composed of a science working group (defining the sample, the goals, and the measurements) and a technology team exploring the instrument solutions. At the same time, the strong international collaborations of the Italian community in atmospheric characterization will be exploited to secure partners and ensure the synergy with the development of the coronagraph and main imager. It should also be explored whether the simultaneous development of the EU-based LIFE mission will be likely, as HWO and LIFE has been shown to yield the maximum science return when flying together and combining optical + IR observations of a common sub-set of the sample of temperate exoplanets. 

Finally, we plan to explore the possibility of linking with a fiber the coronagraph with a high-resolution, near-infrared spectrograph to be proposed for bay E, which we describe in the next section. In the fibre-fed configuration, the bay E spectrograph is fed by the CI-NIR, whose 1.0--1.7~$\mu$m range and optics temperature restrict its coverage to the J and H bands, excluding the K band without further adaptation of the project. The fibre-fed high dispersion coronagraphy variant additionally requires a cross-instrument (from CI to E) architecture not foreseen in the current guidelines, which we would propose as an additional capability to be evaluated. This configuration would respond to a scientific need that has been recently acknowledged by \cite{Jaffe2026}.

\begin{figure}
\includegraphics[width=0.35\linewidth]{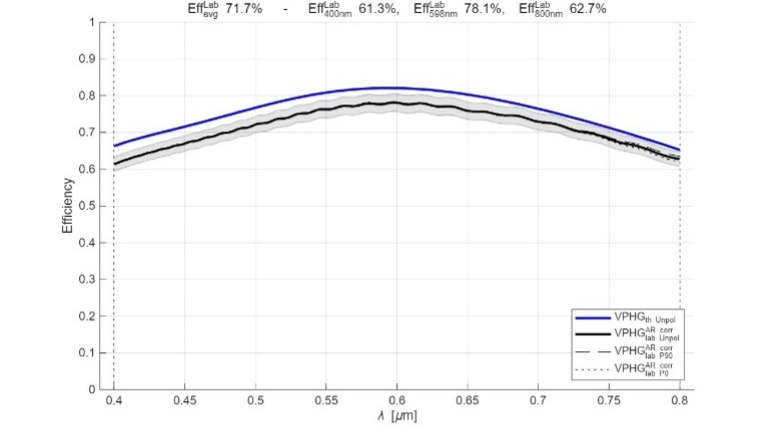}
\includegraphics[width=0.3\linewidth]{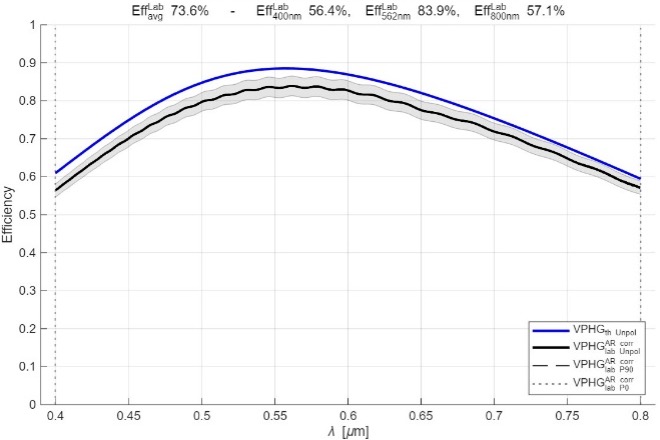}
\includegraphics[width=0.35\linewidth]{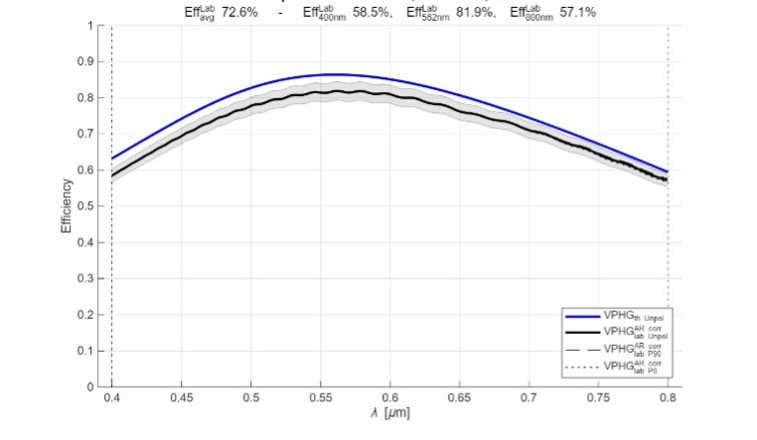}
\caption{Three different VPH simulations (courtesy of the INAF HOLOLAB: Bianco A., Frangiamore M.) showing that despite the different number of lines/mm (300, 400 and 600) there is always a good throughput along the full VIS-NIR band for theoretical R in the order of a few thousand in the case of a compact instrument.
}
\label{fig:CoroSpec1}
\end{figure}

\begin{figure}
\centering
\includegraphics[width=0.45\linewidth,keepaspectratio]{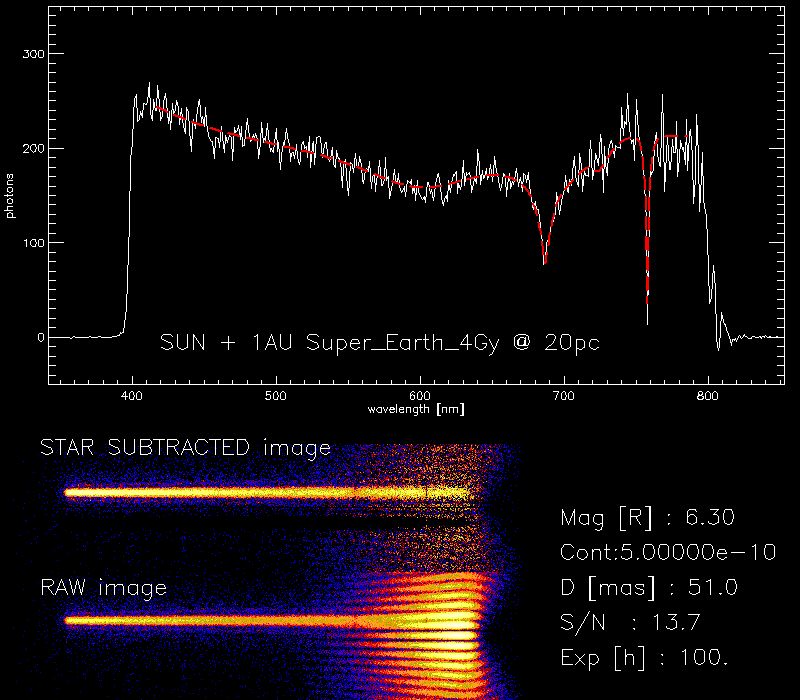}
\includegraphics[width=0.45\linewidth,keepaspectratio]{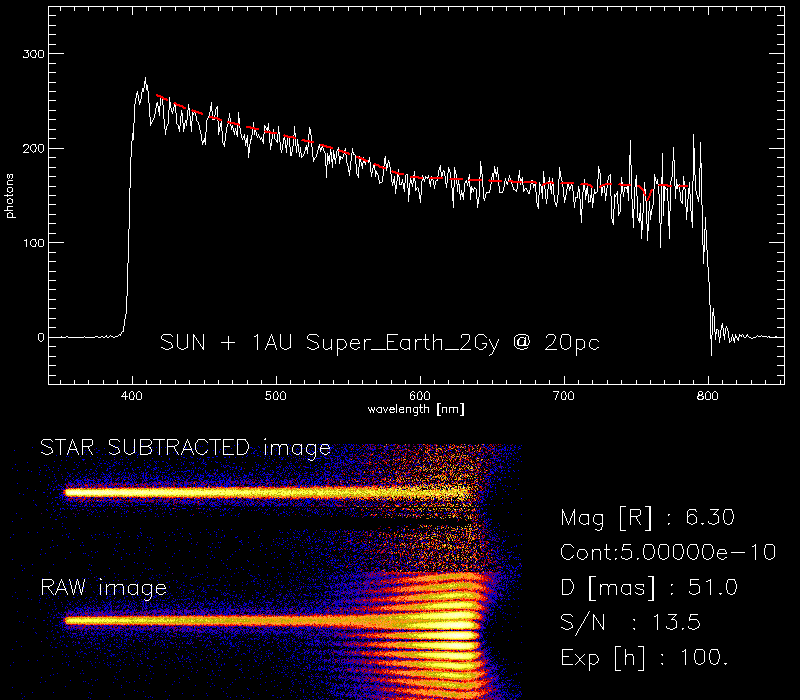}

\caption{Simulated detector images and extracted spectra at a contrast of $5\times10^{-10}$ for long exposures using our Spectro-Coronagraph model, showing a Super-Earth orbiting at 1~AU from a G2V star at a distance of 20~pc. Left: An evolved planet exhibiting a prominent atmospheric blue Rayleigh-scattering slope, $\text{O}_2\text{-B}$ and $\text{O}_2\text{-A}$ absorption bands, and a far-red edge attributed to chlorophyll, providing strong evidence for life. Right: The same Super-Earth at a younger evolutionary stage (2~Gyr), where bio-signatures are not yet present in the spectrum. The input planetary spectra used in the simulations are overlaid in dashed red and served to calculate the signal-to-noise ratio ($S/N$). The raw images—initially degraded by residual stellar halo leakage, particularly at longer wavelengths, have been cleaned by exploiting the spatial symmetry of the instrument point spread function (PSF).}
\label{fig:CoroSpec4}
\end{figure}

%These simulations indicates that low-resolution spectroscopy, performed by this kind of instrument, can analyze the light reflected by exoplanets at contrast levels fainter than $10^{-9}$ discriminating different model of atmospheres. We have recently initiated the design and development phase of this high-contrast spectroscopy concept, where significant optimization remains possible by designing well beyond a basic Lyot coronagraph layout while exploiting the very chromatic dispersed PSF of a spectrograph focal plane. Furthermore upon the availability of next generation detectors as the Superconducting Nanowire Single photon detector (sensitive from 200nm to several micron with dark current of less than 1e-/pixel per day \citep{2025jpl..data.2153W}, \citep{2024SPIE13103E..04H}) and appropriate dispersing elements this concept may be extended to different or larger bands up to a disruptive UV-VIS-NIR-SWIR spectro-coronagraph the Holy Grail for the Exoplanet science. INAF is actively seeking to expand this collaboration to optimize the simulation, layout new design and build prototypes for their verification both within its own laboratories and at other European or international high-contrast testbed facilities.

\subsubsection{The Spectroscopic Coronagraph Concept $^*$}
\label{sec:spectro_coronagraph}
\blfootnote{$^**$For further information feel free to contact Fernando Pedichini  \href{mailto:fernando.pedichini@inaf.it}{\nolinkurl{fernando.pedichini@inaf.it}}}

Within this framework, we propose a novel approach to achieve high-contrast, high-spatial-resolution spectroscopy over broad spectral bandwidths (an octave or more) by introducing the concept of a \textbf{Spectroscopic Coronagraph}. By reversing the traditional architecture—positioning the dispersion element \emph{upstream} of the coronagraphic mask rather than downstream—this instrument aims to overcome the intrinsic bandwidth limitations of conventional coronagraph designs. 

The feasibility of this concept relies on three critical prerequisites:
\begin{enumerate}
    \item \textbf{High-Efficiency Dispersive Optics:} The availability of ultra-wideband dispersion elements, such as the Volume Phase Holographic (VPH) gratings produced at the INAF HoloLab (Fig.~\ref{fig:CoroSpec1}), which can be tailored to meet the stringent requirements of flagship space missions like HWO
    \citep{2025SPIE13529E..0FB}.
    \item \textbf{Extreme Wavefront Control:} An ultra-stable active optics system capable of sub-nanometer wavefront error stability, achieved by integrating an extreme Pyramid Wavefront Sensor (PWFS) as described in Section~\ref{sec:tec.contrast}.
    \item \textbf{End-to-End End Simulations:} Physical optics modeling to evaluate realistic performance limits on challenging exoplanetary targets.
\end{enumerate}

To validate this architecture, we performed preliminary physical optics simulations using the \textsc{PROPER} library \citep{Krist2007}. In our simplified setup, a wideband slit spectrograph ($\lambda = 400\text{--}800~\text{nm}$, $R \approx 600$) is aligned along the star--planet separation vector. The dispersed spectrum of the host star is occulted at the intermediate focal plane by a linear Gaussian transmission mask and subsequently collimated. At the downstream pupil plane, a undersized Lyot stop suppresses the diffracted starlight before a camera lens re-images the off-axis planetary spectrum onto the detector.

Despite its current unoptimized design and modest overall throughput ($\leq 20\%$) due mainly to a simple circular Lyot stop, end-to-end simulations incorporating photon noise demonstrate that a Signal-to-Noise Ratio ($\text{S/N} \geq 10$) is achievable across the full VIS--NIR band in a 50-100 hour integration time (Fig.~\ref{fig:CoroSpec4}). This performance applies to targets at flux contrasts fainter than $10^{-9}$ with angular separations as small as $50~\text{mas}$ from an $R = 6.3~\text{mag}$ host star. Such parameters correspond to a Super-Earth with a contrast of $5x10^{-5}$ at $1.0~\text{AU}$ around a G2V star at $20~\text{pc}$ i.e. 5x enhanced compared to an Earth around a solar-type star.

These preliminary results confirm that low- to medium-resolution spectroscopy with a Spectroscopic Coronagraph can effectively characterize exoplanetary atmospheres at extreme contrast ratios ($< 10^{-9}$). Significant performance gains are expected as we move beyond a basic Lyot mask by taking advantage of the chromatic Point Spread Function (PSF) dispersion at the focal plane. 

Furthermore, integrating next-generation detectors—such as Superconducting Nanowire Single-Photon Detectors (SNSPDs), which operate from $200~\text{nm}$ to the SWIR with dark currents below $10^{-4}~e^-\,\text{pixel}^{-1}\,\text{day}^{-1}$ \citep{2024SPIE13103E..04H, 2025jpl..data.2153W}—will allow this concept to span a continuous UV-to-SWIR coverage. INAF is actively expanding international collaborations to optimize numerical models, design advanced optical layouts, and test physical prototypes both in-house and at European/international high-contrast testbed facilities.

\subsection{Extreme-precision astrometry solutions for HWO$^*$}\label{sec:strum.astrometry}
\label{sec:instr_astrometry}
\blfootnote{$^*$For further information feel free to contact  Deborah Busonero \href{mailto:deborah.busonero@inaf.it}{\nolinkurl{deborah.busonero@inaf.it}} and/or Alessandro Sozzetti \href{mailto:alessandro.sozzetti@inaf.it}{\nolinkurl{alessandro.sozzetti@inaf.it}}  }%\textit{Text by: D. Busonero, A. Sozzetti}

Based on the preliminary specifications of the mission profile, one of the candidate focal-plane instruments for HWO will be a high-angular-resolution imaging camera spanning the UV, VIS, and NIR channels, with a typical field of view of $6-10$ arcmin$^2$. In the event that precursor science efforts do not deliver ahead of time a sample of bona-fide Earth-like planet candidates, it would be necessary for HWO itself to be able to perform observations to obtain (see Section \ref{sec:sci.exoplanets}) precise and accurate orbit and mass determinations for such companions. This could be achieved by either tailoring the design of the imaging camera to accommodate the capability to perform high-precision astrometry (albeit at the cost of the relatively modest field of view) or through addition of a fourth instrument onboard specifically designed for the purpose. 

\begin{figure}[ht]
\begin{center}
\includegraphics[width=0.95\textwidth]{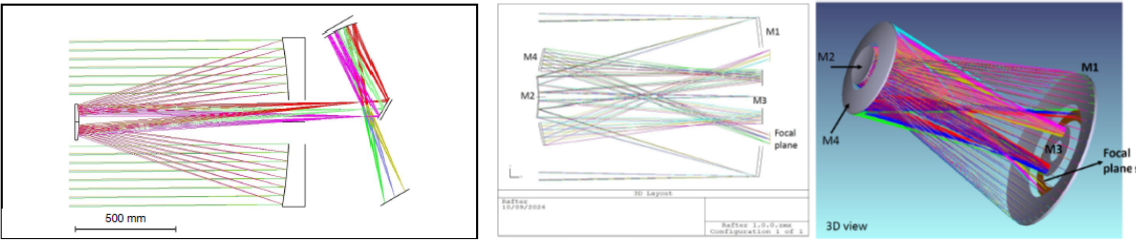}
\caption{\small Left: proposed optical design of the {\it Theia} concept (see \citealt{Malbet2022} for details), featuring a diffraction-limited telescope with a large field of view (FOV) and a metrology system for the focal plane detector array. Similar to Euclid, {\it Theia} uses an off-axis Three-Mirror Anastigmat (TMA) configuration with a primary mirror of 0.8m, an effective focal length (EFL) of 13 m, and a field of fiew (FOV) of $0.60\times0.76$ deg, operating in the 400 - 900 nm wavelength range. Right: proposed optical design for the {\it RAFTER} concept (see \citealt{Riva2020,Gai2022} for details), featuring an on-axis TMA configuration with annular mirrors (annular-field telescope assembly - TA), introducing an optimization approach on the annular area of the focal plane at $1$ deg from the optical axis. This novel configuration reduces the classic TMA complications while maintaining its large (up to $2$ deg radius) corrected field advantage and guarantees a simplified alignment of M1/M3-M2/M4 zones, compensating tilting of single elements in the overall assembly. The baseline design utilizes a 1-m class telescope with EFL $= 15 m$ operating in the 400-900 nm wavelength range, with annular mirrors (rings) and an annular focal plane hosting a ring of detectors.}
\label{fig:theia_rafter_1}
\end{center}
\end{figure}

\begin{figure}[ht]
\begin{center}
\includegraphics[width=0.95\textwidth]{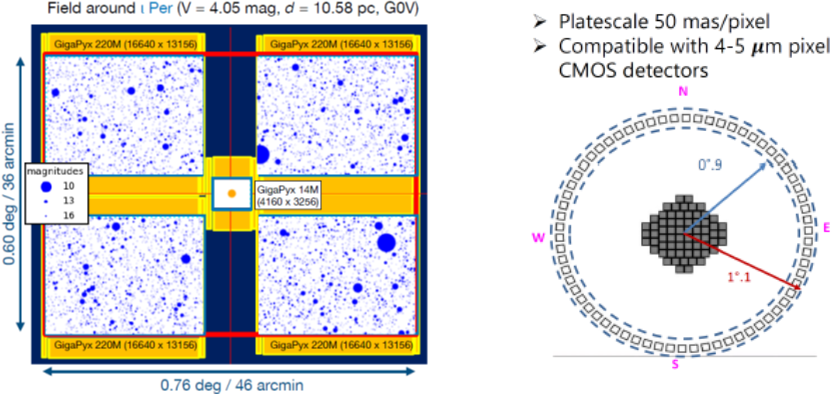}
\caption{\small Left: {\it Theia} focal plane layout with 4 GigaPyx 220M arrays and 1 central GigaPyx 14M array CMOS detectors. The full 0.33 deg$^2$ FOV around the exoplanet host star iota Persei, one of the targets from HWO reference sample \citep{Mamajek2024,Tuchow2025}  has been represented to scale. Right: the {\it RAFTER} focal plane assembly ring with 1 deg radius composed of 66 small-size CMOS detectors, and the same number of devices (in gray) arranged around the optical axis for comparison.}
\label{fig:theia_rafter_2}
\end{center}
\end{figure}

The Italian community is already involved in the development of traditional and novel mission concepts for high precision space-based differential astrometry at the sub-$\mu$as level, in particular the {\it Theia} mission \citep{Malbet2021,Malbet2022} and the {\it RAFTER} mission \citep{Riva2020,Gai2022}, both proposed within the context of ESA's Voyage 2050 program. Fig. \ref{fig:theia_rafter_1} and \ref{fig:theia_rafter_2} illustrate possible the optical design and focal plane detector array solutions for the two proposed missions. In these context we have started tackling some of the core issues in connection with the optimized calibration and correction of the focal plane assembly and optical distortions through hardware and/or software solutions that would relax complexity and cost requirements imposed by standard approaches based on full-fledged laser metrology systems for continuously monitoring the entire payload structure and maximizing instrument stability control. 

We have recently contributed to the definition of the scientific requirements and a preliminary system analysis for the concept of a dedicated astrometric instrument for HWO \citep{Malbet2024,Malbet2026,Amiaux2026,Lizzana2026}. The analysis integrates the definition of the mission profile, the instrument concept architecture, and an error budget that breaks down the key contributors to the sub-$\mu$as precision required for a single measurement. The proposed architecture of the instrument concept is derived from error budget and mission constraints. It would have the capability to produce diffraction-limited images of 0.5 deg$^2$ fields enabled by large visible CMOS-based stitched pixel arrays, achieving a point-spread function (PSF) with a resolution of 20 mas, resulting in an overall gigapixel focal plane. A portion of the error budget addresses photo-center estimation for both the target and calibration stars used in differential astrometry. Other major contributors are related to instrument control of systematics.  A Detector Calibration Unit system using interferometric laser fringes to calibrate pixel positions would enable instrument control of systematics in the reconstruction of differential angle measurements from pixel data (focal plane calibrations) to on-sky line of sight (telescope distortion calibrations). As the field of view requirement would grant access to large numbers of reference stars, the local reference frames could then be used to set up feedback loop procedures that iteratively improve the instrument calibration model through analysis and updating of the reference stars positions used as metrology sources, ultimately materializing the required sensitivity to sub$\mu$as-level astrometric signals produced by true Earth twins around the nearest stars. Finally, we assessed the mission profile to estimate the fraction of survey time required for astrometric survey to achieve the science objective, evaluated the Technology Readiness Level (TRL) and proposed a way forward to reach TRL 5 level for key technologies by the Mission Consolidation Review in 2029.

Thus far, the Italian contribution (within a team that includes French, UK, German, and US colleagues) has priotized the definition of the scientific requirements and on the investigation of detailed simulations of observational scenarios. It would be highly desirable to explore opportunities for further involvement at the technological (hardware/software) level, also taking into account the expertise the Italian community has developed during the decades-long involvement in the Gaia mission.

\subsection{SHOS - A near-IR multi-mode spectrograph to look beyond habitable worlds$^*$}
\label{sec:strum.SHOS}
%\textit{Text by: P. Saracco}
%, P. Conconi, C. Arcidiacono, L. Barbalini, S. Bisogni, R. Bonito, A. Caratti o Garatti, E. Cascone, V. Cianniello,  F. D’Ammando, E. Dalla Bont\`a, V. De Caprio, I. Di Antonio, B. Di Francesco, G. Di Rico, C. Eredia, P. Franzetti, A. Gargiulo, M. G. Guarcello, L. Izzo, F. La Barbera, A. Longobardo, H. Mahmoodzadeh, C. Mancini, M. Mirabile,  E. Molinari, A. Pizzella, E. Portaluri, L. Prisinzano, G. Vietri, H.-F. Wang}

\blfootnote{$^*$For further information feel free to contact Paolo Saracco, \href{mailto:paolo.saracco@inaf.it}{\nolinkurl{paolo.saracco@inaf.it}} }
We propose here SHOS ({\bf SH}ARP {\bf O}n {\bf S}ky), a near-infrared  
spectrograph, as the fourth 
candidate instrument for HWO.
Building on the preliminary specifications and candidate instruments shown 
in the recent presentation of the HWO, %\url{https://assets.science.nasa.gov/content/dam/science/missions/habitable-worlds-observatory/hwo-meetings/aas-245-(jan-2025)/2_%20HWO%20TMPO.pdf},
and on the conceptual design of the SHARP multi-mode spectrograph developed for the Extremely Large Telescope (ELT) \citep{saracco24,mahmoodzadeh25}, SHOS is designed to 
fully exploit the angular resolution at the diffraction limit of HWO.
We assume a telescope's diameter D$\simeq$6.5 m, 
a bandpass 0.1-2.5 $\mu$m, and the following three candidate instruments: 
high-contrast coronagraph (HCC hereafter) in the range 0.35-1.7 $\mu$m, 
high-resolution imager (HRI hereafter) in the range 0.2-2.5  $\mu$m and 
UV Multi-Object spectrograph (UV-MOS hereafter) 
in the range 0.1-1.0 $\mu$m.

While these three candidate instruments for HWO (HCC, HRI, and UV-MOS) fully cover the 
UV and optical bands, SHOS strategically extends the observatory's spectroscopic 
capabilities into the near-infrared range 1.0–2.1(2.5) $\mu$m. 
This allows HWO to explore areas that are poorly or even completely unexplored by the 
three candidate instruments proposed so far, thus maximizing HWO's scientific return.
The main scientific drivers behind SHOS are described in Sec. \ref{sec:sci.galaxiesAGN}.

To tackle the issues described in Sec. \ref{sec:sci.galaxiesAGN}, observations require a specific combination of capabilities: near-IR coverage, high-angular resolution, and multiplexing capabilities:

%To tackle the above issues, observations require a specific combination of capabilities defining the spectrograph SHOS that is discussed in the following.
%Namely:
{\bf Near-IR Coverage (1.0 - 2.5 $\mu$m)} - Essential for:\\
$\bullet$ tracking galaxy properties across cosmic time. This bandpass captures critical redshifted diagnostic lines: Mgb [5120] and Fe lines for total metallicity; the Mg/Fe ratio to gauge star formation duration; the D4000 break for stellar age; Balmer lines, OII [3727], and OIII [5000] for SFR and AGN diagnostics; and HeII [1640] at $z$$>$9 for Pop III stars. \\
$\bullet$ probing the nuclear regions of galaxies, to study the physics of Active Galactic Nuclei (AGN) as well as to detect and study dual AGNs; to probe dust-shrouded regions in the nearby Universe (resolving stars in the centers of Globular Clusters (GCs) and studying 
star and planet formation); \\
$\bullet$ detecting water, organic compounds, and ice signatures in exoplanet atmospheres, and on the surfaces of small bodies and moons within our own Solar System.

%Fig.~\ref{fig:kband_z} shows the observed wavelength of some atomic emission and 
%absorption lines as a function of redshift. 
%The range $\lambda_{obs}$ considered, 0.2-2.5 $\mu$m, is the one covered by the High-Resolution Imager (HRI hereafter) currently planned at HWO, the blue horizontal line 
%marks the limiting wavelength at 1 $\mu$m of the  UV Multi-Object spectrograph (UV-MOS hereafter).

%\begin{wrapfigure}{l}{0.55\textwidth}
%\includegraphics[width=10truecm]{Figures/SHOS_wave_z.png}
%\caption{Observed wavelength of some of the main atomic emission (solid lines) and absorption lines (dotted lines) as a function of redshift: H$\alpha$($\lambda$6563), 
%Mg-b($\lambda$5175), G-band($\lambda$4304), OII($\lambda$3727), MgII($\lambda$2800),
%HeII($\lambda$1640), Ly$\alpha$($\lambda$1216).
%The observed wavelength range $\lambda_{obs}$=0.2-2.5 $\mu$m is the one covered
%by the HRI at HWO. 
%The blue horizontal line marks the limiting wavelength 1.0 $\mu$m of the UV-MOS at HWO.  
%The cyan region is the corresponding redshift range accessible in case of this 
%spectroscopic limit.}
%\label{fig:kband_z}
%\end{wrapfigure}

%A spectrograph covering the same wavelength range as the HRI would also allow HWO 
%to probe the Universe beyond the habitable worlds.

\begin{figure}
\includegraphics[width=8truecm,height=5truecm]{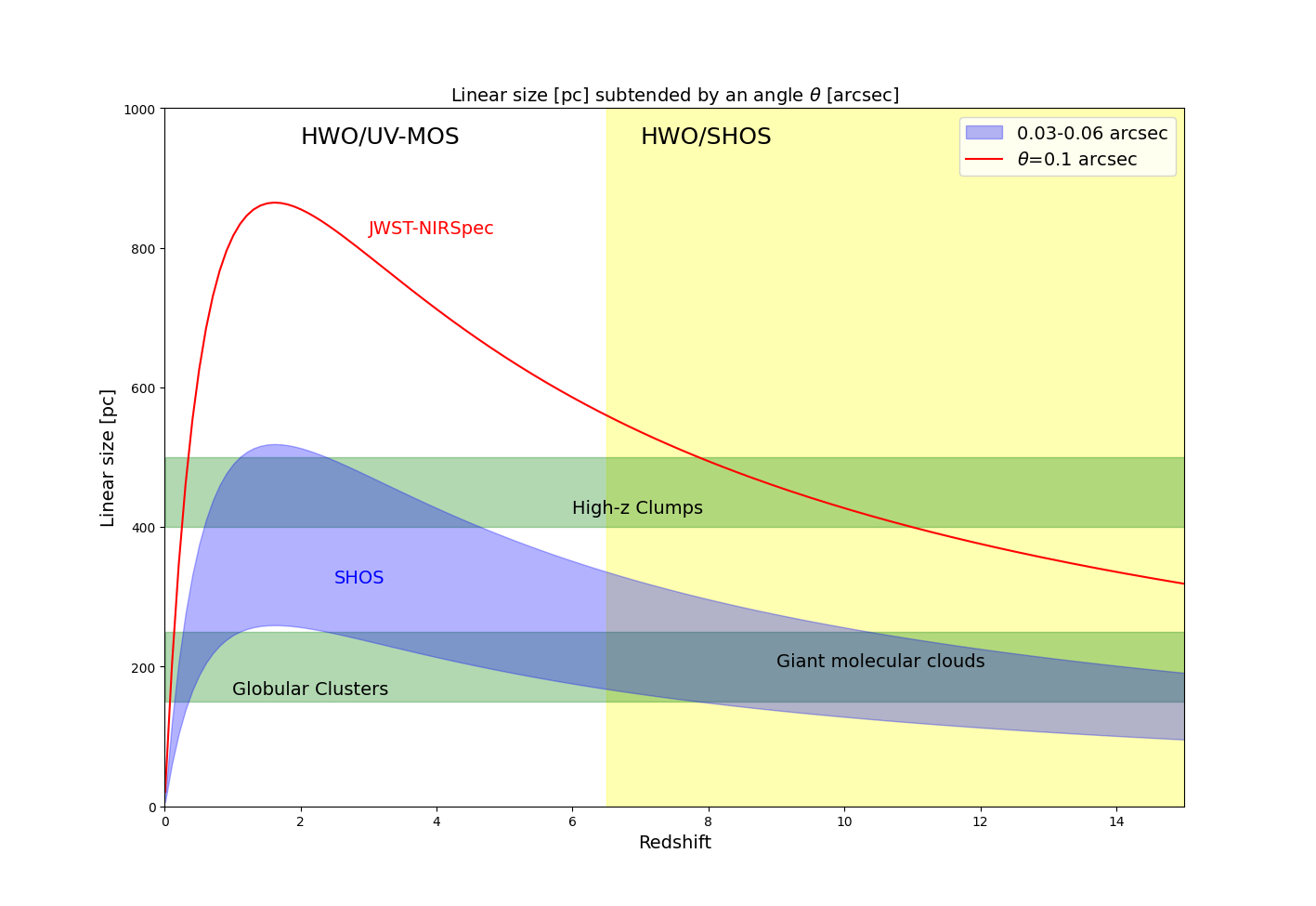}
\includegraphics[width=8truecm,height=5truecm]{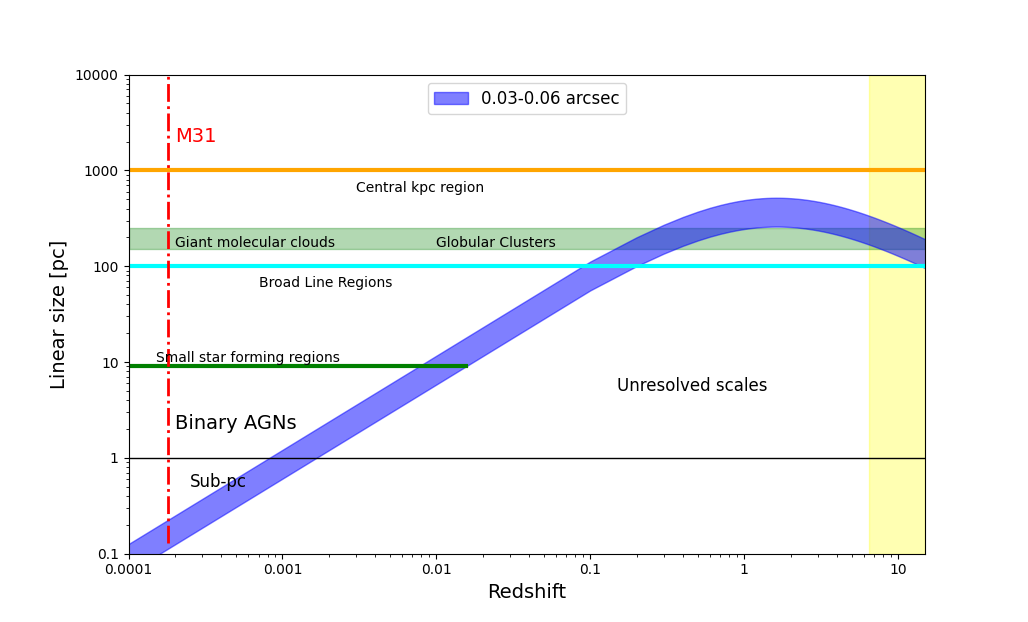}
\caption{\small Left - The linear size [pc] subtended by an angle $\theta$=0.030-0.060" (blue band) is compared to  the pixel scale $\theta$=0.1" of NIRSpec (red curve) at 
JWST as a function of redshift. 
The light-green stripes mark the size of giant molecular gas clouds, 150-250 pc and of high-redshift clumps observed by JWST. 
The yellow region indicates the fraction of the Universe not accessible to the spectrographs currently planned for HWO. Right - Same as left, but on a logarithmic scale. Typical sizes of some dust-shrouded regions of astrophysical interest are marked by horizontal lines.
}
\label{fig:linearscale}
\end{figure}

{\bf Angular Resolution (30-60 mas)} - 
To investigate baryon assembly, observations must resolve regions comparable to Giant Molecular Clouds (GMCs, $\sim$150--250 pc), which are the fundamental units of star formation, chemical enrichment, and galaxy kinematics. Sampling these critical physical scales across all cosmic epochs strictly requires an angular resolution in the range of 30--60 mas.
A $\ge$6.0 m class telescope delivering diffraction-limited performance achieves a nominal angular resolution of 0.035", 0.046", and 0.07" at 1.0 $\mu$m, 1.2 $\mu$m, and 1.8 $\mu$m, respectively. Therefore, by employing an appropriate instrument pixel scale, HWO can successfully resolve $\sim$150--250 pc regions throughout cosmic history. This is illustrated in  Fig. \ref{fig:linearscale} (left panel), where the linear size subtended by a 0.03"--0.06" angle (blue band) is plotted as a function of redshift. 
Crucially, resolving these scales is intrinsically beyond the capabilities of the JWST. Despite its aperture, the  spectroscopic spatial resolution of JWST is limited by the 0.1" pixel scale of the NIRSpec instrument, which does not allow it to detect these subgalactic structures, as shown by the red line in Fig. \ref{fig:linearscale}.

{\bf Multiplexing} - High-resolution observations must be performed simultaneously on multiple targets to reconstruct the dynamic state of high-redshift protoclusters, 
study GC kinematics in UDGs, and survey massive populations of young stellar objects 
in crowded local environments.

%\subsubsection{Summary of Observational Requirements}

%The main scientific drivers behind the development of a spectrograph like SHOS and the resulting main requirements are described in Sec. \ref{sec:drivershos}. 
%To make these studies feasible, observations require a specific combination of capabilities: near-IR coverage, high-angular resolution, and multiplexing capabilities. 
SHOS integrates two core units:
\begin{itemize}
\item 
a slit Multi-Object Spectrograph (MOS) powered by a Configurable Slit System (CSS);
\item
a modular multi-Integral Field Unit (mIFU) fed by a Field Selector System (FSS)
\end{itemize}
both operating in the wavelength range 1.0-2.5 $\mu$m. 
A further extension to longer wavelengths (e.g., 3 $\mu$m) could be considered, 
depending on the thermal background, function of the telescope design.

Considering the angular resolution of a telescope of $\sim$6.5 m 
in the wavelength range 1.0-2.5 $\mu$m,
%is 0.035", 
%0.046", 0.07" and 0.085" at 1 $\mu$m, 1.2 $\mu$m, 1.8 $\mu$m and 2.2 $\mu$m, 
%respectively, 
both units should have a pixel scale of about 35-40 mas,
providing an angular resolution three times better than NIRSpec at JWST (see Fig. \ref{fig:linearscale}).
This pixel scale would allow us to match the diffraction limit of
HWO also at the shortest wavelengths ($\sim$0.035"-0.045" at $\lambda$$<$1.2 $\mu$m) of the range
considered, while satisfying Nyquist sampling at longer wavelengths.
Assuming a similar F/\# to JWST ($\sim$20), for a pixel size 
of about 35 mas the FoV of the SHOS-MOS could be about 1x1 arcmin$^2$.

SHOS-MOS would be powered by a CSS system in which two bars move toward each other to form the slit.
Unlike the MSA system, this system allows for perfect centering of all simultaneously observed targets and adjustment of the slit width.
This system is already operational in a cryogenic environment at the near-infrared spectrographs MOSFIRE at Keck and ERIS at GranTeCan.
The inclusion of an inversion prism system to power the slits would dramatically 
increase the MOS yield \cite[see][]{saracco24}. 

Similarly, a module of the SHOS-mIFU could be composed of 6 probes 
of about 2$\times$2 arcsec$^2$ each.   
The probes would be fed by a system similar to the one adopted for SHARP-VESPER to 
probe an area of about 10"$\times$40" arcsec \citep[see][]{saracco24,mahmoodzadeh25}.

Table~\ref{tab:tab1} summarizes the approximate properties of the two main units of SHOS.

\begin{table*}[ht]
\caption{Main characteristics of SHOS}
\label{tab:tab1}
\smallskip
\begin{center}
{\small
\begin{tabular}{lll}  
\hline
\noalign{\smallskip}
SHOS & MOS &  mIFU  \\
  \noalign{\smallskip}
\hline 
Wavelength coverage & [1.0-2.5(3.0)] $\mu$m & [1.2-2.4] $\mu$m \\
\noalign{\smallskip}
Field of view/Area probed &$\sim$60”x60” &$\sim$10”x40” (one module)\\
 \noalign{\smallskip}
\hline 
\noalign{\smallskip}
Multiplexing & $\sim$30 slits (2” length) & 6 (FoV$\sim$2”x2” each) \\
\noalign{\smallskip}
\hline 
\noalign{\smallskip}
Pixel/spaxel scale &30-40 mas/pixel & 30-40 mas\\
\noalign{\smallskip}
\hline
\noalign{\smallskip}
Spectral resolution& 3000 / 1000/ 200 (0”.2 slit width) & 3000\\
\noalign{\smallskip}
\hline\
\end{tabular}
}
\end{center}
\end{table*}

%\subsubsection{High-angular resolution spectroscopy in the 2030s-2040s }\label{sec:high-angres-2040}
We believe SHOS is competitive in the landscape of near-IR high-angular resolution spectrographs also when HWO will be launched.
Currently, NIRSpec \citep{jakobsen22} at JWST sets the standard for high-angular resolution Near-IR spectroscopy. 
Its strengths include multiplexing for approximately 100 integrated spectra via the Micro-Shutter Assembly (MSA), alongside a single Integral Field Unit (IFU). 
{\it Its angular resolution is limited by the pixel size to 0.1 arcsec, while the MSA system prevents the centering for all the targets.}

While a new generation of ground-based telescopes, the ELT\citep{gilmozzi07}, the Giant Magellan Telescope \citep[GMT,][]{matt06,burgett24}, and the Thirty Meter Telescope (TMT), will become operational during the 2030s, the spectroscopy capabilities will remain largely confined by NIRSpec's current paradigm.
 
The ELT's AO-supported instruments include MICADO \citep{davies21,davies18}, with longslit capability along the parallactic angle and, likely, a single IFU; HARMONI \citep{niranjan24}, a single IFU, and ANDES \citep{marconi24}, a single longlist and IFU with extremely high spectral resolution (R$\sim$100000). 
Similarly, at GMT, the Integral Field Spectrograph (IFS) is expected
\citep[IFS][]{sharp16} 
and, at the TMT IRIS, a comparable instrument.
These instruments will provide spectra for a single object at a time.
The exception is the MOSAIC \citep{pello24} multi-object spectrograph, whose fiber-fed system limits the 
angular resolution to 0.2" (the fiber size) and the wavelength range to 1.8 $\mu$m.
Furthermore, the versatility of systems like ANDES and IRIS is inherently constrained by the requirement for very bright nearby guide stars, which are essential for their Single-Conjugate Adaptive Optics (SCAO) systems.
Consequently, when the JWST mission concludes, likely between 2035-2040, a critical gap will emerge: there will be no spectrographs capable of efficiently pursuing and detailing the scientific pathways pioneered by JWST.
\\
\\
A conceptually similar spectrograph to SHOS is positioned to fill this gap, allowing HWO to address the two scientific questions that will guide research in the coming decades:
\begin{itemize}
\item How matter assembled to form the first stars, galaxies, and the Universe we observe today;
\item Whether we are alone in the Universe or if habitable worlds exist out there.
\end{itemize}

\subsection{A high-resolution near-infrared spectrograph in bay E.$^*$} 
\label{sec:strum.NIRHRspec}
\blfootnote{$^*$For further information feel free to contact Lorenzo Pino  \href{mailto:lorenzo.pino@inaf.it}{\nolinkurl{lorenzo.pino@inaf.it}} and/or Leonardo Testi \href{mailto:leonardo.testi@unibo.it}{\nolinkurl{leonardo.testi@unibo.it}} and/or Andrea Tozzi \href{mailto:andrea.tozzi@inaf.it}{\nolinkurl{andrea.tozzi@inaf.it}}}

%\textit{Text by:L. Pino, L. Testi, A. Tozzi}

The HWO baseline contains no high-resolution spectroscopic capability in the NIR, leaving a clear scientific gap that the Italian community is well placed to fill. We envisage a near-infrared spectrograph in the Extra Instrument bay.
%, to perform high-resolution cross-correlation spectroscopy (HRCCS) and precision radial velocities on bright sources.  The principal challenges are readily identified: an echelle spectrograph at $R\sim10^5$ within the bay's mass and power allocation (200 kg, 200 W for the EAC-5 guidelines) is demanding, and in the fibre-fed configuration the spectrograph is fed by the CI-NIR, whose 1.0--1.7 $\mu$m range and optics temperature restrict its coverage to the J and H bands, excluding the K band without further adaptation of the project. The fibre-fed HDC variant additionally requires a cross-instrument (from CI to E) architecture not foreseen in the current guidelines, which we would propose as an additional capability to be evaluated.

% *** Lorenzo Pino comment *** The next two paragraphs would actually fit better - perhaps in shortened fashion - in Sec. 2.1.

% \textbf{The added value of longer-wavelength HR spectroscopy}
While the question of exoplanet habitability around Solar twins remains the primary focus of the main coronagraphic instruments of HWO, this mission constitutes a unique opportunity to open a new parameter space: high resolution ($R > 50,000$), infrared transmission and emission spectroscopy of transiting planets 
%with periods larger than 6 - 8 days. This parameter space includes temperate planets across all spectral types and the habitable zone of late-type stars with known transiting planets such as K2-18b (Montet et al., 2015) and the TRAPPIST-1 system (Gillon et al., 2017). 
% *** Lorenzo Pino edit on 28th June 2026 *** 
with relatively long periods. These planets have periods too short to effectively separate telluric contamination from the atmospheric signal via its Doppler signature. At the same time, their spatial separations are not large enough to apply direct imaging or molecular mapping techniques, making them elusive targets even for the ELTs. This parameter space includes warm and temperate planets of all sizes and across all spectral types and the habitable zone of late-type stars. As discussed previously, the infrared ($\lambda > 1 \mu \mathrm{m}$) hosts the key atmospheric features needed for a full characterization of a potentially habitable planet. A high resolution spectrograph limited to $\lambda < 2.5 \mu \mathrm{m}$ could unambiguously detect H$_2$O and CO. An extension to the L and M bands ($\lambda < 2.5 \mu \mathrm{m}$) provides the best sensitivity to CH$_4$, CO$_2$ and CO.  
% *** End of edit 28th June 2026 *** 

% *** Lorenzo Pino's comment *** References lost their bibtex format, likely in some copy/paste
While JWST already performs thermal infrared spectroscopy from space, its low resolving power can lead to ambiguous interpretations of atmospheric composition for temperate planets (Madhusudhan et al., 2025; Welbanks et al., 2026), and is susceptible to stellar contamination (Lim et al, 2023; Radica et al., 2024, Espinoza et al., 2025), which can be better diagnosed and mitigated at a high spectral resolution (Guilluy et al., 2020). Indeed, the power of high spectral resolution has been extensively demonstrated with ground-based telescopes (Snellen et al., 2024). Unfortunately, from the ground, aggressive data-reduction techniques are needed to correct for the contamination of telluric lines, prominent in this wavelength range, resulting in a suppression of planetary signals that do not move sufficiently in the Doppler space (Dash et al., 2026). 
% *** Lorenzo Pino edit on 28th June 2026 *** 
%For example, TRAPPIST-1e moves by $2~\mathrm{km~s^{-1}}$ between ingress and regress, which corresponds to less than a resolution element of a $R = 100,000$ spectrograph. Its atmospheric lines therefore appear as stationary even to state-of-the-art high resolution spectrographs, and can not be disentangled from telluric lines. This is a fundamental limitation that even the ELTs can not surpass (Pallè et al., 2024). The only way to overcome this limitation is to perform infrared high resolution spectroscopy from space.
TRAPPIST-1e is probably the best-known transiting planet in the HZ of a late-type host star, and is a clear example of this challenge. With an orbital period of about 6 days, it moves by $2~\mathrm{km~s^{-1}}$ between ingress and regress, which corresponds to less than a resolution element of a $R = 100,000$ spectrograph. While ELT ANDES aims to characterize the atmosphere of this (and similar) planets out to the K-band, the situation is less clear in the L and M-band, where the thermal bacgkround starts ramping up and telluric contamination becomes significantly worse. We therefore argue that exploring the impact of a high-resolution spectrograph covering out to the M-band to be placed in bay E is crucially important in the coming years, when the limitations of the ELT will be clarified with the first on-sky data from METIS first, and ANDES later.

Italy has acquired significant expertise in high resolution infrared spectroscopy, both in terms of instrument design (e.g. GIANO-B, ANDES) and in terms of data exploitation. It is therefore ideally positioned to finalize the key requirements of a high-resolution (R ~ 50,000) infrared ($\lambda > 1 \mu \mathrm{um}$) spectrograph optimized to observe transiting planets, which would constitute a strong candidate as fourth instrument for HWO. In terms of sensitivity, such an instrument should be capable of collecting around $10^{7}--10^{8}$ photons per pixel, which would suffice to reach a contrast of $10^{-4}$ per spectral line in this photon-dominated regime. Using cross-correlation techniques, the contrast can be pushed by a factor of 10 to 100 lower, revealing atmospheric signals as small as 1 part per million, sufficient for an in-depth chemical characterization of target exoplanets.

Our proposal leverages on an existing study of an AO-fed IFU L-M band spectrograph developed for ground based 8m-class telescopes which can deliver full instantaneous spectra at R$\sim$50000 resolution. The instrument is designed to be the first efficient, high-resolution spectrograph for the study of exoplanetary atmospheres in the L and M bands. Crucially, the instrument concept is based on a mirror-based field splitter and a Germanium immersed grating which  offer the unique  possibility of a compact and lightweight design (as compared to traditional infrared Echelle spectrographs), with fixed optics, which might be a key option to develop a space-based instrument.
%UK, French, Irish, and USA colleagues have expressed interest in being involved in the instrument, and especially the development of a potential concept for a space based instrument. The most urgent technical area to address would be to qualify the grating for space use, produce a design of the instrument compatible with the HWO constraints, and provide a quantitative estimate of the performances of the space based instrument. 
We are organizing a scientific and technical working group that would address the suitability of this concept for a HWO instrument proposal.

\subsection{Ultracompact polarimeter using dual-functional metalenses$^*$}\label{sec:strum.polarimetry}
\blfootnote{$^*$For further information feel free to contact  Andrea Vogliardi \href{mailto:andrea.vogliardi@phd.unipd.it}{\nolinkurl{andrea.vogliardi@phd.unipd.it}} and/or Gianluca Ruffato \href{mailto:gianluca.ruffato@unipd.it}{\nolinkurl{gianluca.ruffato@unipd.it}} and/or Filippo Romanato \href{mailto:filippo.romanato@unipd.it }{\nolinkurl{filippo.romanato@unipd.it }}}%\textit{Text by: A.Vogliardi, G.Ruffato and F.Romanato}

The HWO is being developed with the central scientific objective of directly imaging and characterizing rocky planets in the habitable zones of nearby stars, with particular emphasis on the search for atmospheric biosignatures \citep{NASA_HWO,GSFC_HWO,Astro2020}. In this observational regime, the planet is detected in reflected stellar light after extreme starlight suppression by a coronagraph. The measurement is intrinsically difficult because the planetary signal is many orders of magnitude fainter than the host star and because the information carried by the disk-integrated reflected flux is often degenerate: different combinations of atmospheric composition, cloud coverage, aerosol loading, surface properties and planetary phase can produce similar intensity spectra. A polarimeter placed in the HWO optical path can therefore provide an additional observable that is not available from flux-only measurements.

The physical reason is that reflected planetary light is generally polarized, whereas disk-integrated stellar light is approximately unpolarized at the level relevant for exoplanet observations. The polarization state depends on wavelength, phase angle and scattering geometry, and is therefore sensitive to the atmospheric and surface properties of the planet \citep{Berdyugina2016PolarizedScattering,TreesStam2022OceanSigs}. In ground-based high-contrast imaging, polarimetry is often used both as a differential technique to suppress unpolarized stellar residuals and as a diagnostic of scattering material. In the HWO context considered here, the primary role of polarimetry is instead the second one: it is a science observable that complements spectroscopy and photometry after coronagraphic suppression. The coronagraph provides the main starlight rejection, while the polarimeter measures the residual polarization state of the planetary light in order to constrain the physics of the atmosphere and surface.

The polarization state of the light can be described by the Stokes vector but, for reflected-light exoplanet science, the most relevant quantity is often the degree of linear polarization $P_L$. The reason why this parameter is particularly important is that atmospheric scattering and surface reflection processes tend to generate strong linear polarization signatures. Rayleigh scattering, for instance, produces a polarization maximum close to quadrature, where the scattering angle is near \(90^\circ\). This phase dependence can be more diagnostic than the phase dependence of the total reflected flux alone, because it depends directly on the scattering matrix of the atmosphere and therefore on molecular scattering, aerosols, haze and cloud particles \citep{TreesStam2022OceanSigs,Berdyugina2016PolarizedScattering}.

Polarimetry is also sensitive to cloud and surface properties that are difficult to isolate from intensity spectra alone. Liquid-water clouds can produce polarized rainbow features whose angular position and spectral behavior depend on droplet size and refractive index. Ocean surfaces can generate specular reflection, or glint, whose polarization can become particularly strong near geometries related to the Brewster angle \citep{TreesStam2022OceanSigs}. Therefore, a measurement of $P_L$ as a function of wavelength and orbital phase can help distinguish between a dry surface, a cloudy atmosphere, a hazy atmosphere and a planet with surface liquid water. This is especially relevant for HWO because biosignature interpretation will require not only the detection of molecular absorption bands, but also an understanding of the atmospheric and surface context in which those bands appear.

The basic measurement product of an HWO polarimetric channel is thus a coronagraphic image cube containing at least the linear Stokes parameters, either in broad optical bands or in low-resolution spectral channels. By repeating the measurement at multiple orbital phases, one obtains polarimetric phase curves that probe the scattering geometry. Full-Stokes measurementsmay also be useful for calibration or for specific astrophysical cases, but most reflected-light exoplanet applications can already be addressed with high-precision linear polarimetry.

In an ideal linear-analyzer polarimeter, the Stokes parameters can be reconstructed from intensity measurements through four analyzer orientations. Moreover, simultaneous or near-simultaneous acquisition of multiple analyzer states is highly desirable. In high-contrast imaging, the planetary signal is weak and systematic errors can arise from temporal drifts, pointing variations, wavefront evolution, detector effects and non-common-path aberrations. If the different polarization states are measured sequentially with moving optics, any temporal variation between exposures can leak into the reconstructed Stokes parameters. For this reason, the preferred architecture for an HWO polarimeter should minimize moving components and should measure orthogonal or complementary polarization states in a common-path or quasi-common-path configuration.

This is where metalenses and metasurfaces become attractive. Conventional polarimetric systems often rely on combinations of bulk retarders, Wollaston prisms, polarizing beam splitters, rotating waveplates and relay optics. These components can be extremely precise, but they add volume, mass, alignment complexity and possible non-common-path errors. A metasurface, by contrast, can combine phase control, polarization analysis and focusing in a single planar optical element. If placed close to the science detector, a meta-lens polarimeter can reduce the amount of downstream optics, improve mechanical stability and enable snapshot polarimetric measurements. This is particularly relevant for a space telescope such as HWO, where compactness, stability and calibration repeatability are essential.

The meta-lens polarimeter concept considered here is based on dual-functional dielectric metalenses. A metalens is a planar optical element composed of subwavelength dielectric scatterers, usually called meta-atoms, whose geometry and orientation locally control the phase, amplitude and polarization response of the transmitted field.

In particular, dual-functional metalenses implement two different optical functions for two different polarization states. In the DFML approach, the metasurface is designed so that right- and left-handed circularly polarized input light experience two independent phase profiles relying on the combination of dynamic phase and geometric phase.\citep{Vogliardi2023DFML}.

The link between DFMLs and polarimetry is made by recognizing that a polarimeter is not simply a focusing system: it is an optical system that projects the incident Stokes vector onto a set of known analyzer states. The most direct route to single-shot polarimetry with DFMLs is therefore to design a metasurface analyzer whose channels implement known projections of the incident Stokes vector. The DFML provides the polarization-dependent phase profiles that route the selected polarization components to different focal spots; the local retarders or meta-atoms define the analyzer basis; and the detector records all channels simultaneously. In such a device, the focusing and analyzing functions are no longer separated into different bulk optical components. They are co-designed at the metasurface level.

For HWO, this integration is important because the polarimeter would likely be located within, or immediately after, the coronagraphic instrument, preferably close to the science detector. Coronagraphic systems are sensitive to polarization aberrations introduced by reflective optics and coatings, and these aberrations can couple into wavefront errors and contrast leakage \citep{Chipman2018HabExPolAberr}. A compact transmissive meta-lens analyzer placed late in the optical train can reduce additional non-common-path optics and simplify the polarimetric calibration path. At the same time, because HWO aims at extremely high contrast, the DFML polarimeter cannot be treated as an isolated component: its Mueller matrix must be calibrated together with the upstream coronagraphic optics.

To adapt this technology to HWO, the implementation must be on near-visible regime. Silicon, which is useful in the telecom band, is not the ideal material for a broadband visible transmissive metalens because of absorption. For HWO-relevant optical bands, low-loss dielectrics such as TiO\(_2\), SiN or related high-index materials are more appropriate \citep{Chen2018AchromaticMetalens}. The design must also address bandwidth. A simple diffractive metalens is naturally chromatic, because the imposed phase profile depends on wavelength. For HWO, where reflected-light characterization may require broad optical coverage or spectropolarimetry, the meta-atom library should therefore be dispersion-engineered so that both the focusing phase and the polarization response remain controlled over the required bandpass. In practice, this means optimizing not only the phase at a central wavelength, but also the wavelength dependence of the Jones or Mueller response.

The fabrication of such a device follows the general route of dielectric metasurface manufacturing. A transparent substrate, such as fused silica, is prepared and coated with a high-index dielectric layer. The nanostructure pattern is then defined by electron-beam lithography for early prototypes, or by stepper lithography or nanoimprint lithography for larger-scale fabrication. Reactive-ion etching transfers the pattern into the dielectric layer, forming nanopillars or nanofins with the desired lateral dimensions, height and orientation \citep{vogliardi2023all,vogliardi2023silicon}. After etching, the remaining mask is removed, surfaces may be passivated, and antireflection coatings can be added if required by the optical design. For a space-qualified component, this fabrication sequence must be followed by metrology of the geometry, wavefront, efficiency and polarization response, as well as environmental testing for thermal stability, vibration, contamination and radiation compatibility.

\bibliographystyle{mn2e} 
% You should give the same name for your .bbl as your main .tex
% since it is a requirement for posting on ArXiv.
\bibliography{main.bib} 

@PROCEEDINGS{Evansetal2020,
        title = "{Ground-based and Airborne Instrumentation for Astronomy VIII}",
    booktitle = {Ground-based and Airborne Instrumentation for Astronomy VIII},
         year = 2020,
       editor = {{Evans}, Christopher J. and {Bryant}, Julia J. and {Motohara}, Kentaro},
       series = {Society of Photo-Optical Instrumentation Engineers (SPIE) Conference Series},
       volume = {11447},
        month = dec,
          doi = {10.1117/12.2591716},
       adsurl = {https://ui.adsabs.harvard.edu/abs/2020SPIE11447E....E}
}

@INPROCEEDINGS{simonetti25,
       author = {{Simonetti}, Paolo and {Ivanovski}, Stavro and {Biasiotti}, Lorenzo and {Vladilo}, Giovanni and {Monai}, Sergio and {Dogo}, Federico and {Calderone}, Lorenzo and {Politi}, Romolo and {Turrini}, Diego},
        title = "{A new energy balance model to map the habitability of tidally locked rocky planets: application to the Ariel target list}",
    booktitle = {EPSC-DPS Joint Meeting 2025},
         year = 2025,
       volume = {2025},
        month = sep,
          eid = {EPSC-DPS2025-1010},
        pages = {EPSC-DPS2025-1010},
          doi = {10.5194/epsc-dps2025-1010},
       adsurl = {https://ui.adsabs.harvard.edu/abs/2025epsc.conf.1010S}
}

@article{FrebelNorris2015,
   author = "Frebel, Anna and Norris, John E.",
   title = "Near-Field Cosmology with Extremely Metal-Poor Stars", 
   journal= "Annual Review of Astronomy and Astrophysics",
   year = "2015",
   volume = "53",
   number = "Volume 53, 2015",
   pages = "631-688",
   doi = "https://doi.org/10.1146/annurev-astro-082214-122423",
   url = "https://www.annualreviews.org/content/journals/10.1146/annurev-astro-082214-122423",
   publisher = "Annual Reviews",
   issn = "1545-4282",
   type = "Journal Article",
  }

@ARTICLE{Keller2014Natur,
       author = {{Keller}, S.~C. and {Bessell}, M.~S. and {Frebel}, A. and {Casey}, A.~R. and {Asplund}, M. and {Jacobson}, H.~R. and {Lind}, K. and {Norris}, J.~E. and {Yong}, D. and {Heger}, A. and {Magic}, Z. and {da Costa}, G.~S. and {Schmidt}, B.~P. and {Tisserand}, P.},
        title = "{A single low-energy, iron-poor supernova as the source of metals in the star SMSS J031300.36-670839.3}",
      journal = {\nat},
         year = 2014,
        month = feb,
       volume = {506},
       number = {7489},
        pages = {463-466},
          doi = {10.1038/nature12990},
archivePrefix = {arXiv},
       eprint = {1402.1517},
 primaryClass = {astro-ph.SR},
       adsurl = {https://ui.adsabs.harvard.edu/abs/2014Natur.506..463K}
}

@ARTICLE{Bosman22,
       author = {{Bosman}, Sarah E.~I. and {Davies}, Frederick B. and {Becker}, George D. and {Keating}, Laura C. and {Davies}, Rebecca L. and {Zhu}, Yongda and {Eilers}, Anna-Christina and {D'Odorico}, Valentina and {Bian}, Fuyan and {Bischetti}, Manuela and et al.},
        title = "{Hydrogen reionization ends by z = 5.3: Lyman-{\ensuremath{\alpha}} optical depth measured by the XQR-30 sample}",
      journal = {\mnras},
         year = 2022,
        month = jul,
       volume = {514},
       number = {1},
        pages = {55-76},
          doi = {10.1093/mnras/stac1046},
archivePrefix = {arXiv},
       eprint = {2108.03699},
 primaryClass = {astro-ph.CO},
       adsurl = {https://ui.adsabs.harvard.edu/abs/2022MNRAS.514...55B}
}

@ARTICLE{Compostella13,
       author = {{Compostella}, Michele and {Cantalupo}, Sebastiano and {Porciani}, Cristiano},
        title = "{The imprint of inhomogeneous He II reionization on the H I and He II Ly{\ensuremath{\alpha}} forest}",
      journal = {\mnras},
         year = 2013,
        month = nov,
       volume = {435},
       number = {4},
        pages = {3169-3190},
          doi = {10.1093/mnras/stt1510},
archivePrefix = {arXiv},
       eprint = {1306.5745},
 primaryClass = {astro-ph.CO},
       adsurl = {https://ui.adsabs.harvard.edu/abs/2013MNRAS.435.3169C}
}

@ARTICLE{Compostella14,
       author = {{Compostella}, Michele and {Cantalupo}, Sebastiano and {Porciani}, Cristiano},
        title = "{AGN-driven helium reionization and the incidence of extended He III regions at redshift z > 3}",
      journal = {\mnras},
         year = 2014,
        month = dec,
       volume = {445},
       number = {4},
        pages = {4186-4196},
          doi = {10.1093/mnras/stu2035},
archivePrefix = {arXiv},
       eprint = {1407.1316},
 primaryClass = {astro-ph.CO},
       adsurl = {https://ui.adsabs.harvard.edu/abs/2014MNRAS.445.4186C}
}

@ARTICLE{Garaldi19,
       author = {{Garaldi}, Enrico and {Compostella}, Michele and {Porciani}, Cristiano},
        title = "{The Goldilocks problem of the quasar contribution to reionization}",
      journal = {\mnras},
         year = 2019,
        month = mar,
       volume = {483},
       number = {4},
        pages = {5301-5314},
          doi = {10.1093/mnras/sty3414},
archivePrefix = {arXiv},
       eprint = {1809.10144},
 primaryClass = {astro-ph.CO},
       adsurl = {https://ui.adsabs.harvard.edu/abs/2019MNRAS.483.5301G}
}

@ARTICLE{Maiolino24a,
       author = {{Maiolino}, Roberto and {Scholtz}, Jan and {Curtis-Lake}, Emma and {Carniani}, Stefano and {Baker}, William and {de Graaff}, Anna and {Tacchella}, Sandro and {{\"U}bler}, Hannah and {D'Eugenio}, Francesco and {Witstok}, Joris and et al.},
        title = "{JADES: The diverse population of infant black holes at 4 < z < 11: Merging, tiny, poor, but mighty}",
      journal = {\aap},
         year = 2024,
        month = nov,
       volume = {691},
          eid = {A145},
        pages = {A145},
          doi = {10.1051/0004-6361/202347640},
archivePrefix = {arXiv},
       eprint = {2308.01230},
 primaryClass = {astro-ph.GA},
       adsurl = {https://ui.adsabs.harvard.edu/abs/2024A&A...691A.145M}
}

@ARTICLE{Madau24,
       author = {{Madau}, Piero and {Giallongo}, Emanuele and {Grazian}, Andrea and {Haardt}, Francesco},
        title = "{Cosmic Reionization in the JWST Era: Back to AGNs?}",
      journal = {\apj},
         year = 2024,
        month = aug,
       volume = {971},
       number = {1},
          eid = {75},
        pages = {75},
          doi = {10.3847/1538-4357/ad5ce8},
archivePrefix = {arXiv},
       eprint = {2406.18697},
 primaryClass = {astro-ph.CO},
       adsurl = {https://ui.adsabs.harvard.edu/abs/2024ApJ...971...75M}
}

@ARTICLE{Worseck19,
       author = {{Worseck}, G{\'a}bor and {Davies}, Frederick B. and {Hennawi}, Joseph F. and {Prochaska}, J. Xavier},
        title = "{The Evolution of the He II-ionizing Background at Redshifts 2.3 < z < 3.8 Inferred from a Statistical Sample of 24 HST/COS He II Ly{\ensuremath{\alpha}} Absorption Spectra}",
      journal = {\apj},
         year = 2019,
        month = apr,
       volume = {875},
       number = {2},
          eid = {111},
        pages = {111},
          doi = {10.3847/1538-4357/ab0fa1},
archivePrefix = {arXiv},
       eprint = {1808.05247},
 primaryClass = {astro-ph.GA},
       adsurl = {https://ui.adsabs.harvard.edu/abs/2019ApJ...875..111W}
}

@ARTICLE{Worseck11,
       author = {{Worseck}, G{\'a}bor and {Prochaska}, J. Xavier},
        title = "{GALEX Far-ultraviolet Color Selection of UV-bright High-redshift Quasars}",
      journal = {\apj},
         year = 2011,
        month = feb,
       volume = {728},
       number = {1},
          eid = {23},
        pages = {23},
          doi = {10.1088/0004-637X/728/1/23},
archivePrefix = {arXiv},
       eprint = {1004.3347},
 primaryClass = {astro-ph.CO},
       adsurl = {https://ui.adsabs.harvard.edu/abs/2011ApJ...728...23W}
}

@ARTICLE{Syphers12,
       author = {{Syphers}, David and {Anderson}, Scott F. and {Zheng}, Wei and {Meiksin}, Avery and {Schneider}, Donald P. and {York}, Donald G.},
        title = "{HST/COS Observations of Thirteen New He II Quasars}",
      journal = {\aj},
         year = 2012,
        month = apr,
       volume = {143},
       number = {4},
          eid = {100},
        pages = {100},
          doi = {10.1088/0004-6256/143/4/100},
archivePrefix = {arXiv},
       eprint = {1202.0236},
 primaryClass = {astro-ph.CO},
       adsurl = {https://ui.adsabs.harvard.edu/abs/2012AJ....143..100S}
}

@ARTICLE{Zhu2015,
       author = {{Zhu}, Zhaohuan},
        title = "{Accreting Circumplanetary Disks: Observational Signatures}",
      journal = {\apj},
         year = 2015,
        month = jan,
       volume = {799},
       number = {1},
          eid = {16},
        pages = {16},
          doi = {10.1088/0004-637X/799/1/16},
archivePrefix = {arXiv},
       eprint = {1408.6554},
 primaryClass = {astro-ph.EP},
       adsurl = {https://ui.adsabs.harvard.edu/abs/2015ApJ...799...16Z}
}

@ARTICLE{Aoyama2018,
       author = {{Aoyama}, Yuhiko and {Ikoma}, Masahiro and {Tanigawa}, Takayuki},
        title = "{Theoretical Model of Hydrogen Line Emission from Accreting Gas Giants}",
      journal = {\apj},
         year = 2018,
        month = oct,
       volume = {866},
       number = {2},
          eid = {84},
        pages = {84},
          doi = {10.3847/1538-4357/aadc11},
archivePrefix = {arXiv},
       eprint = {1808.06776},
 primaryClass = {astro-ph.EP},
       adsurl = {https://ui.adsabs.harvard.edu/abs/2018ApJ...866...84A}
}

@ARTICLE{Marleau2019,
       author = {{Marleau}, Gabriel-Dominique and {Mordasini}, Christoph and {Kuiper}, Rolf},
        title = "{The Planetary Accretion Shock. II. Grid of Postshock Entropies and Radiative Shock Efficiencies for Nonequilibrium Radiation Transport}",
      journal = {\apj},
         year = 2019,
        month = aug,
       volume = {881},
       number = {2},
          eid = {144},
        pages = {144},
          doi = {10.3847/1538-4357/ab245b},
archivePrefix = {arXiv},
       eprint = {1906.05869},
 primaryClass = {astro-ph.EP},
       adsurl = {https://ui.adsabs.harvard.edu/abs/2019ApJ...881..144M}
}

@ARTICLE{Wagner2018,
       author = {{Wagner}, Kevin and {Follete}, Katherine B. and {Close}, Laird M. and {Apai}, D{\'a}niel and {Gibbs}, Aidan and {Keppler}, Miriam and {M{\"u}ller}, Andr{\'e} and {Henning}, Thomas and {Kasper}, Markus and {Wu}, Ya-Lin and {Long}, Joseph and {Males}, Jared and {Morzinski}, Katie and {McClure}, Melissa},
        title = "{Magellan Adaptive Optics Imaging of PDS 70: Measuring the Mass Accretion Rate of a Young Giant Planet within a Gapped Disk}",
      journal = {\apjl},
         year = 2018,
        month = aug,
       volume = {863},
       number = {1},
          eid = {L8},
        pages = {L8},
          doi = {10.3847/2041-8213/aad695},
archivePrefix = {arXiv},
       eprint = {1807.10766},
 primaryClass = {astro-ph.EP},
       adsurl = {https://ui.adsabs.harvard.edu/abs/2018ApJ...863L...8W}
}

@ARTICLE{Haffert2019,
       author = {{Haffert}, S.~Y. and {Bohn}, A.~J. and {de Boer}, J. and {Snellen}, I.~A.~G. and {Brinchmann}, J. and {Girard}, J.~H. and {Keller}, C.~U. and {Bacon}, R.},
        title = "{Two accreting protoplanets around the young star PDS 70}",
      journal = {Nature Astronomy},
         year = 2019,
        month = jun,
       volume = {3},
        pages = {749-754},
          doi = {10.1038/s41550-019-0780-5},
archivePrefix = {arXiv},
       eprint = {1906.01486},
 primaryClass = {astro-ph.EP},
       adsurl = {https://ui.adsabs.harvard.edu/abs/2019NatAs...3..749H}
}

@ARTICLE{Rigby2018,
       author = {{Rigby}, J.~R. and {Bayliss}, M.~B. and {Chisholm}, J. and {Bordoloi}, R. and {Sharon}, K. and {Gladders}, M.~D. and {Johnson}, T. and {Paterno-Mahler}, R. and {Wuyts}, E. and {Dahle}, H. and {Acharyya}, A.},
        title = "{The Magellan Evolution of Galaxies Spectroscopic and Ultraviolet Reference Atlas (MegaSaura). II. Stacked Spectra}",
      journal = {\apj},
         year = 2018,
        month = jan,
       volume = {853},
       number = {1},
          eid = {87},
        pages = {87},
          doi = {10.3847/1538-4357/aaa2fc},
archivePrefix = {arXiv},
       eprint = {1710.07499},
 primaryClass = {astro-ph.GA},
       adsurl = {https://ui.adsabs.harvard.edu/abs/2018ApJ...853...87R}
}

@INPROCEEDINGS{Benisty2023,
       author = {{Benisty}, M. and {Dominik}, C. and {Follette}, K. and {Garufi}, A. and {Ginski}, C. and {Hashimoto}, J. and {Keppler}, M. and {Kley}, W. and {Monnier}, J.},
        title = "{Optical and Near-infrared View of Planet-forming Disks and Protoplanets}",
    booktitle = {Astronomical Society of the Pacific Conference Series},
         year = 2023,
       editor = {{Inutsuka}, S. and {Aikawa}, Y. and {Muto}, T. and {Tomida}, K. and {Tamura}, M.},
       series = {Astronomical Society of the Pacific Conference Series},
       volume = {534},
        month = jul,
        pages = {605},
       adsurl = {https://ui.adsabs.harvard.edu/abs/2023ASPC..534..605B}
}

@article{Avenhaus2018,
	Adsurl = {https://ui.adsabs.harvard.edu/abs/2018ApJ...863...44A},
	Archiveprefix = {arXiv},
	Author = {{Avenhaus}, Henning and {Quanz}, Sascha P. and {Garufi}, Antonio and {Perez}, Sebastian and {Casassus}, Simon and {Pinte}, Christophe and {Bertrang}, Gesa H. -M. and {Caceres}, Claudio and {Benisty}, Myriam and {Dominik}, Carsten},
	Doi = {10.3847/1538-4357/aab846},
	Eid = {44},
	Eprint = {1803.10882},
	Journal = {The Astrophysical Journal},
	Month = {Aug},
	Number = {1},
	Pages = {44},
	Primaryclass = {astro-ph.SR},
	Title = {{Disks around T Tauri Stars with SPHERE (DARTTS-S). I. SPHERE/IRDIS Polarimetric Imaging of Eight Prominent T Tauri Disks}},
	Volume = {863},
	Year = {2018}}

@ARTICLE{Garufi2024,
       author = {{Garufi}, A. and {Ginski}, C. and {van Holstein}, R.~G. and {Benisty}, M. and {Manara}, C.~F. and {P{\'e}rez}, S. and {Pinilla}, P. and {Ribas}, {\'A}. and {Weber}, P. and {Williams}, J. and {Cieza}, L. and {Dominik}, C. and {Facchini}, S. and {Huang}, J. and {Zurlo}, A. and {Bae}, J. and {Hagelberg}, J. and {Henning}, Th. and {Hogerheijde}, M.~R. and {Janson}, M. and {M{\'e}nard}, F. and {Messina}, S. and {Meyer}, M.~R. and {Pinte}, C. and {Quanz}, S.~P. and {Rigliaco}, E. and {Roccatagliata}, V. and {Schmid}, H.~M. and {Szul{\'a}gyi}, J. and {van Boekel}, R. and {Wahhaj}, Z. and {Antichi}, J. and {Baruffolo}, A. and {Moulin}, T.},
        title = "{The SPHERE view of the Taurus star-forming region. The full census of planet-forming disks with GTO and DESTINYS programs}",
      journal = {\aap},
         year = 2024,
        month = may,
       volume = {685},
          eid = {A53},
        pages = {A53},
          doi = {10.1051/0004-6361/202347586},
archivePrefix = {arXiv},
       eprint = {2403.02158},
 primaryClass = {astro-ph.GA},
       adsurl = {https://ui.adsabs.harvard.edu/abs/2024A&A...685A..53G}
}

@ARTICLE{Duchene2024,
       author = {{Duch{\^e}ne}, Gaspard and {M{\'e}nard}, Fran{\c{c}}ois and {Stapelfeldt}, Karl R. and {Villenave}, Marion and {Wolff}, Schuyler G. and {Perrin}, Marshall D. and {Pinte}, Christophe and {Tazaki}, Ryo and {Padgett}, Deborah L.},
        title = "{JWST Imaging of Edge-on Protoplanetary Disks. I. Fully Vertically Mixed 10 {\ensuremath{\mu}}m Grains in the Outer Regions of a 1000 au Disk}",
      journal = {\aj},
         year = 2024,
        month = feb,
       volume = {167},
       number = {2},
          eid = {77},
        pages = {77},
          doi = {10.3847/1538-3881/acf9a7},
archivePrefix = {arXiv},
       eprint = {2309.07040},
 primaryClass = {astro-ph.EP},
       adsurl = {https://ui.adsabs.harvard.edu/abs/2024AJ....167...77D}
}

@ARTICLE{Perrin2015,
       author = {{Perrin}, Marshall D. and {Duchene}, Gaspard and {Millar-Blanchaer}, Max and {Fitzgerald}, Michael P. and {Graham}, James R. and {Wiktorowicz}, Sloane J. and {Kalas}, Paul G. and {Macintosh}, Bruce and {Bauman}, Brian and {Cardwell}, Andrew and {Chilcote}, Jeffrey and {De Rosa}, Robert J. and {Dillon}, Daren and {Doyon}, Ren{\'e} and {Dunn}, Jennifer and {Erikson}, Darren and {Gavel}, Donald and {Goodsell}, Stephen and {Hartung}, Markus and {Hibon}, Pascale and {Ingraham}, Patrick and {Kerley}, Daniel and {Konapacky}, Quinn and {Larkin}, James E. and {Maire}, J{\'e}r{\^o}me and {Marchis}, Franck and {Marois}, Christian and {Mittal}, Tushar and {Morzinski}, Katie M. and {Oppenheimer}, B.~R. and {Palmer}, David W. and {Patience}, Jennifer and {Poyneer}, Lisa and {Pueyo}, Laurent and {Rantakyr{\"o}}, Fredrik T. and {Sadakuni}, Naru and {Saddlemyer}, Leslie and {Savransky}, Dmitry and {Soummer}, R{\'e}mi and {Sivaramakrishnan}, Anand and {Song}, Inseok and {Thomas}, Sandrine and {Wallace}, J. Kent and {Wang}, Jason J. and {Wolff}, Schuyler G.},
        title = "{Polarimetry with the Gemini Planet Imager: Methods, Performance at First Light, and the Circumstellar Ring around HR 4796A}",
      journal = {\apj},
         year = 2015,
        month = feb,
       volume = {799},
       number = {2},
          eid = {182},
        pages = {182},
          doi = {10.1088/0004-637X/799/2/182},
archivePrefix = {arXiv},
       eprint = {1407.2495},
 primaryClass = {astro-ph.EP},
       adsurl = {https://ui.adsabs.harvard.edu/abs/2015ApJ...799..182P}
}

@ARTICLE{Min2016,
       author = {{Min}, M. and {Rab}, Ch. and {Woitke}, P. and {Dominik}, C. and {M{\'e}nard}, F.},
        title = "{Multiwavelength optical properties of compact dust aggregates in protoplanetary disks}",
      journal = {\aap},
         year = 2016,
        month = jan,
       volume = {585},
          eid = {A13},
        pages = {A13},
          doi = {10.1051/0004-6361/201526048},
archivePrefix = {arXiv},
       eprint = {1510.05426},
 primaryClass = {astro-ph.EP},
       adsurl = {https://ui.adsabs.harvard.edu/abs/2016A&A...585A..13M}
}

@ARTICLE{Tazaki2022,
       author = {{Tazaki}, R. and {Dominik}, C.},
        title = "{The size of monomers of dust aggregates in planet-forming disks. Insights from quantitative optical and near-infrared polarimetry}",
      journal = {\aap},
         year = 2022,
        month = jul,
       volume = {663},
          eid = {A57},
        pages = {A57},
          doi = {10.1051/0004-6361/202243485},
archivePrefix = {arXiv},
       eprint = {2204.08506},
 primaryClass = {astro-ph.EP},
       adsurl = {https://ui.adsabs.harvard.edu/abs/2022A&A...663A..57T}
}

@ARTICLE{Breuval25,
       author = {{Breuval}, Louise and {Anand}, Gagandeep S. and {Anderson}, Richard I. and {Beaton}, Rachael and {Bhardwaj}, Anupam and {Casertano}, Stefano and {Clementini}, Gisella and {Cruz Reyes}, Mauricio and {De Somma}, Giulia and {Groenewegen}, Martin A.~T. and {Huang}, Caroline D. and {Kervella}, Pierre and {Khan}, Saniya and {Macri}, Lucas M. and {Marconi}, Marcella and {Minniti}, Javier H. and {Riess}, Adam G. and {Ripepi}, Vincenzo and {Romaniello}, Martino and {Scolnic}, Daniel and {Trentin}, Erasmo and {Wielg{\'o}rski}, Piotr and {Yuan}, Wenlong},
        title = "{Converging on the Cepheid Metallicity Dependence: Implications of Nonstandard Gaia Parallax Recalibration on Distance Measures}",
      journal = {\apj},
         year = 2025,
        month = nov,
       volume = {994},
       number = {1},
          eid = {111},
        pages = {111},
          doi = {10.3847/1538-4357/ae0cb9},
archivePrefix = {arXiv},
       eprint = {2507.15936},
 primaryClass = {astro-ph.GA},
       adsurl = {https://ui.adsabs.harvard.edu/abs/2025ApJ...994..111B}
}

@ARTICLE{Kasper2021,
       author = {{Kasper}, M. and {Cerpa Urra}, N. and {Pathak}, P. and {Bonse}, M. and {Nousiainen}, J. and {Engler}, B. and {Heritier}, C.~T. and {Kammerer}, J. and {Leveratto}, S. and {Rajani}, C. and {Bristow}, P. and {Le Louarn}, M. and {Madec}, P.-Y. and {Str{\"o}bele}, S. and {Verinaud}, C. and {Glauser}, A. and {Quanz}, S.~P. and {Helin}, T. and {Keller}, C. and {Snik}, F. and {Boccaletti}, A. and {Chauvin}, G. and {Mouillet}, D. and {Kulcs{\'a}r}, C. and {Raynaud}, H.-F.},
        title = "{PCS {\textemdash} A Roadmap for Exoearth Imaging with the ELT}",
      journal = {The Messenger},
         year = 2021,
        month = mar,
       volume = {182},
        pages = {38-43},
          doi = {10.18727/0722-6691/5221},
archivePrefix = {arXiv},
       eprint = {2103.11196},
 primaryClass = {astro-ph.IM},
       adsurl = {https://ui.adsabs.harvard.edu/abs/2021Msngr.182...38K}
}

@ARTICLE{Ripepi26,
       author = {{Ripepi}, V. and {Trentin}, E. and {Catanzaro}, G. and {Marconi}, M. and {Bhardwaj}, A. and {Clementini}, G. and {Cusano}, F. and {De Somma}, G. and {Molinaro}, R. and {Sicignano}, T. and {Storm}, J.},
        title = "{Cepheid Metallicity in the Leavitt Law (C─MetaLL) survey: IX. Metallicity dependence of period-Wesenheit relations based on a homogeneous spectroscopic sample}",
      journal = {\aap},
         year = 2026,
        month = apr,
       volume = {708},
          eid = {A216},
        pages = {A216},
          doi = {10.1051/0004-6361/202556963},
archivePrefix = {arXiv},
       eprint = {2508.17447},
 primaryClass = {astro-ph.SR},
       adsurl = {https://ui.adsabs.harvard.edu/abs/2026A&A...708A.216R}
}

@ARTICLE{Marconi24cep,
       author = {{Marconi}, Marcella and {De Somma}, Giulia and {Molinaro}, Roberto and {Bhardwaj}, Anupam and {Ripepi}, Vincenzo and {Musella}, Ilaria and {Sicignano}, Teresa and {Trentin}, Erasmo and {Leccia}, Silvio},
        title = "{The Hertzsprung progression of classical Cepheids in the Gaia era}",
      journal = {\mnras},
         year = 2024,
        month = apr,
       volume = {529},
       number = {4},
        pages = {4210-4233},
          doi = {10.1093/mnras/stae734},
archivePrefix = {arXiv},
       eprint = {2403.05699},
 primaryClass = {astro-ph.SR},
       adsurl = {https://ui.adsabs.harvard.edu/abs/2024MNRAS.529.4210M}
}

@ARTICLE{DeSomma24,
       author = {{De Somma}, Giulia and {Marconi}, Marcella and {Cassisi}, Santi and {Molinaro}, Roberto and {Bhardwaj}, Anupam and {Ripepi}, Vincenzo and {Musella}, Ilaria and {Pietrinferni}, Adriano and {Sicignano}, Teresa and {Trentin}, Erasmo and {Leccia}, Silvio},
        title = "{Classical Cepheid pulsation properties in the Rubin-LSST filters}",
      journal = {\mnras},
         year = 2024,
        month = mar,
       volume = {528},
       number = {4},
        pages = {6637-6659},
          doi = {10.1093/mnras/stae450},
archivePrefix = {arXiv},
       eprint = {2402.05721},
 primaryClass = {astro-ph.SR},
       adsurl = {https://ui.adsabs.harvard.edu/abs/2024MNRAS.528.6637D}
}

@article{birkby2018,
	title = {Exoplanet {Atmospheres} at {High} {Spectral} {Resolution}},
	url = {http://arxiv.org/abs/1806.04617},
	urldate = {2020-04-24},
	journal = {arXiv:1806.04617 [astro-ph]},
	author = {Birkby, J. L.},
	month = jun,
	year = {2018},
	note = {arXiv: 1806.04617},
}

@article{brogi2017,
	title = {A {Framework} to {Combine} {Low}- and {High}-resolution {Spectroscopy} for the {Atmospheres} of {Transiting} {Exoplanets}},
	volume = {839},
	issn = {0004-637X},
	url = {http://adsabs.harvard.edu/abs/2017ApJ...839L...2B},
	doi = {10.3847/2041-8213/aa6933},
	urldate = {2021-05-04},
	journal = {The Astrophysical Journal Letters},
	author = {Brogi, M. and Line, M. and Bean, J. and Désert, J.-M. and Schwarz, H.},
	month = apr,
	year = {2017},
	pages = {L2},
}

@article{snellen2010,
	title = {The orbital motion, absolute mass and high-altitude winds of exoplanet {HD} 209458b},
	volume = {465},
	issn = {0028-0836, 1476-4687},
	url = {http://www.nature.com/articles/nature09111},
	doi = {10.1038/nature09111},
	language = {en},
	number = {7301},
	urldate = {2020-06-03},
	journal = {Nature},
	author = {Snellen, Ignas A. G. and de Kok, Remco J. and de Mooij, Ernst J. W. and Albrecht, Simon},
	month = jun,
	year = {2010},
	note = {Number: 7301},
	pages = {1049--1051}
}

@article{snellen2014,
	title = {The fast spin-rotation of a young extrasolar planet},
	volume = {509},
	issn = {0028-0836, 1476-4687},
	url = {http://arxiv.org/abs/1404.7506},
	doi = {10.1038/nature13253},
	number = {7498},
	urldate = {2026-04-23},
	journal = {Nature},
	author = {Snellen, Ignas and Brandl, Bernhard and Kok, Remco de and Brogi, Matteo and Birkby, Jayne and Schwarz, Henriette},
	month = may,
	year = {2014},
	note = {arXiv:1404.7506 [astro-ph]},
	pages = {63--65},
}

@ARTICLE{Schwieterman2018,
       author = {{Schwieterman}, Edward W. and {Kiang}, Nancy Y. and {Parenteau}, Mary N. and {Harman}, Chester E. and {DasSarma}, Shiladitya and {Fisher}, Theresa M. and {Arney}, Giada N. and {Hartnett}, Hilairy E. and {Reinhard}, Christopher T. and {Olson}, Stephanie L. and {Meadows}, Victoria S. and {Cockell}, Charles S. and {Walker}, Sara I. and {Grenfell}, John Lee and {Hegde}, Siddharth and {Rugheimer}, Sarah and {Hu}, Renyu and {Lyons}, Timothy W.},
        title = "{Exoplanet Biosignatures: A Review of Remotely Detectable Signs of Life}",
      journal = {Astrobiology},
         year = 2018,
        month = jun,
       volume = {18},
       number = {6},
        pages = {663-708},
          doi = {10.1089/ast.2017.1729},
archivePrefix = {arXiv},
       eprint = {1705.05791},
 primaryClass = {astro-ph.EP},
       adsurl = {https://ui.adsabs.harvard.edu/abs/2018AsBio..18..663S}
}

@ARTICLE{Meadows2018,
       author = {{Meadows}, Victoria S. and {Reinhard}, Christopher T. and {Arney}, Giada N. and {Parenteau}, Mary N. and {Schwieterman}, Edward W. and {Domagal-Goldman}, Shawn D. and {Lincowski}, Andrew P. and {Stapelfeldt}, Karl R. and {Rauer}, Heike and {DasSarma}, Shiladitya and {Hegde}, Siddharth and {Narita}, Norio and {Deitrick}, Russell and {Lustig-Yaeger}, Jacob and {Lyons}, Timothy W. and {Siegler}, Nicholas and {Grenfell}, J. Lee},
        title = "{Exoplanet Biosignatures: Understanding Oxygen as a Biosignature in the Context of Its Environment}",
      journal = {Astrobiology},
         year = 2018,
        month = jun,
       volume = {18},
       number = {6},
        pages = {630-662},
          doi = {10.1089/ast.2017.1727},
archivePrefix = {arXiv},
       eprint = {1705.07560},
 primaryClass = {astro-ph.EP},
       adsurl = {https://ui.adsabs.harvard.edu/abs/2018AsBio..18..630M}
}

@ARTICLE{Alei2024,
       author = {{Alei}, E. and {Quanz}, S.~P. and {Konrad}, B.~S. and {Garvin}, E.~O. and {Kofman}, V. and {Mandell}, A. and {Angerhausen}, D. and {Molli{\`e}re}, P. and {Meyer}, M.~R. and {Robinson}, T. and {Rugheimer}, S. and {the LIFE Collaboration}},
        title = "{Large Interferometer For Exoplanets (LIFE): XIII. The value of combining thermal emission and reflected light for the characterization of Earth twins}",
      journal = {\aap},
         year = 2024,
        month = sep,
       volume = {689},
          eid = {A245},
        pages = {A245},
          doi = {10.1051/0004-6361/202450320},
archivePrefix = {arXiv},
       eprint = {2406.13037},
 primaryClass = {astro-ph.EP},
       adsurl = {https://ui.adsabs.harvard.edu/abs/2024A&A...689A.245A}
}

@ARTICLE{Plavchan2024,
       author = {{Plavchan}, Peter and {Berberian}, Jr, John E. and {Kane}, Stephen R and {Morgan}, Rhonda and {Peretz}, Eliad and {Economon}, Sophia},
        title = "{Analytic relations assessing the impact of precursor knowledge and key mission parameters on direct imaging survey yield}",
      journal = {arXiv e-prints},
         year = 2024,
        month = jan,
          eid = {arXiv:2401.02039},
        pages = {arXiv:2401.02039},
          doi = {10.48550/arXiv.2401.02039},
archivePrefix = {arXiv},
       eprint = {2401.02039},
 primaryClass = {astro-ph.EP},
       adsurl = {https://ui.adsabs.harvard.edu/abs/2024arXiv240102039P}
}

@ARTICLE{Kane2024,
       author = {{Kane}, Stephen R. and {Li}, Zhexing and {Turnbull}, Margaret C. and {Dressing}, Courtney D. and {Harada}, Caleb K.},
        title = "{Dynamical Viability Assessment for Habitable Worlds Observatory Targets}",
      journal = {\aj},
         year = 2024,
        month = nov,
       volume = {168},
       number = {5},
          eid = {195},
        pages = {195},
          doi = {10.3847/1538-3881/ad6a50},
archivePrefix = {arXiv},
       eprint = {2408.00263},
 primaryClass = {astro-ph.EP},
       adsurl = {https://ui.adsabs.harvard.edu/abs/2024AJ....168..195K}
}

@ARTICLE{Painter2025,
       author = {{Painter}, Katie E. and {Bowler}, Brendan P. and {Franson}, Kyle and {Becker}, Juliette C. and {Burt}, Jennifer A.},
        title = "{Astrometric Accelerations of Provisional Targets for the Habitable Worlds Observatory}",
      journal = {\aj},
         year = 2025,
        month = sep,
       volume = {170},
       number = {3},
          eid = {147},
        pages = {147},
          doi = {10.3847/1538-3881/ade442},
archivePrefix = {arXiv},
       eprint = {2506.21768},
 primaryClass = {astro-ph.EP},
       adsurl = {https://ui.adsabs.harvard.edu/abs/2025AJ....170..147P}
}

@ARTICLE{Damiano2025,
       author = {{Damiano}, Mario and {Burr}, Zachary and {Hu}, Renyu and {Burt}, Jennifer and {Kataria}, Tiffany},
        title = "{Effects of Planetary Mass Uncertainties on the Interpretation of the Reflectance Spectra of Earth-like Exoplanets}",
      journal = {\aj},
         year = 2025,
        month = feb,
       volume = {169},
       number = {2},
          eid = {97},
        pages = {97},
          doi = {10.3847/1538-3881/ada610},
archivePrefix = {arXiv},
       eprint = {2502.01513},
 primaryClass = {astro-ph.EP},
       adsurl = {https://ui.adsabs.harvard.edu/abs/2025AJ....169...97D}
}

@ARTICLE{Stark2024,
       author = {{Stark}, Christopher C. and {Latouf}, Natasha and {Mandell}, Avi M. and {Young}, Amber},
        title = "{Optimized bandpasses for the Habitable Worlds Observatory's exoEarth survey}",
      journal = {Journal of Astronomical Telescopes, Instruments, and Systems},
         year = 2024,
        month = jan,
       volume = {10},
          eid = {014005},
        pages = {014005},
          doi = {10.1117/1.JATIS.10.1.014005},
archivePrefix = {arXiv},
       eprint = {2404.05654},
 primaryClass = {astro-ph.EP},
       adsurl = {https://ui.adsabs.harvard.edu/abs/2024JATIS..10a4005S}
}

@ARTICLE{Stark2014,
       author = {{Stark}, Christopher C. and {Roberge}, Aki and {Mandell}, Avi and {Robinson}, Tyler D.},
        title = "{Maximizing the ExoEarth Candidate Yield from a Future Direct Imaging Mission}",
      journal = {\apj},
         year = 2014,
        month = nov,
       volume = {795},
       number = {2},
          eid = {122},
        pages = {122},
          doi = {10.1088/0004-637X/795/2/122},
archivePrefix = {arXiv},
       eprint = {1409.5128},
 primaryClass = {astro-ph.SR},
       adsurl = {https://ui.adsabs.harvard.edu/abs/2014ApJ...795..122S}
}

@ARTICLE{Heller2022,
       author = {{Heller}, Ren{\'e} and {Harre}, Jan-Vincent and {Samadi}, R{\'e}za},
        title = "{Transit least-squares survey. IV. Earth-like transiting planets expected from the PLATO mission}",
      journal = {\aap},
         year = 2022,
        month = sep,
       volume = {665},
          eid = {A11},
        pages = {A11},
          doi = {10.1051/0004-6361/202141640},
archivePrefix = {arXiv},
       eprint = {2206.02071},
 primaryClass = {astro-ph.EP},
       adsurl = {https://ui.adsabs.harvard.edu/abs/2022A&A...665A..11H}
}

@ARTICLE{Matuszewski2023,
       author = {{Matuszewski}, F. and {Nettelmann}, N. and {Cabrera}, J. and {B{\"o}rner}, A. and {Rauer}, H.},
        title = "{Estimating the number of planets that PLATO can detect}",
      journal = {\aap},
         year = 2023,
        month = sep,
       volume = {677},
          eid = {A133},
        pages = {A133},
          doi = {10.1051/0004-6361/202245287},
archivePrefix = {arXiv},
       eprint = {2307.12163},
 primaryClass = {astro-ph.EP},
       adsurl = {https://ui.adsabs.harvard.edu/abs/2023A&A...677A.133M}
}

@ARTICLE{Bryson2021,
       author = {{Bryson}, Steve and {Kunimoto}, Michelle and {Kopparapu}, Ravi K. and {Coughlin}, Jeffrey L. and {Borucki}, William J. and {Koch}, David and {Aguirre}, Victor Silva and {Allen}, Christopher and {Barentsen}, Geert and {Batalha}, Natalie M. and {Berger}, Travis and {Boss}, Alan and {Buchhave}, Lars A. and {Burke}, Christopher J. and {Caldwell}, Douglas A. and {Campbell}, Jennifer R. and {Catanzarite}, Joseph and {Chandrasekaran}, Hema and {Chaplin}, William J. and {Christiansen}, Jessie L. and {Christensen-Dalsgaard}, J{\o}rgen and {Ciardi}, David R. and {Clarke}, Bruce D. and {Cochran}, William D. and {Dotson}, Jessie L. and {Doyle}, Laurance R. and {Duarte}, Eduardo Seperuelo and {Dunham}, Edward W. and {Dupree}, Andrea K. and {Endl}, Michael and {Fanson}, James L. and {Ford}, Eric B. and {Fujieh}, Maura and {Gautier}, III, Thomas N. and {Geary}, John C. and {Gilliland}, Ronald L. and {Girouard}, Forrest R. and {Gould}, Alan and {Haas}, Michael R. and {Henze}, Christopher E. and {Holman}, Matthew J. and {Howard}, Andrew W. and {Howell}, Steve B. and {Huber}, Daniel and {Hunter}, Roger C. and {Jenkins}, Jon M. and {Kjeldsen}, Hans and {Kolodziejczak}, Jeffery and {Larson}, Kipp and {Latham}, David W. and {Li}, Jie and {Mathur}, Savita and {Meibom}, S{\o}ren and {Middour}, Chris and {Morris}, Robert L. and {Morton}, Timothy D. and {Mullally}, Fergal and {Mullally}, Susan E. and {Pletcher}, David and {Prsa}, Andrej and {Quinn}, Samuel N. and {Quintana}, Elisa V. and {Ragozzine}, Darin and {Ramirez}, Solange V. and {Sanderfer}, Dwight T. and {Sasselov}, Dimitar and {Seader}, Shawn E. and {Shabram}, Megan and {Shporer}, Avi and {Smith}, Jeffrey C. and {Steffen}, Jason H. and {Still}, Martin and {Torres}, Guillermo and {Troeltzsch}, John and {Twicken}, Joseph D. and {Uddin}, Akm Kamal and {Van Cleve}, Jeffrey E. and {Voss}, Janice and {Weiss}, Lauren M. and {Welsh}, William F. and {Wohler}, Bill and {Zamudio}, Khadeejah A.},
        title = "{The Occurrence of Rocky Habitable-zone Planets around Solar-like Stars from Kepler Data}",
      journal = {\aj},
         year = 2021,
        month = jan,
       volume = {161},
       number = {1},
          eid = {36},
        pages = {36},
          doi = {10.3847/1538-3881/abc418},
archivePrefix = {arXiv},
       eprint = {2010.14812},
 primaryClass = {astro-ph.EP},
       adsurl = {https://ui.adsabs.harvard.edu/abs/2021AJ....161...36B}
}

@ARTICLE{Meunier2024,
       author = {{Meunier}, Nad{\`e}ge},
        title = "{Impact of stellar variability on exoplanet detectability and characterisation}",
      journal = {Comptes Rendus Physique},
         year = 2024,
        month = jan,
       volume = {24},
       number = {S2},
          eid = {140},
        pages = {140},
          doi = {10.5802/crphys.140},
       adsurl = {https://ui.adsabs.harvard.edu/abs/2024CRPhy..24S.140M}
}

@ARTICLE{Meunier2022,
       author = {{Meunier}, N. and {Lagrange}, A.-M.},
        title = "{A new estimation of astrometric exoplanet detection limits in the habitable zone around nearby stars}",
      journal = {\aap},
         year = 2022,
        month = mar,
       volume = {659},
          eid = {A104},
        pages = {A104},
          doi = {10.1051/0004-6361/202142702},
archivePrefix = {arXiv},
       eprint = {2202.06301},
 primaryClass = {astro-ph.EP},
       adsurl = {https://ui.adsabs.harvard.edu/abs/2022A&A...659A.104M}
}

@ARTICLE{Sozzetti2005,
       author = {{Sozzetti}, Alessandro},
        title = "{Astrometric Methods and Instrumentation to Identify and Characterize Extrasolar Planets: A Review}",
      journal = {\pasp},
         year = 2005,
        month = oct,
       volume = {117},
       number = {836},
        pages = {1021-1048},
          doi = {10.1086/444487},
archivePrefix = {arXiv},
       eprint = {astro-ph/0507115},
 primaryClass = {astro-ph},
       adsurl = {https://ui.adsabs.harvard.edu/abs/2005PASP..117.1021S}
}

@ARTICLE{Makarov2009,
       author = {{Makarov}, V.~V. and {Beichman}, C.~A. and {Catanzarite}, J.~H. and {Fischer}, D.~A. and {Lebreton}, J. and {Malbet}, F. and {Shao}, M.},
        title = "{Starspot Jitter in Photometry, Astrometry, and Radial Velocity Measurements}",
      journal = {\apjl},
         year = 2009,
        month = dec,
       volume = {707},
       number = {1},
        pages = {L73-L76},
          doi = {10.1088/0004-637X/707/1/L73},
archivePrefix = {arXiv},
       eprint = {0911.2008},
 primaryClass = {astro-ph.SR},
       adsurl = {https://ui.adsabs.harvard.edu/abs/2009ApJ...707L..73M}
}

@ARTICLE{GravityCollaboration2021,
       author = {{GRAVITY Collaboration} and {Abuter}, R. and {Amorim}, A. and {Baub{\"o}ck}, M. and {Berger}, J.~P. and {Bonnet}, H. and {Brandner}, W. and {Cl{\'e}net}, Y. and {Davies}, R. and {de Zeeuw}, P.~T. and {Dexter}, J. and {Dallilar}, Y. and {Drescher}, A. and {Eckart}, A. and {Eisenhauer}, F. and {F{\"o}rster Schreiber}, N.~M. and {Garcia}, P. and {Gao}, F. and {Gendron}, E. and {Genzel}, R. and {Gillessen}, S. and {Habibi}, M. and {Haubois}, X. and {Hei{\ss}el}, G. and {Henning}, T. and {Hippler}, S. and {Horrobin}, M. and {Jim{\'e}nez-Rosales}, A. and {Jochum}, L. and {Jocou}, L. and {Kaufer}, A. and {Kervella}, P. and {Lacour}, S. and {Lapeyr{\`e}re}, V. and {Le Bouquin}, J.-B. and {L{\'e}na}, P. and {Lutz}, D. and {Nowak}, M. and {Ott}, T. and {Paumard}, T. and {Perraut}, K. and {Perrin}, G. and {Pfuhl}, O. and {Rabien}, S. and {Rodr{\'\i}guez-Coira}, G. and {Shangguan}, J. and {Shimizu}, T. and {Scheithauer}, S. and {Stadler}, J. and {Straub}, O. and {Straubmeier}, C. and {Sturm}, E. and {Tacconi}, L.~J. and {Vincent}, F. and {von Fellenberg}, S. and {Waisberg}, I. and {Widmann}, F. and {Wieprecht}, E. and {Wiezorrek}, E. and {Woillez}, J. and {Yazici}, S. and {Young}, A. and {Zins}, G.},
        title = "{Improved GRAVITY astrometric accuracy from modeling optical aberrations}",
      journal = {\aap},
         year = 2021,
        month = mar,
       volume = {647},
          eid = {A59},
        pages = {A59},
          doi = {10.1051/0004-6361/202040208},
archivePrefix = {arXiv},
       eprint = {2101.12098},
 primaryClass = {astro-ph.GA},
       adsurl = {https://ui.adsabs.harvard.edu/abs/2021A&A...647A..59G}
}

@ARTICLE{Sagynbayeva2025,
       author = {{Sagynbayeva}, Sabina and {Abbas}, Asif and {Kane}, Stephen R. and {Nielsen}, Eric L. and {Thompson}, William and {Blunt}, Sarah and {Rice}, Malena and {Christiansen}, Jessie L. and {Harada}, Caleb K. and {Newton}, Elisabeth R. and {Hasegawa}, Yasuhiro and {Armitage}, Philip J. and {Daylan}, Tansu},
        title = "{Requirements for Joint Orbital Characterization of Cold Giants and Habitable Worlds with Habitable Worlds Observatory}",
      journal = {\aj},
         year = 2025,
        month = oct,
       volume = {170},
       number = {4},
          eid = {208},
        pages = {208},
          doi = {10.3847/1538-3881/adf84d},
archivePrefix = {arXiv},
       eprint = {2507.21443},
 primaryClass = {astro-ph.EP},
       adsurl = {https://ui.adsabs.harvard.edu/abs/2025AJ....170..208S}
}

@ARTICLE{Spohn2022,
       author = {{Spohn}, Corey and {Savransky}, Dmitry and {Morgan}, Rhonda},
        title = "{Scheduling Direct Imaging Observations Based on Radial Velocity Orbital Fits: Best Practices for Translating Orbits and Failure Modes}",
      journal = {\aj},
         year = 2022,
        month = apr,
       volume = {163},
       number = {4},
          eid = {163},
        pages = {163},
          doi = {10.3847/1538-3881/ac5049},
       adsurl = {https://ui.adsabs.harvard.edu/abs/2022AJ....163..163S}
}

@INPROCEEDINGS{Spohn2024,
       author = {{Spohn}, Corey and {Stark}, Chris and {Savransky}, Dmitry},
        title = "{How the Habitable Worlds Observatory's field of regard will impact the use of precursor science}",
    booktitle = {Space Telescopes and Instrumentation 2024: Optical, Infrared, and Millimeter Wave},
         year = 2024,
       editor = {{Coyle}, Laura E. and {Matsuura}, Shuji and {Perrin}, Marshall D.},
       series = {Society of Photo-Optical Instrumentation Engineers (SPIE) Conference Series},
       volume = {13092},
        month = aug,
          eid = {130925L},
        pages = {130925L},
          doi = {10.1117/12.3020689},
       adsurl = {https://ui.adsabs.harvard.edu/abs/2024SPIE13092E..5LS}
}

@INPROCEEDINGS{Biazzo2022,
       author = {{Biazzo}, K. and {Bozza}, V. and {Mancini}, L. and {Sozzetti}, A.},
        title = "{The Demographics of Close-In Planets}",
    booktitle = {Demographics of Exoplanetary Systems, Lecture Notes of the 3rd Advanced School on Exoplanetary Science},
         year = 2022,
       editor = {{Biazzo}, Katia and {Bozza}, Valerio and {Mancini}, Luigi and {Sozzetti}, Alessandro},
       series = {Astrophysics and Space Science Library},
       volume = {466},
        month = jan,
        pages = {143-234},
          doi = {10.1007/978-3-030-88124-5_3},
       adsurl = {https://ui.adsabs.harvard.edu/abs/2022ASSL..466..143B}
}

@ARTICLE{Spite2025,
author = {Spite, F. and Barbuy, B. and Tan, K.},
title = {Precise boron abundance in a sample of metal-poor stars from far-ultraviolet lines},
journal = {Astronomy \& Astrophysics},
year = {2025},
volume = {702},
pages = {A217},
bibcode = {2025A&A...702A.217S},
adsurl = {https://ui.adsabs.harvard.edu/abs/2025A%26A...702A.217S}
}

@ARTICLE{Smiljanic2014,
       author = {{Smiljanic}, R.},
        title = "{Stellar abundances of beryllium and CUBES}",
      journal = {\apss},
         year = 2014,
        month = nov,
       volume = {354},
       number = {1},
        pages = {55-64},
          doi = {10.1007/s10509-014-1916-9},
archivePrefix = {arXiv},
       eprint = {1403.6276},
 primaryClass = {astro-ph.SR},
       adsurl = {https://ui.adsabs.harvard.edu/abs/2014Ap&SS.354...55S}
}

@ARTICLE{Pepe2021,
       author = {{Pepe}, F. and {Cristiani}, S. and {Rebolo}, R. and {Santos}, N.~C. and {Dekker}, H. and {Cabral}, A. and {Di Marcantonio}, P. and {Figueira}, P. and {Lo Curto}, G. and {Lovis}, C. and {Mayor}, M. and {M{\'e}gevand}, D. and {Molaro}, P. and {Riva}, M. and {Zapatero Osorio}, M.~R. and {Amate}, M. and {Manescau}, A. and {Pasquini}, L. and {Zerbi}, F.~M. and {Adibekyan}, V. and {Abreu}, M. and {Affolter}, M. and {Alibert}, Y. and {Aliverti}, M. and {Allart}, R. and {Allende Prieto}, C. and {{\'A}lvarez}, D. and {Alves}, D. and {Avila}, G. and {Baldini}, V. and {Bandy}, T. and {Barros}, S.~C.~C. and {Benz}, W. and {Bianco}, A. and {Borsa}, F. and {Bourrier}, V. and {Bouchy}, F. and {Broeg}, C. and {Calderone}, G. and {Cirami}, R. and {Coelho}, J. and {Conconi}, P. and {Coretti}, I. and {Cumani}, C. and {Cupani}, G. and {D'Odorico}, V. and {Damasso}, M. and {Deiries}, S. and {Delabre}, B. and {Demangeon}, O.~D.~S. and {Dumusque}, X. and {Ehrenreich}, D. and {Faria}, J.~P. and {Fragoso}, A. and {Genolet}, L. and {Genoni}, M. and {G{\'e}nova Santos}, R. and {Gonz{\'a}lez Hern{\'a}ndez}, J.~I. and {Hughes}, I. and {Iwert}, O. and {Kerber}, F. and {Knudstrup}, J. and {Landoni}, M. and {Lavie}, B. and {Lillo-Box}, J. and {Lizon}, J.-L. and {Maire}, C. and {Martins}, C.~J.~A.~P. and {Mehner}, A. and {Micela}, G. and {Modigliani}, A. and {Monteiro}, M.~A. and {Monteiro}, M.~J.~P.~F.~G. and {Moschetti}, M. and {Murphy}, M.~T. and {Nunes}, N. and {Oggioni}, L. and {Oliveira}, A. and {Oshagh}, M. and {Pall{\'e}}, E. and {Pariani}, G. and {Poretti}, E. and {Rasilla}, J.~L. and {Rebord{\~a}o}, J. and {Redaelli}, E.~M. and {Santana Tschudi}, S. and {Santin}, P. and {Santos}, P. and {S{\'e}gransan}, D. and {Schmidt}, T.~M. and {Segovia}, A. and {Sosnowska}, D. and {Sozzetti}, A. and {Sousa}, S.~G. and {Span{\`o}}, P. and {Su{\'a}rez Mascare{\~n}o}, A. and {Tabernero}, H. and {Tenegi}, F. and {Udry}, S. and {Zanutta}, A.},
        title = "{ESPRESSO at VLT. On-sky performance and first results}",
      journal = {\aap},
         year = 2021,
        month = jan,
       volume = {645},
          eid = {A96},
        pages = {A96},
          doi = {10.1051/0004-6361/202038306},
archivePrefix = {arXiv},
       eprint = {2010.00316},
 primaryClass = {astro-ph.IM},
       adsurl = {https://ui.adsabs.harvard.edu/abs/2021A&A...645A..96P}
}

@INPROCEEDINGS{Malbet2024,
author = {{Malbet}, Fabien and {Lizzana}, Manon and {Pancher}, Fabrice and {Soler}, Sebastien and {Leger}, Alain and {Thierry}, Lepine and {Mamon}, Gary A. and {Sozzetti}, Alessandro and {Riva}, Alberto and {Busonero}, Deborah and {Labadie}, Lucas and {Lagage}, Pierre-Olivier and {Goullioud}, Renaud.},
title = "{Challenges in focal plane and telescope calibration for high-precision space astrometry}",
booktitle = {Space Telescopes and Instrumentation 2024: Optical, Infrared, and Millimeter Wave},
         year = 2024,
       editor = {{Coyle}, Laura E. and {Matsuura}, Shuji and {Perrin}, Marshall D.},
       series = {Society of Photo-Optical Instrumentation Engineers (SPIE) Conference Series},
       volume = {13092},
        month = aug,
          eid = {130920B},
        pages = {130920B},
          doi = {10.1117/12.3019233},
archivePrefix = {arXiv},
       eprint = {2207.12540},
 primaryClass = {astro-ph.IM},
       adsurl = {https://ui.adsabs.harvard.edu/abs/2024SPIE13092E..0BM}
}

@INPROCEEDINGS{Malbet2022,
       author = {{Malbet}, Fabien and {Labadie}, Lucas and {Sozzetti}, Alessandro and {Mamon}, Gary A. and {Shao}, Mike and {Goullioud}, Renaud and {L{\'e}ger}, Alain and {Gai}, Mario and {Riva}, Alberto and {Busonero}, Deborah and {L{\'e}pine}, Thierry and {Lizzana}, Manon and {Brandeker}, Alexis and {Villaver}, Eva},
        title = "{Theia: science cases and mission profiles for high precision astrometry in the future}",
    booktitle = {Space Telescopes and Instrumentation 2022: Optical, Infrared, and Millimeter Wave},
         year = 2022,
       editor = {{Coyle}, Laura E. and {Matsuura}, Shuji and {Perrin}, Marshall D.},
       series = {Society of Photo-Optical Instrumentation Engineers (SPIE) Conference Series},
       volume = {12180},
        month = aug,
          eid = {121801F},
        pages = {121801F},
          doi = {10.1117/12.2629927},
archivePrefix = {arXiv},
       eprint = {2207.12540},
 primaryClass = {astro-ph.IM},
       adsurl = {https://ui.adsabs.harvard.edu/abs/2022SPIE12180E..1FM}
}

@ARTICLE{Malbet2021,
       author = {{Malbet}, Fabien and {Boehm}, C{\'e}line and {Krone-Martins}, Alberto and {Amorim}, Antonio and {Anglada-Escud{\'e}}, Guillem and {Brandeker}, Alexis and {Courbin}, Fr{\'e}d{\'e}ric and {En{\ss}lin}, Torsten and {Falc{\~a}o}, Antonio and {Freese}, Katherine and {Holl}, Berry and {Labadie}, Lucas and {L{\'e}ger}, Alain and {Mamon}, Gary A. and {McArthur}, Barbara and {Mora}, Alcione and {Shao}, Mike and {Sozzetti}, Alessandro and {Spolyar}, Douglas and {Villaver}, Eva and {Abbas}, Ummi and {Albertus}, Conrado and {Alves}, Jo{\~a}o and {Barnes}, Rory and {Bonomo}, Aldo Stefano and {Bouy}, Herv{\'e} and {Brown}, Warren R. and {Cardoso}, Vitor and {Castellani}, Marco and {Chemin}, Laurent and {Clark}, Hamish and {Correia}, Alexandre C.~M. and {Crosta}, Mariateresa and {Crouzier}, Antoine and {Damasso}, Mario and {Darling}, Jeremy and {Davies}, Melvyn B. and {Diaferio}, Antonaldo and {Fortin}, Morgane and {Fridlund}, Malcolm and {Gai}, Mario and {Garcia}, Paulo and {Gnedin}, Oleg and {Goobar}, Ariel and {Gordo}, Paulo and {Goullioud}, Renaud and {Hall}, David and {Hambly}, Nigel and {Harrison}, Diana and {Hobbs}, David and {Holland}, Andrew and {H{\o}g}, Erik and {Jordi}, Carme and {Klioner}, Sergei and {Lan{\c{c}}on}, Ariane and {Laskar}, Jacques and {Lattanzi}, Mario and {Le Poncin-Lafitte}, Christophe and {Luri}, Xavier and {Michalik}, Daniel and {Moitinho de Almeida}, Andr{\'e} and {Mour{\~a}o}, Ana and {Moustakas}, Leonidas and {Murray}, Neil J. and {Muterspaugh}, Matthew and {Oertel}, Micaela and {Ostorero}, Luisa and {Portell}, Jordi and {Prost}, Jean-Pierre and {Quirrenbach}, Andreas and {Schneider}, Jean and {Scott}, Pat and {Siebert}, Arnaud and {Silva}, Antonio da and {Silva}, Manuel and {Th{\'e}bault}, Philippe and {Tomsick}, John and {Traub}, Wesley and {de Val-Borro}, Miguel and {Valluri}, Monica and {Walton}, Nicholas A. and {Watkins}, Laura L. and {White}, Glenn and {Wyrzykowski}, Lukasz and {Wyse}, Rosemary and {Yamada}, Yoshiyuki},
        title = "{Faint objects in motion: the new frontier of high precision astrometry}",
      journal = {Experimental Astronomy},
         year = 2021,
        month = jun,
       volume = {51},
       number = {3},
        pages = {845-886},
          doi = {10.1007/s10686-021-09781-1},
archivePrefix = {arXiv},
       eprint = {2111.08709},
 primaryClass = {astro-ph.IM},
       adsurl = {https://ui.adsabs.harvard.edu/abs/2021ExA....51..845M}
}

@INPROCEEDINGS{Riva2020,
       author = {{Riva}, Alberto and {Gai}, Mario and {Vecchiato}, Alberto and {Busonero}, Deborah and {Lattanzi}, Mario Glilberto and {Landini}, Federico and {Qi}, Zhaoxiang and {Tang}, Zhenghong},
        title = "{RAFTER: Ring Astrometric Field Telescope for Exo-planets and Relativity}",
    booktitle = {Space Telescopes and Instrumentation 2020: Optical, Infrared, and Millimeter Wave},
         year = 2020,
       editor = {{Lystrup}, Makenzie and {Perrin}, Marshall D.},
       series = {Society of Photo-Optical Instrumentation Engineers (SPIE) Conference Series},
       volume = {11443},
        month = dec,
          eid = {114430P},
        pages = {114430P},
          doi = {10.1117/12.2576806},
archivePrefix = {arXiv},
       eprint = {2104.03003},
 primaryClass = {astro-ph.IM},
       adsurl = {https://ui.adsabs.harvard.edu/abs/2020SPIE11443E..0PR}
}

@ARTICLE{Gai2022,
       author = {{Gai}, M. and {Vecchiato}, A. and {Riva}, A. and {Butkevich}, A.~G. and {Busonero}, D. and {Qi}, Z. and {Lattanzi}, M.~G.},
        title = "{Relative Astrometry in an Annular Field}",
      journal = {\pasp},
         year = 2022,
        month = mar,
       volume = {134},
       number = {1033},
          eid = {035001},
        pages = {035001},
          doi = {10.1088/1538-3873/ac50a1},
archivePrefix = {arXiv},
       eprint = {2212.03001},
 primaryClass = {astro-ph.IM},
       adsurl = {https://ui.adsabs.harvard.edu/abs/2022PASP..134c5001G}
}

@INPROCEEDINGS{Malbet2026,
       author = {{Malbet}, Fabien and {Amiaux}, J{\'e}r{\^o}me and {Ardellier-Desages}, Florence and {Goullioud}, Renaud and {Greene}, Thomas and {Labadie}, Lucas and {Lagage}, Pierre-Olivier and {Lizzana}, Manon and {L{\'e}ger}, Alain and {L{\'e}pine}, Thierry and {Mamon}, Gary and {Martignac}, J{\'e}r{\^o}me and {Pancher}, Fabrice and {Pichon}, Thibault and {Roberge}, Aki and {Ronayette}, Samuel and {Rousset}, Hugo and {Soler}, S{\'e}bastien and {Sozzetti}, Alessandro and {Tourette}, Thierry},
        title = "{Very High Precision Astrometry for Exoplanets and Dark Matter with the Habitable Worlds Observatory}",
    booktitle = {HWO25 Proceedings Part II: Mission Framework, Technology, and Broader Contributions},
         year = 2026,
       editor = {{Lee}, Janice C. and {Noviello}, Jessica and {LaMassa}, Stephanie and {Postman}, Marc},
       series = {Astronomical Society of the Pacific Conference Series},
       volume = {543},
        month = nov,
    publisher = {ASP},
        pages = {65},
          doi = {10.48550/arXiv.2510.18920},
archivePrefix = {arXiv},
       eprint = {2510.18920},
 primaryClass = {astro-ph.IM},
       adsurl = {https://ui.adsabs.harvard.edu/abs/2026ASPC..543...65M}
}

@INPROCEEDINGS{Amiaux2026,
       author = {{Amiaux}, J{\'e}r{\^o}me and {Malbet}, Fabien and {Ardellier-Desages}, Florence and {Doumayrou}, Eric and {Frugier}, Pierre-Antoine and {Goullioud}, Renaud and {Greene}, Thomas and {Labadie}, Lucas and {Lagage}, Pierre-Olivier and {Lizzana}, Manon and {Leger}, Alain and {Lepine}, Thierry and {Mamon}, Gary and {Martignac}, J{\'e}r{\^o}me and {Michelot}, Julien and {Pancher}, Fabrice and {Pichon}, Thibault and {Roberge}, Aki and {Ronayette}, Samuel and {Rousset}, Hugo and {Sitarski}, Breann and {Sozzetti}, Alessandro and {Tourette}, Thierry},
        title = "{System Analysis for a High-Precision High-Accuracy Astrometric Instrument for HWO}",
    booktitle = {Astronomical Society of the Pacific Conference Series},
         year = 2026,
       editor = {{Lee}, Janice C. and {Noviello}, Jessica and {LaMassa}, Stephanie and {Postman}, Marc},
       series = {Astronomical Society of the Pacific Conference Series},
       volume = {543},
        month = nov,
        pages = {327},
       adsurl = {https://ui.adsabs.harvard.edu/abs/2026ASPC..543..327A}
}

@INPROCEEDINGS{Lizzana2026,
       author = {{Lizzana}, Manon and {Malbet}, Fabien and {Leger}, Alain and {Pancher}, Fabrice and {Soler}, S{\'e}bastien and {Rousset}, Hugo and {Lepine}, Thierry and {Michelot}, Julien and {Er-Rahmaouy}, Yahya and {Bakka}, Youssef},
        title = "{Experimental Tests of the Calibration of High Precision Differential Astrometry for HWO}",
    booktitle = {Astronomical Society of the Pacific Conference Series},
         year = 2026,
       editor = {{Lee}, Janice C. and {Noviello}, Jessica and {LaMassa}, Stephanie and {Postman}, Marc},
       series = {Astronomical Society of the Pacific Conference Series},
       volume = {543},
        month = nov,
        pages = {335},
       adsurl = {https://ui.adsabs.harvard.edu/abs/2026ASPC..543..335L}
}

@BOOK{Villanueva2022,
       author = {{Villanueva}, Geronimo Luis and {Liuzzi}, Giuliano and {Faggi}, Sara and {Protopapa}, Silvia and {Kofman}, Vincent and {Fauchez}, Thomas and {Stone}, Shane Wesley and {Mandell}, Avi Max},
        title = "{Fundamentals of the Planetary Spectrum Generator}",
         year = 2022,
       adsurl = {https://ui.adsabs.harvard.edu/abs/2022fpsg.book.....V}
}

@software{MacDonald2024,
       author = {{MacDonald}, Ryan J. and {Madhusudhan}, Nikku},
        title = "{POSEIDON: Multidimensional atmospheric retrieval of exoplanet spectra}",
 howpublished = {Astrophysics Source Code Library, record ascl:2412.028},
         year = 2024,
        month = dec,
          eid = {ascl:2412.028},
archivePrefix = {ascl},
       eprint = {2412.028},
       adsurl = {https://ui.adsabs.harvard.edu/abs/2024ascl.soft12028M}
}

@INPROCEEDINGS{Barbisan2026,
       author = {{Barbisan}, Diego and {Barbato}, Marco and {Maris}, Michele and {Silva}, Laura and {Boccia}, Beatrice and {La Rocca}, Nicoletta and {Coccola}, Lorenzo and {Poletto}, Luca and {Trivellin}, Nicola and {Peron}, Fabio},
        title = "{UV LED-based solar flare simulator for space environment studies}",
    booktitle = {Society of Photo-Optical Instrumentation Engineers (SPIE) Conference Series},
         year = 2026,
       editor = {{Kim}, Jong Kyu and {Krames}, Michael R. and {Strassburg}, Martin},
       series = {Society of Photo-Optical Instrumentation Engineers (SPIE) Conference Series},
       volume = {13913},
        month = mar,
          eid = {139130G},
        pages = {139130G},
          doi = {10.1117/12.3077482},
       adsurl = {https://ui.adsabs.harvard.edu/abs/2026SPIE13913E..0GB}
}

@INPROCEEDINGS{Bellucci2024,
       author = {{Bellucci}, Micoli and {Pedone}, Maria and {Pezzilli}, Serena and {Pacelli}, Claudia and {Gangi}, Manuele and {Billi}, Daniela and {Selbmann}, Laura and {Cavalazzi}, Barbara and {Negri}, Barbara},
        title = "{a Comparison Between Three Projects Exploring how Life Might Origin and Perpetuate on Mars Using Planetary Field Analogues}",
    booktitle = {45th COSPAR Scientific Assembly},
         year = 2024,
       volume = {45},
        month = jul,
        pages = {2248},
       adsurl = {https://ui.adsabs.harvard.edu/abs/2024cosp...45.2248B}
}

@ARTICLE{murante20,
       author = {{Murante}, Giuseppe and {Provenzale}, Antonello and {Vladilo}, Giovanni and {Taffoni}, Giuliano and {Silva}, Laura and {Palazzi}, Elisa and {Hardenberg}, Jost von and {Maris}, Michele and {Londero}, Elisa and {Knapic}, Cristina and {Zorba}, Sonia},
        title = "{Climate bistability of Earth-like exoplanets}",
      journal = {\mnras},
         year = 2020,
        month = feb,
       volume = {492},
       number = {2},
        pages = {2638-2650},
          doi = {10.1093/mnras/stz3529},
archivePrefix = {arXiv},
       eprint = {1912.05392},
 primaryClass = {astro-ph.EP},
       adsurl = {https://ui.adsabs.harvard.edu/abs/2020MNRAS.492.2638M}
}

@ARTICLE{bisesi26a,
       author = {{Bisesi}, E. and {Murante}, G. and {Provenzale}, A. and {von Hardenberg}, J. and {Maris}, M. and {Silva}, L.},
        title = "{Interaction between vegetation and Snowball phases in the late Proterozoic Earth}",
      journal = {arXiv e-prints},
         year = 2026,
        month = mar,
          eid = {arXiv:2603.25321},
        pages = {arXiv:2603.25321},
          doi = {10.48550/arXiv.2603.25321},
archivePrefix = {arXiv},
       eprint = {2603.25321},
 primaryClass = {astro-ph.EP},
       adsurl = {https://ui.adsabs.harvard.edu/abs/2026arXiv260325321B}
}

@ARTICLE{silva17a,
       author = {{Silva}, Laura and {Vladilo}, Giovanni and {Schulte}, Patricia M. and {Murante}, Giuseppe and {Provenzale}, Antonello},
        title = "{From climate models to planetary habitability: temperature constraints for complex life}",
      journal = {International Journal of Astrobiology},
         year = 2017,
        month = jul,
       volume = {16},
       number = {3},
        pages = {244-265},
          doi = {10.1017/S1473550416000215},
archivePrefix = {arXiv},
       eprint = {1604.08864},
 primaryClass = {astro-ph.EP},
       adsurl = {https://ui.adsabs.harvard.edu/abs/2017IJAsB..16..244S}
}

@ARTICLE{silva17b,
       author = {{Silva}, Laura and {Vladilo}, Giovanni and {Murante}, Giuseppe and {Provenzale}, Antonello},
        title = "{Quantitative estimates of the surface habitability of Kepler-452b}",
      journal = {\mnras},
         year = 2017,
        month = sep,
       volume = {470},
       number = {2},
        pages = {2270-2282},
          doi = {10.1093/mnras/stx1396},
archivePrefix = {arXiv},
       eprint = {1706.01224},
 primaryClass = {astro-ph.EP},
       adsurl = {https://ui.adsabs.harvard.edu/abs/2017MNRAS.470.2270S}
}

@Article{fraedrich05,
author = "Fraedrich, Klaus and Jansen, Heiko and Kirk, Edilbert and Lunkeit, Frank",
journal = "Meteorologische Zeitschrift",
month = 07,
year = 2005,
title = "The Planet Simulator: Green planet and desert world",
number = "3",
volume = "14",
pages = {305-314},
url = "http://dx.doi.org/10.1127/0941-2948/2005/0044",
doi = "10.1127/0941-2948/2005/0044",
publisher = "Schweizerbart Science Publishers",
address = "Stuttgart, Germany"
}

@ARTICLE{paradise22,
       author = {{Paradise}, Adiv and {Macdonald}, Evelyn and {Menou}, Kristen and {Lee}, Christopher and {Fan}, Bo Lin},
        title = "{ExoPlaSim: Extending the Planet Simulator for exoplanets}",
      journal = {\mnras},
         year = 2022,
        month = apr,
       volume = {511},
       number = {3},
        pages = {3272-3303},
          doi = {10.1093/mnras/stac172},
archivePrefix = {arXiv},
       eprint = {2107.07685},
 primaryClass = {astro-ph.EP},
       adsurl = {https://ui.adsabs.harvard.edu/abs/2022MNRAS.511.3272P}
}

@ARTICLE{bisesi24,
       author = {{Bisesi}, E. and {Murante}, G. and {Provenzale}, A. and {Biasiotti}, L. and {von Hardenberg}, J. and {Ivanovski}, S. and {Maris}, M. and {Monai}, S. and {Silva}, L. and {Simonetti}, P. and {Vladilo}, G.},
        title = "{Impact of vegetation albedo on the habitability of Earth-like exoplanets}",
      journal = {\mnras},
         year = 2024,
        month = oct,
       volume = {534},
       number = {1},
        pages = {1-11},
          doi = {10.1093/mnras/stae2016},
archivePrefix = {arXiv},
       eprint = {2409.01746},
 primaryClass = {astro-ph.EP},
       adsurl = {https://ui.adsabs.harvard.edu/abs/2024MNRAS.534....1B}
}

@ARTICLE{bisesi26b,
       author = {{Bisesi}, E. and {Murante}, G. and {von Hardenberg}, J. and {Caballero}, J. A. and {Maris}, M. and {Billi}, D. and {La Rocca}, N. and {Silva}, L. },
        title = "{Assessing the Climate and Habitability of 
Tidally Locked Rocky Exoplanets}",
      journal = {Astrobiology},
         year = 2026,
        month = apr,
       volume = {accepted},
       number = {},
        pages = {},
          doi = {},
archivePrefix = {},
       eprint = {},
 primaryClass = {},
       adsurl = {}
}

@ARTICLE{charney75,
       author = {{Charney}, J. and {Stone}, P.~H. and {Quirk}, W.~J.},
        title = "{Drought in the Sahara - A biogeophysical feedback mechanism}",
      journal = {Science},
         year = 1975,
        month = feb,
       volume = {187},
        pages = {434},
          doi = {10.1126/science.187.4175.434},
       adsurl = {https://ui.adsabs.harvard.edu/abs/1975Sci...187..434C}
}

@ARTICLE{aleina13,
       author = {{Aleina}, Fabio Cresto and {Baudena}, Mara and {D'Andrea}, Fabio and {Provenzale}, Antonello},
        title = "{Multiple equilibria on planet Dune: climate-vegetation dynamics on a sandy planet}",
      journal = {Tellus Series B Chemical and Physical Meteorology B},
         year = 2013,
        month = jan,
       volume = {65},
          eid = {17662},
        pages = {17662},
          doi = {10.3402/tellusb.v65i0.17662},
       adsurl = {https://ui.adsabs.harvard.edu/abs/2013TellB..6517662C}
}

@ARTICLE{molliere19,
       author = {{Molli{\`e}re}, P. and {Wardenier}, J.~P. and {van Boekel}, R. and {Henning}, Th. and {Molaverdikhani}, K. and {Snellen}, I.~A.~G.},
        title = "{petitRADTRANS. A Python radiative transfer package for exoplanet characterization and retrieval}",
      journal = {\aap},
         year = 2019,
        month = jul,
       volume = {627},
          eid = {A67},
        pages = {A67},
          doi = {10.1051/0004-6361/201935470},
archivePrefix = {arXiv},
       eprint = {1904.11504},
 primaryClass = {astro-ph.EP},
       adsurl = {https://ui.adsabs.harvard.edu/abs/2019A&A...627A..67M}
}

@ARTICLE{Hall:etal:2023,
       author = {{Hall}, C. and {Stancil}, P.~C. and {Terry}, J.~P. and {Ellison}, C.~K.},
        title = "{A New Definition of Exoplanet Habitability: Introducing the Photosynthetic Habitable Zone}",
      journal = {{\rm{Astrophsyical Journal Letters}}},
         year = 2023,
        month = may,
       volume = {948},
       number = {2},
          eid = {L26},
        pages = {L26},
          doi = {10.3847/2041-8213/acccfb},
archivePrefix = {arXiv},
       eprint = {2301.13836},
 primaryClass = {astro-ph.EP}
}

@ARTICLE{Abbott2017,
       author = {{Abbott}, B.~P. and {Abbott}, R. and {Abbott}, T.~D. and {Acernese}, F. and {Ackley}, K. and {Adams}, C. and {Adams}, T. and {Addesso}, P. and {Adhikari}, R.~X. and {Adya}, V.~B. and {Affeldt}, C. and {Afrough}, M. and {Agarwal}, B. and {Agathos}, M. and {Agatsuma}, K. and {Aggarwal}, N. and {Aguiar}, O.~D. and {Aiello}, L. and {Ain}, A. and {Ajith}, P. and {Allen}, B. and {Allen}, G. and {Allocca}, A. and {Altin}, P.~A. and {Amato}, A. and {Ananyeva}, A. and {Anderson}, S.~B. and {Anderson}, W.~G. and {Angelova}, S.~V. and {Antier}, S. and {Appert}, S. and {Arai}, K. and {Araya}, M.~C. and {Areeda}, J.~S. and {Arnaud}, N. and {Arun}, K.~G. and {Ascenzi}, S. and {Ashton}, G. and {Ast}, M. and {Aston}, S.~M. and {Astone}, P. and {Atallah}, D.~V. and {Aufmuth}, P. and {Aulbert}, C. and {AultONeal}, K. and {Austin}, C. and {Avila-Alvarez}, A. and {Babak}, S. and {Bacon}, P. and {Bader}, M.~K.~M. and {Bae}, S. and {Bailes}, M. and {Baker}, P.~T. and {Baldaccini}, F. and {Ballardin}, G. and {Ballmer}, S.~W. and {Banagiri}, S. and {Barayoga}, J.~C. and {Barclay}, S.~E. and {Barish}, B.~C. and {Barker}, D. and {Barkett}, K. and {Barone}, F. and {Barr}, B. and {Barsotti}, L. and {Barsuglia}, M. and {Barta}, D. and {Barthelmy}, S.~D. and {Bartlett}, J. and {Bartos}, I. and {Bassiri}, R. and {Basti}, A. and {Batch}, J.~C. and {Bawaj}, M. and {Bayley}, J.~C. and {Bazzan}, M. and {B{\'e}csy}, B. and {Beer}, C. and {Bejger}, M. and {Belahcene}, I. and {Bell}, A.~S. and {Berger}, B.~K. and {Bergmann}, G. and {Bernuzzi}, S. and {Bero}, J.~J. and {Berry}, C.~P.~L. and {Bersanetti}, D. and {Bertolini}, A. and {Betzwieser}, J. and {Bhagwat}, S. and {Bhandare}, R. and {Bilenko}, I.~A. and {Billingsley}, G. and {Billman}, C.~R. and {Birch}, J. and {Birney}, R. and {Birnholtz}, O. and {Biscans}, S. and {Biscoveanu}, S. and {Bisht}, A. and {Bitossi}, M. and {Biwer}, C. and {Bizouard}, M.~A. and {Blackburn}, J.~K. and {Blackman}, J. and {Blair}, C.~D. and {Blair}, D.~G. and {Blair}, R.~M. and {Bloemen}, S. and {Bock}, O. and {Bode}, N. and {Boer}, M. and {Bogaert}, G. and {Bohe}, A. and {Bondu}, F. and {Bonilla}, E. and {Bonnand}, R. and {Boom}, B.~A. and {Bork}, R. and {Boschi}, V. and {Bose}, S. and {Bossie}, K. and {Bouffanais}, Y. and {Bozzi}, A. and {Bradaschia}, C. and {Brady}, P.~R. and {Branchesi}, M. and {Brau}, J.~E. and {Briant}, T. and {Brillet}, A. and {Brinkmann}, M. and {Brisson}, V. and {Brockill}, P. and {Broida}, J.~E. and {Brooks}, A.~F. and {Brown}, D.~A. and {Brown}, D.~D. and {Brunett}, S. and {Buchanan}, C.~C. and {Buikema}, A. and {Bulik}, T. and {Bulten}, H.~J. and {Buonanno}, A. and {Buskulic}, D. and {Buy}, C. and {Byer}, R.~L. and {Cabero}, M. and {Cadonati}, L. and {Cagnoli}, G. and {Cahillane}, C. and {Calder{\'o}n Bustillo}, J. and {Callister}, T.~A. and {Calloni}, E. and {Camp}, J.~B. and {Canepa}, M. and {Canizares}, P. and {Cannon}, K.~C. and {Cao}, H. and {Cao}, J. and {Capano}, C.~D. and {Capocasa}, E. and {Carbognani}, F. and {Caride}, S. and {Carney}, M.~F. and {Carullo}, G. and {Casanueva Diaz}, J. and {Casentini}, C. and {Caudill}, S. and {Cavagli{\`a}}, M. and {Cavalier}, F. and {Cavalieri}, R. and {Cella}, G. and {Cepeda}, C.~B. and {Cerd{\'a}-Dur{\'a}n}, P. and {Cerretani}, G. and {Cesarini}, E. and {Chamberlin}, S.~J. and {Chan}, M. and {Chao}, S. and {Charlton}, P. and {Chase}, E. and {Chassande-Mottin}, E. and {Chatterjee}, D. and {Chatziioannou}, K. and {Cheeseboro}, B.~D. and {Chen}, H.~Y. and {Chen}, X. and {Chen}, Y. and {Cheng}, H.-P. and {Chia}, H. and {Chincarini}, A. and {Chiummo}, A. and {Chmiel}, T. and {Cho}, H.~S. and {Cho}, M. and {Chow}, J.~H. and {Christensen}, N. and {Chu}, Q. and {Chua}, A.~J.~K. and {Chua}, S.},
        title = "{GW170817: Observation of Gravitational Waves from a Binary Neutron Star Inspiral}",
      journal = {\prl},
         year = 2017,
        month = oct,
       volume = {119},
       number = {16},
          eid = {161101},
        pages = {161101},
          doi = {10.1103/PhysRevLett.119.161101},
archivePrefix = {arXiv},
       eprint = {1710.05832},
 primaryClass = {gr-qc},
       adsurl = {https://ui.adsabs.harvard.edu/abs/2017PhRvL.119p1101A}
}

@ARTICLE{Bailes2021,
       author = {{Bailes}, M. and {Berger}, B.~K. and {Brady}, P.~R. and {Branchesi}, M. and {Danzmann}, K. and {Evans}, M. and {Holley-Bockelmann}, K. and {Iyer}, B.~R. and {Kajita}, T. and {Katsanevas}, S. and {Kramer}, M. and {Lazzarini}, A. and {Lehner}, L. and {Losurdo}, G. and {L\"uck}, H. and {McClelland}, D.~E. and {McLaughlin}, M.~A. and {Punturo}, M. and {Ransom}, S. and {Raychaudhury}, S. and {Reitze}, D. and {Ricci}, F. and {Rowan}, S. and {Saito}, Y. and {Sanders}, G.~H. and {Sathyaprakash}, B.~S. and {Schutz}, B.~F. and {Sesana}, A. and {Shinkai}, H. and {Siemens}, X. and {Shoemaker}, D.~H. and {Thorpe}, J. and {van den Brand}, J.~F.~J. and {Vitale}, S.},
        title = "{Gravitational-wave physics and astronomy in the 2020s and 2030s}",
      journal = {Nature Reviews Physics},
         year = 2021,
       volume = {3},
        pages = {344-366},
          doi = {10.1038/s42254-021-00303-8},
       adsurl = {https://ui.adsabs.harvard.edu/abs/2021NatRP...3..344B}
}

@ARTICLE{Bonifacio2025,
       author = {{Bonifacio}, Piercarlo and {Caffau}, Elisabetta and {Fran{\c{c}}ois}, Patrick and {Spite}, Monique},
        title = "{The most metal-poor stars}",
      journal = {\aapr},
         year = 2025,
        month = jul,
       volume = {33},
       number = {1},
          eid = {2},
        pages = {2},
          doi = {10.1007/s00159-025-00159-2},
archivePrefix = {arXiv},
       eprint = {2504.06335},
 primaryClass = {astro-ph.GA},
       adsurl = {https://ui.adsabs.harvard.edu/abs/2025A&ARv..33....2B}
}

@ARTICLE{LiPaczynski1998,
       author = {{Li}, L.-X. and {Paczy\'nski}, B.},
        title = "{Transient Events from Neutron Star Mergers}",
      journal = {\apjl},
         year = 1998,
       volume = {507},
        pages = {L59},
       adsurl = {https://ui.adsabs.harvard.edu/abs/1998ApJ...507L..59L}
}

@ARTICLE{Law2022,
       author = {{Law}, Nicholas and {Vasquez Soto}, Alan and {Corbett}, Hank and
                 {Galliher}, Nathan and {Gonzalez}, Ramses and {Machia}, Lawrence and
                 {Walters}, Glenn},
        title = "{The inside-out, upside-down telescope: the Argus Array's new pseudofocal design}",
         year = 2022,
        month = jul,
archivePrefix = {arXiv},
       eprint = {2207.14318},
 primaryClass = {astro-ph.IM},
       adsurl = {https://ui.adsabs.harvard.edu/abs/2022arXiv220714318L}
}

@ARTICLE{Law2024,
       author = {{Law}, C.~J. and {Sharma}, K. and {Ravi}, V. and {Chen}, G. and
                 {Catha}, M. and {Connor}, L. and {Faber}, J.~T. and {Hallinan}, G. and
                 {Harnach}, C. and {Hellbourg}, G. and {Hobbs}, R. and {Hodge}, D. and
                 {Hodges}, M. and {Lamb}, J.~W. and {Rasmussen}, P. and
                 {Sherman}, M.~B. and {Shi}, J. and {Simard}, D. and
                 {Squillace}, R. and {Weinreb}, S. and {Woody}, D.~P. and
                 {Yadlapalli}, N.},
        title = "{Deep Synoptic Array Science: First FRB and Host Galaxy Catalog}",
      journal = {\apj},
         year = 2024,
        month = jan,
archivePrefix = {arXiv},
       eprint = {2307.03344},
 primaryClass = {astro-ph.HE},
       adsurl = {https://ui.adsabs.harvard.edu/abs/2024arXiv230703344L}
}

@ARTICLE{Metzger2010,
       author = {{Metzger}, B.~D. and {Mart{\'\i}nez-Pinedo}, G. and {Darbha}, S. and {Quataert}, E. and {Arcones}, A. and {Kasen}, D. and {Thomas}, R. and {Nugent}, P. and {Panov}, I.~V. and {Zinner}, N.~T.},
        title = "{Electromagnetic counterparts of compact object mergers powered by the radioactive decay of r-process nuclei}",
      journal = {\mnras},
         year = 2010,
        month = aug,
       volume = {406},
       number = {4},
        pages = {2650-2662},
          doi = {10.1111/j.1365-2966.2010.16864.x},
archivePrefix = {arXiv},
       eprint = {1001.5029},
 primaryClass = {astro-ph.HE},
       adsurl = {https://ui.adsabs.harvard.edu/abs/2010MNRAS.406.2650M}
}

@ARTICLE{Metzger2019,
       author = {{Metzger}, B.~D.},
        title = "{Kilonovae}",
      journal = {Living Reviews in Relativity},
         year = 2019,
       volume = {23},
        pages = {1},
       adsurl = {https://ui.adsabs.harvard.edu/abs/2019LRR....23....1M}
}

@INPROCEEDINGS{Angel2022,
       author = {{Angel}, J.~Roger and {Bender}, Chad and {Berkson}, Joel and
                 {Didato}, Nick and {Ford}, J. and {Gray}, Peter and
                 {Jannuzi}, Buell and {Ketelsen}, Dean and {Kim}, Daewook and
                 {Chavez Lopez}, G. and {Monson}, Andrew and {Oh}, Chang Jin and
                 {Patrou}, J. and {Rademacher}, M. and {Schwab}, Christian and
                 {Sisco}, M. and {Wortley}, R. and {Young}, Andrew},
        title = "{LFAST, the Large Fiber Array Spectroscopic Telescope}",
    booktitle = {Ground-Based and Airborne Telescopes IX},
         year = 2022,
       series = {Proc.\ SPIE},
       volume = {12182},
          eid = {121821U},
        pages = {121821U},
          doi = {10.1117/12.2629655},
       adsurl = {https://ui.adsabs.harvard.edu/abs/2022SPIE12182E..1UA}
}

@ARTICLE{Roy2026,
       author = {{Roy}, Arpita and {Feldman}, Stuart and {Klupar}, Pete and
                 {DiPalma}, John and {Perlmutter}, Saul and {Douglas}, Ewan S. and
                 {Aldering}, Greg and {Furesz}, Gabor and {Ingraham}, Patrick and
                 {Stefansson}, Gudmundur and {Kelly}, Douglas and {Yang}, Fan Yang and
                 {Wevers}, Thomas and {Arulanantham}, Nicole and {Lasker}, James and
                 {Rigault}, Mickael and {Schlawin}, Everett and
                 {Zandbergen}, Sander R. and {Worden}, S.~Pete},
        title = "{The Lazuli Space Observatory: Architecture \& Capabilities}",
         year = 2026,
        month = jan,
          doi = {10.48550/arXiv.2601.02556},
archivePrefix = {arXiv},
       eprint = {2601.02556},
 primaryClass = {astro-ph.IM},
       adsurl = {https://ui.adsabs.harvard.edu/abs/2026arXiv260102556R}
}

@ARTICLE{Belladitta2025,
       author = {{Belladitta}, Silvia and {Ba{\~n}ados}, Eduardo and {Xie}, Zhang-Liang and {Decarli}, Roberto and {Onorato}, Silvia and {Yang}, Jinyi and {Bischetti}, Manuela and {Onoue}, Masafusa and {Loiacono}, Federica and {Mart{\'\i}nez-Ram{\'\i}rez}, Laura N. and {Mazzucchelli}, Chiara and {Davies}, Frederick B. and {Wolf}, Julien and {Schindler}, Jan-Torge and {Fan}, Xiaohui and {Wang}, Feige and {Walter}, Fabian and {Mkrtchyan}, Tatevik and {Stern}, Daniel and {Farina}, Emanuele P. and {Venemans}, Bram P.},
        title = "{Discovery and characterization of 25 new quasars at 4.6 < z < 6.9 from wide-field multiband surveys}",
      journal = {\aap},
         year = 2025,
        month = jul,
       volume = {699},
          eid = {A335},
        pages = {A335},
          doi = {10.1051/0004-6361/202554859},
archivePrefix = {arXiv},
       eprint = {2505.15923},
 primaryClass = {astro-ph.GA},
       adsurl = {https://ui.adsabs.harvard.edu/abs/2025A&A...699A.335B}
}

@ARTICLE{Berger2014,
       author = {{Berger}, E.},
        title = "{Short-Duration Gamma-Ray Bursts}",
      journal = {\araa},
         year = 2014,
       volume = {52},
        pages = {43},
       adsurl = {https://ui.adsabs.harvard.edu/abs/2014ARA&A..52...43B}
}

@ARTICLE{Christensen2023,
       author = {{Christensen}, L. and {Jakobsen}, P. and {Willott}, C. and {Arribas}, S. and {Bunker}, A. and {Charlot}, S. and {Maiolino}, R. and {Marshall}, M. and {Perna}, M. and {{\"U}bler}, H.},
        title = "{Metal enrichment and evolution in four z > 6.5 quasar sightlines observed with JWST/NIRSpec}",
      journal = {\aap},
         year = 2023,
        month = dec,
       volume = {680},
          eid = {A82},
        pages = {A82},
          doi = {10.1051/0004-6361/202347943},
archivePrefix = {arXiv},
       eprint = {2309.06470},
 primaryClass = {astro-ph.GA},
       adsurl = {https://ui.adsabs.harvard.edu/abs/2023A&A...680A..82C}
}

@ARTICLE{DAvanzo2015,
       author = {{D'Avanzo}, P.},
        title = "{Short gamma-ray bursts: A review}",
      journal = {Journal of High Energy Astrophysics},
         year = 2015,
        month = sep,
       volume = {7},
        pages = {73-80},
          doi = {10.1016/j.jheap.2015.07.002},
       adsurl = {https://ui.adsabs.harvard.edu/abs/2015JHEAp...7...73D}
}

@ARTICLE{Dodorico2024,
       author = {{D'Odorico}, Valentina and {Bolton}, James S. and {Christensen}, Lise and {De Cia}, Annalisa and {Zackrisson}, Erik and {Kordt}, Aron and {Izzo}, Luca and {Li}, Jiangtao and {Maiolino}, Roberto and {Marconi}, Alessandro and {Richter}, Philipp and {Saccardi}, Andrea and {Salvadori}, Stefania and {Vanni}, Irene and {Feruglio}, Chiara and {Fumagalli}, Michele and {Fynbo}, Johan P.~U. and {Noterdaeme}, Pasquier and {Papaderos}, Polychronis and {P{\'e}roux}, C{\'e}line and {Verma}, Aprajita and {Di Marcantonio}, Paolo and {Origlia}, Livia and {Zanutta}, Alessio},
        title = "{Galaxy formation and symbiotic evolution with the inter-galactic medium in the age of ELT-ANDES}",
      journal = {Experimental Astronomy},
         year = 2024,
        month = dec,
       volume = {58},
       number = {3},
          eid = {21},
        pages = {21},
          doi = {10.1007/s10686-024-09967-3},
archivePrefix = {arXiv},
       eprint = {2311.16803},
 primaryClass = {astro-ph.GA},
       adsurl = {https://ui.adsabs.harvard.edu/abs/2024ExA....58...21D}
}

@ARTICLE{Dodorico2023,
       author = {{D'Odorico}, Valentina and {Ba{\~n}ados}, E. and {Becker}, G.~D. and {Bischetti}, M. and {Bosman}, S.~E.~I. and {Cupani}, G. and {Davies}, R. and {Farina}, E.~P. and {Ferrara}, A. and {Feruglio}, C. and {Mazzucchelli}, C. and {Ryan-Weber}, E. and {Schindler}, J.-T. and {Sodini}, A. and {Venemans}, B.~P. and {Walter}, F. and {Chen}, H. and {Lai}, S. and {Zhu}, Y. and {Bian}, F. and {Campo}, S. and {Carniani}, S. and {Cristiani}, S. and {Davies}, F. and {Decarli}, R. and {Drake}, A. and {Eilers}, A.-C. and {Fan}, X. and {Gaikwad}, P. and {Gallerani}, S. and {Greig}, B. and {Haehnelt}, M.~G. and {Hennawi}, J. and {Keating}, L. and {Kulkarni}, G. and {Mesinger}, A. and {Meyer}, R.~A. and {Neeleman}, M. and {Onoue}, M. and {Pallottini}, A. and {Qin}, Y. and {Rojas-Ruiz}, S. and {Satyavolu}, S. and {Sebastian}, A. and {Tripodi}, R. and {Wang}, F. and {Wolfson}, M. and {Yang}, J. and {Zanchettin}, M.~V.},
        title = "{XQR-30: The ultimate XSHOOTER quasar sample at the reionization epoch}",
      journal = {\mnras},
         year = 2023,
        month = jul,
       volume = {523},
       number = {1},
        pages = {1399-1420},
          doi = {10.1093/mnras/stad1468},
archivePrefix = {arXiv},
       eprint = {2305.05053},
 primaryClass = {astro-ph.GA},
       adsurl = {https://ui.adsabs.harvard.edu/abs/2023MNRAS.523.1399D}
}

@ARTICLE{Fan2023,
       author = {{Fan}, Xiaohui and {Ba{\~n}ados}, Eduardo and {Simcoe}, Robert A.},
        title = "{Quasars and the Intergalactic Medium at Cosmic Dawn}",
      journal = {\araa},
         year = 2023,
        month = aug,
       volume = {61},
        pages = {373-426},
          doi = {10.1146/annurev-astro-052920-102455},
archivePrefix = {arXiv},
       eprint = {2212.06907},
 primaryClass = {astro-ph.GA},
       adsurl = {https://ui.adsabs.harvard.edu/abs/2023ARA&A..61..373F}
}

@ARTICLE{Ferrara2026,
       author = {{Ferrara}, Andrea and {Carniani}, Stefano and {Morishita}, Takahiro and {Stiavelli}, Massimo},
        title = "{Possible evidence for a pair-instability supernova nature of ultra-early JWST sources}",
      journal = {arXiv e-prints},
         year = 2026,
        month = jan,
          eid = {arXiv:2601.07374},
        pages = {arXiv:2601.07374},
          doi = {10.48550/arXiv.2601.07374},
archivePrefix = {arXiv},
       eprint = {2601.07374},
 primaryClass = {astro-ph.GA},
       adsurl = {https://ui.adsabs.harvard.edu/abs/2026arXiv260107374F}
}

@ARTICLE{flury22,
       author = {{Flury}, Sophia R. and {Jaskot}, Anne E. and {Ferguson}, Harry C. and {Worseck}, G{\'a}bor and {Makan}, Kirill and {Chisholm}, John and {Saldana-Lopez}, Alberto and {Schaerer}, Daniel and {McCandliss}, Stephan and {Wang}, Bingjie and {Ford}, N.~M. and {Heckman}, Timothy and {Ji}, Zhiyuan and {Giavalisco}, Mauro and {Amorin}, Ricardo and {Atek}, Hakim and {Blaizot}, Jeremy and {Borthakur}, Sanchayeeta and {Carr}, Cody and {Castellano}, Marco and {Cristiani}, Stefano and {De Barros}, Stephane and {Dickinson}, Mark and {Finkelstein}, Steven L. and {Fleming}, Brian and {Fontanot}, Fabio and {Garel}, Thibault and {Grazian}, Andrea and {Hayes}, Matthew and {Henry}, Alaina and {Mauerhofer}, Valentin and {Micheva}, Genoveva and {Oey}, M.~S. and {Ostlin}, Goran and {Papovich}, Casey and {Pentericci}, Laura and {Ravindranath}, Swara and {Rosdahl}, Joakim and {Rutkowski}, Michael and {Santini}, Paola and {Scarlata}, Claudia and {Teplitz}, Harry and {Thuan}, Trinh and {Trebitsch}, Maxime and {Vanzella}, Eros and {Verhamme}, Anne and {Xu}, Xinfeng},
        title = "{The Low-redshift Lyman Continuum Survey. I. New, Diverse Local Lyman Continuum Emitters}",
      journal = {\apjs},
         year = 2022,
        month = may,
       volume = {260},
       number = {1},
          eid = {1},
        pages = {1},
          doi = {10.3847/1538-4365/ac5331},
archivePrefix = {arXiv},
       eprint = {2201.11716},
 primaryClass = {astro-ph.GA},
       adsurl = {https://ui.adsabs.harvard.edu/abs/2022ApJS..260....1F}
}

@ARTICLE{Goldstein2017,
       author = {{Goldstein}, A. and {Veres}, P. and {Burns}, E. and {Briggs}, M.~S. and {Hamburg}, R. and {Kocevski}, D. and {Wilson-Hodge}, C.~A. and {Preece}, R.~D. and {Poolakkil}, S. and {Roberts}, O.~J. and {Hui}, C.~M. and {Connaughton}, V. and {Racusin}, J. and {von Kienlin}, A. and {Dal Canton}, T. and {Christensen}, N. and {Littenberg}, T. and {Siellez}, K. and {Blackburn}, L. and {Broida}, J. and {Bissaldi}, E. and {Cleveland}, W.~H. and {Gibby}, M.~H. and {Giles}, M.~M. and {Kippen}, R.~M. and {McBreen}, S. and {McEnery}, J. and {Meegan}, C.~A. and {Paciesas}, W.~S. and {Stanbro}, M.},
        title = "{An Ordinary Short Gamma-Ray Burst with Extraordinary Implications: Fermi-GBM Detection of GRB 170817A}",
      journal = {\apjl},
         year = 2017,
        month = oct,
       volume = {848},
       number = {2},
          eid = {L14},
        pages = {L14},
          doi = {10.3847/2041-8213/aa8f41},
archivePrefix = {arXiv},
       eprint = {1710.05446},
 primaryClass = {astro-ph.HE},
       adsurl = {https://ui.adsabs.harvard.edu/abs/2017ApJ...848L..14G}
}

@ARTICLE{inoue2014,
       author = {{Inoue}, Akio K. and {Shimizu}, Ikkoh and {Iwata}, Ikuru and {Tanaka}, Masayuki},
        title = "{An updated analytic model for attenuation by the intergalactic medium}",
      journal = {\mnras},
         year = 2014,
        month = aug,
       volume = {442},
       number = {2},
        pages = {1805-1820},
          doi = {10.1093/mnras/stu936},
archivePrefix = {arXiv},
       eprint = {1402.0677},
 primaryClass = {astro-ph.CO},
       adsurl = {https://ui.adsabs.harvard.edu/abs/2014MNRAS.442.1805I}
}

@ARTICLE{jaskot25,
       author = {{Jaskot}, Anne E.},
        title = "{Ionizing Radiation Escape from Low-Redshift Galaxies and Its Connection to Cosmic Reionization}",
      journal = {\araa},
         year = 2025,
        month = aug,
       volume = {63},
       number = {1},
        pages = {45-82},
          doi = {10.1146/annurev-astro-111324-074935},
archivePrefix = {arXiv},
       eprint = {2508.18411},
 primaryClass = {astro-ph.GA},
       adsurl = {https://ui.adsabs.harvard.edu/abs/2025ARA&A..63...45J}
}

@ARTICLE{Kim2023,
       author = {{Kim}, Keunho J. and {Bayliss}, Matthew B. and {Rigby}, Jane R. and {Gladders}, Michael D. and {Chisholm}, John and {Sharon}, Keren and {Dahle}, H{\r{a}}kon and {Rivera-Thorsen}, T. Emil and {Florian}, Michael K. and {Khullar}, Gourav and {Mahler}, Guillaume and {Mainali}, Ramesh and {Napier}, Kate A. and {Navarre}, Alexander and {Owens}, M. Riley and {Roberson}, Joshua},
        title = "{Small Region, Big Impact: Highly Anisotropic Lyman-continuum Escape from a Compact Starburst Region with Extreme Physical Properties}",
      journal = {\apjl},
         year = 2023,
        month = sep,
       volume = {955},
       number = {1},
          eid = {L17},
        pages = {L17},
          doi = {10.3847/2041-8213/acf0c5},
archivePrefix = {arXiv},
       eprint = {2305.13405},
 primaryClass = {astro-ph.GA},
       adsurl = {https://ui.adsabs.harvard.edu/abs/2023ApJ...955L..17K}
}

@ARTICLE{Koutsouridou2024,
       author = {{Koutsouridou}, Ioanna and {Salvadori}, Stefania and {Sk{\'u}lad{\'o}ttir}, {\'A}sa},
        title = "{True Pair-instability Supernova Descendant: Implications for the First Stars' Mass Distribution}",
      journal = {\apjl},
         year = 2024,
        month = feb,
       volume = {962},
       number = {2},
          eid = {L26},
        pages = {L26},
          doi = {10.3847/2041-8213/ad2466},
archivePrefix = {arXiv},
       eprint = {2312.05309},
 primaryClass = {astro-ph.GA},
       adsurl = {https://ui.adsabs.harvard.edu/abs/2024ApJ...962L..26K}
}

@ARTICLE{liu25,
       author = {{Liu}, Y. and {Mascia}, S. and {Pentericci}, L. and {Watson}, P. and {Alavi}, A. and {Bergamini}, P. and {Brada{\v{c}}}, M. and {Calabr{\`o}}, A. and {Glazebrook}, K. and {Henry}, A. and {Llerena}, M. and {Merlin}, E. and {Metha}, B. and {Nanayakkara}, T. and {Napolitano}, L. and {Roy}, N. and {Siana}, B. and {Vanzella}, E. and {Vulcani}, B. and {Wang}, X.},
        title = "{A Lyman continuum analysis of {\ensuremath{\sim}}100 galaxies at z$_{spec}${\ensuremath{\sim}} 3 in the Abell 2744 cluster field}",
      journal = {\aap},
         year = 2025,
        month = dec,
       volume = {704},
          eid = {A328},
        pages = {A328},
          doi = {10.1051/0004-6361/202556410},
archivePrefix = {arXiv},
       eprint = {2507.11045},
 primaryClass = {astro-ph.GA},
       adsurl = {https://ui.adsabs.harvard.edu/abs/2025A&A...704A.328L}
}

@ARTICLE{Martinez2026,
       author = {{Mart{\'\i}nez-Ram{\'\i}rez}, L.~N. and {Wolf}, Julien and {Belladitta}, Silvia and {Ba{\~n}ados}, Eduardo and {Bauer}, F.~E. and {Hviding}, Raphael E. and {Stern}, Daniel and {Mazzucchelli}, Chiara and {Meyer}, Romain A. and {Treister}, Ezequiel and {Loiacono}, Federica},
        title = "{16 new quasars at the end of the reionization unveiled by self-supervised learning}",
      journal = {arXiv e-prints},
         year = 2026,
        month = mar,
          eid = {arXiv:2603.08830},
        pages = {arXiv:2603.08830},
          doi = {10.48550/arXiv.2603.08830},
archivePrefix = {arXiv},
       eprint = {2603.08830},
 primaryClass = {astro-ph.GA},
       adsurl = {https://ui.adsabs.harvard.edu/abs/2026arXiv260308830M}
}

@ARTICLE{mascia,
       author = {{Mascia}, S. and {Pentericci}, L. and {Calabr{\`o}}, A. and {Santini}, P. and {Napolitano}, L. and {Arrabal Haro}, P. and {Castellano}, M. and {Dickinson}, M. and {Ocvirk}, P. and {Lewis}, J.~S.~W. and {Amor{\'\i}n}, R. and {Bagley}, M. and {Bhatawdekar}, R. and {Cleri}, N.~J. and {Costantin}, L. and {Dekel}, A. and {Finkelstein}, S.~L. and {Fontana}, A. and {Giavalisco}, M. and {Grogin}, N.~A. and {Hathi}, N.~P. and {Hirschmann}, M. and {Holwerda}, B.~W. and {Jung}, I. and {Kartaltepe}, J.~S. and {Koekemoer}, A.~M. and {Lucas}, R.~A. and {Papovich}, C. and {P{\'e}rez-Gonz{\'a}lez}, P.~G. and {Pirzkal}, N. and {Trump}, J.~R. and {Wilkins}, S.~M. and {Yung}, L.~Y.~A.},
        title = "{New insight on the nature of cosmic reionizers from the CEERS survey}",
      journal = {\aap},
         year = 2024,
        month = may,
       volume = {685},
          eid = {A3},
        pages = {A3},
          doi = {10.1051/0004-6361/202347884},
archivePrefix = {arXiv},
       eprint = {2309.02219},
 primaryClass = {astro-ph.GA},
       adsurl = {https://ui.adsabs.harvard.edu/abs/2024A&A...685A...3M}
}

@ARTICLE{Rastinejad2025,
       author = {{Rastinejad}, J.~C. and {Fong}, W. and {Kilpatrick}, C.~D. and {Nicholl}, M. and {Metzger}, B.~D.},
        title = "{Uniform Modeling of Observed Kilonovae: Implications for Diversity and the Progenitors of Merger-driven Long Gamma-Ray Bursts}",
      journal = {\apj},
         year = 2025,
        month = feb,
       volume = {979},
       number = {2},
          eid = {190},
        pages = {190},
          doi = {10.3847/1538-4357/ad9c77},
archivePrefix = {arXiv},
       eprint = {2409.02158},
 primaryClass = {astro-ph.HE},
       adsurl = {https://ui.adsabs.harvard.edu/abs/2025ApJ...979..190R}
}

@ARTICLE{Saccardi2023,
       author = {{Saccardi}, Andrea and {Salvadori}, Stefania and {D'Odorico}, Valentina and {Cupani}, Guido and {Fumagalli}, Michele and {Berg}, Trystyn A.~M. and {Becker}, George D. and {Ellison}, Sara and {Lopez}, Sebastian},
        title = "{Evidence of First Stars-enriched Gas in High-redshift Absorbers}",
      journal = {\apj},
         year = 2023,
        month = may,
       volume = {948},
       number = {1},
          eid = {35},
        pages = {35},
          doi = {10.3847/1538-4357/acc39f},
archivePrefix = {arXiv},
       eprint = {2305.02346},
 primaryClass = {astro-ph.GA},
       adsurl = {https://ui.adsabs.harvard.edu/abs/2023ApJ...948...35S}
}

@ARTICLE{Savchenko2017,
       author = {{Savchenko}, V. and {Ferrigno}, C. and {Kuulkers}, E. and {Bazzano}, A. and {Bozzo}, E. and {Brandt}, S. and {Chenevez}, J. and {Courvoisier}, T.~J.-L. and {Diehl}, R. and {Domingo}, A. and {Hanlon}, L. and {Jourdain}, E. and {von Kienlin}, A. and {Laurent}, P. and {Lebrun}, F. and {Lutovinov}, A. and {Martin-Carrillo}, A. and {Mereghetti}, S. and {Natalucci}, L. and {Rodi}, J. and {Roques}, J.-P. and {Sunyaev}, R. and {Ubertini}, P.},
        title = "{INTEGRAL Detection of the First Prompt Gamma-Ray Signal Coincident with the Gravitational-wave Event GW170817}",
      journal = {\apjl},
         year = 2017,
        month = oct,
       volume = {848},
       number = {2},
          eid = {L15},
        pages = {L15},
          doi = {10.3847/2041-8213/aa8f94},
archivePrefix = {arXiv},
       eprint = {1710.05449},
 primaryClass = {astro-ph.HE},
       adsurl = {https://ui.adsabs.harvard.edu/abs/2017ApJ...848L..15S}
}

@ARTICLE{Sodini2024,
       author = {{Sodini}, Alessio and {D'Odorico}, Valentina and {Salvadori}, Stefania and {Vanni}, Irene and {Bischetti}, Manuela and {Cupani}, Guido and {Davies}, Rebecca and {Becker}, George D. and {Ba{\~n}ados}, Eduardo and {Bosman}, Sarah and {Davies}, Frederick and {Paolo Farina}, Emanuele and {Ferrara}, Andrea and {Keating}, Laura and {Kulkarni}, Girish and {Lai}, Samuel and {Ryan-Weber}, Emma and {Maria Sebastian}, Alma and {Walter}, Fabian},
        title = "{Evidence of Pop III stars' chemical signature in neutral gas at z {\ensuremath{\sim}} 6. A study based on the E-XQR-30 spectroscopic sample}",
      journal = {\aap},
         year = 2024,
        month = jul,
       volume = {687},
          eid = {A314},
        pages = {A314},
          doi = {10.1051/0004-6361/202349062},
archivePrefix = {arXiv},
       eprint = {2404.10722},
 primaryClass = {astro-ph.GA},
       adsurl = {https://ui.adsabs.harvard.edu/abs/2024A&A...687A.314S}
}

@ARTICLE{Vanni2024,
       author = {{Vanni}, Irene and {Salvadori}, Stefania and {D'Odorico}, Valentina and {Becker}, George D. and {Cupani}, Guido},
        title = "{Chemical Diagnostics to Unveil Environments Enriched by First Stars}",
      journal = {\apjl},
         year = 2024,
        month = jun,
       volume = {967},
       number = {2},
          eid = {L22},
        pages = {L22},
          doi = {10.3847/2041-8213/ad46fa},
archivePrefix = {arXiv},
       eprint = {2402.18640},
 primaryClass = {astro-ph.GA},
       adsurl = {https://ui.adsabs.harvard.edu/abs/2024ApJ...967L..22V}
}

@ARTICLE{Pian2017,
       author = {{Pian}, E. and {D'Avanzo}, P. and {Benetti}, S. and {Branchesi}, M. and {Brocato}, E. and {Campana}, S. and {Cappellaro}, E. and {Covino}, S. and {D'Elia}, V. and {Fynbo}, J.~P.~U. and {Getman}, F. and {Ghirlanda}, G. and {Ghisellini}, G. and {Grado}, A. and {Greco}, G. and {Hjorth}, J. and {Kouveliotou}, C. and {Levan}, A. and {Limatola}, L. and {Malesani}, D. and {Mazzali}, P.~A. and {Melandri}, A. and {M{\o}ller}, P. and {Nicastro}, L. and {Palazzi}, E. and {Piranomonte}, S. and {Rossi}, A. and {Salafia}, O.~S. and {Selsing}, J. and {Stratta}, G. and {Tanaka}, M. and {Tanvir}, N.~R. and {Tomasella}, L. and {Watson}, D. and {Yang}, S. and {Amati}, L. and {Antonelli}, L.~A. and {Ascenzi}, S. and {Bernardini}, M.~G. and {Bo{\"e}r}, M. and {Bufano}, F. and {Bulgarelli}, A. and {Capaccioli}, M. and {Casella}, P. and {Castro-Tirado}, A.~J. and {Chassande-Mottin}, E. and {Ciolfi}, R. and {Copperwheat}, C.~M. and {Dadina}, M. and {De Cesare}, G. and {di Paola}, A. and {Fan}, Y.~Z. and {Gendre}, B. and {Giuffrida}, G. and {Giunta}, A. and {Hunt}, L.~K. and {Israel}, G.~L. and {Jin}, Z.-P. and {Kasliwal}, M.~M. and {Klose}, S. and {Lisi}, M. and {Longo}, F. and {Maiorano}, E. and {Mapelli}, M. and {Masetti}, N. and {Nava}, L. and {Patricelli}, B. and {Perley}, D. and {Pescalli}, A. and {Piran}, T. and {Possenti}, A. and {Pulone}, L. and {Razzano}, M. and {Salvaterra}, R. and {Schipani}, P. and {Spera}, M. and {Stamerra}, A. and {Stella}, L. and {Tagliaferri}, G. and {Testa}, V. and {Troja}, E. and {Turatto}, M. and {Vergani}, S.~D. and {Vergani}, D.},
        title = "{Spectroscopic identification of r-process nucleosynthesis in a double neutron-star merger}",
      journal = {\nat},
         year = 2017,
        month = nov,
       volume = {551},
       number = {7678},
        pages = {67-70},
          doi = {10.1038/nature24298},
archivePrefix = {arXiv},
       eprint = {1710.05858},
 primaryClass = {astro-ph.HE},
       adsurl = {https://ui.adsabs.harvard.edu/abs/2017Natur.551...67P}
}

@ARTICLE{Hallinan2017,
       author = {{Hallinan}, G. and {Corsi}, A. and {Mooley}, K.~P. and {Hotokezaka}, K. and {Nakar}, E. and {Kasliwal}, M.~M. and {Kaplan}, D.~L. and {Frail}, D.~A. and {Myers}, S.~T. and {Murphy}, T. and {De}, K. and {Dobie}, D. and {Allison}, J.~R. and {Bannister}, K.~W. and {Bhalerao}, V. and {Chandra}, P. and {Clarke}, T.~E. and {Giacintucci}, S. and {Ho}, A.~Y.~Q. and {Horesh}, A. and {Kassim}, N.~E. and {Kulkarni}, S.~R. and {Lenc}, E. and {Lockman}, F.~J. and {Lynch}, C. and {Nichols}, D. and {Nissanke}, S. and {Palliyaguru}, N. and {Peters}, W.~M. and {Piran}, T. and {Rana}, J. and {Sadler}, E.~M. and {Singer}, L.~P.},
        title = "{A radio counterpart to a neutron star merger}",
      journal = {Science},
         year = 2017,
        month = dec,
       volume = {358},
       number = {6370},
        pages = {1579-1583},
          doi = {10.1126/science.aap9855},
archivePrefix = {arXiv},
       eprint = {1710.05435},
 primaryClass = {astro-ph.HE},
       adsurl = {https://ui.adsabs.harvard.edu/abs/2017Sci...358.1579H}
}

@ARTICLE{Troja2017,
       author = {{Troja}, E. and {Piro}, L. and {van Eerten}, H. and {Wollaeger}, R.~T. and {Im}, M. and {Fox}, O.~D. and {Butler}, N.~R. and {Cenko}, S.~B. and {Sakamoto}, T. and {Fryer}, C.~L. and {Ricci}, R. and {Lien}, A. and {Ryan}, R.~E. and {Korobkin}, O. and {Lee}, S.-K. and {Burgess}, J.~M. and {Lee}, W.~H. and {Watson}, A.~M. and {Choi}, C. and {Covino}, S. and {D'Avanzo}, P. and {Fontes}, C.~J. and {Gonz{\'a}lez}, J. Becerra and {Khandrika}, H.~G. and {Kim}, J. and {Kim}, S.-L. and {Lee}, C.-U. and {Lee}, H.~M. and {Kutyrev}, A. and {Lim}, G. and {S{\'a}nchez-Ram{\'\i}rez}, R. and {Veilleux}, S. and {Wieringa}, M.~H. and {Yoon}, Y.},
        title = "{The X-ray counterpart to the gravitational-wave event GW170817}",
      journal = {\nat},
         year = 2017,
        month = nov,
       volume = {551},
       number = {7678},
        pages = {71-74},
          doi = {10.1038/nature24290},
archivePrefix = {arXiv},
       eprint = {1710.05433},
 primaryClass = {astro-ph.HE},
       adsurl = {https://ui.adsabs.harvard.edu/abs/2017Natur.551...71T}
}

@ARTICLE{DAvanzo2018,
       author = {{D'Avanzo}, P. and {Campana}, S. and {Salafia}, O.~S. and {Ghirlanda}, G. and {Ghisellini}, G. and {Melandri}, A. and {Bernardini}, M.~G. and {Branchesi}, M. and {Chassande-Mottin}, E. and {Covino}, S. and {D'Elia}, V. and {Nava}, L. and {Salvaterra}, R. and {Tagliaferri}, G. and {Vergani}, S.~D.},
        title = "{The evolution of the X-ray afterglow emission of GW 170817/ GRB 170817A in XMM-Newton observations}",
      journal = {\aap},
         year = 2018,
        month = may,
       volume = {613},
          eid = {L1},
        pages = {L1},
          doi = {10.1051/0004-6361/201832664},
archivePrefix = {arXiv},
       eprint = {1801.06164},
 primaryClass = {astro-ph.HE},
       adsurl = {https://ui.adsabs.harvard.edu/abs/2018A&A...613L...1D}
}

@ARTICLE{Ghirlanda2019,
       author = {{Ghirlanda}, G. and {Salafia}, O.~S. and {Paragi}, Z. and {Giroletti}, M. and {Yang}, J. and {Marcote}, B. and {Blanchard}, J. and {Agudo}, I. and {An}, T. and {Bernardini}, M.~G. and {Beswick}, R. and {Branchesi}, M. and {Campana}, S. and {Casadio}, C. and {Chassande-Mottin}, E. and {Colpi}, M. and {Covino}, S. and {D'Avanzo}, P. and {D'Elia}, V. and {Frey}, S. and {Gawronski}, M. and {Ghisellini}, G. and {Gurvits}, L.~I. and {Jonker}, P.~G. and {van Langevelde}, H.~J. and {Melandri}, A. and {Moldon}, J. and {Nava}, L. and {Perego}, A. and {Perez-Torres}, M.~A. and {Reynolds}, C. and {Salvaterra}, R. and {Tagliaferri}, G. and {Venturi}, T. and {Vergani}, S.~D. and {Zhang}, M.},
        title = "{Compact radio emission indicates a structured jet was produced by a binary neutron star merger}",
      journal = {Science},
         year = 2019,
        month = mar,
       volume = {363},
       number = {6430},
        pages = {968-971},
          doi = {10.1126/science.aau8815},
archivePrefix = {arXiv},
       eprint = {1808.00469},
 primaryClass = {astro-ph.HE},
       adsurl = {https://ui.adsabs.harvard.edu/abs/2019Sci...363..968G}
}

@ARTICLE{Gillanders2022,
       author = {{Gillanders}, J.~H. and {Smartt}, S.~J. and {Sim}, S.~A. and {Bauswein}, A. and {Goriely}, S.},
        title = "{Modelling the spectra of the kilonova AT2017gfo - I. The photospheric epochs}",
      journal = {\mnras},
         year = 2022,
        month = sep,
       volume = {515},
       number = {1},
        pages = {631-651},
          doi = {10.1093/mnras/stac1258},
archivePrefix = {arXiv},
       eprint = {2202.01786},
 primaryClass = {astro-ph.HE},
       adsurl = {https://ui.adsabs.harvard.edu/abs/2022MNRAS.515..631G}
}

@ARTICLE{Bulla2023,
       author = {{Bulla}, Mattia},
        title = "{The critical role of nuclear heating rates, thermalization efficiencies, and opacities for kilonova modelling and parameter inference}",
      journal = {\mnras},
         year = 2023,
        month = apr,
       volume = {520},
       number = {2},
        pages = {2558-2570},
          doi = {10.1093/mnras/stad232},
archivePrefix = {arXiv},
       eprint = {2211.14348},
 primaryClass = {astro-ph.HE},
       adsurl = {https://ui.adsabs.harvard.edu/abs/2023MNRAS.520.2558B}
}

@ARTICLE{McKernan2019,
       author = {{McKernan}, B. and {Ford}, K.~E.~S. and {Bartos}, I. and {Graham}, M.~J. and {Lyra}, W. and {Marka}, S. and {Marka}, Z. and {Ross}, N.~P. and {Stern}, D. and {Yang}, Y.},
        title = "{Ram-pressure Stripping of a Kicked Hill Sphere: Prompt Electromagnetic Emission from the Merger of Stellar Mass Black Holes in an AGN Accretion Disk}",
      journal = {\apjl},
         year = 2019,
        month = oct,
       volume = {884},
       number = {2},
          eid = {L50},
        pages = {L50},
          doi = {10.3847/2041-8213/ab4886},
archivePrefix = {arXiv},
       eprint = {1907.03746},
 primaryClass = {astro-ph.HE},
       adsurl = {https://ui.adsabs.harvard.edu/abs/2019ApJ...884L..50M}
}

@ARTICLE{Antoni2019,
       author = {{Antoni}, Andrea and {MacLeod}, Morgan and {Ramirez-Ruiz}, Enrico},
        title = "{The Evolution of Binaries in a Gaseous Medium: Three-dimensional Simulations of Binary Bondi-Hoyle-Lyttleton Accretion}",
      journal = {\apj},
         year = 2019,
        month = oct,
       volume = {884},
       number = {1},
          eid = {22},
        pages = {22},
          doi = {10.3847/1538-4357/ab3466},
archivePrefix = {arXiv},
       eprint = {1901.07572},
 primaryClass = {astro-ph.HE},
       adsurl = {https://ui.adsabs.harvard.edu/abs/2019ApJ...884...22A}
}

@ARTICLE{Tagawa2024,
       author = {{Tagawa}, Hiromichi and {Kimura}, Shigeo S. and {Haiman}, Zolt{\'a}n and {Perna}, Rosalba and {Bartos}, Imre},
        title = "{Shock Cooling and Breakout Emission for Optical Flares Associated with Gravitational-wave Events}",
      journal = {\apj},
         year = 2024,
        month = may,
       volume = {966},
       number = {1},
          eid = {21},
        pages = {21},
          doi = {10.3847/1538-4357/ad2e0b},
archivePrefix = {arXiv},
       eprint = {2310.18392},
 primaryClass = {astro-ph.HE},
       adsurl = {https://ui.adsabs.harvard.edu/abs/2024ApJ...966...21T}
}

@ARTICLE{Wang2022,
       author = {{Wang}, Yi-Han and {Lazzati}, Davide and {Perna}, Rosalba},
        title = "{The emergence of diffused gamma-ray burst afterglows from the discs of active galactic nuclei}",
      journal = {\mnras},
         year = 2022,
        month = nov,
       volume = {516},
       number = {4},
        pages = {5935-5944},
          doi = {10.1093/mnras/stac1968},
archivePrefix = {arXiv},
       eprint = {2207.05084},
 primaryClass = {astro-ph.HE},
       adsurl = {https://ui.adsabs.harvard.edu/abs/2022MNRAS.516.5935W}
}

@ARTICLE{Anderson2000,
       author = {{Anderson}, Jay and {King}, Ivan R.},
        title = "{Toward High-Precision Astrometry with WFPC2. I. Deriving an Accurate Point-Spread Function}",
      journal = {\pasp},
         year = 2000,
        month = oct,
       volume = {112},
       number = {776},
        pages = {1360-1382},
          doi = {10.1086/316632},
archivePrefix = {arXiv},
       eprint = {astro-ph/0006325},
 primaryClass = {astro-ph},
       adsurl = {https://ui.adsabs.harvard.edu/abs/2000PASP..112.1360A}
}

@ARTICLE{Magrini2022,
       author = {{Magrini}, L. and {Danielski}, C. and {Bossini}, D. and {Rainer}, M. and {Turrini}, D. and {Benatti}, S. and {Brucalassi}, A. and {Tsantaki}, M. and {Delgado Mena}, E. and {Sanna}, N. and {Biazzo}, K. and {Campante}, T.~L. and {Van der Swaelmen}, M. and {Sousa}, S.~G. and {He{\l}miniak}, K.~G. and {Neitzel}, A.~W. and {Adibekyan}, V. and {Bruno}, G. and {Casali}, G.},
        title = "{Ariel stellar characterisation. I. Homogeneous stellar parameters of 187 FGK planet host stars: Description and validation of the method}",
      journal = {\aap},
         year = 2022,
        month = jul,
       volume = {663},
          eid = {A161},
        pages = {A161},
          doi = {10.1051/0004-6361/202243405},
archivePrefix = {arXiv},
       eprint = {2204.08825},
 primaryClass = {astro-ph.SR},
       adsurl = {https://ui.adsabs.harvard.edu/abs/2022A&A...663A.161M}
}

@ARTICLE{Tsantaki2025,
       author = {{Tsantaki}, M. and {Magrini}, L. and {Danielski}, C. and {Bossini}, D. and {Turrini}, D. and {Moedas}, N. and {Folsom}, C.~P. and {Ramler}, H. and {Biazzo}, K. and {Campante}, T.~L. and {Delgado-Mena}, E. and {da Silva}, R. and {Sousa}, S.~G. and {Benatti}, S. and {Casali}, G. and {He{\l}miniak}, K.~G. and {Rainer}, M. and {Sanna}, N.},
        title = "{Ariel stellar characterisation: III. Fast rotators and new FGK stars in the Ariel mission candidate sample}",
      journal = {\aap},
         year = 2025,
        month = may,
       volume = {697},
          eid = {A102},
        pages = {A102},
          doi = {10.1051/0004-6361/202453059},
archivePrefix = {arXiv},
       eprint = {2502.20950},
 primaryClass = {astro-ph.SR},
       adsurl = {https://ui.adsabs.harvard.edu/abs/2025A&A...697A.102T}
}

@ARTICLE{daSilva2024,
       author = {{da Silva}, R. and {Danielski}, C. and {Delgado Mena}, E. and {Magrini}, L. and {Turrini}, D. and {Biazzo}, K. and {Tsantaki}, M. and {Rainer}, M. and {Helminiak}, K.~G. and {Benatti}, S. and {Adibekyan}, V. and {Sanna}, N. and {Sousa}, S. and {Casali}, G. and {Van der Swaelmen}, M.},
        title = "{Ariel stellar characterisation. II. Chemical abundances of carbon, nitrogen, and oxygen for 181 planet-host FGK dwarf stars}",
      journal = {\aap},
         year = 2024,
        month = aug,
       volume = {688},
          eid = {A193},
        pages = {A193},
          doi = {10.1051/0004-6361/202450604},
archivePrefix = {arXiv},
       eprint = {2406.12393},
 primaryClass = {astro-ph.SR},
       adsurl = {https://ui.adsabs.harvard.edu/abs/2024A&A...688A.193D}
}

@ARTICLE{Peacock2025,
       author = {{Peacock}, Sarah and {Wilson}, David J. and {Richey-Yowell}, Tyler and {Tuchow}, Noah W. and {France}, Kevin and {Caballero}, Jos{\'e} A. and {Spinelli}, Riccardo and {Corrales}, L{\'\i}a and {Zelakiewicz}, Aiden S. and {Redfield}, Seth and {Rockcliffe}, Keighley and {Youngblood}, Allison and {Froning}, Cynthia S. and {Duvvuri}, Girish M. and {Binder}, Breanna A. and {Hinkel}, Natalie R. and {Mamajek}, Eric E.},
        title = "{HWO Target Stars and Systems: A Survey of Archival UV and X-Ray Data}",
      journal = {\aj},
         year = 2025,
        month = nov,
       volume = {170},
       number = {5},
          eid = {293},
        pages = {293},
          doi = {10.3847/1538-3881/ae0a2f},
archivePrefix = {arXiv},
       eprint = {2509.08999},
 primaryClass = {astro-ph.SR},
       adsurl = {https://ui.adsabs.harvard.edu/abs/2025AJ....170..293P}
}

@ARTICLE{Tuchow2025,
       author = {{Tuchow}, Noah W. and {Harada}, Caleb K. and {Mamajek}, Eric E. and {Tanner}, Angelle and {Hinkel}, Natalie R. and {Belikov}, Ruslan and {Sirbu}, Dan and {Ciardi}, David R. and {Stark}, Christopher C. and {Morgan}, Rhonda M. and {Savransky}, Dmitry and {Turmon}, Michael},
        title = "{HWO Target Stars and Systems: A Prioritized Community List of Potential Stellar Targets for the Habitable Worlds Observatory's ExoEarth Survey}",
      journal = {\pasp},
         year = 2025,
        month = oct,
       volume = {137},
       number = {10},
          eid = {104402},
        pages = {104402},
          doi = {10.1088/1538-3873/ae0a81},
archivePrefix = {arXiv},
       eprint = {2509.20544},
 primaryClass = {astro-ph.SR},
       adsurl = {https://ui.adsabs.harvard.edu/abs/2025PASP..137j4402T}
}

@ARTICLE{Tuchow2024,
       author = {{Tuchow}, Noah W. and {Stark}, Christopher C. and {Mamajek}, Eric},
        title = "{HPIC: The Habitable Worlds Observatory Preliminary Input Catalog}",
      journal = {\aj},
         year = 2024,
        month = mar,
       volume = {167},
       number = {3},
          eid = {139},
        pages = {139},
          doi = {10.3847/1538-3881/ad25ec},
       adsurl = {https://ui.adsabs.harvard.edu/abs/2024AJ....167..139T}
}

@ARTICLE{Mamajek2024,
       author = {{Mamajek}, Eric and {Stapelfeldt}, Karl},
        title = "{NASA Exoplanet Exploration Program (ExEP) Mission Star List for the Habitable Worlds Observatory (2023)}",
      journal = {arXiv e-prints},
         year = 2024,
        month = feb,
          eid = {arXiv:2402.12414},
        pages = {arXiv:2402.12414},
          doi = {10.48550/arXiv.2402.12414},
archivePrefix = {arXiv},
       eprint = {2402.12414},
 primaryClass = {astro-ph.IM},
       adsurl = {https://ui.adsabs.harvard.edu/abs/2024arXiv240212414M}
}

@INPROCEEDINGS{Tinetti2022,
       author = {{Tinetti}, Giovanna and {Eccleston}, Paul and {Lueftinger}, Theresa and {Salvignol}, Jean-Christophe and {Fahmy}, Salma and {Alves de Oliveira}, Caterina},
        title = "{Ariel: Enabling planetary science across light-years}",
    booktitle = {European Planetary Science Congress},
         year = 2022,
        month = sep,
          eid = {EPSC2022-1114},
        pages = {EPSC2022-1114},
          doi = {10.5194/epsc2022-1114},
archivePrefix = {arXiv},
       eprint = {2104.04824},
 primaryClass = {astro-ph.IM},
       adsurl = {https://ui.adsabs.harvard.edu/abs/2022EPSC...16.1114T}
}

@ARTICLE{Tinetti2018,
       author = {{Tinetti}, Giovanna and {Drossart}, Pierre and {Eccleston}, Paul and {Hartogh}, Paul and {Heske}, Astrid and {Leconte}, J{\'e}r{\'e}my and {Micela}, Giusi and {Ollivier}, Marc and {Pilbratt}, G{\"o}ran and {Puig}, Ludovic and {Turrini}, Diego and {Vandenbussche}, Bart and {Wolkenberg}, Paulina and {Beaulieu}, Jean-Philippe and {Buchave}, Lars A. and {Ferus}, Martin and {Griffin}, Matt and {Guedel}, Manuel and {Justtanont}, Kay and {Lagage}, Pierre-Olivier and {Machado}, Pedro and {Malaguti}, Giuseppe and {Min}, Michiel and {N{\o}rgaard-Nielsen}, Hans Ulrik and {Rataj}, Mirek and {Ray}, Tom and {Ribas}, Ignasi and {Swain}, Mark and {Szabo}, Robert and {Werner}, Stephanie and {Barstow}, Joanna and {Burleigh}, Matt and {Cho}, James and {Coud{\'e} du Foresto}, Vincent and {Coustenis}, Athena and {Decin}, Leen and {Encrenaz}, Therese and {Galand}, Marina and {Gillon}, Michael and {Helled}, Ravit and {Morales}, Juan Carlos and {Garc{\'\i}a Mu{\~n}oz}, Antonio and {Moneti}, Andrea and {Pagano}, Isabella and {Pascale}, Enzo and {Piccioni}, Giuseppe and {Pinfield}, David and {Sarkar}, Subhajit and {Selsis}, Franck and {Tennyson}, Jonathan and {Triaud}, Amaury and {Venot}, Olivia and {Waldmann}, Ingo and {Waltham}, David and {Wright}, Gillian and {Amiaux}, Jerome and {Augu{\`e}res}, Jean-Louis and {Berth{\'e}}, Michel and {Bezawada}, Naidu and {Bishop}, Georgia and {Bowles}, Neil and {Coffey}, Deirdre and {Colom{\'e}}, Josep and {Crook}, Martin and {Crouzet}, Pierre-Elie and {Da Peppo}, Vania and {Sanz}, Isabel Escudero and {Focardi}, Mauro and {Frericks}, Martin and {Hunt}, Tom and {Kohley}, Ralf and {Middleton}, Kevin and {Morgante}, Gianluca and {Ottensamer}, Roland and {Pace}, Emanuele and {Pearson}, Chris and {Stamper}, Richard and {Symonds}, Kate and {Rengel}, Miriam and {Renotte}, Etienne and {Ade}, Peter and {Affer}, Laura and {Alard}, Christophe and {Allard}, Nicole and {Altieri}, Francesca and {Andr{\'e}}, Yves and {Arena}, Claudio and {Argyriou}, Ioannis and {Aylward}, Alan and {Baccani}, Cristian and {Bakos}, Gaspar and {Banaszkiewicz}, Marek and {Barlow}, Mike and {Batista}, Virginie and {Bellucci}, Giancarlo and {Benatti}, Serena and {Bernardi}, Pernelle and {B{\'e}zard}, Bruno and {Blecka}, Maria and {Bolmont}, Emeline and {Bonfond}, Bertrand and {Bonito}, Rosaria and {Bonomo}, Aldo S. and {Brucato}, John Robert and {Brun}, Allan Sacha and {Bryson}, Ian and {Bujwan}, Waldemar and {Casewell}, Sarah and {Charnay}, Bejamin and {Pestellini}, Cesare Cecchi and {Chen}, Guo and {Ciaravella}, Angela and {Claudi}, Riccardo and {Cl{\'e}dassou}, Rodolphe and {Damasso}, Mario and {Damiano}, Mario and {Danielski}, Camilla and {Deroo}, Pieter and {Di Giorgio}, Anna Maria and {Dominik}, Carsten and {Doublier}, Vanessa and {Doyle}, Simon and {Doyon}, Ren{\'e} and {Drummond}, Benjamin and {Duong}, Bastien and {Eales}, Stephen and {Edwards}, Billy and {Farina}, Maria and {Flaccomio}, Ettore and {Fletcher}, Leigh and {Forget}, Fran{\c{c}}ois and {Fossey}, Steve and {Fr{\"a}nz}, Markus and {Fujii}, Yuka and {Garc{\'\i}a-Piquer}, {\'A}lvaro and {Gear}, Walter and {Geoffray}, Herv{\'e} and {G{\'e}rard}, Jean Claude and {Gesa}, Lluis and {Gomez}, H. and {Graczyk}, Rafa{\l} and {Griffith}, Caitlin and {Grodent}, Denis and {Guarcello}, Mario Giuseppe and {Gustin}, Jacques and {Hamano}, Keiko and {Hargrave}, Peter and {Hello}, Yann and {Heng}, Kevin and {Herrero}, Enrique and {Hornstrup}, Allan and {Hubert}, Benoit and {Ida}, Shigeru and {Ikoma}, Masahiro and {Iro}, Nicolas and {Irwin}, Patrick and {Jarchow}, Christopher and {Jaubert}, Jean and {Jones}, Hugh and {Julien}, Queyrel and {Kameda}, Shingo and {Kerschbaum}, Franz and {Kervella}, Pierre and {Koskinen}, Tommi and {Krijger}, Matthijs and {Krupp}, Norbert and {Lafarga}, Marina and {Landini}, Federico and {Lellouch}, Emanuel and {Leto}, Giuseppe and {Luntzer}, A. and {Rank-L{\"u}ftinger}, Theresa and {Maggio}, Antonio and {Maldonado}, Jesus and {Maillard}, Jean-Pierre and {Mall}, Urs and {Marquette}, Jean-Baptiste and {Mathis}, Stephane and {Maxted}, Pierre and {Matsuo}, Taro and {Medvedev}, Alexander and {Miguel}, Yamila and {Minier}, Vincent and {Morello}, Giuseppe and {Mura}, Alessandro and {Narita}, Norio and {Nascimbeni}, Valerio and {Nguyen Tong}, N. and {Noce}, Vladimiro and {Oliva}, Fabrizio and {Palle}, Enric and {Palmer}, Paul and {Pancrazzi}, Maurizio and {Papageorgiou}, Andreas and {Parmentier}, Vivien and {Perger}, Manuel and {Petralia}, Antonino and {Pezzuto}, Stefano and {Pierrehumbert}, Ray and {Pillitteri}, Ignazio},
        title = "{A chemical survey of exoplanets with ARIEL}",
      journal = {Experimental Astronomy},
         year = 2018,
        month = nov,
       volume = {46},
       number = {1},
        pages = {135-209},
          doi = {10.1007/s10686-018-9598-x},
       adsurl = {https://ui.adsabs.harvard.edu/abs/2018ExA....46..135T}
}

@ARTICLE{Soubiran2016,
       author = {{Soubiran}, Caroline and {Le Campion}, Jean-Fran{\c{c}}ois and {Brouillet}, Nathalie and {Chemin}, Laurent},
        title = "{The PASTEL catalogue: 2016 version}",
      journal = {\aap},
         year = 2016,
        month = jun,
       volume = {591},
          eid = {A118},
        pages = {A118},
          doi = {10.1051/0004-6361/201628497},
archivePrefix = {arXiv},
       eprint = {1605.07384},
 primaryClass = {astro-ph.SR},
       adsurl = {https://ui.adsabs.harvard.edu/abs/2016A&A...591A.118S}
}

@ARTICLE{Lazar2020,
       author = {{Lazar}, Alexandres and {Bullock}, James S.},
        title = "{Accurate mass estimates from the proper motions of dispersion-supported galaxies}",
      journal = {\mnras},
         year = 2020,
        month = apr,
       volume = {493},
       number = {4},
        pages = {5825-5837},
          doi = {10.1093/mnras/staa692},
archivePrefix = {arXiv},
       eprint = {1907.08841},
 primaryClass = {astro-ph.GA},
       adsurl = {https://ui.adsabs.harvard.edu/abs/2020MNRAS.493.5825L}
}

@ARTICLE{Vitral2026,
       author = {{Vitral}, Eduardo and {van der Marel}, Roeland P. and {Sohn}, Sangmo Tony and {Pe{\~n}arrubia}, Jorge and {Patel}, Ekta and {Watkins}, Laura L. and {Libralato}, Mattia and {McKinnon}, Kevin A. and {Bellini}, Andrea and {del Pino}, Andr{\'e}s and {Bennet}, Paul},
        title = "{HSTPROMO Internal Proper-motion Kinematics of Dwarf Spheroidal Galaxies. II. Velocity Anisotropy and Dark Matter Cusp Slope of Sculptor}",
      journal = {\apj},
         year = 2026,
        month = feb,
       volume = {998},
       number = {2},
          eid = {206},
        pages = {206},
          doi = {10.3847/1538-4357/ae1f8a},
archivePrefix = {arXiv},
       eprint = {2508.20711},
 primaryClass = {astro-ph.GA},
       adsurl = {https://ui.adsabs.harvard.edu/abs/2026ApJ...998..206V}
}

@ARTICLE{Vitral2024,
       author = {{Vitral}, Eduardo and {van der Marel}, Roeland P. and {Sohn}, Sangmo Tony and {Libralato}, Mattia and {del Pino}, Andr{\'e}s and {Watkins}, Laura L. and {Bellini}, Andrea and {Walker}, Matthew G. and {Besla}, Gurtina and {Pawlowski}, Marcel S. and {Mamon}, Gary A.},
        title = "{HSTPROMO Internal Proper-motion Kinematics of Dwarf Spheroidal Galaxies. I. Velocity Anisotropy and Dark Matter Cusp Slope of Draco}",
      journal = {\apj},
         year = 2024,
        month = jul,
       volume = {970},
       number = {1},
          eid = {1},
        pages = {1},
          doi = {10.3847/1538-4357/ad571c},
archivePrefix = {arXiv},
       eprint = {2407.07769},
 primaryClass = {astro-ph.GA},
       adsurl = {https://ui.adsabs.harvard.edu/abs/2024ApJ...970....1V}
}

@ARTICLE{Massari2020,
       author = {{Massari}, D. and {Helmi}, A. and {Mucciarelli}, A. and {Sales}, L.~V. and {Spina}, L. and {Tolstoy}, E.},
        title = "{Stellar 3D kinematics in the Draco dwarf spheroidal galaxy}",
      journal = {\aap},
         year = 2020,
        month = jan,
       volume = {633},
          eid = {A36},
        pages = {A36},
          doi = {10.1051/0004-6361/201935613},
archivePrefix = {arXiv},
       eprint = {1904.04037},
 primaryClass = {astro-ph.GA},
       adsurl = {https://ui.adsabs.harvard.edu/abs/2020A&A...633A..36M}
}

@ARTICLE{Massari2018,
       author = {{Massari}, D. and {Breddels}, M.~A. and {Helmi}, A. and {Posti}, L. and {Brown}, A.~G.~A. and {Tolstoy}, E.},
        title = "{Three-dimensional motions in the Sculptor dwarf galaxy as a glimpse of a new era}",
      journal = {Nature Astronomy},
         year = 2018,
        month = nov,
       volume = {2},
        pages = {156-161},
          doi = {10.1038/s41550-017-0322-y},
archivePrefix = {arXiv},
       eprint = {1711.08945},
 primaryClass = {astro-ph.GA},
       adsurl = {https://ui.adsabs.harvard.edu/abs/2018NatAs...2..156M}
}

@INCOLLECTION{Walker2013,
       author = {{Walker}, Matthew},
        title = "{Dark Matter in the Galactic Dwarf Spheroidal Satellites}",
    booktitle = {Planets, Stars and Stellar Systems. Volume 5: Galactic Structure and Stellar Populations},
         year = 2013,
       editor = {{Oswalt}, Terry D. and {Gilmore}, Gerard},
       volume = {5},
        pages = {1039},
          doi = {10.1007/978-94-007-5612-0_20},
       adsurl = {https://ui.adsabs.harvard.edu/abs/2013pss5.book.1039W}
}

@ARTICLE{Binney1982,
       author = {{Binney}, J. and {Mamon}, G.~A.},
        title = "{M/L and velocity anisotropy from observations of spherical galaxies, or must M87 have a massive black hole ?}",
      journal = {\mnras},
         year = 1982,
        month = jul,
       volume = {200},
        pages = {361-375},
          doi = {10.1093/mnras/200.2.361},
       adsurl = {https://ui.adsabs.harvard.edu/abs/1982MNRAS.200..361B}
}

@ARTICLE{Read2005,
       author = {{Read}, J.~I. and {Gilmore}, G.},
        title = "{Mass loss from dwarf spheroidal galaxies: the origins of shallow dark matter cores and exponential surface brightness profiles}",
      journal = {\mnras},
         year = 2005,
        month = jan,
       volume = {356},
       number = {1},
        pages = {107-124},
          doi = {10.1111/j.1365-2966.2004.08424.x},
archivePrefix = {arXiv},
       eprint = {astro-ph/0409565},
 primaryClass = {astro-ph},
       adsurl = {https://ui.adsabs.harvard.edu/abs/2005MNRAS.356..107R}
}

@ARTICLE{Navarro1996,
       author = {{Navarro}, Julio F. and {Eke}, Vincent R. and {Frenk}, Carlos S.},
        title = "{The cores of dwarf galaxy haloes}",
      journal = {\mnras},
         year = 1996,
        month = dec,
       volume = {283},
       number = {3},
        pages = {L72-L78},
          doi = {10.1093/mnras/283.3.L72},
archivePrefix = {arXiv},
       eprint = {astro-ph/9610187},
 primaryClass = {astro-ph},
       adsurl = {https://ui.adsabs.harvard.edu/abs/1996MNRAS.283L..72N}
}

@ARTICLE{Moore1994,
       author = {{Moore}, Ben},
        title = "{Evidence against dissipation-less dark matter from observations of galaxy haloes}",
      journal = {\nat},
         year = 1994,
        month = aug,
       volume = {370},
       number = {6491},
        pages = {629-631},
          doi = {10.1038/370629a0},
       adsurl = {https://ui.adsabs.harvard.edu/abs/1994Natur.370..629M}
}

@article{battistuzzi2023a,
  title={Oxygenic photosynthetic responses of cyanobacteria exposed under an M-dwarf starlight simulator: Implications for exoplanet’s habitability},
  author={Battistuzzi, Mariano and Cocola, Lorenzo and Claudi, Riccardo and Pozzer, Anna Caterina and Segalla, Anna and Simionato, Diana and Morosinotto, Tomas and Poletto, Luca and La Rocca, Nicoletta},
  doi = {10.3389/fpls.2023.1070359},
  journal={Frontiers in plant science},
  volume={14},
  pages={1070359},
  year={2023},
  publisher={Frontiers Media SA}
}

@article{battistuzzi2023b,
  title={Growth and photosynthetic efficiency of microalgae and plants with different levels of complexity exposed to a simulated m-dwarf starlight},
  author={Battistuzzi, Mariano and Cocola, Lorenzo and Liistro, Elisabetta and Claudi, Riccardo and Poletto, Luca and La Rocca, Nicoletta},
  doi = {10.3390/life13081641},
  journal={Life},
  volume={13},
  number={8},
  pages={1641},
  year={2023},
  publisher={MDPI}
}

@ARTICLE{Ragazzoni1996,
       author = {{Ragazzoni}, Roberto},
        title = "{Pupil plane wavefront sensing with an oscillating prism}",
      journal = {Journal of Modern Optics},
         year = 1996,
        month = feb,
       volume = {43},
       number = {2},
        pages = {289-293},
          doi = {10.1080/09500349608232742},
       adsurl = {https://ui.adsabs.harvard.edu/abs/1996JMOp...43..289R}
}

@INPROCEEDINGS{Pinna2015,
       author = {{Pinna}, Enrico and {Pedichini}, Fernando and {Esposito}, Simone and {Centrone}, Mauro and {Puglisi}, Alfio and {Farinato}, Jacopo and {Carbonaro}, Luca and {Agapito}, Guido and {Stangalini}, Marco and {Riccardi}, Amando and {Xompero}, Marco and {Briguglio}, R. and {Hinz}, Philip and {Bayley}, Vanessa and {Montoya}, Manny},
        title = "{XAO at LBT: current performances in the visible and upcoming upgrade}",
    booktitle = {Adaptive Optics for Extremely Large Telescopes IV (AO4ELT4)},
         year = 2015,
        month = oct,
          eid = {E58},
        pages = {E58},
       adsurl = {https://ui.adsabs.harvard.edu/abs/2015aoel.confE..58P}
}

@INPROCEEDINGS{Simioni2024,
       author = {{Simioni}, Matteo and {Jodlbauer}, Daniel and {Arcidiacono}, Carmelo and {Grazian}, Andrea and {Gullieuszik}, Marco and {Portaluri}, Elisa and {Vulcani}, Benedetta and {Wagner}, Roland and {Zanella}, Anita and {Hartke}, Johanna and {Helin}, Tapio and {Kuncarayakti}, Hanindyo and {Masciadri}, Elena and {Pedichini}, Fernando and {Piazzesi}, Roberto and {Turchi}, Alessio and {Vaccari}, Piero},
        title = "{The MICADO first light imager for the ELT: off-axis performance of PSF reconstruction}",
    booktitle = {Adaptive Optics Systems IX},
         year = 2024,
       editor = {{Jackson}, Kathryn J. and {Schmidt}, Dirk and {Vernet}, Elise},
       series = {Society of Photo-Optical Instrumentation Engineers (SPIE) Conference Series},
       volume = {13097},
        month = aug,
          eid = {130977C},
        pages = {130977C},
          doi = {10.1117/12.3017375},
archivePrefix = {arXiv},
       eprint = {2407.09201},
 primaryClass = {astro-ph.IM},
       adsurl = {https://ui.adsabs.harvard.edu/abs/2024SPIE13097E..7CS}
}

@INPROCEEDINGS{Pinna2016,
       author = {{Pinna}, E. and {Esposito}, S. and {Hinz}, P. and {Agapito}, G. and {Bonaglia}, M. and {Puglisi}, A. and {Xompero}, M. and {Riccardi}, A. and {Briguglio}, R. and {Arcidiacono}, C. and {Carbonaro}, L. and {Fini}, L. and {Montoya}, M. and {Durney}, O.},
        title = "{SOUL: the Single conjugated adaptive Optics Upgrade for LBT}",
    booktitle = {Adaptive Optics Systems V},
         year = 2016,
       editor = {{Marchetti}, Enrico and {Close}, Laird M. and {V{\'e}ran}, Jean-Pierre},
       series = {Society of Photo-Optical Instrumentation Engineers (SPIE) Conference Series},
       volume = {9909},
        month = jul,
          eid = {99093V},
        pages = {99093V},
          doi = {10.1117/12.2234444},
       adsurl = {https://ui.adsabs.harvard.edu/abs/2016SPIE.9909E..3VP}
}

@ARTICLE{TauREx2021,
       author = {{Al-Refaie}, A.~F. and {Changeat}, Q. and {Waldmann}, I.~P. and {Tinetti}, G.},
        title = "{TauREx 3: A Fast, Dynamic, and Extendable Framework for Retrievals}",
      journal = {\apj},
         year = 2021,
        month = aug,
       volume = {917},
       number = {1},
          eid = {37},
        pages = {37},
          doi = {10.3847/1538-4357/ac0252},
archivePrefix = {arXiv},
       eprint = {1912.07759},
 primaryClass = {astro-ph.IM},
       adsurl = {https://ui.adsabs.harvard.edu/abs/2021ApJ...917...37A}
}

@ARTICLE{Whiteford2023,
       author = {{Whiteford}, Niall and {Glasse}, Alistair and {Chubb}, Katy L. and {Kitzmann}, Daniel and {Ray}, Shrishmoy and {Phillips}, Mark W. and {Biller}, Beth A. and {Palmer}, Paul I. and {Rice}, Ken and {Waldmann}, Ingo P. and {Changeat}, Quentin and {Skaf}, Nour and {Wang}, Jason and {Edwards}, Billy and {Al-Refaie}, Ahmed},
        title = "{Retrieval study of cool, directly imaged exoplanet 51 Eri b}",
      journal = {\mnras},
         year = 2023,
        month = oct,
       volume = {525},
       number = {1},
        pages = {1375-1400},
          doi = {10.1093/mnras/stad670},
archivePrefix = {arXiv},
       eprint = {2302.07939},
 primaryClass = {astro-ph.EP},
       adsurl = {https://ui.adsabs.harvard.edu/abs/2023MNRAS.525.1375W}
}

@ARTICLE{GrecoBrandt2016,
       author = {{Greco}, Johnny P. and {Brandt}, Timothy D.},
        title = "{The Measurement, Treatment, and Impact of Spectral Covariance and Bayesian Priors in Integral-field Spectroscopy of Exoplanets}",
      journal = {\apj},
         year = 2016,
        month = dec,
       volume = {833},
       number = {2},
          eid = {134},
        pages = {134},
          doi = {10.3847/1538-4357/833/2/134},
archivePrefix = {arXiv},
       eprint = {1602.00691},
 primaryClass = {astro-ph.EP},
       adsurl = {https://ui.adsabs.harvard.edu/abs/2016ApJ...833..134G}
}

@ARTICLE{Vaswani2017,
       author = {{Vaswani}, Ashish and {Shazeer}, Noam and {Parmar}, Niki and {Uszkoreit}, Jakob and {Jones}, Llion and {Gomez}, Aidan N. and {Kaiser}, Lukasz and {Polosukhin}, Illia},
        title = "{Attention Is All You Need}",
      journal = {arXiv e-prints},
         year = 2017,
        month = jun,
          eid = {arXiv:1706.03762},
        pages = {arXiv:1706.03762},
          doi = {10.48550/arXiv.1706.03762},
archivePrefix = {arXiv},
       eprint = {1706.03762},
 primaryClass = {cs.CL},
       adsurl = {https://ui.adsabs.harvard.edu/abs/2017arXiv170603762V}
}

@ARTICLE{Zingales2018,
       author = {{Zingales}, Tiziano and {Waldmann}, Ingo P.},
        title = "{ExoGAN: Retrieving Exoplanetary Atmospheres Using Deep Convolutional Generative Adversarial Networks}",
      journal = {\aj},
         year = 2018,
        month = dec,
       volume = {156},
       number = {6},
          eid = {268},
        pages = {268},
          doi = {10.3847/1538-3881/aae77c},
archivePrefix = {arXiv},
       eprint = {1806.02906},
 primaryClass = {astro-ph.IM},
       adsurl = {https://ui.adsabs.harvard.edu/abs/2018AJ....156..268Z}
}

@ARTICLE{Robinson2015PASP,
       author = {{Robinson}, Tyler D. and {Stapelfeldt}, Karl R. and {Marley}, Mark S.},
        title = "{Characterizing Rocky and Gaseous Exoplanets with 2 m Class Space-based Coronagraphs}",
      journal = {\pasp},
         year = 2016,
        month = feb,
       volume = {128},
       number = {960},
        pages = {025003},
          doi = {10.1088/1538-3873/128/960/025003},
archivePrefix = {arXiv},
       eprint = {1507.00777},
 primaryClass = {astro-ph.EP},
       adsurl = {https://ui.adsabs.harvard.edu/abs/2016PASP..128b5003R}
}

@ARTICLE{CoronagraphPkg2019,
       author = {{Lustig-Yaeger}, Jacob and {Robinson}, Tyler and {Arney}, Giada},
        title = "{coronagraph: Telescope Noise Modeling for Exoplanets in Python}",
      journal = {The Journal of Open Source Software},
         year = 2019,
        month = aug,
       volume = {4},
       number = {40},
          eid = {1387},
        pages = {1387},
          doi = {10.21105/joss.01387},
       adsurl = {https://ui.adsabs.harvard.edu/abs/2019JOSS....4.1387L}
}

@ARTICLE{FujiiNakajima2017,
       author = {{Fujii}, Keisuke and {Nakajima}, Kohei},
        title = "{Harnessing Disordered-Ensemble Quantum Dynamics for Machine Learning}",
      journal = {Physical Review Applied},
         year = 2017,
        month = aug,
       volume = {8},
       number = {2},
          eid = {024030},
        pages = {024030},
          doi = {10.1103/PhysRevApplied.8.024030},
archivePrefix = {arXiv},
       eprint = {1602.08159},
 primaryClass = {quant-ph},
       adsurl = {https://ui.adsabs.harvard.edu/abs/2017PhRvP...8b4030F}
}

@ARTICLE{Mujal2021,
       author = {{Mujal}, Pere and {Mart{\'\i}nez-Pe{\~n}a}, Rodrigo and {Nokkala}, Johannes and {Garc{\'\i}a-Beni}, Jorge and {Giorgi}, Gian Luca and {Soriano}, Miguel C. and {Zambrini}, Roberta},
        title = "{Opportunities in Quantum Reservoir Computing and Extreme Learning Machines}",
      journal = {arXiv e-prints},
         year = 2021,
        month = feb,
          eid = {arXiv:2102.11831},
        pages = {arXiv:2102.11831},
          doi = {10.48550/arXiv.2102.11831},
archivePrefix = {arXiv},
       eprint = {2102.11831},
 primaryClass = {quant-ph},
       adsurl = {https://ui.adsabs.harvard.edu/abs/2021arXiv210211831M}
}

@ARTICLE{LloydWeedbrook2018,
       author = {{Lloyd}, Seth and {Weedbrook}, Christian},
        title = "{Quantum Generative Adversarial Learning}",
      journal = {\prl},
         year = 2018,
        month = jul,
       volume = {121},
       number = {4},
          eid = {040502},
        pages = {040502},
          doi = {10.1103/PhysRevLett.121.040502},
archivePrefix = {arXiv},
       eprint = {1804.09139},
 primaryClass = {quant-ph},
       adsurl = {https://ui.adsabs.harvard.edu/abs/2018PhRvL.121d0502L}
}

@article{DallaireKilloran2018,
  title = {Quantum generative adversarial networks},
  author = {Dallaire-Demers, Pierre-Luc and Killoran, Nathan},
  journal = {Phys. Rev. A},
  volume = {98},
  issue = {1},
  pages = {012324},
  numpages = {8},
  year = {2018},
  month = {Jul},
  publisher = {American Physical Society},
  doi = {10.1103/PhysRevA.98.012324},
  url = {https://link.aps.org/doi/10.1103/PhysRevA.98.012324}
}

@article{Hu2018qGAN,
author = {Ling Hu  and Shu-Hao Wu  and Weizhou Cai  and Yuwei Ma  and Xianghao Mu  and Yuan Xu  and Haiyan Wang  and Yipu Song  and Dong-Ling Deng  and Chang-Ling Zou  and Luyan Sun },
title = {Quantum generative adversarial learning in a superconducting quantum circuit},
journal = {Science Advances},
volume = {5},
number = {1},
pages = {eaav2761},
year = {2019},
doi = {10.1126/sciadv.aav2761},
URL = {https://www.science.org/doi/abs/10.1126/sciadv.aav2761},
eprint = {https://www.science.org/doi/pdf/10.1126/sciadv.aav2761}}

@misc{Brassard2002QAE,
   title={Quantum amplitude amplification and estimation},
   ISBN={9780821878958},
   ISSN={0271-4132},
   url={http://dx.doi.org/10.1090/conm/305/05215},
   DOI={10.1090/conm/305/05215},
   journal={Quantum Computation and Information},
   publisher={American Mathematical Society},
   author={Brassard, Gilles and Høyer, Peter and Mosca, Michele and Tapp, Alain},
   year={2002},
   pages={53–74} }

@ARTICLE{flasseur24,
       author = {{Flasseur}, Olivier and {Bodrito}, Th{\'e}o and {Mairal}, Julien and {Ponce}, Jean and {Langlois}, Maud and {Lagrange}, Anne-Marie},
        title = "{deep PACO: combining statistical models with deep learning for exoplanet detection and characterization in direct imaging at high contrast}",
      journal = {\mnras},
         year = 2024,
        month = jan,
       volume = {527},
       number = {1},
        pages = {1534-1562},
          doi = {10.1093/mnras/stad3143},
archivePrefix = {arXiv},
       eprint = {2303.02461},
 primaryClass = {astro-ph.IM},
       adsurl = {https://ui.adsabs.harvard.edu/abs/2024MNRAS.527.1534F}
}

@ARTICLE{Pagliaro2026,
       author = {{Pagliaro}, L. and {Zingales}, T. and {Piotto}, G. and {Giovannini}, I. and {Mantovan}, G.},
        title = "{$\texttt{Exoformer}$: Accelerating Bayesian atmospheric retrievals with transformer neural networks}",
      journal = {arXiv e-prints},
         year = 2026,
        month = mar,
          eid = {arXiv:2603.27623},
        pages = {arXiv:2603.27623},
          doi = {10.48550/arXiv.2603.27623},
archivePrefix = {arXiv},
       eprint = {2603.27623},
 primaryClass = {astro-ph.EP},
       adsurl = {https://ui.adsabs.harvard.edu/abs/2026arXiv260327623P}
}

@article{Vetrano2026,
	author = {Vetrano, Marco and Zingales, Tiziano and Palma, G. Massimo and Lorenzo, Salvatore},
	date = {2026/06/22},
	doi = {10.1007/s42484-026-00414-4},
	id = {Vetrano2026},
	isbn = {2524-4914},
	journal = {Quantum Machine Intelligence},
	number = {2},
	pages = {69},
	title = {Exoplanetary atmospheres retrieval via a quantum extreme learning machine},
	url = {https://doi.org/10.1007/s42484-026-00414-4},
	volume = {8},
	year = {2026}}

@ARTICLE{antwi25,
       author = {{Antwi-Danso}, Jacqueline and {Papovich}, Casey and {Esdaile}, James and {Nanayakkara}, Themiya and {Glazebrook}, Karl and {Hutchison}, Taylor A. and {Whitaker}, Katherine E. and {Marsan}, Z. Cemile and {Diaz}, Ruben J. and {Marchesini}, Danilo and {Muzzin}, Adam and {Tran}, Kim-Vy H. and {Setton}, David J. and {Kaushal}, Yasha and {Speagle}, Joshua S. and {Cole}, Justin},
        title = "{The FENIKS Survey: Spectroscopic Confirmation of Massive Quiescent Galaxies at z {\ensuremath{\sim}} 3--5}",
      journal = {\apj},
         year = 2025,
        month = jan,
       volume = {978},
       number = {1},
          eid = {90},
        pages = {90},
          doi = {10.3847/1538-4357/ad8b30},
archivePrefix = {arXiv},
       eprint = {2307.09590},
 primaryClass = {astro-ph.GA},
       adsurl = {https://ui.adsabs.harvard.edu/abs/2025ApJ...978...90A}
}

@ARTICLE{boylan23,
       author = {{Boylan-Kolchin}, Michael},
        title = "{Stress testing {\ensuremath{\Lambda}}CDM with high-redshift galaxy candidates}",
      journal = {Nature Astronomy},
         year = 2023,
        month = jun,
       volume = {7},
        pages = {731-735},
          doi = {10.1038/s41550-023-01937-7},
archivePrefix = {arXiv},
       eprint = {2208.01611},
 primaryClass = {astro-ph.CO},
       adsurl = {https://ui.adsabs.harvard.edu/abs/2023NatAs...7..731B}
}

@INPROCEEDINGS{burgett24,
       author = {{Burgett}, W. and {Bernstein}, R. and {Ashby}, D. and {Bigelow}, B. and {Brossus}, G. and {Cox}, M. and {Demers}, R. and {Figueroa}, F. and {Fischer}, B. and {Groark}, F. and {Laskin}, R. and {Millan-Gabet}, R. and {Park}, S. and {Turner}, R. and {Walls}, B.},
        title = "{The Giant Magellan Telescope project in 2024: status and look ahead}",
    booktitle = {Ground-based and Airborne Telescopes X},
         year = 2024,
       editor = {{Marshall}, Heather K. and {Spyromilio}, Jason and {Usuda}, Tomonori},
       series = {Society of Photo-Optical Instrumentation Engineers (SPIE) Conference Series},
       volume = {13094},
        month = aug,
          eid = {1309417},
        pages = {1309417},
          doi = {10.1117/12.3020733},
       adsurl = {https://ui.adsabs.harvard.edu/abs/2024SPIE13094E..17B}
}

@ARTICLE{carleton19,
       author = {{Carleton}, Timothy and {Errani}, Rapha{\"e}l and {Cooper}, Michael and {Kaplinghat}, Manoj and {Pe{\~n}arrubia}, Jorge and {Guo}, Yicheng},
        title = "{The formation of ultra-diffuse galaxies in cored dark matter haloes through tidal stripping and heating}",
      journal = {\mnras},
         year = 2019,
        month = may,
       volume = {485},
       number = {1},
        pages = {382-395},
          doi = {10.1093/mnras/stz383},
archivePrefix = {arXiv},
       eprint = {1805.06896},
 primaryClass = {astro-ph.GA},
       adsurl = {https://ui.adsabs.harvard.edu/abs/2019MNRAS.485..382C}
}

@ARTICLE{carnall24,
       author = {{Carnall}, A.~C. and {Cullen}, F. and {McLure}, R.~J. and {McLeod}, D.~J. and {Begley}, R. and {Donnan}, C.~T. and {Dunlop}, J.~S. and {Shapley}, A.~E. and {Rowlands}, K. and {Almaini}, O. and {Arellano-C{\'o}rdova}, K.~Z. and {Barrufet}, L. and {Cimatti}, A. and {Ellis}, R.~S. and {Grogin}, N.~A. and {Hamadouche}, M.~L. and {Illingworth}, G.~D. and {Koekemoer}, A.~M. and {Leung}, H.-H. and {Lovell}, C.~C. and {P{\'e}rez-Gonz{\'a}lez}, P.~G. and {Santini}, P. and {Stanton}, T.~M. and {Wild}, V.},
        title = "{The JWST EXCELS survey: too much, too young, too fast? Ultra-massive quiescent galaxies at 3 < z < 5}",
      journal = {\mnras},
         year = 2024,
        month = oct,
       volume = {534},
       number = {1},
        pages = {325-348},
          doi = {10.1093/mnras/stae2092},
archivePrefix = {arXiv},
       eprint = {2405.02242},
 primaryClass = {astro-ph.GA},
       adsurl = {https://ui.adsabs.harvard.edu/abs/2024MNRAS.534..325C}
}

@ARTICLE{chworowsky24,
       author = {{Chworowsky}, Katherine and {Finkelstein}, Steven L. and {Boylan-Kolchin}, Michael and {McGrath}, Elizabeth J. and {Iyer}, Kartheik G. and {Papovich}, Casey and {Dickinson}, Mark and {Taylor}, Anthony J. and {Yung}, L.~Y. Aaron and {Arrabal Haro}, Pablo and {Bagley}, Micaela B. and {Backhaus}, Bren E. and {Bhatawdekar}, Rachana and {Cheng}, Yingjie and {Cleri}, Nikko J. and {Cole}, Justin W. and {Cooper}, M.~C. and {Costantin}, Luca and {Dekel}, Avishai and {Franco}, Maximilien and {Fujimoto}, Seiji and {Hayward}, Christopher C. and {Holwerda}, Benne W. and {Huertas-Company}, Marc and {Hirschmann}, Michaela and {Hutchison}, Taylor A. and {Koekemoer}, Anton M. and {Larson}, Rebecca L. and {Li}, Zhaozhou and {Long}, Arianna S. and {Lucas}, Ray A. and {Pirzkal}, Nor and {Rodighiero}, Giulia and {Somerville}, Rachel S. and {Vanderhoof}, Brittany N. and {de la Vega}, Alexander and {Wilkins}, Stephen M. and {Yang}, Guang and {Zavala}, Jorge A.},
        title = "{Evidence for a Shallow Evolution in the Volume Densities of Massive Galaxies at z = 4{\textendash}8 from CEERS}",
      journal = {\aj},
         year = 2024,
        month = sep,
       volume = {168},
       number = {3},
          eid = {113},
        pages = {113},
          doi = {10.3847/1538-3881/ad57c1},
archivePrefix = {arXiv},
       eprint = {2311.14804},
 primaryClass = {astro-ph.GA},
       adsurl = {https://ui.adsabs.harvard.edu/abs/2024AJ....168..113C}
}

@ARTICLE{conroy12,
       author = {{Conroy}, Charlie and {van Dokkum}, Pieter G.},
        title = "{The Stellar Initial Mass Function in Early-type Galaxies From Absorption Line Spectroscopy. II. Results}",
      journal = {\apj},
         year = 2012,
        month = nov,
       volume = {760},
       number = {1},
          eid = {71},
        pages = {71},
          doi = {10.1088/0004-637X/760/1/71},
archivePrefix = {arXiv},
       eprint = {1205.6473},
 primaryClass = {astro-ph.CO},
       adsurl = {https://ui.adsabs.harvard.edu/abs/2012ApJ...760...71C}
}

@ARTICLE{davies21,
       author = {{Davies}, R. and {H{\"o}rmann}, V. and {Rabien}, S. and {Sturm}, E. and {Alves}, J. and {Cl{\'e}net}, Y. and {Kotilainen}, J. and {Lang-Bardl}, F. and {Nicklas}, H. and {Pott}, J. -U. and {Tolstoy}, E. and {Vulcani}, B. and {MICADO Consortium}},
        title = "{MICADO: The Multi-Adaptive Optics Camera for Deep Observations}",
      journal = {The Messenger},
         year = 2021,
        month = mar,
       volume = {182},
        pages = {17-21},
          doi = {10.18727/0722-6691/5217},
archivePrefix = {arXiv},
       eprint = {2103.11631},
 primaryClass = {astro-ph.IM},
       adsurl = {https://ui.adsabs.harvard.edu/abs/2021Msngr.182...17D}
}

@INPROCEEDINGS{davies18,
       author = {{Davies}, R. and {Alves}, J. and {Cl{\'e}net}, Y. and {Lang-Bardl}, F. and {Nicklas}, H. and {Pott}, J. -U. and {Ragazzoni}, R. and {Tolstoy}, E. and {Amico}, P. and {Anwand-Heerwart}, H. and {Barboza}, S. and {Barl}, L. and {Baudoz}, P. and {Bender}, R. and {Bezawada}, N. and {Bizenberger}, P. and {Boland}, W. and {Bonifacio}, P. and {Borgo}, B. and {Buey}, T. and {Chapron}, F. and {Chemla}, F. and {Cohen}, M. and {Czoske}, O. and {D{\'e}o}, V. and {Disseau}, K. and {Dreizler}, S. and {Dupuis}, O. and {Fabricius}, M. and {Falomo}, R. and {Fedou}, P. and {F{\"o}rster Schreiber}, N. and {Garrel}, V. and {Geis}, N. and {Gemperlein}, H. and {Gendron}, E. and {Genzel}, R. and {Gillessen}, S. and {Gl{\"u}ck}, M. and {Grupp}, F. and {Hartl}, M. and {H{\"a}user}, M. and {Hess}, H. -J. and {Hofferbert}, R. and {Hopp}, U. and {H{\"o}rmann}, V. and {Hubert}, Z. and {Huby}, E. and {Huet}, J. -M. and {Hutterer}, V. and {Ives}, D. and {Janssen}, A. and {Jellema}, W. and {Kausch}, W. and {Kerber}, F. and {Kravcar}, H. and {Le Ruyet}, B. and {Leschinski}, K. and {Mandla}, C. and {Manhart}, M. and {Massari}, D. and {Mei}, S. and {Merlin}, F. and {Mohr}, L. and {Monna}, A. and {Muench}, N. and {M{\"u}ller}, F. and {Musters}, G. and {Navarro}, R. and {Neumann}, U. and {Neumayer}, N. and {Niebsch}, J. and {Plattner}, M. and {Przybilla}, N. and {Rabien}, S. and {Ramlau}, R. and {Ramos}, J. and {Ramsay}, S. and {Rhode}, P. and {Richter}, A. and {Richter}, J. and {Rix}, H. -W. and {Rodeghiero}, G. and {Rohloff}, R. -R. and {Rosensteiner}, M. and {Rousset}, G. and {Schlichter}, J. and {Schubert}, J. and {Sevin}, A. and {Stuik}, R. and {Sturm}, E. and {Thomas}, J. and {Tromp}, N. and {Verdoes-Kleijn}, G. and {Vidal}, F. and {Wagner}, R. and {Wegner}, M. and {Zeilinger}, W. and {Ziegleder}, J. and {Ziegler}, B. and {Zins}, G.},
        title = "{The MICADO first light imager for the ELT: overview, operation, simulation}",
    booktitle = {Ground-based and Airborne Instrumentation for Astronomy VII},
         year = 2018,
       editor = {{Evans}, Christopher J. and {Simard}, Luc and {Takami}, Hideki},
       series = {Society of Photo-Optical Instrumentation Engineers (SPIE) Conference Series},
       volume = {10702},
        month = jul,
          eid = {107021S},
        pages = {107021S},
          doi = {10.1117/12.2311483},
archivePrefix = {arXiv},
       eprint = {1807.10003},
 primaryClass = {astro-ph.IM},
       adsurl = {https://ui.adsabs.harvard.edu/abs/2018SPIE10702E..1SD}
}

@ARTICLE{deugenio20,
       author = {{D'Eugenio}, C. and {Daddi}, E. and {Gobat}, R. and {Strazzullo}, V. and {Lustig}, P. and {Delvecchio}, I. and {Jin}, S. and {Puglisi}, A. and {Calabr{\'o}}, A. and {Mancini}, C. and {Dickinson}, M. and {Cimatti}, A. and {Onodera}, M.},
        title = "{The Typical Massive Quiescent Galaxy at z {\ensuremath{\sim}} 3 is a Post-starburst}",
      journal = {\apjl},
         year = 2020,
        month = mar,
       volume = {892},
       number = {1},
          eid = {L2},
        pages = {L2},
          doi = {10.3847/2041-8213/ab7a96},
archivePrefix = {arXiv},
       eprint = {2003.04342},
 primaryClass = {astro-ph.GA},
       adsurl = {https://ui.adsabs.harvard.edu/abs/2020ApJ...892L...2D}
}

@ARTICLE{djorgovski87,
       author = {{Djorgovski}, S. and {Davis}, Marc},
        title = "{Fundamental Properties of Elliptical Galaxies}",
      journal = {\apj},
         year = 1987,
        month = feb,
       volume = {313},
        pages = {59},
          doi = {10.1086/164948},
       adsurl = {https://ui.adsabs.harvard.edu/abs/1987ApJ...313...59D}
}

@ARTICLE{dressler87,
       author = {{Dressler}, Alan and {Lynden-Bell}, Donald and {Burstein}, David and {Davies}, Roger L. and {Faber}, S.~M. and {Terlevich}, Roberto and {Wegner}, Gary},
        title = "{Spectroscopy and Photometry of Elliptical Galaxies. I. New Distance Estimator}",
      journal = {\apj},
         year = 1987,
        month = feb,
       volume = {313},
        pages = {42},
          doi = {10.1086/164947},
       adsurl = {https://ui.adsabs.harvard.edu/abs/1987ApJ...313...42D}
}

@ARTICLE{forrest22,
       author = {{Forrest}, Ben and {Wilson}, Gillian and {Muzzin}, Adam and {Marchesini}, Danilo and {Cooper}, M.~C. and {Marsan}, Z. Cemile and {Annunziatella}, Marianna and {McConachie}, Ian and {Zaidi}, Kumail and {Gomez}, Percy and {Urbano Stawinski}, Stephanie M. and {Chang}, Wenjun and {de Lucia}, Gabriella and {La Barbera}, Francesco and {Lubin}, Lori and {Nantais}, Julie and {Pe{\~n}a}, Theodore and {Saracco}, Paolo and {Surace}, Jason and {Stefanon}, Mauro},
        title = "{MAGAZ3NE: High Stellar Velocity Dispersions for Ultramassive Quiescent Galaxies at z {\ensuremath{\gtrsim}} 3}",
      journal = {\apj},
         year = 2022,
        month = oct,
       volume = {938},
       number = {2},
          eid = {109},
        pages = {109},
          doi = {10.3847/1538-4357/ac8747},
archivePrefix = {arXiv},
       eprint = {2208.04329},
 primaryClass = {astro-ph.GA},
       adsurl = {https://ui.adsabs.harvard.edu/abs/2022ApJ...938..109F}
}

@ARTICLE{gallazzi05,
       author = {{Gallazzi}, Anna and {Charlot}, St{\'e}phane and {Brinchmann}, Jarle and {White}, Simon D.~M. and {Tremonti}, Christy A.},
        title = "{The ages and metallicities of galaxies in the local universe}",
      journal = {\mnras},
         year = 2005,
        month = sep,
       volume = {362},
       number = {1},
        pages = {41-58},
          doi = {10.1111/j.1365-2966.2005.09321.x},
archivePrefix = {arXiv},
       eprint = {astro-ph/0506539},
 primaryClass = {astro-ph},
       adsurl = {https://ui.adsabs.harvard.edu/abs/2005MNRAS.362...41G}
}

@ARTICLE{gilmozzi07,
       author = {{Gilmozzi}, R. and {Spyromilio}, J.},
        title = "{The European Extremely Large Telescope (E-ELT)}",
      journal = {The Messenger},
         year = 2007,
        month = mar,
       volume = {127},
        pages = {11},
       adsurl = {https://ui.adsabs.harvard.edu/abs/2007Msngr.127...11G}
}

@ARTICLE{glazebrook17,
   author = {{Glazebrook}, K. and {Schreiber}, C. and {Labb{\'e}}, I. and 
	{Nanayakkara}, T. and {Kacprzak}, G.~G. and {Oesch}, P.~A. and 
	{Papovich}, C. and {Spitler}, L.~R. and {Straatman}, C.~M.~S. and 
	{Tran}, K.-V.~H. and {Yuan}, T.},
    title = "{A massive, quiescent galaxy at a redshift of 3.717}",
  journal = {\nat},
archivePrefix = "arXiv",
   eprint = {1702.01751},
     year = 2017,
    month = apr,
   volume = 544,
    pages = {71-74},
      doi = {10.1038/nature21680}
}

@ARTICLE{glazebrook24,
       author = {{Glazebrook}, Karl and {Nanayakkara}, Themiya and {Schreiber}, Corentin and {Lagos}, Claudia and {Kawinwanichakij}, Lalitwadee and {Jacobs}, Colin and {Chittenden}, Harry and {Brammer}, Gabriel and {Kacprzak}, Glenn G. and {Labbe}, Ivo and {Marchesini}, Danilo and {Marsan}, Z. Cemile and {Oesch}, Pascal A. and {Papovich}, Casey and {Remus}, Rhea-Silvia and {Tran}, Kim-Vy H. and {Esdaile}, James and {Chandro-Gomez}, Angel},
        title = "{A massive galaxy that formed its stars at z {\ensuremath{\approx}} 11}",
      journal = {\nat},
         year = 2024,
        month = apr,
       volume = {628},
       number = {8007},
        pages = {277-281},
          doi = {10.1038/s41586-024-07191-9},
archivePrefix = {arXiv},
       eprint = {2308.05606},
 primaryClass = {astro-ph.GA},
       adsurl = {https://ui.adsabs.harvard.edu/abs/2024Natur.628..277G}
}

@ARTICLE{Holmbeck23,
       author = {{Holmbeck}, Erika M. and {Barnes}, Jennifer and {Lund}, Kelsey A. and {Sprouse}, Trevor M. and {McLaughlin}, G.~C. and {Mumpower}, Matthew R.},
        title = "{Superheavy Elements in Kilonovae}",
      journal = {\apjl},
         year = 2023,
        month = jul,
       volume = {951},
       number = {1},
          eid = {L13},
        pages = {L13},
          doi = {10.3847/2041-8213/acd9cb},
archivePrefix = {arXiv},
       eprint = {2304.02125},
 primaryClass = {astro-ph.HE},
       adsurl = {https://ui.adsabs.harvard.edu/abs/2023ApJ...951L..13H}
}

@ARTICLE{jakobsen22,
       author = {{Jakobsen}, P. and {Ferruit}, P. and {Alves de Oliveira}, C. and {Arribas}, S. and {Bagnasco}, G. and {Barho}, R. and {Beck}, T.~L. and {Birkmann}, S. and {B{\"o}ker}, T. and {Bunker}, A.~J. and {Charlot}, S. and {de Jong}, P. and {de Marchi}, G. and {Ehrenwinkler}, R. and {Falcolini}, M. and {Fels}, R. and {Franx}, M. and {Franz}, D. and {Funke}, M. and {Giardino}, G. and {Gnata}, X. and {Holota}, W. and {Honnen}, K. and {Jensen}, P.~L. and {Jentsch}, M. and {Johnson}, T. and {Jollet}, D. and {Karl}, H. and {Kling}, G. and {K{\"o}hler}, J. and {Kolm}, M. -G. and {Kumari}, N. and {Lander}, M.~E. and {Lemke}, R. and {L{\'o}pez-Caniego}, M. and {L{\"u}tzgendorf}, N. and {Maiolino}, R. and {Manjavacas}, E. and {Marston}, A. and {Maschmann}, M. and {Maurer}, R. and {Messerschmidt}, B. and {Moseley}, S.~H. and {Mosner}, P. and {Mott}, D.~B. and {Muzerolle}, J. and {Pirzkal}, N. and {Pittet}, J. -F. and {Plitzke}, A. and {Posselt}, W. and {Rapp}, B. and {Rauscher}, B.~J. and {Rawle}, T. and {Rix}, H. -W. and {R{\"o}del}, A. and {Rumler}, P. and {Sabbi}, E. and {Salvignol}, J. -C. and {Schmid}, T. and {Sirianni}, M. and {Smith}, C. and {Strada}, P. and {te Plate}, M. and {Valenti}, J. and {Wettemann}, T. and {Wiehe}, T. and {Wiesmayer}, M. and {Willott}, C.~J. and {Wright}, R. and {Zeidler}, P. and {Zincke}, C.},
        title = "{The Near-Infrared Spectrograph (NIRSpec) on the James Webb Space Telescope. I. Overview of the instrument and its capabilities}",
      journal = {A\& A},
         year = 2022,
        month = may,
       volume = {661},
          eid = {A80},
        pages = {A80},
          doi = {10.1051/0004-6361/202142663},
archivePrefix = {arXiv},
       eprint = {2202.03305},
 primaryClass = {astro-ph.IM},
       adsurl = {https://ui.adsabs.harvard.edu/abs/2022A&A...661A..80J}
}

@ARTICLE{hopkins18,
       author = {{Hopkins}, A.~M.},
        title = "{The Dawes Review 8: Measuring the Stellar Initial Mass Function}",
      journal = {\pasa},
         year = 2018,
        month = nov,
       volume = {35},
          eid = {e039},
        pages = {e039},
          doi = {10.1017/pasa.2018.29},
archivePrefix = {arXiv},
       eprint = {1807.09949},
 primaryClass = {astro-ph.GA},
       adsurl = {https://ui.adsabs.harvard.edu/abs/2018PASA...35...39H}
}

@ARTICLE{katz23,
       author = {{Katz}, Harley and {Kimm}, Taysun and {Ellis}, Richard S. and {Devriendt}, Julien and {Slyz}, Adrianne},
        title = "{The challenges of identifying Population III stars in the early Universe}",
      journal = {\mnras},
         year = 2023,
        month = sep,
       volume = {524},
       number = {1},
        pages = {351-360},
          doi = {10.1093/mnras/stad1903},
archivePrefix = {arXiv},
       eprint = {2207.04751},
 primaryClass = {astro-ph.GA},
       adsurl = {https://ui.adsabs.harvard.edu/abs/2023MNRAS.524..351K}
}

@ARTICLE{Ji&Frebel18,
       author = {{Ji}, Alexander P. and {Frebel}, Anna},
        title = "{From Actinides to Zinc: Using the Full Abundance Pattern of the Brightest Star in Reticulum II to Distinguish between Different r-process Sites}",
      journal = {\apj},
         year = 2018,
        month = apr,
       volume = {856},
       number = {2},
          eid = {138},
        pages = {138},
          doi = {10.3847/1538-4357/aab14a},
archivePrefix = {arXiv},
       eprint = {1802.07272},
 primaryClass = {astro-ph.SR},
       adsurl = {https://ui.adsabs.harvard.edu/abs/2018ApJ...856..138J}
}

@ARTICLE{labarbera13,
       author = {{La Barbera}, F. and {Ferreras}, I. and {Vazdekis}, A. and {de la Rosa}, I.~G. and {de Carvalho}, R.~R. and {Trevisan}, M. and {Falc{\'o}n-Barroso}, J. and {Ricciardelli}, E.},
        title = "{SPIDER VIII - constraints on the stellar initial mass function of early-type galaxies from a variety of spectral features}",
      journal = {\mnras},
         year = 2013,
        month = aug,
       volume = {433},
       number = {4},
        pages = {3017-3047},
          doi = {10.1093/mnras/stt943},
archivePrefix = {arXiv},
       eprint = {1305.2273},
 primaryClass = {astro-ph.CO},
       adsurl = {https://ui.adsabs.harvard.edu/abs/2013MNRAS.433.3017L}
}

@ARTICLE{labarbera19,
       author = {{La Barbera}, F. and {Vazdekis}, A. and {Ferreras}, I. and {Pasquali}, A. and {Allende Prieto}, C. and {Mart{\'\i}n-Navarro}, I. and {Aguado}, D.~S. and {de Carvalho}, R.~R. and {Rembold}, S. and {Falc{\'o}n-Barroso}, J. and {van de Ven}, G.},
        title = "{IMF radial gradients in most massive early-type galaxies}",
      journal = {\mnras},
         year = 2019,
        month = nov,
       volume = {489},
       number = {3},
        pages = {4090-4110},
          doi = {10.1093/mnras/stz2192},
archivePrefix = {arXiv},
       eprint = {1909.01382},
 primaryClass = {astro-ph.GA},
       adsurl = {https://ui.adsabs.harvard.edu/abs/2019MNRAS.489.4090L}
}

@ARTICLE{mahmoodzadeh25,
       author = {{Mahmoodzadeh}, Hossein and {Saracco}, Paolo and {Conconi}, Paolo and {Saggin}, Bortolino and {Scaccabarozzi}, Diego and {Di Antonio}, Ivan and {Riva}, Marco and {Molinari}, Emilio and {Arcidiacono}, Carmelo and {Arosio}, Ilaria and {Cascone}, Enrico and {Cianniello}, Vincenzo and {De Caprio}, Vincenzo and {Di Rico}, Gianluca and {Di Francesco}, Benedetta and {Eredia}, Christian and {Franzetti}, Paolo and {Fumana}, Marco and {Greggio}, Davide and {Portaluri}, Elisa and {Scalera}, Marcello},
        title = "{Conceptual opto-mechanical design of SHARP: a near-infrared multi-mode spectrograph conceived for the next-generation telescopes}",
      journal = {Journal of Astronomical Telescopes, Instruments, and Systems},
         year = 2025,
        month = jul,
       volume = {11},
          eid = {035002},
        pages = {035002},
          doi = {10.1117/1.JATIS.11.3.035002},
archivePrefix = {arXiv},
       eprint = {2509.07057},
 primaryClass = {astro-ph.IM},
       adsurl = {https://ui.adsabs.harvard.edu/abs/2025JATIS..11c5002M}
}

@INPROCEEDINGS{marconi24,
       author = {{Marconi}, A. and {Abreu}, M. and {Adibekyan}, V. and {Alberti}, V. and {Albrecht}, S. and {Alcaniz}, J. and {Aliverti}, M. and {Allende Prieto}, C. and {Alvarado-Gomez}, J.~D. and {Alves}, C.~S. and {Amado}, P.~J. and {Amate}, M. and {Andersen}, M.~I. and {Antoniucci}, S. and {Artigau}, E. and {Bailet}, C. and {Baker}, C. and {Baldini}, V. and {Balestra}, A. and {Barnes}, S.~A. and {Baron}, F. and {Barros}, S.~C.~C. and {Bauer}, S.~M. and {Beaulieu}, M. and {Bellido-Tirado}, O. and {Benneke}, B. and {Bensby}, T. and {Bergin}, E.~A. and {Berio}, P. and {Biazzo}, K. and {Bigot}, L. and {Bik}, A. and {Birkby}, J.~L. and {Blind}, N. and {Boebion}, O. and {Boisse}, I. and {Bolmont}, E. and {Bolton}, J.~S. and {Bonaglia}, M. and {Bonfils}, X. and {Bonhomme}, L. and {Borsa}, F. and {Bouret}, J.-C. and {Brandeker}, A. and {Brandner}, W. and {Broeg}, C.~H. and {Brogi}, M. and {Brousseau}, D. and {Brucalassi}, A. and {Brynnel}, J. and {Buchhave}, L.~A. and {Buscher}, D.~F. and {Cabona}, L. and {Cabral}, A. and {Calderone}, G. and {Calvo-Ortega}, R. and {Cantalloube}, F. and {Canto Martins}, B.~L. and {Carbonaro}, L. and {Caujolle}, Y. and {Chauvin}, G. and {Chazelas}, B. and {Cheffot}, A.-L. and {Cheng}, Y.~S. and {Chiavassa}, A. and {Christensen}, L. and {Cirami}, R. and {Cirasuolo}, M. and {Cook}, N.~J. and {Cooke}, R.~J. and {Coretti}, I. and {Covino}, S. and {Cowan}, N. and {Cresci}, G. and {Cristiani}, S. and {Cunha Parro}, V. and {Cupani}, G. and {D'Odorico}, V. and {Dadi}, K. and {de Castro Le{\~a}o}, I. and {De Cia}, A. and {De Medeiros}, J.~R. and {Debras}, F. and {Debus}, M. and {Delorme}, A. and {Demangeon}, O. and {Derie}, F. and {Dessauges-Zavadsky}, M. and {Di Marcantonio}, P. and {Di Stefano}, S. and {Dionies}, F. and {Domiciano de Souza}, A. and {Doyon}, R. and {Dunn}, J. and {Egner}, S. and {Ehrenreich}, D. and {Faria}, J.~P. and {Ferruzzi}, D. and {Feruglio}, C. and {Fisher}, M. and {Fontana}, A. and {Frank}, B.~S. and {Fuesslein}, C. and {Fumagalli}, M. and {Fusco}, T. and {Fynbo}, J. and {Gabella}, O. and {Gaessler}, W. and {Gallo}, E. and {Gao}, X. and {Genolet}, L. and {Genoni}, M. and {Giacobbe}, P. and {Giro}, E. and {Gon{\c{c}}alves}, R.~S. and {Gonzalez}, O.~A. and {Gonz{\'a}lez-Hern{\'a}ndez}, J.~I. and {Gouvret}, C. and {Gracia T{\'e}mich}, F. and {Haehnelt}, M.~G. and {Haniff}, C. and {Hatzes}, A. and {Helled}, R. and {Hoeijmakers}, H.~J. and {Hughes}, I. and {Huke}, P. and {Ivanisenko}, Y. and {J{\"a}rvinen}, A.~S. and {J{\"a}rvinen}, S.~P. and {Kaminski}, A. and {Kern}, J. and {Knoche}, J. and {Kordt}, A. and {Korhonen}, H. and {Korn}, A.~J. and {Kouach}, D. and {Kowzan}, G. and {Kreidberg}, L. and {Landoni}, M. and {Lanotte}, A.~A. and {Lavail}, A. and {Lavie}, B. and {Lee}, D. and {Lehmitz}, M. and {Li}, J. and {Li}, W. and {Liske}, J. and {Lovis}, C. and {Lucatello}, S. and {Lunney}, D. and {MacIntosh}, M.~J. and {Madhusudhan}, N. and {Magrini}, L. and {Maiolino}, R. and {Maldonado}, J. and {Malo}, L. and {Man}, A.~W.~S. and {Marquart}, T. and {Marques}, C.~M.~J. and {Marques}, E.~L. and {Martinez}, P. and {Martins}, A. and {Martins}, C.~J.~A.~P. and {Martins}, J.~H.~C. and {Maslowski}, P. and {Mason}, C. and {Mason}, E. and {McCracken}, R.~A. and {Melo e Sousa}, M.~A.~F. and {Mergo}, P. and {Micela}, G. and {Milakovi{\'c}}, D. and {Molli{\`e}re}, P. and {Monteiro}, M.~A. and {Montgomery}, D. and {Mordasini}, C. and {Morin}, J. and {Mucciarelli}, A. and {Murphy}, M.~T. and {N'Diaye}, M. and {Nardetto}, N. and {Neichel}, B. and {Neri}, N. and {Niedzielski}, A.~T. and {Niemczura}, E. and {Nisini}, B. and {Nortmann}, L. and {Noterdaeme}, P. and {Nunes}, N.~J. and {Oggioni}, L. and {Olchewsky}, F. and {Oliva}, E. and {{\"O}nel}, H. and {Origlia}, L. and {{\"O}stlin}, G. and {Ouellette}, N.~N.-Q. and {Pall{\'e}}, E. and {Papaderos}, P. and {Pariani}, G. and {Pasquini}, L.},
        title = "{ANDES, the high resolution spectrograph for the ELT: science goals, project overview, and future developments}",
    booktitle = {Ground-based and Airborne Instrumentation for Astronomy X},
         year = 2024,
       editor = {{Bryant}, Julia J. and {Motohara}, Kentaro and {Vernet}, Jo{\"e}l. R.~D.},
       series = {Society of Photo-Optical Instrumentation Engineers (SPIE) Conference Series},
       volume = {13096},
        month = jul,
          eid = {1309613},
        pages = {1309613},
          doi = {10.1117/12.3017966},
archivePrefix = {arXiv},
       eprint = {2407.14601},
 primaryClass = {astro-ph.IM},
       adsurl = {https://ui.adsabs.harvard.edu/abs/2024SPIE13096E..13M}
}

@INPROCEEDINGS{matt06,
       author = {{Johns}, Matt},
        title = "{The Giant Magellan Telescope (GMT)}",
    booktitle = {Ground-based and Airborne Telescopes},
         year = 2006,
       editor = {{Stepp}, Larry M.},
       series = {Society of Photo-Optical Instrumentation Engineers (SPIE) Conference Series},
       volume = {6267},
        month = jun,
          eid = {626729},
        pages = {626729},
          doi = {10.1117/12.670839},
       adsurl = {https://ui.adsabs.harvard.edu/abs/2006SPIE.6267E..29J}
}

@INPROCEEDINGS{niranjan24,
       author = {{Thatte}, Niranjan A. and {Melotte}, Dave and {Neichel}, Benoit and {Le Mignant}, David and {Rees}, Phil and {Clarke}, Fraser and {Ferraro-Wood}, Vanessa and {Gonzalez}, Oscar and {Jones}, Maia and {{\'A}lvarez Urue{\~n}a}, Alonso and {Argelaguet Vilaseca}, Heribert and {Arribas Mocoroa}, Santiago and {Caballero}, Jos{\'e} Antonio and {Carracedo Carballal}, Gonzalo Jos{\'e} and {Estrada Piqueras}, Alberto and {Ferro}, Irene and {Garc{\'\i}a Garc{\'\i}a}, Miriam and {Lamperti}, Isabella and {Pereira Santaella}, Miguel and {Perna}, Michele and {Piqueras Lopez}, Javier and {Bouch{\'e}}, Nicolas and {Boudon}, Didier and {Daguise}, Eric and {Domenis}, Nicola and {Fensch}, J{\'e}r{\'e}my and {Olivier Flasseur}, Olivier and {Giroud}, R{\'e}mi and {Guibert}, Matthieu and {Jarno}, Aurelien and {Jeanneau}, Alexandre and {Krogager}, Jens-Kristian and {Langlois}, Maud and {Laurent}, Florence and {Loupias}, Magali and {Migniau}, Jean-Emmanuel and {Nguyen}, Dieu and {Piqueras}, Laure and {Remillieux}, Alban and {Richard}, Johan and {Pecontal}, Arlette and {Bardou}, Lisa and {Barr}, David and {Cetre}, Sylvain and {Dimoudi}, Sofia and {Dubbeldam}, Marc and {Dunn}, Andrew and {Gadotti}, Dimitri and {Guy}, Joss and {King}, David and {McLeod}, Anna and {Morris}, Simon and {Morris}, Tim and {O'Brien}, Kieran and {Ronson}, Emily and {Smith}, Russell and {Staykov}, Lazar and {Swinbank}, Mark and {Accardo}, Matteo and {Alvarez Mendez}, Domingo and {Fuerte Rodriguez}, Pablo Alberto and {George}, Elizabeth and {Ives}, Derek and {Mehrgan}, Leander and {Mueller}, Eric and {Reyes}, Javier and {Conzelmann}, Ralf and {Gutierrez Cheetham}, Pablo and {Alonso Sanchez}, Angel and {Battaglia}, Giuseppina and {Cagigas}, Miguel and {Castro-Almaz{\'a}n}, Julio A. and {Chulani}, Haresh and {Delgado-Garc{\'\i}a}, Graciela and {Fernandez Izquierdo}, Patricia and {Esparza-Arredondo}, Donaji and {Garc{\'\i}a-Lorenzo}, Bego{\~n}a. and {Hern{\'a}ndez Gonz{\'a}lez}, Alberto and {Hern{\'a}ndez Su{\'a}rez}, Elvio and {Licandro}, Javier and {Joven}, Enrique and {L{\'o}pez L{\'o}pez}, Roberto and {Lujan Gonzalez}, Alejandro Antonio and {Mart{\'\i}n Hernando}, Yolanda and {Mart{\'\i}n-Navarro}, Ignacio and {Mediavilla}, Evencio and {Men{\'e}ndez Mendoza}, Sa{\'u}l and {Montoya Mart{\'\i}nez}, Luz Maria and {Pe{\~n}ate Castro}, Jos{\'e} and {Murgas}, Felipe and {Pall{\'e}}, Enric and {P{\'e}rez}, {\'A}lvaro and {Rasilla}, Jose Luis and {Rebolo}, Rafael and {Rodr{\'\i}guez}, Horacio and {Rodr{\'\i}guez Ramos}, Luis Fernando and {S{\'a}nchez B{\'e}jar}, Victor and {Shahbaz}, Tariq and {Vega Moreno}, Afrodisio and {Viera}, Teodora and {Bonnefoy}, Micka{\"e}l. and {Bret}, Tony and {Carlotti}, Alexis and {Correia}, Jean-Jacques and {Curaba}, St{\'e}phane and {Delboulb{\'e}}, Alain and {Guieu}, Sylvain and {Hours}, Adrien and {Hubert}, Zoltan and {Jocou}, Laurent and {Magnard}, Yves and {Michaud}, Laurence and {Moulin}, Thibaut and {Pancher}, Fabrice and {Rabou}, Patrick and {Rochat}, Sylvain and {Stadler}, Eric and {Contini}, Thierry and {Larrieu}, Marie and {Mamessier}, S{\'e}bastien and {Boebion}, Olivier and {Fantei-Caujolle}, Yan and {Lecron}, Daniel and {Amram}, Philippe and {Blanchard}, Patrick and {Bon}, William and {Bonnefoi}, Anne and {Bozier}, Alexandre and {Ceria}, William and {Challita}, Zalpha and {Charles}, Yannick and {Choquet}, Elodie and {Costille}, Anne and {Delsanti}, Audrey and {Dohlen}, Kjetil and {Ducret}, Franck and {El Hadi}, Kacem and {Foulon}, Benjamin and {Gimenez}, Jean-Luc and {Groussin}, Olivier and {Jaquet}, Marc and {Renault}, Edgard and {Rouquette}, Paul and {Sanchez}, Patrice and {Vigan}, Arthur and {Zavagno}, Annie and {F{\'e}tick}, Romain and {Fusco}, Thierry and {H{\'e}ritier}, Cedric and {Sauvage}, Jean-Francois and {Vedrenne}, Nicolas and {Aksoy}, Demet and {Caldwell}, Martin and {Fitzpatrick}, Ann and {Geddert}, Carl and {Hiscock}, Peter and {Johnson}, Emma and {Nalagatla}, Murali and {Saraff}, Louise and {Shreeves}, Joe and {Tildesley}, Matthew and {Wells}, Mark and {Aretos}, Anastasios and {Barrett}, Lee and {Black}, Martin and {Bond}, Charlotte and {Brierley}, Saskia and {Bryson}, Ian and {Calderhead}, Amelia and {Campbell}, Kenny and {Carruthers}, James and {Chapman}, Lee and {Cochrane}, William and {Gillespie}, Rory and {Harman}, Joel and {Harvey}, Douglas and {Harvey}, Eamonn and {Johnson}, Bethany and {Louth}, Tom and {MacIntosh}, Mike and {MacIver}, Anna and {Miller}, Chris and {Montgomery}, David and {Murali}, Meenu and {Murray}, John and {O'Malley}, Norman and {Sanchez-Janssen}, Ruben and {Schwartz}, Noah and {Smith}, Patrick and {Strachan}, Jonathan and {Todd}, Stephen and {Wasley}, Dawn and {Wilson}, Sandi and {Zhou}, Junyi and {Bell}, Eric and {Gnedin}, Oleg and {Gultekin}, Kayhan and {Mateo}, Mario and {Meyer}, Michael and {Birkby}, Jayne},
        title = "{HARMONI at ELT: project status and instrument overview}",
    booktitle = {Ground-based and Airborne Instrumentation for Astronomy X},
         year = 2024,
       editor = {{Bryant}, Julia J. and {Motohara}, Kentaro and {Vernet}, Jo{\"e}l. R.~D.},
       series = {Society of Photo-Optical Instrumentation Engineers (SPIE) Conference Series},
       volume = {13096},
        month = jul,
          eid = {1309614},
        pages = {1309614},
          doi = {10.1117/12.3018520},
       adsurl = {https://ui.adsabs.harvard.edu/abs/2024SPIE13096E..14T}
}

@ARTICLE{parikh18,
       author = {{Parikh}, Taniya and {Thomas}, Daniel and {Maraston}, Claudia and {Westfall}, Kyle B. and {Goddard}, Daniel and {Lian}, Jianhui and {Meneses-Goytia}, Sofia and {Jones}, Amy and {Vaughan}, Sam and {Andrews}, Brett H. and {Bershady}, Matthew and {Bizyaev}, Dmitry and {Brinkmann}, Jonathan and {Brownstein}, Joel R. and {Bundy}, Kevin and {Drory}, Niv and {Emsellem}, Eric and {Law}, David R. and {Newman}, Jeffrey A. and {Roman-Lopes}, Alexandre and {Wake}, David and {Yan}, Renbin and {Zheng}, Zheng},
        title = "{SDSS-IV MaNGA: the spatially resolved stellar initial mass function in {\ensuremath{\sim}}400 early-type galaxies}",
      journal = {\mnras},
         year = 2018,
        month = jul,
       volume = {477},
       number = {3},
        pages = {3954-3982},
          doi = {10.1093/mnras/sty785},
archivePrefix = {arXiv},
       eprint = {1803.08515},
 primaryClass = {astro-ph.GA},
       adsurl = {https://ui.adsabs.harvard.edu/abs/2018MNRAS.477.3954P}
}

@INPROCEEDINGS{pello24,
       author = {{Pell{\'o}}, Roser and {Puech}, Mathieu and {Prieto}, {\'E}ric and {Rodrigues}, Myriam and {Sanchez-Janssen}, Rub{\'e}n. and {Dalton}, Gavin B. and {Ducret}, Franck and {El Hadi}, Kacem and {Garc{\'\i}a-Vargas}, Mar{\'\i}a. L. and {Lynn}, Jeff and {Bharmal}, Nazim A. and {Chapuis}, Diane and {Dupieux}, Michel and {Hottier}, Cl{\'e}ment and {Larrieu}, Marie and {Martin}, Laurent and {Mohamed}, Meghna and {Morris}, Tim and {P{\'e}rez}, Ana and {Seifert}, Walter and {Xu}, Wenli and {Morris}, Simon and {Kaper}, Lex and {Gallego}, Jes{\'u}s and {Afonso}, Jose and {Barbuy}, Beatriz and {Contini}, Thierry and {Finoguenov}, Alexis and {Kassin}, Susan and {Miller}, Christopher and {Ostlin}, G{\"o}ran and {Pentericci}, Laura and {Schaerer}, Daniel and {Steinmetz}, Matthias and {Ziegler}, Bodo and {Araujo}, Ricardo and {Brynnel}, Joar and {Castilho}, Bruno and {Conselice}, Christopher J. and {Cvetojevic}, Nick and {Davison}, Christopher and {Dejonghe}, Julien and {Dessauges-Zavadsky}, Mirka and {Dohlen}, Kjetil and {Ferreira}, D{\'e}cio and {Gil de Paz}, Armando and {Gon{\c{c}}alves}, Thiago S. and {Guinouard}, Isabelle and {Hayes}, Matthew J. and {Ives}, Derek and {Janssen}, Annemieke and {Kehrig}, Carol and {Kelz}, Andreas and {Krajnovi{\'c}}, Davor and {Lanotte}, Audrey A. and {Laporte}, Nicolas and {Laporte}, Philippe and {Larsen}, Soren and {Lemasle}, Bertran and {Lewis}, Ian and {Li}, Jiang-Tao and {Pancino}, Elena and {Pieri}, Matthew M. and {Surace}, Christian and {Thurneysen}, Markus and {Vergani}, Susanna and {Wildi}, Fran{\c{c}}ois and {{\'A}lvarez Moreno}, Fernando and {Artan}, Raziye and {Beaulieu}, Mathilde and {Besada}, Eva and {Bik}, Arjan and {Bond}, Charlotte and {Bouri}, Mohamed and {Boy}, J{\'e}r{\'e}mie and {Bramall}, David and {Brands}, Sarah and {Braulio}, Antonio and {Butterley}, Tim and {Cabello}, Cristina and {Calero de Ory}, Marina and {Calvo}, Rocio and {Castillo Morales}, Africa and {Challita}, Zalpha and {Chittik}, Stephen and {Curto Maldonado}, Andr{\'e}s. and {De Frontat}, F{\'a}tima and {Dijkstra}, Elfi and {Elswijk}, Eddy and {Fasola}, Gilles and {Feiz}, Carmen and {Fialho}, Fabio and {Floriot}, Johan and {Franzetti}, Paolo and {Fumana}, Marco and {Gabarra}, Luis and {Garcia}, Lia and {Gargiulo}, Adriana and {Gaudemard}, Julien and {Giannone}, Domenico and {Gill}, Polly and {Gomez-Gutierrez}, Alicia and {Gouvret}, Carole and {Guenther}, Alan and {Harvey}, Douglas and {Ib{\'a}{\~n}ez Mengual}, Jos{\'e} Miguel and {Iglesias}, Jorge and {Ivanisenko}, Yevgeniy and {Kunst}, Peter and {Kwast}, Sander and {Leschinski}, Kieran and {Licausi}, Gianluca and {Ligori}, Sebastiano and {L{\'o}pez Orozco}, Juan Antonio and {Lowe}, Adam and {Macintosh}, Mike and {Magan}, H{\'e}ctor and {Maldonado}, Manuel and {Marquart}, Thomas and {Martins}, Lucimara and {Melara}, Marisole and {Melinder}, Jens and {Molema}, Jeannet and {Montgomery}, David and {Morales}, Mar{\'\i}a. and {Najarro}, Francisco and {Nardetto}, Nicolas and {Navarro}, Ramon and {Ottomani}, Antoine and {Pamplona}, Tony and {Pannetier}, Cyril and {Parr-Burman}, Phil and {Pascual}, Sergio and {Pe{\~n}ataro}, Mar{\'\i}a. and {P{\'e}rez Grande}, Isabel and {Peterzon}, Jan Rinze and {Piqueras}, Javier and {Piskunov}, Nicolai and {Rodriguez Cardoso}, Ramon and {Rodriguez Venzal}, Sergio and {Romp}, Rick and {Rostami}, Hossein and {Royer}, Fr{\'e}d{\'e}ric and {Sablowski}, Daniel and {Sanchez}, Ainhoa and {S{\'a}nchez Blanco}, Ernesto and {Schalling}, Ellen and {Schmoll}, Jurgen and {Schwartz}, Noah and {Stephan}, Jay and {Taburet}, Sylvestre and {Terrett}, David and {Torralbo}, Ignacio and {Tromp}, Niels and {Veredas}, Gerardo and {Yang}, Yanbin and {York}, Alec and {Zeilinger}, Werner},
        title = "{MOSAIC at the ELT: a unique instrument for the largest ground-based telescope}",
    booktitle = {Ground-based and Airborne Instrumentation for Astronomy X},
         year = 2024,
       editor = {{Bryant}, Julia J. and {Motohara}, Kentaro and {Vernet}, Jo{\"e}l. R.~D.},
       series = {Society of Photo-Optical Instrumentation Engineers (SPIE) Conference Series},
       volume = {13096},
        month = jul,
          eid = {1309615},
        pages = {1309615},
          doi = {10.1117/12.3019047},
       adsurl = {https://ui.adsabs.harvard.edu/abs/2024SPIE13096E..15P}
}

@ARTICLE{Roederer22,
       author = {{Roederer}, Ian U. and {Lawler}, James E. and {Den Hartog}, Elizabeth A. and {Placco}, Vinicius M. and {Surman}, Rebecca and {Beers}, Timothy C. and {Ezzeddine}, Rana and {Frebel}, Anna and {Hansen}, Terese T. and {Hattori}, Kohei and {Holmbeck}, Erika M. and {Sakari}, Charli M.},
        title = "{The R-process Alliance: A Nearly Complete R-process Abundance Template Derived from Ultraviolet Spectroscopy of the R-process-enhanced Metal-poor Star HD 222925}",
      journal = {\apjs},
         year = 2022,
        month = jun,
       volume = {260},
       number = {2},
          eid = {27},
        pages = {27},
          doi = {10.3847/1538-4365/ac5cbc},
archivePrefix = {arXiv},
       eprint = {2205.03426},
 primaryClass = {astro-ph.SR},
       adsurl = {https://ui.adsabs.harvard.edu/abs/2022ApJS..260...27R}
}

@ARTICLE{saracco20apj,
       author = {{Saracco}, Paolo and {Marchesini}, Danilo and {La Barbera}, Francesco and {Gargiulo}, Adriana and {Annunziatella}, Marianna and {Forrest}, Ben and {Lange Vagle}, Daniel J. and {Marsan}, Z. Cemile and {Muzzin}, Adam and {Stefanon}, Mauro and {Wilson}, Gillian},
        title = "{The Rapid Buildup of Massive Early-type Galaxies: Supersolar Metallicity, High Velocity Dispersion, and Young Age for an Early-type Galaxy at z = 3.35}",
      journal = {\apj},
         year = 2020,
        month = dec,
       volume = {905},
       number = {1},
          eid = {40},
        pages = {40},
          doi = {10.3847/1538-4357/abc7c4},
archivePrefix = {arXiv},
       eprint = {2011.04657},
 primaryClass = {astro-ph.GA}
}

@INPROCEEDINGS{saracco24,
       author = {{Saracco}, P. and {Conconi}, P. and {Arcidiacono}, C. and {Portaluri}, E. and {Mahmoodzadeh}, H. and {D'Orazi}, V. and {Fedele}, D. and {Gargiulo}, A. and {Vanzella}, E. and {Franzetti}, P. and {Arosio}, I. and {Barbalini}, L. and {Lops}, G. and {Molinari}, E. and {Cascone}, E. and {Cianniello}, V. and {D'Auria}, D. and {De Caprio}, V. and {Di Antonio}, I. and {Di Francesco}, B. and {Di Rico}, G. and {Eredia}, C. and {Fumana}, M. and {Greggio}, D. and {Rodeghiero}, G. and {Scalera}, M. and {Alcal{\`a}}, J.~M. and {Bisogni}, S. and {Bonito}, R. and {Bono}, G. and {Caratti o Garatti}, A. and {Dalla Bont{\`a}}, E. and {Dall'Ora}, M. and {Fiorentino}, G. and {Gallazzi}, A.~R. and {Guarcello}, M. and {Izzo}, L. and {La Barbera}, F. and {Lardo}, C. and {Longhetti}, M. and {Longobardo}, A. and {Magrini}, L. and {Mancini}, C. and {Mura}, A. and {Piconcelli}, E. and {Pizzella}, A. and {Podio}, L. and {Polletta}, M. and {Prisinzano}, L. and {Ricci}, F. and {Ripepi}, V. and {Roccatagliata}, V. and {Vietri}, G.},
        title = "{SHARP: a near-IR multi-mode spectrograph conceived for MORFEO@ELT}",
    booktitle = {Ground-based and Airborne Instrumentation for Astronomy X},
         year = 2024,
       editor = {{Bryant}, Julia J. and {Motohara}, Kentaro and {Vernet}, Jo{\"e}l. R.~D.},
       series = {Society of Photo-Optical Instrumentation Engineers (SPIE) Conference Series},
       volume = {13096},
        month = jul,
          eid = {130965I},
        pages = {130965I},
          doi = {10.1117/12.3018977},
       adsurl = {https://ui.adsabs.harvard.edu/abs/2024SPIE13096E..5IS}
}

@INPROCEEDINGS{sharp16,
       author = {{Sharp}, Rob and {Bloxham}, G. and {Boz}, R. and {Bundy}, D. and {Davies}, J. and {Espeland}, B. and {Fordham}, B. and {Hart}, J. and {Herrald}, N. and {Nielsen}, J. and {Vaccarella}, A. and {Vest}, C. and {Young}, P. and {McGregor}, P.},
        title = "{GMTIFS: The Giant Magellan Telescope integral fields spectrograph and imager}",
    booktitle = {Ground-based and Airborne Instrumentation for Astronomy VI},
         year = 2016,
       editor = {{Evans}, Christopher J. and {Simard}, Luc and {Takami}, Hideki},
       series = {Society of Photo-Optical Instrumentation Engineers (SPIE) Conference Series},
       volume = {9908},
        month = aug,
          eid = {99081Y},
        pages = {99081Y},
          doi = {10.1117/12.2231561},
       adsurl = {https://ui.adsabs.harvard.edu/abs/2016SPIE.9908E..1YS}
}

@ARTICLE{smith20,
       author = {{Smith}, Russell J.},
        title = "{Evidence for Initial Mass Function Variation in Massive Early-Type Galaxies}",
      journal = {\araa},
         year = 2020,
        month = aug,
       volume = {58},
        pages = {577-615},
          doi = {10.1146/annurev-astro-032620-020217},
       adsurl = {https://ui.adsabs.harvard.edu/abs/2020ARA&A..58..577S}
}

@ARTICLE{spiniello14,
       author = {{Spiniello}, Chiara and {Trager}, Scott and {Koopmans}, L{\'e}on V.~E. and {Conroy}, Charlie},
        title = "{The stellar IMF in early-type galaxies from a non-degenerate set of optical line indices}",
      journal = {\mnras},
         year = 2014,
        month = feb,
       volume = {438},
       number = {2},
        pages = {1483-1499},
          doi = {10.1093/mnras/stt2282},
archivePrefix = {arXiv},
       eprint = {1305.2873},
 primaryClass = {astro-ph.CO},
       adsurl = {https://ui.adsabs.harvard.edu/abs/2014MNRAS.438.1483S}
}

@ARTICLE{tanaka19,
       author = {{Tanaka}, Masayuki and {Valentino}, Francesco and {Toft}, Sune and {Onodera}, Masato and {Shimakawa}, Rhythm and {Ceverino}, Daniel and {Faisst}, Andreas L. and {Gallazzi}, Anna and {G{\'o}mez-Guijarro}, Carlos and {Kubo}, Mariko and {Magdis}, Georgios E. and {Steinhardt}, Charles L. and {Stockmann}, Mikkel and {Yabe}, Kiyoto and {Zabl}, Johannes},
        title = "{Stellar Velocity Dispersion of a Massive Quenching Galaxy at z = 4.01}",
      journal = {\apjl},
         year = 2019,
        month = nov,
       volume = {885},
       number = {2},
          eid = {L34},
        pages = {L34},
          doi = {10.3847/2041-8213/ab4ff3},
archivePrefix = {arXiv},
       eprint = {1909.10721},
 primaryClass = {astro-ph.GA},
       adsurl = {https://ui.adsabs.harvard.edu/abs/2019ApJ...885L..34T}
}

@ARTICLE{toloba23,
       author = {{Toloba}, Elisa and {Sales}, Laura V. and {Lim}, Sungsoon and {Peng}, Eric W. and {Guhathakurta}, Puragra and {Roediger}, Joel and {Wang}, Kaixiang and {Mihos}, J. Christopher and {C{\^o}t{\'e}}, Patrick and {Durrell}, Patrick R. and {Ferrarese}, Laura},
        title = "{The Next Generation Virgo Cluster Survey (NGVS). XXXV. First Kinematical Clues of Overly Massive Dark Matter Halos in Several Ultradiffuse Galaxies in the Virgo Cluster}",
      journal = {\apj},
         year = 2023,
        month = jul,
       volume = {951},
       number = {1},
          eid = {77},
        pages = {77},
          doi = {10.3847/1538-4357/acd336},
archivePrefix = {arXiv},
       eprint = {2305.06369},
 primaryClass = {astro-ph.GA},
       adsurl = {https://ui.adsabs.harvard.edu/abs/2023ApJ...951...77T}
}

@ARTICLE{Messa2022,
       author = {{Messa}, Matteo and {Dessauges-Zavadsky}, Miroslava and {Richard}, Johan and {Adamo}, Angela and {Nagy}, David and {Combes}, Fran{\c{c}}oise and {Mayer}, Lucio and {Ebeling}, Harald},
        title = "{Multiply lensed star forming clumps in the A521-sys1 galaxy at redshift 1}",
      journal = {\mnras},
         year = 2022,
        month = oct,
       volume = {516},
       number = {2},
        pages = {2420-2443},
          doi = {10.1093/mnras/stac2189},
archivePrefix = {arXiv},
       eprint = {2208.02863},
 primaryClass = {astro-ph.GA},
       adsurl = {https://ui.adsabs.harvard.edu/abs/2022MNRAS.516.2420M}
}

@ARTICLE{Claeyssens2025,
       author = {{Claeyssens}, Ad{\'e}la{\"\i}de and {Adamo}, Angela and {Messa}, Matteo and {Dessauges-Zavadsky}, Miroslava and {Richard}, Johan and {Kramarenko}, Ivan and {Matthee}, Jorryt and {Naidu}, Rohan P.},
        title = "{Tracing star formation across cosmic time at tens of parsec-scales in the lensing cluster field Abell 2744}",
      journal = {\mnras},
         year = 2025,
        month = mar,
       volume = {537},
       number = {3},
        pages = {2535-2558},
          doi = {10.1093/mnras/staf058},
archivePrefix = {arXiv},
       eprint = {2410.10974},
 primaryClass = {astro-ph.GA},
       adsurl = {https://ui.adsabs.harvard.edu/abs/2025MNRAS.537.2535C}
}

@ARTICLE{wisotzki2016,
       author = {{Wisotzki}, L. and {Bacon}, R. and {Blaizot}, J. and {Brinchmann}, J. and {Herenz}, E.~C. and {Schaye}, J. and {Bouch{\'e}}, N. and {Cantalupo}, S. and {Contini}, T. and {Carollo}, C.~M. and {Caruana}, J. and {Courbot}, J. -B. and {Emsellem}, E. and {Kamann}, S. and {Kerutt}, J. and {Leclercq}, F. and {Lilly}, S.~J. and {Patr{\'\i}cio}, V. and {Sandin}, C. and {Steinmetz}, M. and {Straka}, L.~A. and {Urrutia}, T. and {Verhamme}, A. and {Weilbacher}, P.~M. and {Wendt}, M.},
        title = "{Extended Lyman {\ensuremath{\alpha}} haloes around individual high-redshift galaxies revealed by MUSE}",
      journal = {\aap},
         year = 2016,
        month = mar,
       volume = {587},
          eid = {A98},
        pages = {A98},
          doi = {10.1051/0004-6361/201527384},
archivePrefix = {arXiv},
       eprint = {1509.05143},
 primaryClass = {astro-ph.GA},
       adsurl = {https://ui.adsabs.harvard.edu/abs/2016A&A...587A..98W}
}

@ARTICLE{Bhagwat2025,
       author = {{Bhagwat}, Aniket and {Napolitano}, Lorenzo and {Pentericci}, Laura and {Ciardi}, Benedetta and {Costa}, Tiago},
        title = "{Ly {\ensuremath{\alpha}} with SPICE: interpreting Ly {\ensuremath{\alpha}} emission at z > 5}",
      journal = {\mnras},
         year = 2025,
        month = sep,
       volume = {542},
       number = {1},
        pages = {128-135},
          doi = {10.1093/mnras/staf1121},
archivePrefix = {arXiv},
       eprint = {2408.16063},
 primaryClass = {astro-ph.GA},
       adsurl = {https://ui.adsabs.harvard.edu/abs/2025MNRAS.542..128B}
}

@INPROCEEDINGS{Ciliegi2024,
       author = {{Ciliegi}, Paolo and {Agapito}, Guido and {Aliverti}, Matteo and {Annibali}, Francesca and {Aridiacono}, Carmelo and {Azzaroli}, Nicol{\`o} and {Balestra}, Andrea and {Baronchelli}, Ivano and {Ballone}, Alessandro and {Baruffolo}, Andrea and {Battaini}, Federico and {Benedetti}, Simone and {Bergomi}, Maria and {Bianco}, Andrea and {Bonaglia}, Marco and {Briguglio}, Runa and {Busoni}, Lorenzo and {Cantiello}, Michele and {Capasso}, Giulio and {Carl{\`a}}, Giulia and {Carolo}, Elena and {Cascone}, Enrico and {Chauvin}, Ga{\"e}l. and {Chebbo}, Manal and {Chinellato}, Simonetta and {Cianniello}, Vincenzo and {Colapietro}, Mirko and {Correia}, Jean-Jacques and {Cosentino}, Giuseppe and {Costa}, Elia and {D'Auria}, Domenico and {De Caprio}, Vincenzo and {Devaney}, Nicholas and {Di Antonio}, Ivan and {Di Cianno}, Amico and {Di Dato}, Andrea and {Di Filippo}, Simone and {Di Francesco}, Benedetta and {Di Giammatteo}, Ugo and {Di Prospero}, Chiara and {Di Rico}, Gianluca and {Di Rocco}, Andrea and {Diretto}, Daphne and {Dolci}, Mauro and {Eredia}, Christian and {Esposito}, Simone and {Fantinel}, Daniela and {Farinato}, Jacopo and {Feautrier}, Philippe and {Foppiani}, Italo and {Genoni}, Matteo and {Giro}, Enrico and {Gluck}, Laurence and {Goncharov}, Alexander and {Grani}, Paolo and {Greggio}, Davide and {Guieu}, Sylvain and {Gullieuszik}, Marco and {Hubert}, Zoltan and {Jocou}, Laurent and {Lampitelli}, Salvatore and {Lapucci}, Tommaso and {Laudisio}, Fulvio and {Leal}, Vincent and {Magnard}, Yves and {Magrin}, Demetrio and {Malone}, Deborah and {Marafatto}, Luca and {Michel}, Christophe and {Mouillet}, David and {Moulin}, Thibaut and {Munari}, Matteo and {Oberti}, Sylvain and {Pancher}, Fabrice and {Pariani}, Giorgio and {Petrella}, Amedeo and {Pinnard}, Laurent and {Plantet}, Cedric and {Portaluri}, Elisa and {Puglisi}, Alfio and {Rabou}, Patrick and {Radhakrishnan}, Kalyan and {Ragazzoni}, Roberto and {Redaelli}, Edoardo Maria Alberto and {Riva}, Marco and {Rochat}, Sylvain and {Rodeghiero}, Gabriele and {Rosignoli}, Luca and {Salasnich}, Bernardo and {Savarese}, Salvatore and {Scalera}, Marcello and {Schipani}, Pietro and {Selvestrel}, Danilo and {Sassolas}, Benoit and {Sordo}, Rosanna and {Teodori}, Ludovico and {Umbriaco}, Gabriele and {Valentini}, Angelo and {Xompero}, Marco},
        title = "{MORFEO at ELT: the adaptive optics module for ELT}",
    booktitle = {Adaptive Optics Systems IX},
         year = 2024,
       editor = {{Jackson}, Kathryn J. and {Schmidt}, Dirk and {Vernet}, Elise},
       series = {Society of Photo-Optical Instrumentation Engineers (SPIE) Conference Series},
       volume = {13097},
        month = aug,
          eid = {1309722},
        pages = {1309722},
          doi = {10.1117/12.3019058},
       adsurl = {https://ui.adsabs.harvard.edu/abs/2024SPIE13097E..22C}
}

@ARTICLE{Euclid2025d,
       author = {{Euclid Collaboration} and {Bazzanini}, L. and {Angora}, G. and {Bergamini}, P. and {Meneghetti}, M. and {Rosati}, P. and {Acebron}, A. and {Grillo}, C. and {Lombardi}, M. and {Ratta}, R. and {Fogliardi}, M. and {Di Rosa}, G. and {Abriola}, D. and {D'Addona}, M. and {Granata}, G. and {Leuzzi}, L. and {Mercurio}, A. and {Schuldt}, S. and {Vanzella}, E. and {Tortora}, C. and {Altieri}, B. and {Andreon}, S. and {Auricchio}, N. and {Baccigalupi}, C. and {Baldi}, M. and {Balestra}, A. and {Bardelli}, S. and {Battaglia}, P. and {Biviano}, A. and {Branchini}, E. and {Brescia}, M. and {Camera}, S. and {Ca{\~n}as-Herrera}, G. and {Capobianco}, V. and {Carbone}, C. and {Carretero}, J. and {Castellano}, M. and {Castignani}, G. and {Cavuoti}, S. and {Cimatti}, A. and {Colodro-Conde}, C. and {Congedo}, G. and {Conversi}, L. and {Copin}, Y. and {Costille}, A. and {Courbin}, F. and {Courtois}, H.~M. and {Cropper}, M. and {Da Silva}, A. and {Degaudenzi}, H. and {De Lucia}, G. and {Dole}, H. and {Dubath}, F. and {Duncan}, C.~A.~J. and {Dupac}, X. and {Dusini}, S. and {Escoffier}, S. and {Fabricius}, M. and {Farina}, M. and {Farinelli}, R. and {Faustini}, F. and {Ferriol}, S. and {Finelli}, F. and {Frailis}, M. and {Franceschi}, E. and {Fumana}, M. and {Galeotta}, S. and {Gillard}, W. and {Gillis}, B. and {Giocoli}, C. and {Gracia-Carpio}, J. and {Grazian}, A. and {Grupp}, F. and {Guzzo}, L. and {Haugan}, S.~V.~H. and {Hoar}, J. and {Holmes}, W. and {Hook}, I.~M. and {Hormuth}, F. and {Hornstrup}, A. and {Jahnke}, K. and {Jhabvala}, M. and {Joachimi}, B. and {Keih{\"a}nen}, E. and {Kermiche}, S. and {Kiessling}, A. and {Kilbinger}, M. and {Kubik}, B. and {Kunz}, M. and {Kurki-Suonio}, H. and {Laureijs}, R. and {Le Brun}, A.~M.~C. and {Le Mignant}, D. and {Ligori}, S. and {Lilje}, P.~B. and {Lindholm}, V. and {Lloro}, I. and {Mainetti}, G. and {Maino}, D. and {Maiorano}, E. and {Mansutti}, O. and {Marggraf}, O. and {Martinelli}, M. and {Martinet}, N. and {Marulli}, F. and {Massey}, R.~J. and {Medinaceli}, E. and {Mei}, S. and {Melchior}, M. and {Mellier}, Y. and {Merlin}, E. and {Meylan}, G. and {Mora}, A. and {Moresco}, M. and {Moscardini}, L. and {Neissner}, C. and {Niemi}, S.-M. and {Padilla}, C. and {Paltani}, S. and {Pasian}, F. and {Pedersen}, K. and {Percival}, W.~J. and {Pettorino}, V. and {Pires}, S. and {Polenta}, G. and {Poncet}, M. and {Popa}, L.~A. and {Pozzetti}, L. and {Raison}, F. and {Renzi}, A. and {Rhodes}, J. and {Riccio}, G. and {Romelli}, E. and {Roncarelli}, M. and {Saglia}, R. and {Sakr}, Z. and {S{\'a}nchez}, A.~G. and {Sapone}, D. and {Sartoris}, B. and {Schneider}, P. and {Schrabback}, T. and {Secroun}, A. and {Seidel}, G. and {Serrano}, S. and {Simon}, P. and {Sirignano}, C. and {Sirri}, G. and {Stanco}, L. and {Steinwagner}, J. and {Tallada-Cresp{\'\i}}, P. and {Taylor}, A.~N. and {Tereno}, I. and {Tessore}, N. and {Toft}, S. and {Toledo-Moreo}, R. and {Torradeflot}, F. and {Tutusaus}, I. and {Valentijn}, E.~A. and {Valenziano}, L. and {Valiviita}, J. and {Vassallo}, T. and {Verdoes Kleijn}, G. and {Veropalumbo}, A. and {Wang}, Y. and {Weller}, J. and {Zacchei}, A. and {Zamorani}, G. and {Zucca}, E. and {Ballardini}, M. and {Bolzonella}, M. and {Bozzo}, E. and {Burigana}, C. and {Cabanac}, R. and {Calabrese}, M. and {Cappi}, A. and {Di Ferdinando}, D. and {Escartin Vigo}, J.~A. and {Hartley}, W.~G. and {Mart{\'\i}n-Fleitas}, J. and {Matthew}, S. and {Mauri}, N. and {Metcalf}, R.~B. and {Pezzotta}, A. and {P{\"o}ntinen}, M. and {Risso}, I. and {Scottez}, V. and {Sereno}, M. and {Tenti}, M. and {Viel}, M. and {Wiesmann}, M. and {Akrami}, Y. and {Andika}, I.~T. and {Anselmi}, S. and {Archidiacono}, M. and {Atrio-Barandela}, F. and {Aubourg}, E. and {Bertacca}, D. and {Bethermin}, M. and {Blanchard}, A. and {Blot}, L.},
        title = "{Euclid Quick Data Release (Q1). Searching for giant gravitational arcs in galaxy clusters with mask region-based convolutional neural networks}",
      journal = {arXiv e-prints},
         year = 2025,
        month = nov,
          eid = {arXiv:2511.03064},
        pages = {arXiv:2511.03064},
          doi = {10.48550/arXiv.2511.03064},
archivePrefix = {arXiv},
       eprint = {2511.03064},
 primaryClass = {astro-ph.CO},
       adsurl = {https://ui.adsabs.harvard.edu/abs/2025arXiv251103064E}
}

@ARTICLE{Euclid2025c,
       author = {{Euclid Collaboration} and {Lines}, N.~E.~P. and {Collett}, T.~E. and {Walmsley}, M. and {Rojas}, K. and {Li}, T. and {Leuzzi}, L. and {Manj{\'o}n-Garc{\'\i}a}, A. and {Vincken}, S.~H. and {Wilde}, J. and {Holloway}, P. and {Verma}, A. and {Metcalf}, R.~B. and {Andika}, I.~T. and {Melo}, A. and {Melchior}, M. and {Dom{\'\i}nguez S{\'a}nchez}, H. and {D{\'\i}az-S{\'a}nchez}, A. and {Acevedo Barroso}, J.~A. and {Cl{\'e}ment}, B. and {Krawczyk}, C. and {Pearce-Casey}, R. and {Serjeant}, S. and {Courbin}, F. and {Despali}, G. and {Gavazzi}, R. and {Schuldt}, S. and {Degaudenzi}, H. and {Ecker}, L.~R. and {Enzi}, W.~J.~R. and {Finner}, K. and {Galan}, A. and {Giocoli}, C. and {Hogg}, N.~B. and {Jahnke}, K. and {Kruk}, S. and {Mahler}, G. and {More}, A. and {Nagam}, B.~C. and {Pearson}, J. and {Sainz de Murieta}, A. and {Scarlata}, C. and {Sluse}, D. and {Sonnenfeld}, A. and {Spiniello}, C. and {Thai}, T.~T. and {Tortora}, C. and {Ulivi}, L. and {Weisenbach}, L. and {Zumalacarregui}, M. and {Aghanim}, N. and {Altieri}, B. and {Amara}, A. and {Andreon}, S. and {Auricchio}, N. and {Aussel}, H. and {Baccigalupi}, C. and {Baldi}, M. and {Balestra}, A. and {Bardelli}, S. and {Battaglia}, P. and {Bender}, R. and {Bernardeau}, F. and {Biviano}, A. and {Bonchi}, A. and {Bonino}, D. and {Branchini}, E. and {Brescia}, M. and {Brinchmann}, J. and {Camera}, S. and {Ca{\~n}as-Herrera}, G. and {Capobianco}, V. and {Carbone}, C. and {Cardone}, V.~F. and {Carretero}, J. and {Casas}, S. and {Castellano}, M. and {Castignani}, G. and {Cavuoti}, S. and {Chambers}, K.~C. and {Cimatti}, A. and {Colodro-Conde}, C. and {Congedo}, G. and {Conselice}, C.~J. and {Conversi}, L. and {Copin}, Y. and {Costille}, A. and {Courtois}, H.~M. and {Cropper}, M. and {Da Silva}, A. and {De Lucia}, G. and {Di Giorgio}, A.~M. and {Dolding}, C. and {Dole}, H. and {Dubath}, F. and {Duncan}, C.~A.~J. and {Dupac}, X. and {Escoffier}, S. and {Fabricius}, M. and {Farina}, M. and {Farinelli}, R. and {Faustini}, F. and {Ferriol}, S. and {Finelli}, F. and {Fotopoulou}, S. and {Frailis}, M. and {Franceschi}, E. and {Fumana}, M. and {Galeotta}, S. and {George}, K. and {Gillard}, W. and {Gillis}, B. and {G{\'o}mez-Alvarez}, P. and {Gracia-Carpio}, J. and {Granett}, B.~R. and {Grazian}, A. and {Grupp}, F. and {Guzzo}, L. and {Gwyn}, S. and {Haugan}, S.~V.~H. and {Holmes}, W. and {Hook}, I.~M. and {Hormuth}, F. and {Hornstrup}, A. and {Hudelot}, P. and {Jhabvala}, M. and {Keih{\"a}nen}, E. and {Kermiche}, S. and {Kiessling}, A. and {Kubik}, B. and {K{\"u}mmel}, M. and {Kunz}, M. and {Kurki-Suonio}, H. and {Le Boulc'h}, Q. and {Le Brun}, A.~M.~C. and {Le Mignant}, D. and {Ligori}, S. and {Lilje}, P.~B. and {Lindholm}, V. and {Lloro}, I. and {Mainetti}, G. and {Maino}, D. and {Maiorano}, E. and {Mansutti}, O. and {Marcin}, S. and {Marggraf}, O. and {Martinelli}, M. and {Martinet}, N. and {Marulli}, F. and {Massey}, R. and {Maurogordato}, S. and {Medinaceli}, E. and {Mei}, S. and {Mellier}, Y. and {Meneghetti}, M. and {Merlin}, E. and {Meylan}, G. and {Mora}, A. and {Moresco}, M. and {Moscardini}, L. and {Nakajima}, R. and {Neissner}, C. and {Nichol}, R.~C. and {Niemi}, S.-M. and {Nightingale}, J.~W. and {Padilla}, C. and {Paltani}, S. and {Pasian}, F. and {Pedersen}, K. and {Percival}, W.~J. and {Pettorino}, V. and {Pires}, S. and {Polenta}, G. and {Poncet}, M. and {Popa}, L.~A. and {Pozzetti}, L. and {Raison}, F. and {Rebolo}, R. and {Renzi}, A. and {Rhodes}, J. and {Riccio}, G. and {Romelli}, E. and {Roncarelli}, M. and {Saglia}, R. and {Sakr}, Z. and {S{\'a}nchez}, A.~G. and {Sapone}, D. and {Sartoris}, B. and {Schewtschenko}, J.~A. and {Schirmer}, M. and {Schneider}, P. and {Schrabback}, T. and {Secroun}, A. and {Seidel}, G. and {Seiffert}, M. and {Serrano}, S. and {Simon}, P. and {Sirignano}, C. and {Sirri}, G. and {Spurio Mancini}, A.},
        title = "{Euclid Quick Data Release (Q1). The Strong Lensing Discovery Engine C: Finding lenses with machine learning}",
      journal = {arXiv e-prints},
         year = 2025,
        month = mar,
          eid = {arXiv:2503.15326},
        pages = {arXiv:2503.15326},
          doi = {10.48550/arXiv.2503.15326},
archivePrefix = {arXiv},
       eprint = {2503.15326},
 primaryClass = {astro-ph.GA},
       adsurl = {https://ui.adsabs.harvard.edu/abs/2025arXiv250315326E}
}

@ARTICLE{Euclid2025b,
       author = {{Euclid Collaboration} and {Rojas}, K. and {Collett}, T.~E. and {Acevedo Barroso}, J.~A. and {Nightingale}, J.~W. and {Stern}, D. and {Moustakas}, L.~A. and {Schuldt}, S. and {Despali}, G. and {Melo}, A. and {Walmsley}, M. and {Ballard}, D.~J. and {Enzi}, W.~J.~R. and {Li}, T. and {Sainz de Murieta}, A. and {Andika}, I.~T. and {Cl{\'e}ment}, B. and {Courbin}, F. and {Ecker}, L.~R. and {Gavazzi}, R. and {Jackson}, N. and {Kov{\'a}cs}, A. and {Matavulj}, P. and {Meneghetti}, M. and {Serjeant}, S. and {Sluse}, D. and {Tortora}, C. and {Verma}, A. and {Marchetti}, L. and {O'Riordan}, C.~M. and {McCarthy}, K. and {Suyu}, S.~H. and {Metcalf}, R.~B. and {Aghanim}, N. and {Altieri}, B. and {Amara}, A. and {Andreon}, S. and {Auricchio}, N. and {Aussel}, H. and {Baccigalupi}, C. and {Baldi}, M. and {Balestra}, A. and {Bardelli}, S. and {Battaglia}, P. and {Bender}, R. and {Biviano}, A. and {Bonchi}, A. and {Branchini}, E. and {Brescia}, M. and {Brinchmann}, J. and {Camera}, S. and {Ca{\~n}as-Herrera}, G. and {Capobianco}, V. and {Carbone}, C. and {Cardone}, V.~F. and {Carretero}, J. and {Casas}, S. and {Castellano}, M. and {Castignani}, G. and {Cavuoti}, S. and {Chambers}, K.~C. and {Cimatti}, A. and {Colodro-Conde}, C. and {Congedo}, G. and {Conselice}, C.~J. and {Conversi}, L. and {Copin}, Y. and {Courtois}, H.~M. and {Cropper}, M. and {Da Silva}, A. and {Degaudenzi}, H. and {De Lucia}, G. and {Di Giorgio}, A.~M. and {Dolding}, C. and {Dole}, H. and {Dubath}, F. and {Dupac}, X. and {Escoffier}, S. and {Fabricius}, M. and {Farina}, M. and {Farinelli}, R. and {Faustini}, F. and {Ferriol}, S. and {Finelli}, F. and {Fotopoulou}, S. and {Frailis}, M. and {Franceschi}, E. and {Galeotta}, S. and {George}, K. and {Gillard}, W. and {Gillis}, B. and {Giocoli}, C. and {G{\'o}mez-Alvarez}, P. and {Gracia-Carpio}, J. and {Granett}, B.~R. and {Grazian}, A. and {Grupp}, F. and {Guzzo}, L. and {Gwyn}, S. and {Haugan}, S.~V.~H. and {Holmes}, W. and {Hook}, I.~M. and {Hormuth}, F. and {Hornstrup}, A. and {Hudelot}, P. and {Jahnke}, K. and {Jhabvala}, M. and {Keih{\"a}nen}, E. and {Kermiche}, S. and {Kiessling}, A. and {Kubik}, B. and {Kuijken}, K. and {K{\"u}mmel}, M. and {Kunz}, M. and {Kurki-Suonio}, H. and {Le Boulc'h}, Q. and {Le Brun}, A.~M.~C. and {Le Mignant}, D. and {Liebing}, P. and {Ligori}, S. and {Lilje}, P.~B. and {Lindholm}, V. and {Lloro}, I. and {Mainetti}, G. and {Maino}, D. and {Maiorano}, E. and {Mansutti}, O. and {Marcin}, S. and {Marggraf}, O. and {Martinelli}, M. and {Martinet}, N. and {Marulli}, F. and {Massey}, R. and {Maurogordato}, S. and {McCracken}, H.~J. and {Medinaceli}, E. and {Mei}, S. and {Melchior}, M. and {Mellier}, Y. and {Merlin}, E. and {Meylan}, G. and {Mora}, A. and {Moresco}, M. and {Moscardini}, L. and {Nakajima}, R. and {Neissner}, C. and {Nichol}, R.~C. and {Niemi}, S.-M. and {Padilla}, C. and {Paltani}, S. and {Pasian}, F. and {Pedersen}, K. and {Percival}, W.~J. and {Pettorino}, V. and {Pires}, S. and {Polenta}, G. and {Poncet}, M. and {Popa}, L.~A. and {Pozzetti}, L. and {Raison}, F. and {Rebolo}, R. and {Renzi}, A. and {Rhodes}, J. and {Riccio}, G. and {Romelli}, E. and {Roncarelli}, M. and {Saglia}, R. and {Sakr}, Z. and {S{\'a}nchez}, A.~G. and {Sapone}, D. and {Sartoris}, B. and {Schewtschenko}, J.~A. and {Schirmer}, M. and {Schneider}, P. and {Schrabback}, T. and {Secroun}, A. and {Seidel}, G. and {Seiffert}, M. and {Serrano}, S. and {Simon}, P. and {Sirignano}, C. and {Sirri}, G. and {Stanco}, L. and {Steinwagner}, J. and {Tallada-Cresp{\'\i}}, P. and {Taylor}, A.~N. and {Tereno}, I. and {Toft}, S. and {Toledo-Moreo}, R. and {Torradeflot}, F. and {Tutusaus}, I. and {Valenziano}, L. and {Valiviita}, J. and {Vassallo}, T. and {Verdoes Kleijn}, G. and {Veropalumbo}, A. and {Wang}, Y. and {Weller}, J. and {Zacchei}, A. and {Zamorani}, G.},
        title = "{Euclid Quick Data Release (Q1) The Strong Lensing Discovery Engine B -- Early strong lens candidates from visual inspection of high velocity dispersion galaxies}",
      journal = {arXiv e-prints},
         year = 2025,
        month = mar,
          eid = {arXiv:2503.15325},
        pages = {arXiv:2503.15325},
          doi = {10.48550/arXiv.2503.15325},
archivePrefix = {arXiv},
       eprint = {2503.15325},
 primaryClass = {astro-ph.GA},
       adsurl = {https://ui.adsabs.harvard.edu/abs/2025arXiv250315325E}
}

@ARTICLE{Euclid2025a,
       author = {{Euclid Collaboration} and {Walmsley}, M. and {Holloway}, P. and {Lines}, N.~E.~P. and {Rojas}, K. and {Collett}, T.~E. and {Verma}, A. and {Li}, T. and {Nightingale}, J.~W. and {Despali}, G. and {Schuldt}, S. and {Gavazzi}, R. and {Melo}, A. and {Metcalf}, R.~B. and {Andika}, I.~T. and {Leuzzi}, L. and {Manj{\'o}n-Garc{\'\i}a}, A. and {Pearce-Casey}, R. and {Vincken}, S.~H. and {Wilde}, J. and {Busillo}, V. and {Tortora}, C. and {Acevedo Barroso}, J.~A. and {Dole}, H. and {Ecker}, L.~R. and {Pearson}, J. and {Marshall}, P.~J. and {More}, A. and {Saifollahi}, T. and {Gracia-Carpio}, J. and {Baeten}, E. and {Cornen}, C. and {Johnson}, L.~C. and {Macmillan}, C. and {Kruk}, S. and {Remmelgas}, K.~A. and {Cl{\'e}ment}, B. and {Degaudenzi}, H. and {Courbin}, F. and {Bovy}, J. and {Casas}, S. and {Dannerbauer}, H. and {Diego}, J.~M. and {Finner}, K. and {Galan}, A. and {Giocoli}, C. and {Hogg}, N.~B. and {Jahnke}, K. and {Katona}, J. and {Kov{\'a}cs}, A. and {De Leo}, C. and {Mahler}, G. and {Millon}, M. and {Nagam}, B.~C. and {Nugent}, P. and {Sainz de Murieta}, A. and {O'Riordan}, C.~M. and {Sluse}, D. and {Sonnenfeld}, A. and {Spiniello}, C. and {Serjeant}, S. and {Thai}, T.~T. and {Ulivi}, L. and {Walth}, G.~L. and {Weisenbach}, L. and {Zumalacarregui}, M. and {Aghanim}, N. and {Altieri}, B. and {Amara}, A. and {Andreon}, S. and {Auricchio}, N. and {Aussel}, H. and {Baccigalupi}, C. and {Baldi}, M. and {Balestra}, A. and {Bardelli}, S. and {Battaglia}, P. and {Bernardeau}, F. and {Biviano}, A. and {Bonchi}, A. and {Bonino}, D. and {Branchini}, E. and {Brescia}, M. and {Brinchmann}, J. and {Camera}, S. and {Ca{\~n}as-Herrera}, G. and {Capobianco}, V. and {Carbone}, C. and {Cardone}, V.~F. and {Carretero}, J. and {Castander}, F.~J. and {Castellano}, M. and {Castignani}, G. and {Cavuoti}, S. and {Chambers}, K.~C. and {Cimatti}, A. and {Colodro-Conde}, C. and {Congedo}, G. and {Conselice}, C.~J. and {Conversi}, L. and {Copin}, Y. and {Corcione}, L. and {Courtois}, H.~M. and {Cropper}, M. and {Da Silva}, A. and {De Lucia}, G. and {Di Giorgio}, A.~M. and {Dolding}, C. and {Dubath}, F. and {Duncan}, C.~A.~J. and {Dupac}, X. and {Ealet}, A. and {Escoffier}, S. and {Fabricius}, M. and {Farina}, M. and {Farinelli}, R. and {Faustini}, F. and {Finelli}, F. and {Fotopoulou}, S. and {Frailis}, M. and {Franceschi}, E. and {Fumana}, M. and {Galeotta}, S. and {George}, K. and {Gillard}, W. and {Gillis}, B. and {G{\'o}mez-Alvarez}, P. and {Granett}, B.~R. and {Grazian}, A. and {Grupp}, F. and {Guzzo}, L. and {Gwyn}, S. and {Haugan}, S.~V.~H. and {Hoekstra}, H. and {Holmes}, W. and {Hook}, I.~M. and {Hormuth}, F. and {Hornstrup}, A. and {Hudelot}, P. and {Jhabvala}, M. and {Joachimi}, B. and {Keih{\"a}nen}, E. and {Kermiche}, S. and {Kiessling}, A. and {Kubik}, B. and {K{\"u}mmel}, M. and {Kunz}, M. and {Kurki-Suonio}, H. and {Lahav}, O. and {Le Boulc'h}, Q. and {Le Brun}, A.~M.~C. and {Le Mignant}, D. and {Ligori}, S. and {Lilje}, P.~B. and {Lindholm}, V. and {Lloro}, I. and {Mainetti}, G. and {Maino}, D. and {Maiorano}, E. and {Mansutti}, O. and {Marcin}, S. and {Marggraf}, O. and {Martinelli}, M. and {Martinet}, N. and {Marulli}, F. and {Massey}, R. and {Maurogordato}, S. and {McCracken}, H.~J. and {Medinaceli}, E. and {Mei}, S. and {Mellier}, Y. and {Meneghetti}, M. and {Merlin}, E. and {Meylan}, G. and {Mora}, A. and {Moresco}, M. and {Moscardini}, L. and {Nakajima}, R. and {Neissner}, C. and {Nichol}, R.~C. and {Niemi}, S.-M. and {Padilla}, C. and {Paltani}, S. and {Pasian}, F. and {Pedersen}, K. and {Percival}, W.~J. and {Pettorino}, V. and {Pires}, S. and {Polenta}, G. and {Poncet}, M. and {Popa}, L.~A. and {Pozzetti}, L. and {Raison}, F. and {Rebolo}, R. and {Renzi}, A. and {Rhodes}, J. and {Riccio}, G. and {Romelli}, E. and {Roncarelli}, M. and {Saglia}, R.},
        title = "{Euclid Quick Data Release (Q1): The Strong Lensing Discovery Engine A -- System overview and lens catalogue}",
      journal = {arXiv e-prints},
         year = 2025,
        month = mar,
          eid = {arXiv:2503.15324},
        pages = {arXiv:2503.15324},
          doi = {10.48550/arXiv.2503.15324},
archivePrefix = {arXiv},
       eprint = {2503.15324},
 primaryClass = {astro-ph.GA},
       adsurl = {https://ui.adsabs.harvard.edu/abs/2025arXiv250315324E}
}

@ARTICLE{Claeyssens2023,
       author = {{Claeyssens}, Ad{\'e}la{\"\i}de and {Adamo}, Angela and {Richard}, Johan and {Mahler}, Guillaume and {Messa}, Matteo and {Dessauges-Zavadsky}, Miroslava},
        title = "{Star formation at the smallest scales: a JWST study of the clump populations in SMACS0723}",
      journal = {\mnras},
         year = 2023,
        month = apr,
       volume = {520},
       number = {2},
        pages = {2180-2203},
          doi = {10.1093/mnras/stac3791},
archivePrefix = {arXiv},
       eprint = {2208.10450},
 primaryClass = {astro-ph.GA},
       adsurl = {https://ui.adsabs.harvard.edu/abs/2023MNRAS.520.2180C}
}

@ARTICLE{Vanzella2018,
       author = {{Vanzella}, E. and {Nonino}, M. and {Cupani}, G. and {Castellano}, M. and {Sani}, E. and {Mignoli}, M. and {Calura}, F. and {Meneghetti}, M. and {Gilli}, R. and {Comastri}, A. and {Mercurio}, A. and {Caminha}, G.~B. and {Caputi}, K. and {Rosati}, P. and {Grillo}, C. and {Cristiani}, S. and {Balestra}, I. and {Fontana}, A. and {Giavalisco}, M.},
        title = "{Direct Lyman continuum and Ly {\ensuremath{\alpha}} escape observed at redshift 4}",
      journal = {\mnras},
         year = 2018,
        month = may,
       volume = {476},
       number = {1},
        pages = {L15-L19},
          doi = {10.1093/mnrasl/sly023},
archivePrefix = {arXiv},
       eprint = {1712.07661},
 primaryClass = {astro-ph.GA},
       adsurl = {https://ui.adsabs.harvard.edu/abs/2018MNRAS.476L..15V}
}

@article{Filiberto2024,
author = {Filiberto, Justin and Mccanta, Molly},
year = {2024},
month = {05},
pages = {805-813},
title = {Characterizing basalt-atmosphere interactions on Venus: A review of thermodynamic and experimental results},
volume = {109},
journal = {American Mineralogist},
doi = {10.2138/am-2023-9015}
}

\end{document}